\documentclass{article}

\usepackage{arxiv}

\usepackage[utf8]{inputenc} 
\usepackage[T1]{fontenc}    
\usepackage{hyperref}       
\usepackage{url}            
\usepackage{booktabs}       
\usepackage{amsfonts}       
\usepackage{nicefrac}       
\usepackage{microtype}      
\usepackage{lipsum}
\usepackage{graphicx}
\usepackage{amssymb}
\usepackage{amsmath}
\usepackage{color}
\usepackage{multicol,multicap,graphicx}
\usepackage[mathlines]{lineno}
\title{An immersed boundary method for fluid--structure--acoustics interactions involving large deformations and complex geometries}

\author{
  Li Wang \\
  School of Engineering and Information Technology\\
  University of New South Wales\\
  Canberra ACT, 2600, Australia\\
  \texttt{l.wang@unsw.edu.au} \\
   \And
 Fang-Bao Tian \\
  School of Engineering and Information Technology\\
  University of New South Wales\\
  Canberra ACT, 2600, Australia\\
  \texttt{f.tian@adfa.edu.au; onetfbao@gmail.com} \\
  \And
 Joseph C. S. Lai \\
  School of Engineering and Information Technology\\
  University of New South Wales\\
  Canberra ACT, 2600, Australia\\
  \texttt{j.lai@adfa.edu.au} \\
}

\begin{document}
\maketitle

\begin{abstract}
This paper presents an immersed boundary (IB) method for fluid--structure--acoustics interactions involving large deformations and complex geometries. In this method, the fluid dynamics is solved by a finite difference method where the temporal, viscous and convective terms are respectively discretized by the third-order Runge-Kutta scheme, the fourth-order central difference scheme and a fifth-order W/TENO (Weighted/Targeted Essentially Non-oscillation) scheme. Without loss of generality, a nonlinear flexible plate is considered here, and is solved by a finite element method based on the absolute nodal coordinate formulation. The no-slip boundary condition at the fluid--structure interface is achieved by using a diffusion-interface penalty IB method. With the above proposed method, the aeroacoustics field generated by the moving boundaries and the associated flows are inherently solved. In order to validate and verify the current method, several benchmark cases are conducted: acoustic waves scattered from a stationary cylinder in a quiescent flow, sound generation by a stationary and a rotating cylinder in a uniform flow, sound generation by an insect in hovering flight, deformation of a red blood cell induced by acoustic waves and acoustic waves scattered by a stationary sphere. The comparison of the sound scattered by a cylinder shows that the present IB--WENO scheme, a simple approach, has an excellent performance which is even better than the implicit IB--lattice Boltzmann method. For the sound scattered by a sphere, the IB--TENO scheme has a lower dissipation compared with the IB--WENO scheme. Applications of this technique to model fluid-structure-acoustics interactions of flapping foils mimicking an insect wing section during forward flight and flapping foil energy harvester are also presented, considering the effects of foil shape and flexibility. The difference of the force and sound generations of the foils are compared. For wing during forward flight, the results show that flexible wing generates larger thrust with higher acoustic pressure. In terms of the energy harvester, the current results show that the geometrical shape has no significant effects on the force and sound generation, and the flexibility of the plate tends to deteriorate the power extraction efficiency. The flexible plate also induces larger fluctuating pressure at the frequency of $2f$ ($f$ is the flapping frequency) and weaker sound at the frequencies of $f$ and $3f$. The successful validations and applications show that the IB method handled by delta function, whose accuracy is generally lower than that of the internal flow solver, is accurate for predicting the dilatation and acoustics, and thus is an attractive alternative for modelling fluid--structure--acoustics interactions involving large deformations and complex geometries.

\end{abstract}


\section{Introduction}
Flow-induced pressure fluctuations extensively exist in engineering applications, such as aviation technologies~\cite{colonius2004computational}, ventilation~\cite{waye1997effects} and biomechanics~\cite{seo2011high}. In many of these applications, the major task is to identify the noise source and then to reduce or eliminate it (see e.g. rotorcrafts~\cite{colonius2004computational} and ventilations~\cite{waye1997effects}). Flow-induced sound also plays a crucial role in the biomechanics, such as phonation (flow-induced sound in the larynx~\cite{zhao2002computational}) and heart murmurs (blood-flow-induced noise~\cite{el2005computer}). Accurate prediction of the flow-induced noise in these systems can be utilized to optimize low noise engineering design. The acoustic prediction in biomechanics also has potential in medical diagnosis~\cite{seo2011high}. However, in most of the engineering applications, fluid-structure-acoustics interactions involving complex geometries are involved, making the accurate prediction of the sound generation extremely challenging.

Numerical methods for computational aeroacoustics (CAA) can be categorized into three groups~\cite{inoue2002sound}. The first group of methods calculate the near field flow dynamics by using computational fluid dynamics (CFD) techniques, and predict the far-field sound generation by employing acoustic analogies, such as the pioneering analogy proposed by Lighthill~\cite{lighthill1952sound} and the extension of this analogy to include the influence of the solid boundaries in the sound field~\cite{curle1955influence}. Since Lighthill's work, great efforts have been made to improve the method~\cite{schram2009boundary,farassat1988extension,di1997new}. The second group of methods, also called acoustic/viscous splitting methods, decompose the compressible viscous equations into incompressible and perturbed compressible parts. The perturbation in the far-field represents the acoustic quantities~\cite{hardin1994acoustic}. The third group of methods employ direct numerical simulation (DNS) to solve the compressible Navier-Stokes equations. By using DNS, the generation and propagation processes of the sound in the near and intermediate fields can be calculated, without suffering from the limitations such as the low Mach number and compactness of the source region~\cite{inoue2002sound,schlanderer2017boundary}.

In order to resolve the sound pressure which is much smaller compared to the ambient pressure~\cite{colonius2004computational}, high-order finite difference methods (e.g. compact schemes~\cite{lele1992compact} and Weighted/Targeted Essentially Non-oscillation schemes~\cite{liu1994weighted,fu2016family}) on body-conformal meshes are widely used. In these methods, it is difficult to generate high-quality structured meshes around the complex and moving boundaries. The finite volume method has the ability to handle complex boundaries. However, it is still challenging to handle moving boundaries. In addition, it suffers from the low-order accuracy and introduces extra dissipation and dispersion~\cite{sun2012immersed}. The high-order overset grid method~\cite{sherer2005high} has been successfully applied for CAA in complex geometries. But, the applications of this type of methods are limited due to their complexity of implementation and low efficiency at low Mach numbers~\cite{seo2011high}. As an efficient method for fluid--structure interaction (FSI), the immersed boundary (IB) method, first developed by Peskin~\cite{peskin1977numerical,peskin2002immersed}, has been extended to acoustics~\cite{seo2011high,sun2012immersed}. In this method, structured mesh is generated initially and fixed during the computation. Therefore, the regeneration of mesh is avoided. Because of its simple boundary treatment, IB method has gained popularity for a wide range of applications \cite{mittal2005immersed,huang2007simulation,tian2014fluid,sotiropoulos2014immersed}. In the hybrid method proposed by Seo and Mittal~\cite{seo2011high}, the IB method and the hydrodynamic/acoustic splitting technique are combined to handle the acoustic problems involving complex geometries in low-Mach number flow. It encounters challenges when the Mach number rises and the complex geometries move. Penalty IB (pIB) is a typical IB method~\cite{kim2007penalty}, where the IB is conceptually split into two Lagrangian components: one component is massless and interacts with the fluid exactly as the traditional IB method, and the other component carrying mass is connected to the massless component by virtual stiff springs. However, most of the previous studies based on pIB method focus on FSI problems without acoustics~\cite{mittal2005immersed,tian2011efficient,huang2010three,qiu2016boundary,ghias2007sharp,wang2017immersed}. Sun et al. presented an IB method which considers the linear Euler equations for acoustic scattering modeling~\cite{sun2012immersed}. In their work, the moving boundary and fluid viscosity are not explored. The ability of the pIB method for fluid--structure--acoustics interactions involving complex geometries has not been well explored. This is the motivation for this work.

Flapping foil has drawn growing attention recently, due to its high performance in micro aerial vehicles (MAVs)~\cite{tian2013force,shahzad2016effects,shahzad2018effects,shahzad2018effectspof,tian2018aerodynamic} and power generators~\cite{liu2016discrete,liu2017flapping,tian2014improving,liu2019kinematic}. The aerodynamic characteristics of the flapping foil including its thrust, lift and power efficiency have been extensively studied by researchers~\cite{tian2013force,yin2010effect}. However, the aeroacoustics induced by the flapping foil has not been well understood~\cite{geng2017effect}. Although the high performance of insect flight can be achieved with low noises, the flapping sound generated by the foil may have potential effects on biological functions, such as the communication using aposematic signals with the locomotion-induced sound~\cite{geng2017effect}. Moreover, the study on the sound induced by flapping foil may also have potential applications in medical inspections, such as the phonation of human which performs like a flapping foil induced sound~\cite{zhao2002computational}. Due to its importance in many areas, the sound generation by flapping foils is numerically studied by the present method. The effects of both the elasticity and the geometrical shape of the foil on the force and sound generation process are analyzed. The numerical examples presented here can also enrich the limited database of fluid--structure--acoustics interactions.

In this paper, an immersed boundary method introduced in our previous work~\cite{wang2017immersed} is extended to fluid--structure--acoustics interactions involving large deformations and complex geometries. The organization of the paper is as follows. The numerical approach is briefly introduced in Section 2. Several validations including acoustic waves scattered by a stationary cylinder, sound generation by a stationary and a rotating cylinder, sound generation by an insect in hovering flight, deformation of a red blood cell induced by acoustic waves and acoustic waves scattered by a stationary sphere are presented in Section 3. Application of the current numerical method in modelling flapping foil induced acoustics is presented in Section 4. Finally, conclusions are given in Section 5.

\section{Numerical method}
The current numerical method includes three important components: the structure, compressible fluid and fluid--structure interaction. Without loss of generality, a flexible plate immersed in the two-dimensional fluid is used as an example to introduce the structure dynamics. The plate is assumed to be elastic and its dynamics is governed by the following nonlinear equation~\cite{huang2007simulation,tian2010interaction,wang2017immersed}
\begin{equation}
\rho_s \frac{\partial^2 \boldsymbol{X}}{\partial t^2}+ \frac{\partial}{\partial s} \left[(K_S |{\frac{\partial \boldsymbol{X}}{\partial s}}|-1) {\frac{\partial \boldsymbol{X}}{\partial s}}\right] + K_B {\frac{\partial^4 \boldsymbol{X}}{\partial s^4}}= \boldsymbol{F_f},
\label{eq:beamequation}
\end{equation}
where $\boldsymbol{X}$ is the Lagrangian coordinates of the flexible beam, $\rho_s$ is linear density, $K_S$ and $K_B$ are respectively the stretching and bending rigidity, $s$ is the arc coordinate, and $\boldsymbol{F}_f$ is the external force acting on the beam. The absolute nodal coordinate formulation (ANCF) proposed by Shabana~\cite{shabana1997flexible,shabana1998application,shabana2013dynamics}
is adopted to solve Eq.~\ref{eq:beamequation}. This method was combined with the IB method by Wang et al.~\cite{wang2017immersed}.

The flow dynamics considered here are governed by the compressible viscous Navier--Stokes equations
\begin{eqnarray}
&&\frac{\partial Q}{\partial t} + \frac{\partial F}{\partial x} + \frac{\partial G}{\partial y}+ \frac{\partial H}{\partial z}-
\frac{1}{\Re}(\frac{\partial F_u}{\partial x}+\frac{\partial G_v}{\partial y}+\frac{\partial H_v}{\partial z})= S,\\
&&Q=[\rho,\rho u,\rho v,\rho w, E]^T,\quad F=[\rho u,\rho u^2+P,\rho u v,\rho u w, (E+P) u]^T,\\
&&G=[\rho v,\rho u v, \rho v^2+ P,\rho v w, (E+P)v]^T,\quad H=[\rho w,\rho u w,\rho v w, \rho w^2+ P,(E+P)w]^T,\\
&&F_u=[0,\tau_{xx},\tau_{xy},\tau_{xz},b_x]^T,\quad G_v=[0,\tau_{xy},\tau_{yy},\tau_{yz},b_y]^T,\quad H_v=[0,\tau_{xz},\tau_{yz},\tau_{zz},b_z]^T,\\
&&b_x=u \tau_{xx}+ v \tau_{xy}+w \tau_{xz},\quad b_y=u \tau_{xy}+ v \tau_{yy}+ w \tau_{yz},\quad b_z=u \tau_{xz}+ v \tau_{yz}+ w \tau_{zz},
\end{eqnarray}
where $\rho$ is the fluid density, $u$, $v$ and $w$ are respectively the three velocity components, $P$ is the pressure, $E$ is the total energy, $S$ is a general source term including the IB-imposed Eulerian force and other body forces, Re is the Reynolds number, and $\tau_{ij}$ is the shear stress.

In the fluid solver, the fifth-order Weighted Essentially Non-oscillation (WENO) scheme proposed by Liu et al.~\cite{liu1994weighted} is used for the spatial discretization of the convective term. We also introduce a recently developed scheme, the Target Essentially Non-oscillation (TENO)~\cite{fu2016family}, to discretize the convective term, to compare the performance of the IB--WENO with that of the IB--TENO. For the viscous terms, a fourth-order central difference scheme is used to discretize the spatial derivatives. For all unsteady equations involved in flow solver, the third-order TVD Runge-Kutta method is used for temporal discretization~\cite{shu1988efficient}. The dynamics of the fluid and flexible structures are solved independently. The interaction force is calculated explicitly using a feedback law~\cite{goldstein1993modeling} based on the pIB method~\cite{kim2007penalty}. The interaction force between the fluid and the structure can be determined by the feedback law \cite{kim2007penalty}
\begin{equation}
\boldsymbol{F}_f = \alpha \int_0^t (\boldsymbol{U}_{ib} - \boldsymbol{U}) dt + \beta (\boldsymbol{U}_{ib} - \boldsymbol{U}),
\label{eq:penatly}
\end{equation}
where $\boldsymbol{U}_{ib}$ is the boundary velocity obtained by interpolation at the IB, $\boldsymbol{U}$ is the structure velocity, and $\alpha$ and $\beta$ are large positive constants. It is noted that this method does not transform the Lagrangian density of structure into the Eulerian density, as did in Zhu and Peskin~\cite{zhu2002simulation}. Therefore, it is not necessary to apply the incompressible limitation or to modify the continuity equation~\cite{zhu2002simulation,wang2017immersed}. Eq.~\ref{eq:penatly} is the Lagrangian force acting on the structure. The Lagrangian force acting on the fluid by the immersed boundary is -$\boldsymbol{F}_f$, which is spread onto fluid nodes to achieve the boundary condition. The interpolation of the velocity and the spreading of the Lagrange force to the adjacent grid points are expressed as
\begin{equation}
\boldsymbol{U}_{ib} (s, t) = \int_{V} \boldsymbol{u} (x, t) \delta_h (\boldsymbol{X}(s,t) - \boldsymbol{x}) d \boldsymbol{x},
\label{eq:ibvelocity}
\end{equation}
\begin{equation}
\boldsymbol{f} (\boldsymbol{x}, t) = -\int_{\Gamma} \boldsymbol{F}_f (s, t) \delta_h (\boldsymbol{X}(s,t) - \boldsymbol{x}) ds,
\label{eq:fluidforce}
\end{equation}
where $\boldsymbol{u}$ is the fluid velocity, $\boldsymbol{X}$ is the coordinates of structural nodes, $\boldsymbol{x}$ is the coordinates of fluid, $s$ is the arc coordinate for a two dimensional domain, $V$ is the fluid domain, $\Gamma$ is the structure domain and $\delta_h$ is the smoothed Dirac delta function~\cite{peskin2002immersed}, expressed as
\begin{equation}
\delta_h(x,y,z) = \frac{1}{h^3} \lambda(\frac{x}{h}) \lambda(\frac{y}{h}) \lambda(\frac{z}{h}).
\label{eq:phifun}
\end{equation}

In this paper, the four-point delta function introduced by Peskin~\cite{peskin2002immersed} is used
\begin{equation}
\lambda (r) =
\begin{cases}
\frac{1}{8} (3 - 2 |r| + \sqrt{1 + 4|r| - 4|r|^2}), & 0\leq|r|<1 \\
\frac{1}{8} (5 - 2 |r| - \sqrt{-7 + 12|r| - 4|r|^2}), & 1\leq|r|<2 \\
0, & 2\leq|r|.
\end{cases}
\label{eq:deltfun}
\end{equation}

Instead of using uniform mesh which is time consuming, a non-uniform mesh is used here to improve the computational efficiency~\cite{tian2014fluid}. The mesh is uniform in both directions within a small inner box containing the solid for achieving good accuracy and interpolation in the IB method, and it is stretched in the remaining of the computational domain. Non-reflecting boundary conditions are applied on the boundaries~\cite{thompson1987time,thompson1990time}. In addition, the solver is parallelized using hybrid OpenMP and MPI~\cite{rabenseifner2006hybrid}.

\section{Validations}
Validations of the present solver for fluid--structure interactions including flow over a stationary cylinder, structure dynamics, deformation of a flexible panel induced by shock waves in a shock tube, an inclined flexible plate in a hypersonic flow, and shock-induced collapse of a cylindrical helium cavity in the air have been conducted in our previous work~\cite{wang2017immersed}. Here, we focus on validations of acoustics modelling including acoustic waves scattered from a stationary cylinder in a quiescent flow, sound generation by a stationary and a rotating cylinder in a uniform flow, sound generation by an insect in hovering flight, deformation of a red blood cell induced by acoustic waves and acoustic waves scattered by a stationary sphere.

\subsection{Acoustic waves scattered by a stationary cylinder}
In order to verify the accuracy of the present numerical method on uniform Cartesian meshes in simulating the acoustic problem, acoustic waves scattered by a stationary cylinder is validated against the published data~\cite{liu2007brinkman,bailoor2017fluid}. In this simulation, a rigid cylinder with a diameter of $D=1.0$ is fixed with its center at (0, 0). The initial conditions are localized pressure perturbations
of Gaussian distribution~\cite{liu2007brinkman}
\begin{equation}
P_a=\epsilon {\rm exp}[-{\rm ln}(2)\frac{(x-4)^2+y^2}{0.04}]
\end{equation}
where $\epsilon=10^{-3}$. The initial density and pressure are respectively 1.0 and $1/\gamma$ ($\gamma=1.4$), and the initial velocity of the fluid is zero. 

\begin{figure}
 \begin{center}
  \hskip-3.0in (a) \hskip3.0in (b)

  \includegraphics[width=3.2in]{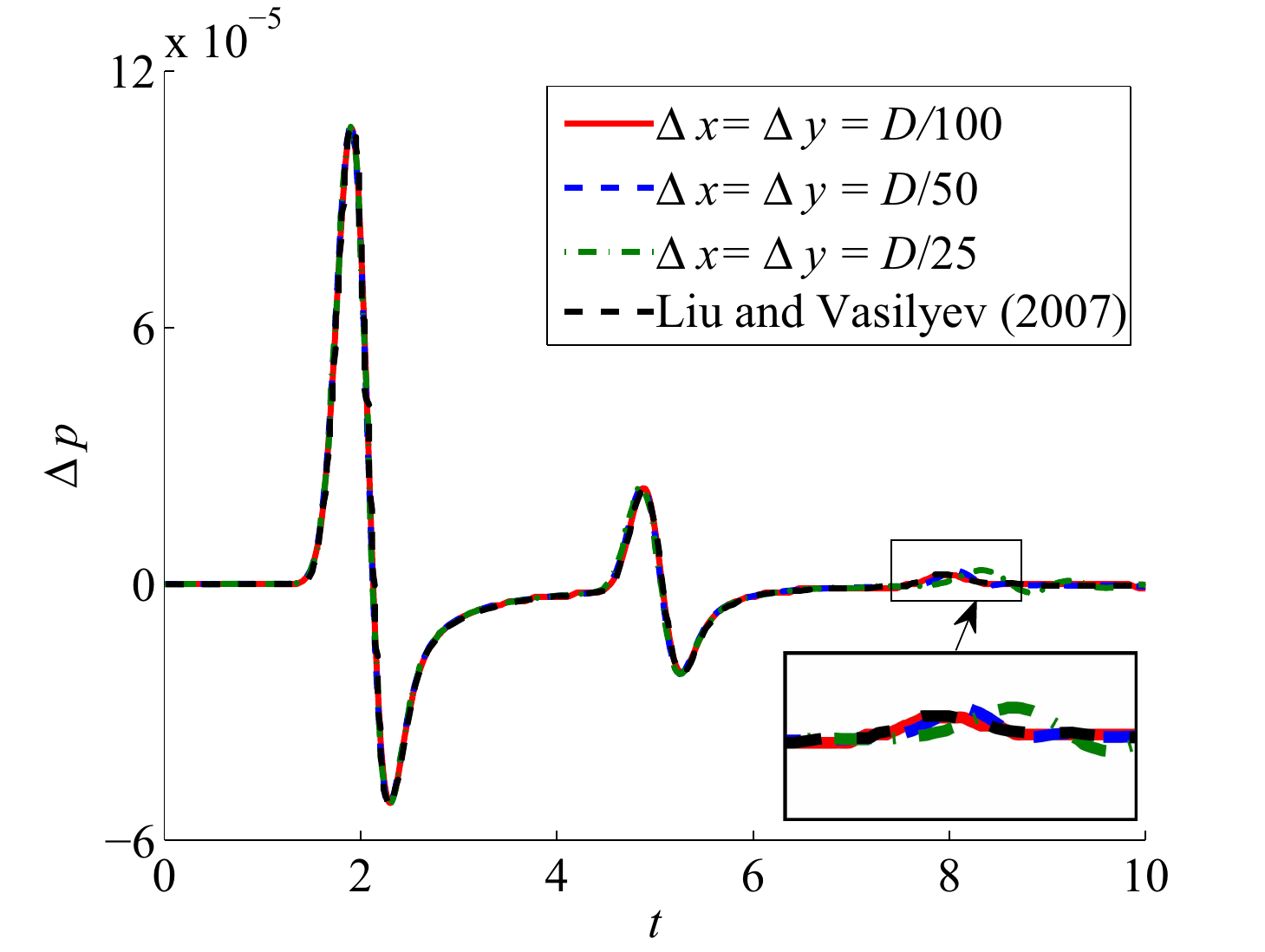}
  \hskip0.1in
  \includegraphics[width=3.2in]{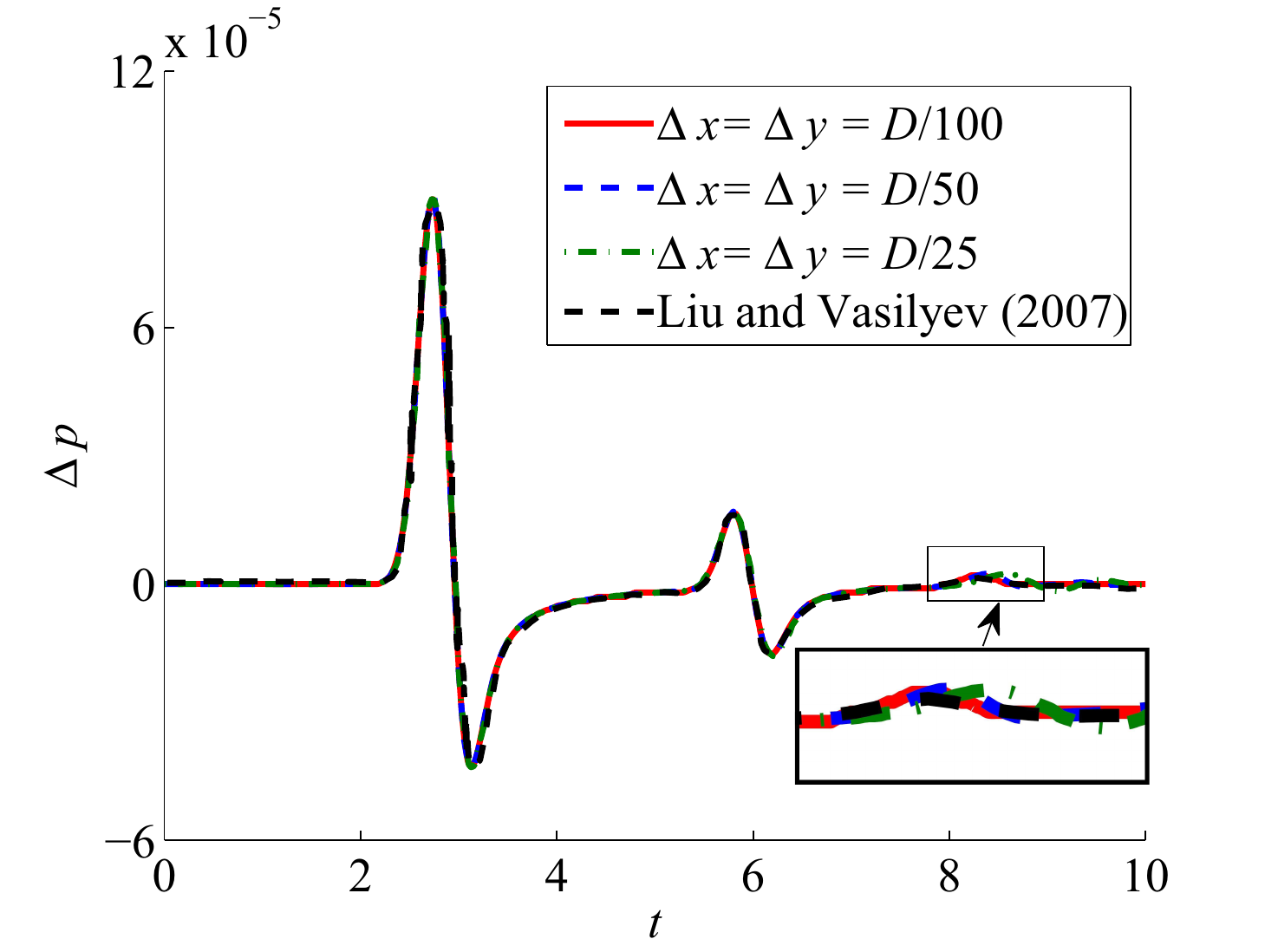}
  \end{center}
\caption{Acoustic waves scattered by a stationary cylinder: comparison of the time histories of pressure fluctuation with available data from Ref.~\cite{liu2007brinkman}. (a) (2, 0), (b) (2, 2).}
\label{Fig:acoustic-pt}
\end{figure}

We used a uniform Cartesian mesh over a computational domain of ($20D\times 20D$) with three different mesh spacings: $D/100$, $D/50$ and $D/25$. The time histories of the fluctuating pressure, $\Delta p=P-p_\infty$ where $P$ and $p_\infty$ are respectively the current and ambient pressure, at points (2, 0) and (2, 2) are plotted in Fig.~\ref{Fig:acoustic-pt} along with the DNS results reported in Ref.~\cite{liu2007brinkman}. According to the comparison in Fig.~\ref{Fig:acoustic-pt}, all the three mesh spacings are able to capture the fluctuating pressure. The small discrepancies depicted in the scaled up figures (see insets in Fig.~\ref{Fig:acoustic-pt}) can be diminished when the mesh size is $D/100$. 

\begin{figure}
 \begin{center}
 \includegraphics[width=3.2in]{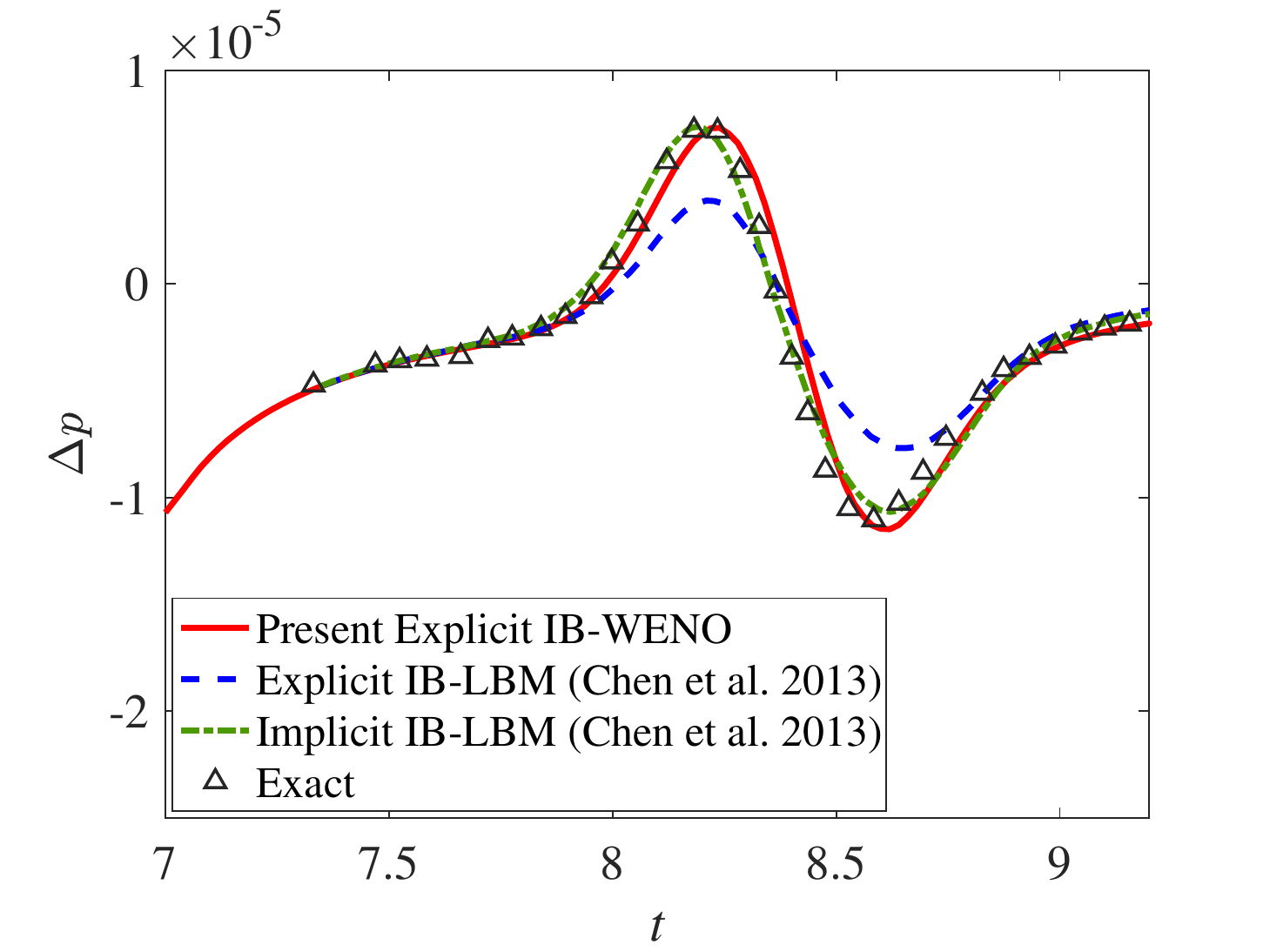}
   \end{center}
\caption{Acoustic waves scattered by a stationary cylinder: comparison of the time histories of fluctuating pressure with available data from Ref.~\cite{chen2014comparative} at (0, 5).}
\label{Fig:acoustic-cy-iblbmcomp}
\end{figure}

Fig.~\ref{Fig:acoustic-cy-iblbmcomp} presents the comparison of present results with those calculated by the immersed boundary--lattice Boltzmann method (IB-LBM) in Ref.~\cite{chen2014comparative}. The mesh spacing for the fluid is $D/40$, same as that used in Ref.~\cite{chen2014comparative}. The results show that the present IB--WENO scheme has an excellent performance on capturing the scattered acoustic pressure. Specifically, the sound pressure predicted by the current explicit IB is significantly better than that predicted by the explicit IB--LBM, as reported in Ref.~\cite{chen2014comparative}.

\begin{figure}
 \begin{center}
  \hskip-5.0in (a)

  \includegraphics[width=3.0in]{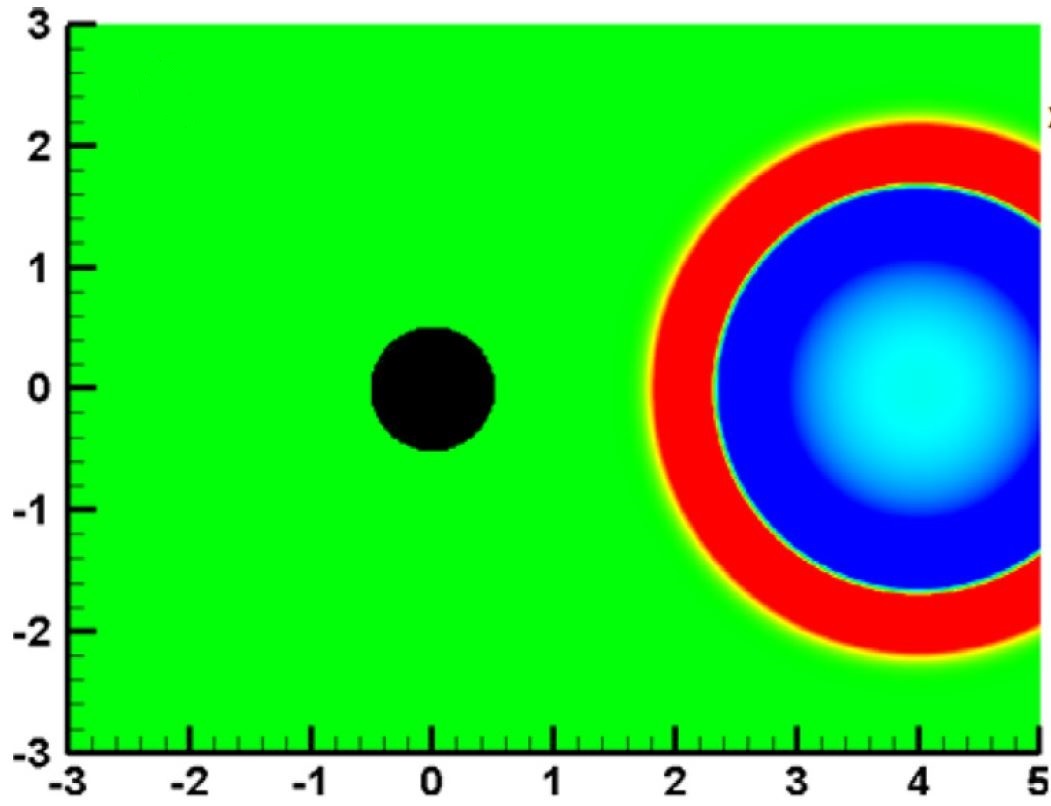}
  \hskip0.1in
  \includegraphics[width=3.2in]{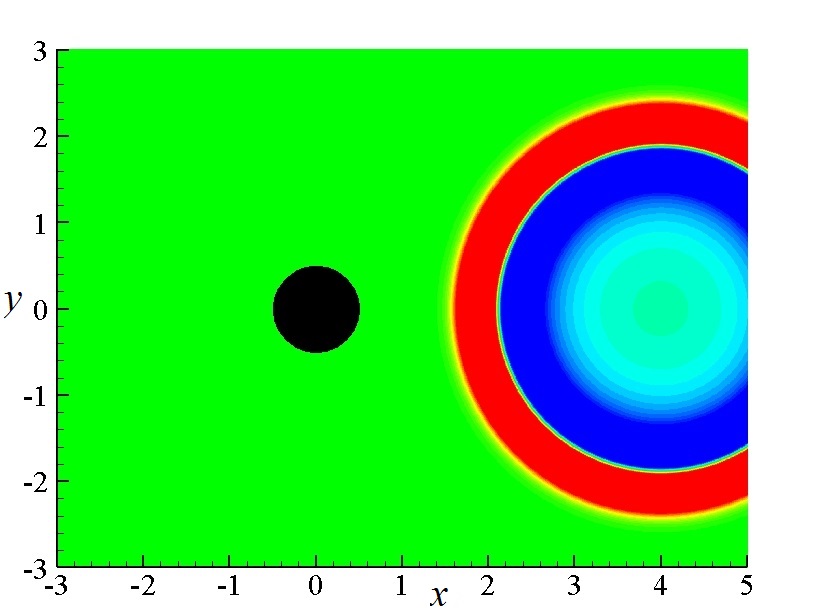}\\
  
  \hskip-5.0in (b)

  \includegraphics[width=3.0in]{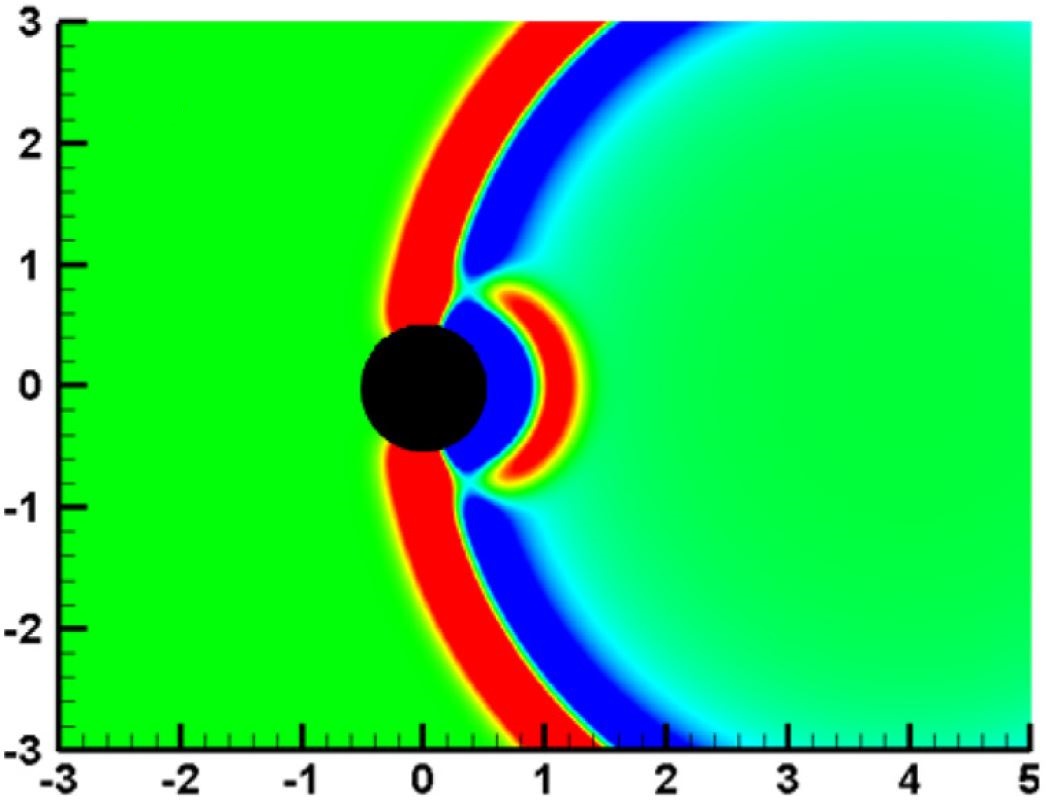}
  \hskip0.1in
  \includegraphics[width=3.2in]{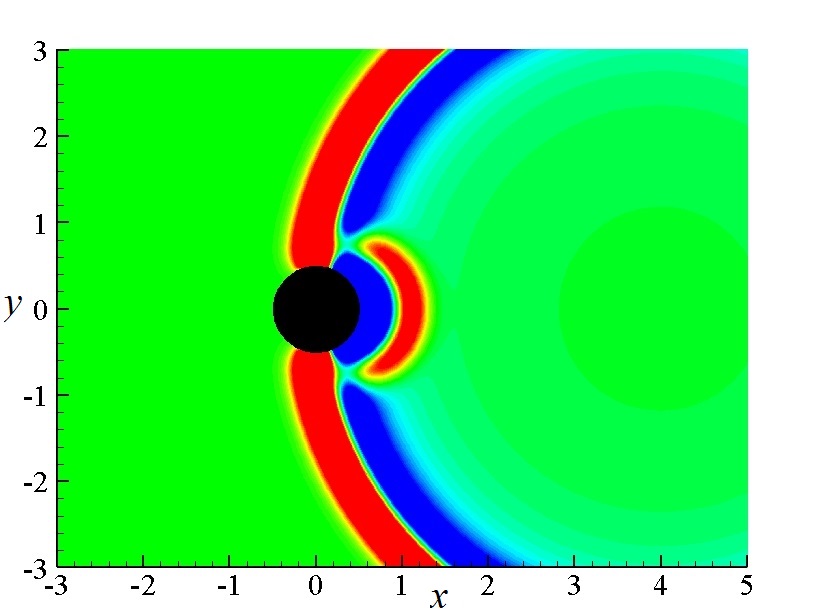}\\
  
  \hskip-5.0in (c)

  \includegraphics[width=3.0in]{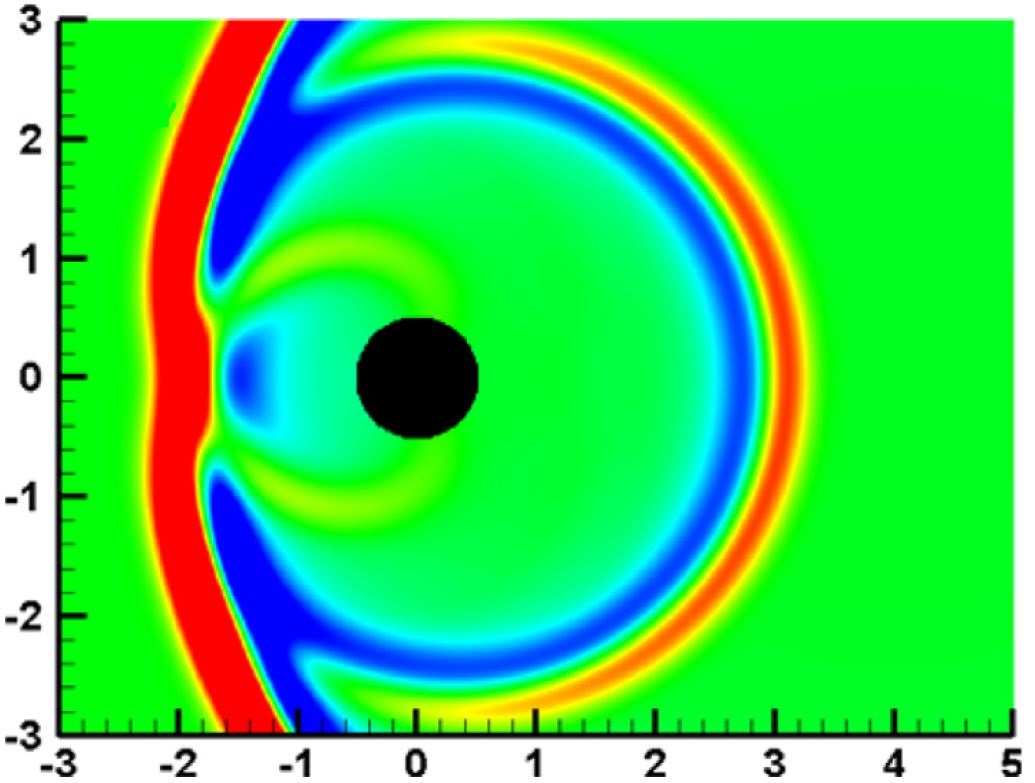}
  \hskip0.1in
  \includegraphics[width=3.2in]{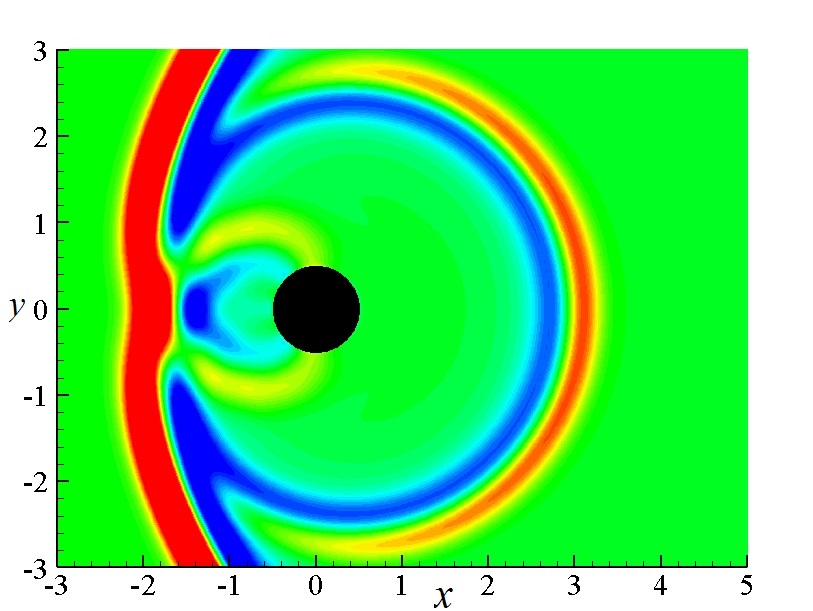}\\
  \end{center}
\caption{Acoustic waves scattered by a stationary cylinder: qualitative comparison of the pressure perturbation contours calculated by Shantanu~\cite{bailoor2017fluid} (left column) and the present method (right column) at times (a) $t$ = 2.0, (b) $t$ = 4.0 and (c) $t$ = 6.0. The contour level ranges from $-2.0\times10^{-5}$ to $2.0\times10^{-5}$.}
\label{Fig:acoustic-p-contour}
\end{figure}

We further present the fluctuating pressure contour for mesh spacing of $D/50$ at three instants in Fig.~\ref{Fig:acoustic-p-contour} along with figures from Ref.~\cite{bailoor2017fluid} to illustrate the propagation of the acoustic wave. As shown in Fig.~\ref{Fig:acoustic-p-contour}(a), a principal pulse is generated due to the initial pressure perturbation. When the acoustic wave impacts on the cylinder, a reflected wave off from the cylinder surface is generated and yields to a secondary acoustic wave, as demonstrated by Fig.~\ref{Fig:acoustic-p-contour}(b). As reported by Liu and Vasilyes~\cite{liu2007brinkman}, two parts of the principal wave front split by the cylinder traverse its span, collide, and merge, thereby generating a third acoustic wave front (see Fig.~\ref{Fig:acoustic-p-contour}(c)). These three acoustic wave fronts are all well captured according to the time history of fluctuating pressure presented in Fig.~\ref{Fig:acoustic-pt}. This benchmark validation shows that the present method has a good accuracy in capturing acoustic waves.

\subsection{Sound generation by a stationary cylinder in a uniform flow}
Flow around a stationary cylinder has been extensively studied theoretically, experimentally and numerically~\cite{son1969numerical,graf1998experiments,he1997lattice}, since the work of Strouhal on aeolian tones. Most of previous studies on the sound generation due to flow past a circular cylinder was done by using hybrid or acoustic/viscous splitting methods to reduce the computational expense~\cite{inoue2002sound}. However, direct numerical simulation (DNS) is an effective way to identify both the aerodynamics and characteristic features of the sound accurately. Inoue and Hatakeyama studied the sound generation by a stationary~\cite{inoue2002sound} and rotating~\cite{inoue2003control} cylinder in a uniform flow using DNS, and clarified the relation between the vortex/flow dynamics and the sound pressure.

In this section, a stationary cylinder in a uniform flow is considered. The cylinder is located at the origin. In order to discuss the sound pressure, we define $r$ (nondimensionalized by the diameter of the cylinder) as the distance from the origin and $\theta$ as the circumferential angle. The non-dimensional parameters that govern this problem are defined as follows
\begin{equation}
{\rm Re}=\rho_f U_0 D/\mu, \quad M=U_0/c,
\label{eq:cy_para}
\end{equation}
where $D$ is the diameter of the cylinder, $U_0$ is the inlet velocity, and $\rho_f$, $\mu$ and $c$ are respectively the density, viscosity and sound speed of the fluid in the far field. In the current stationary case, Re=150 and $M=0.2$. Three mesh regions are used in the simulations to improve the computational efficiency, including a cylinder occupied region, a sound region and a sponge region (similar as that in Ref.~\cite{inoue2002sound}) around the cylinder. The cylinder occupied region extends from ($-1.25D, -1.25D$) to ($1.25D, 1.25D$). The sound region is $212D$ in width and length. The sponge region is as large as $400D$ in both width and length to diminish the reflections from the boundary. Non-reflecting boundary are applied on the external boundaries of the sponge region. In the cylinder occupied region, the mesh spacing is $D/40$. The mesh spacing in the sound region increases to $D/5$. There is a sinusoidal transition between the cylinder occupied region and sound region. The maximum mesh spacing in the sponge region is $12.5D$. In order to achieve a fast transition to the asymmetric K\'arm\'an vortex street, an initial velocity perturbation is applied in the near wake.

\begin{figure}
 \begin{center}
  \hskip-3.0in (a) \hskip3.0in (b)

  \includegraphics[width=3.2in]{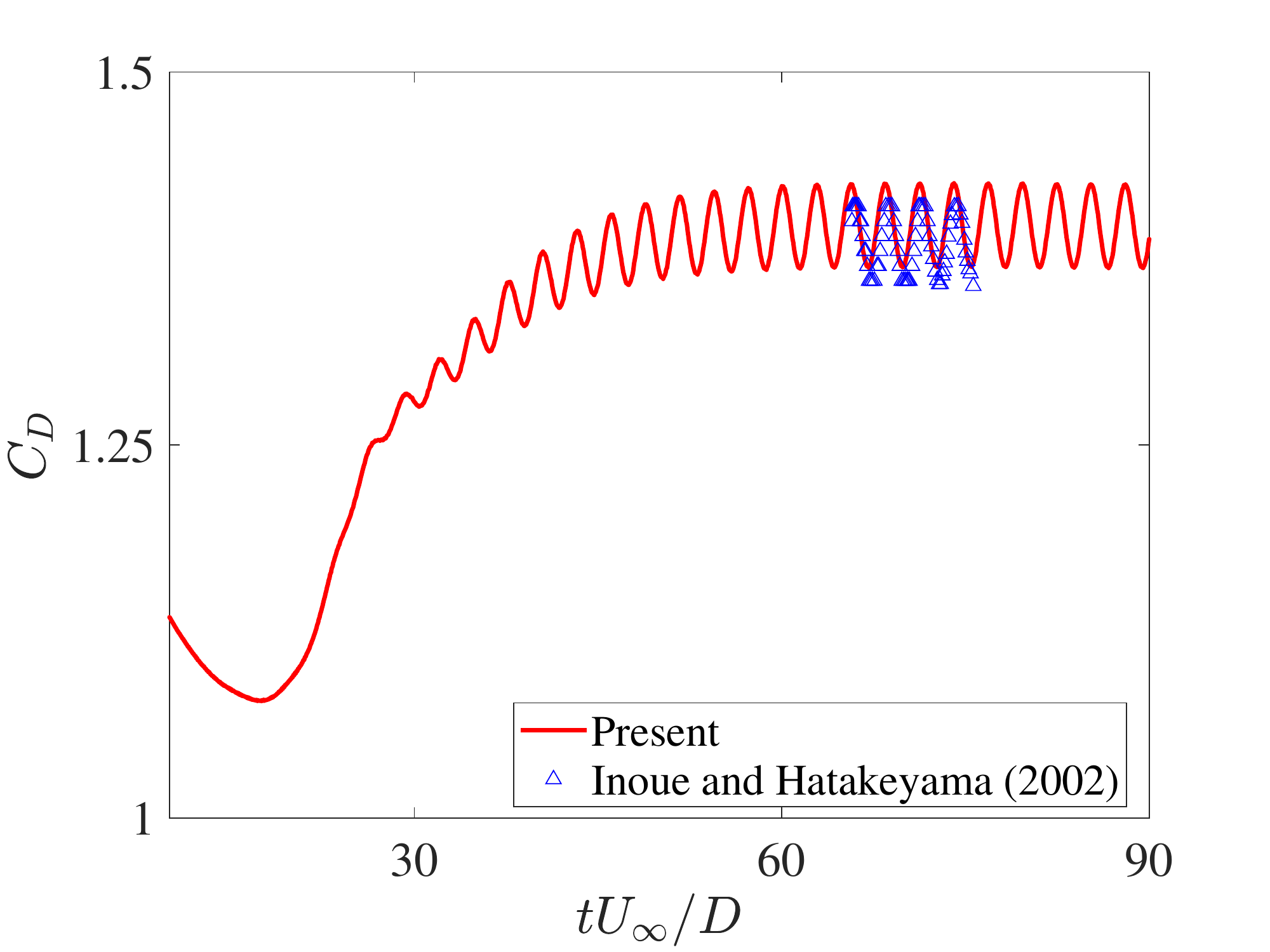}
  \hskip0.1in
  \includegraphics[width=3.2in]{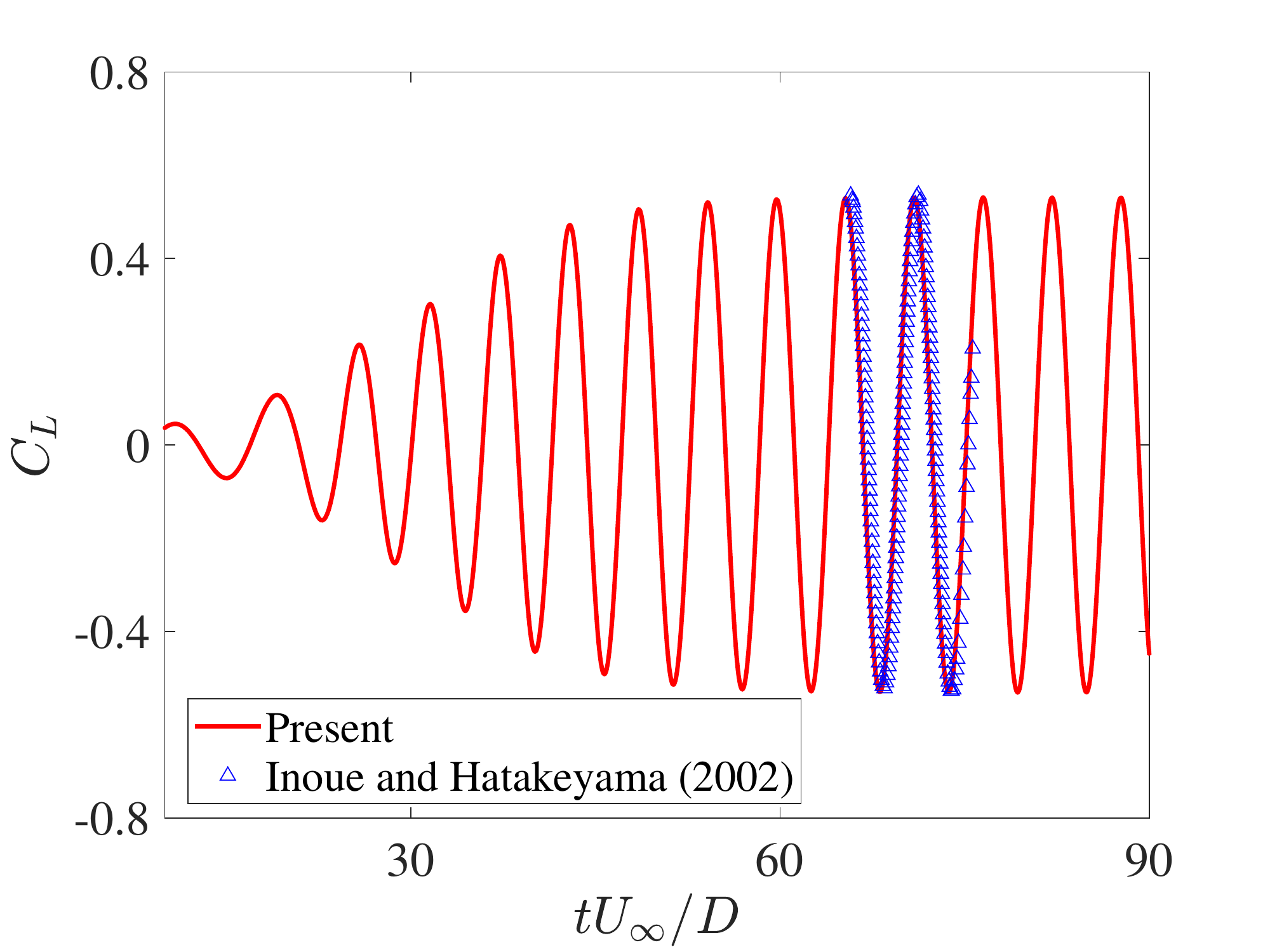}
  \end{center}
\caption{Sound generation by a stationary cylinder in a uniform flow: time histories of $C_D$ (a) and $C_L$ (b) at Re=150 and $M=0.2$.}
\label{Fig:Re150M0_2clcd}
\end{figure}

The time histories of the drag ($C_D$) and lift ($C_L$) coefficients scaled by the $0.5\rho_f U_0^2 D$ are presented in Fig.~\ref{Fig:Re150M0_2clcd}, with data reported in Ref.~\cite{inoue2002sound} for comparison. The amplitudes of $C_L$ and mean value of $C_D$ are $0.525$ and $1.40$, respectively, which agree well with the results ($C_L=0.520$ and $C_{D,m}=1.39$) from Ref.~\cite{inoue2002sound}. The Strouhal number calculated by the present method is about 0.178, which is also close to 0.175 in Ref.~\cite{dumbser2016high} and 0.183 in Ref.~\cite{inoue2002sound}.

\begin{figure}
  \begin{center}
  \includegraphics[width=3.5in]{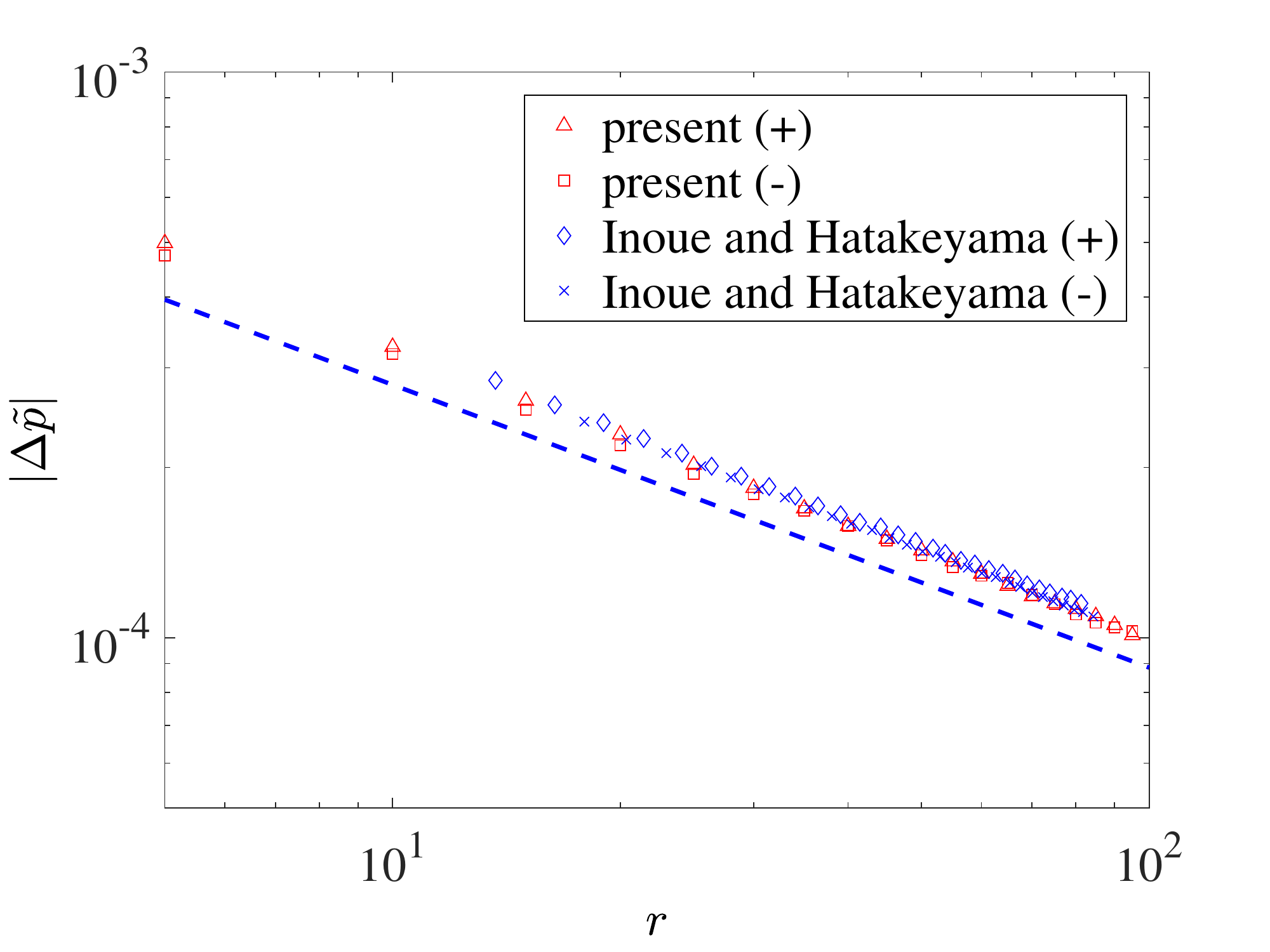}
  \end{center}
\caption{Sound generation by a stationary cylinder in a uniform flow: comparison of the decay of pressure peaks measured at $\theta=90^o$ with available data from Ref.~\cite{inoue2002sound}. $+$ and $-$ denote the positive and negative peaks. The dashed line is the theoretical prediction by Ref.~\cite{Landau1987fluidmech} showing that the pressure peaks tend to decay in proportion to $r^{-\frac{1}{2}}$.}
\label{Fig:Re150M0_2decay}
\end{figure}

The fluctuating pressure $\Delta \tilde{p}$ is defined by $\Delta \tilde{p}(x,y,t)=\Delta p(x,y,t)-\Delta \bar{p}(x,y)$, where $\Delta p$ denotes the total fluctuating pressure and $\bar{p}$ is the time average pressure. $\Delta p=p-p_{\infty}$ with $p_{\infty}$ being the ambient pressure~\cite{inoue2002sound}. In present paper, the fluid density $\rho_f$ and sound speed $c$ are used to nondimensionalize the pressure. Fig.~\ref{Fig:Re150M0_2decay} shows decay of the pressure peaks (include both positive and negative peaks) measured at $\theta=90^o$, which agrees well with results from Ref.~\cite{inoue2002sound}. The pressure peaks of the cylindric pressure waves generated by the unsteady flow tend to decay in proportion to $r^{-\frac{1}{2}}$, which agrees well with the theoretical prediction by Landau and Lifshitz~\cite{Landau1987fluidmech}, as indicated by the dashed line in Fig.~\ref{Fig:Re150M0_2decay}.

\begin{figure}
 \begin{center}
  \hskip-3.0in (a) \hskip3.0in (b)

  \includegraphics[width=3.2in]{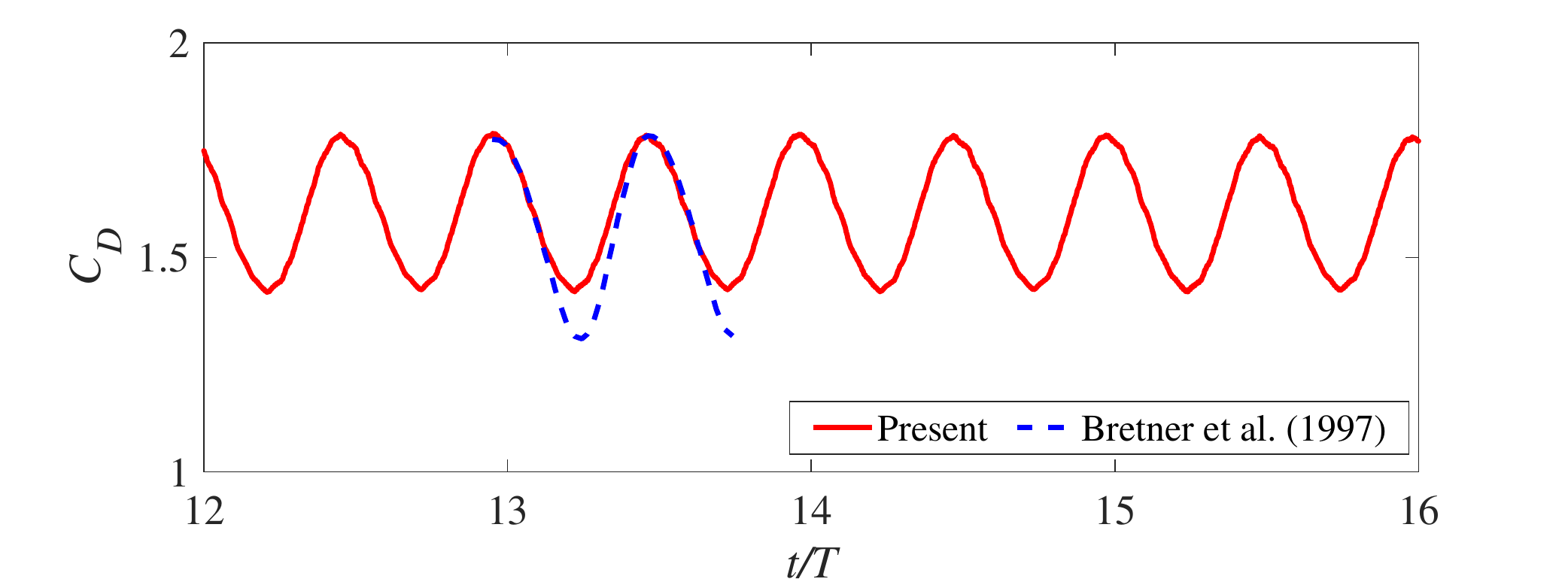}
  \hskip0.1in
  \includegraphics[width=3.2in]{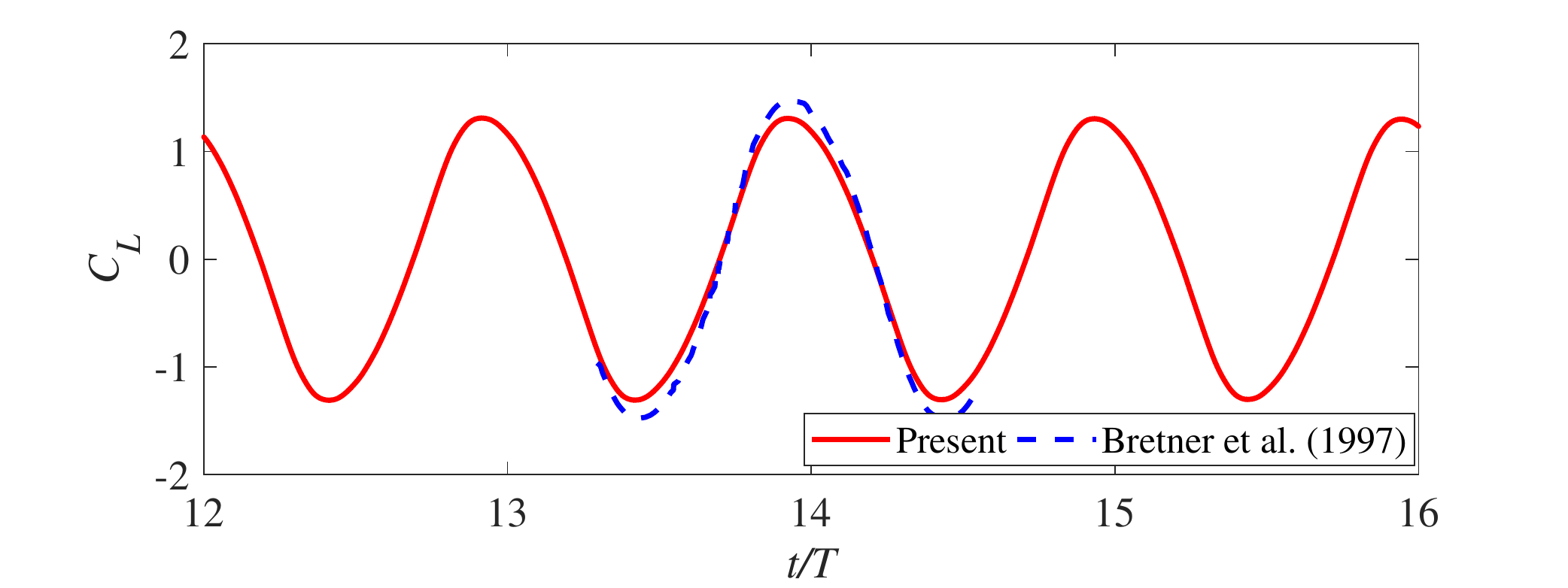}
  \end{center}
\caption{Sound generation by a stationary cylinder in a uniform flow: time histories of $C_D$ (a) and $C_L$ (b) at Re=1000 and $M=0.2$.}
\label{Fig:cyRe1000clcd}
\end{figure}

We further increase the Reynolds number to 1000 (Ma=0.2), and assume it is still in laminar flow regime. The same problem has also been examined by Brentner et al.~\cite{brentner1997computation}, who used the Lighthill acoustic analogy to separate flow dynamics and acoustics calculations. Fig.~\ref{Fig:cyRe1000clcd} presents the comparison of the time histories of $C_D$ and $C_L$ with the data from Ref.~\cite{brentner1997computation}. The present drag coefficient and St number are respectively 1.60 and 0.215, which agree with the numerical results (1.56 and 0.238) in Ref.~\cite{brentner1997computation}. The discrepancy  (2.6\% in the drag coefficient and 9.7\% in St) is probably due to the compressibility of the fluid neglected in the reference, as indicated in Ref.~\cite{brentner1997computation} where the compressible solver predicts lower frequency than the incompressible one.

\begin{figure}
  \begin{center}
  \includegraphics[width=4.5in]{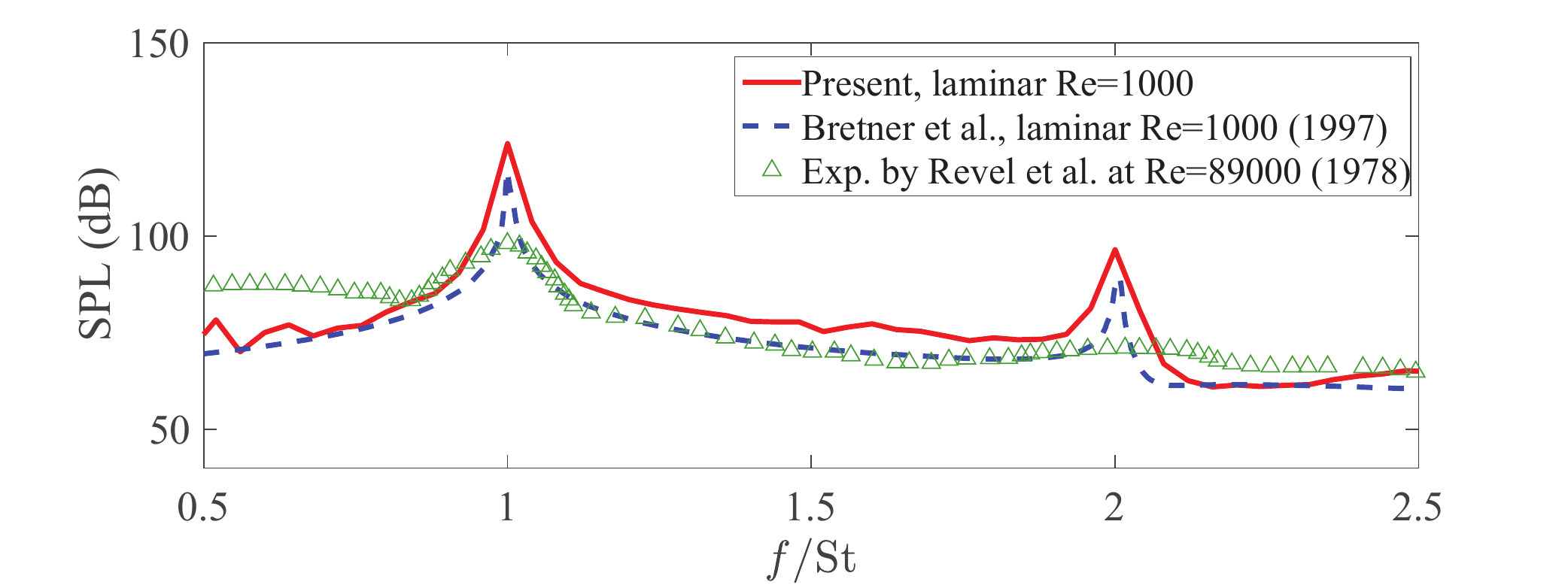}
  \end{center}
\caption{Sound generation by a stationary cylinder in a uniform flow: comparison of the sound pressure level measured at a distance of $128D$ and $90^o$ from the inlet flow: Re=1000 and $M=0.2$.}
\label{Fig:cyRe1000_spl_st}
\end{figure}

\begin{figure}
 \begin{center}
  \includegraphics[width=3.05in]{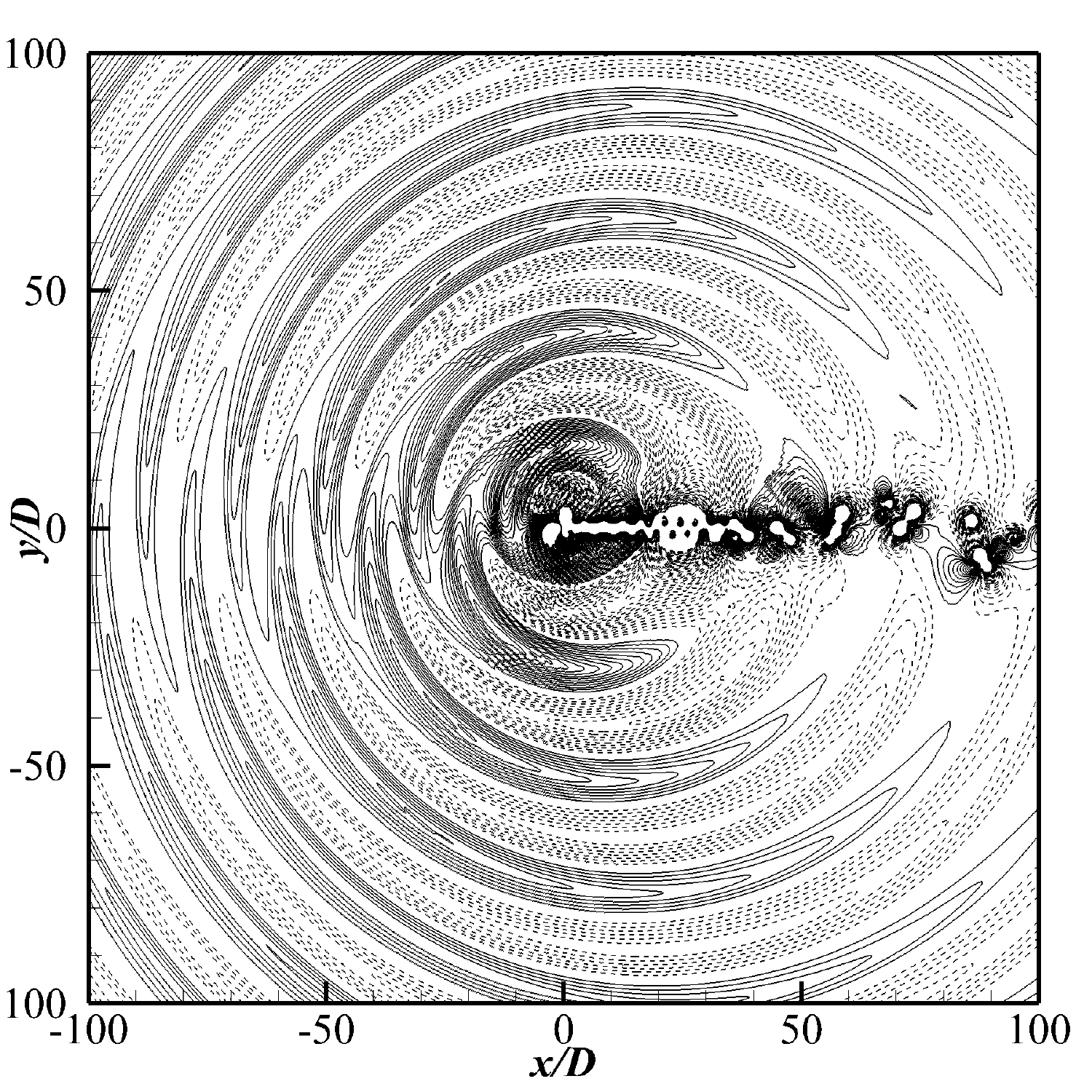}
  \end{center}
\caption{Sound generation by a stationary cylinder in a uniform flow: a snapshot of the total pressure fluctuation $\Delta p$ contours, Re=1000 and $M=0.2$. The contour level ranges from $-0.1M^{2.5}$ to $0.1M^{2.5}$ with an interval of $0.0025M^{2.5}$, the dashed and solid lines indicate the negative and positive pressure, respectively.}
\label{Fig:cyRe1000M0_2_dpcontour}
\end{figure}

Fig.~\ref{Fig:cyRe1000_spl_st} shows the SPL (reference pressure $20 uPa$) in the frequency space obtained by the Fast Fourier transform (FFT). For comparison, data from in Ref~\cite{brentner1997computation,revell1978experimental} are shown in the same figure. The present results agree well with those from the references. The maximum discrepancies of the peak values are about 5\%, which is probably induced by the neglected quadrupole source in FW-H. It should be noted that the results from Ref.~\cite{brentner1997computation} were calculated in two steps: the incompressible N-S equations were first solved, then the acoustics was computed by solving Ffowcs Williams--Hawkings (FW-H) equation using the resolved unsteady flow as the input with the quadrupole source neglected, while, the present method obtains the sound pressure from DNS. The experimental results from Ref.~\cite{revell1978experimental} show much lower peaks at high Reynolds number, but the frequency space still coincides with that at low Reynolds number. According to Ref.~\cite{brentner1997computation}, it is still challenging to accurately predict the SPL in the frequency space by using turbulence models such as $k-\omega$ and SST. In the future, our effort will be made to incorporate large eddy simulation and wall models into the current solver to address this challenge. Fig.~\ref{Fig:cyRe1000M0_2_dpcontour} is a snapshot of the fluctuating pressure contours for future validation of newly developed method.

\subsection{Sound generation by a rotating cylinder in a uniform flow}
Here, we consider the sound generated by a rotating cylinder in a uniform flow, which was studied by Inoue et al.~\cite{inoue2003control}. The non-dimensional parameters governing this problem are Re, $M$ (defined by Eq.~\ref{eq:cy_para}) and the velocity ratio, defined by  $\alpha=M_\theta/M$, where $M_\theta$ is the anticlockwise angular velocity of the rotating cylinder. In the current rotating case, Re=160, $M=0.2$ and $\alpha=1.5$. The mesh strategy is same as that in Section 3.2.

\begin{figure}
  \begin{center}
  \includegraphics[width=3.5in]{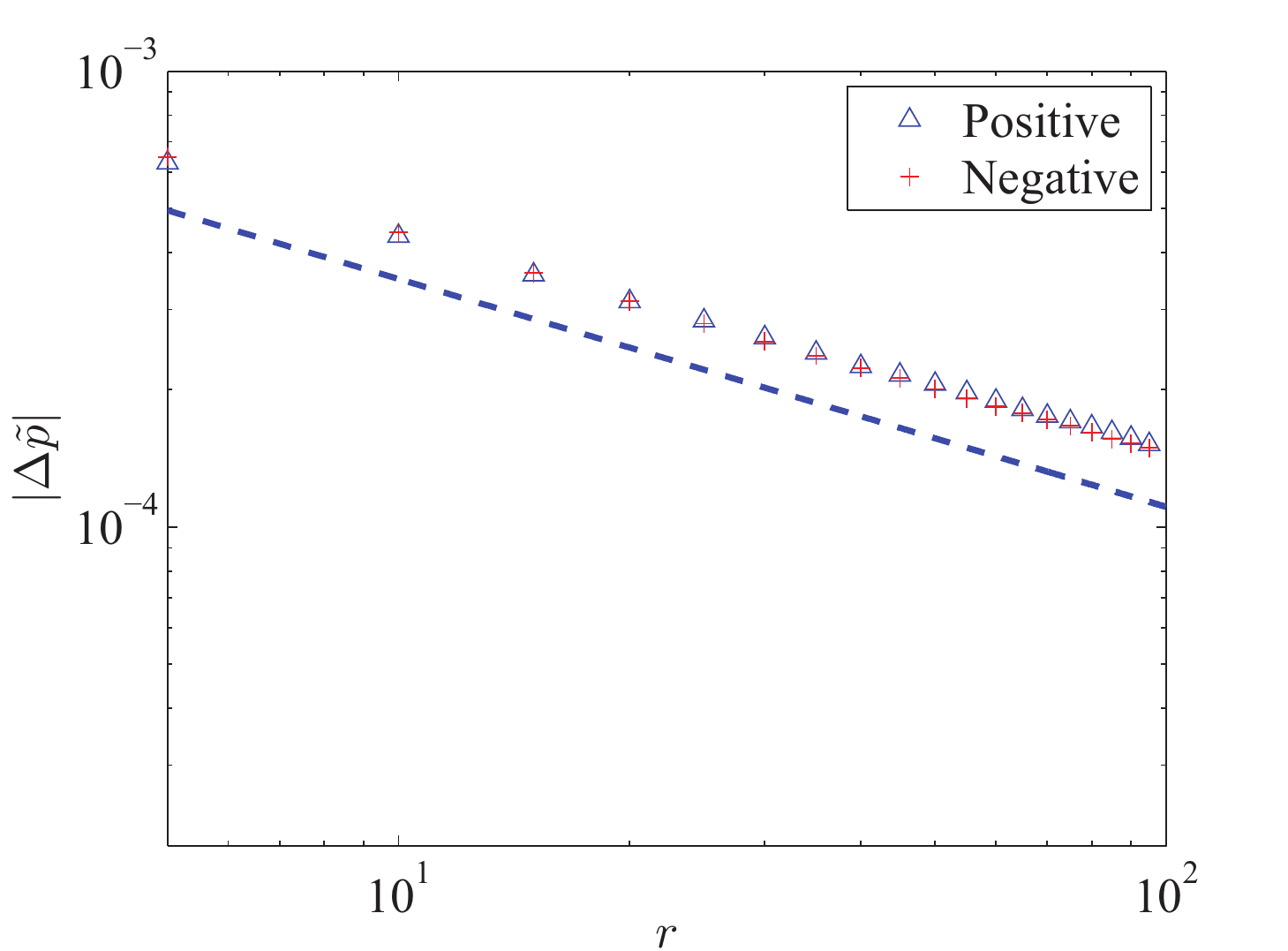}
  \end{center}
\caption{Sound generation by a rotating cylinder in a uniform flow: decay of pressure peaks measured at $\theta=90^o$. The dashed line indicates that the pressure peaks tend to decay in proportion to $r^{-\frac{1}{2}}$.}
\label{Fig:Re160M0_2Rotdecay}
\end{figure}

\begin{figure}
  \begin{center}
  \includegraphics[width=3.5in]{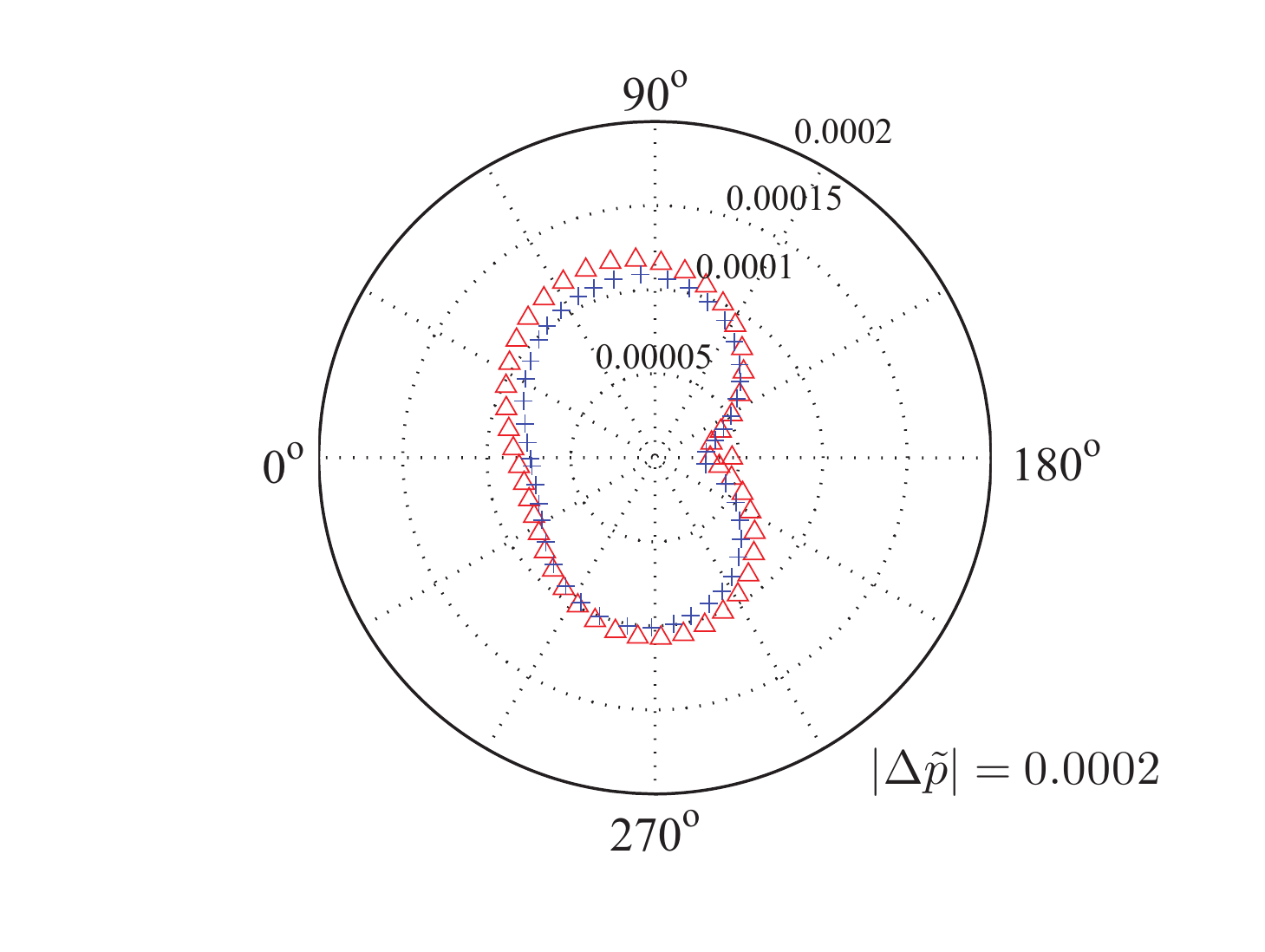}
  \end{center}
\caption{Sound generation by a rotating cylinder in a uniform flow: comparison of the root-mean-square of $\Delta \tilde{p}$ (measured at at $r=r'(1-M {\rm cos}\theta)$ with $r'=75$) from the present computation ($\Delta$) and that obtained from Ref.~\cite{inoue2003control} ($+$).}
\label{Fig:Re160M0_2Rotdp}
\end{figure}

The decay of the fluctuating pressure peaks measured at $\theta=90^o$ is plotted in Fig.~\ref{Fig:Re160M0_2Rotdecay}. The pressure peaks also tend to decay in proportion to $r^{-\frac{1}{2}}$, similar to that discussed in the stationary case. The polar plots in Fig.~\ref{Fig:Re160M0_2Rotdp} are the root-mean-square of the fluctuating pressure $\Delta \tilde{p}$ measured at $r=r'(1-M {\rm cos}\theta)$ with $r'=75$, taking the Doppler effect into account. The data from Ref.~\cite{inoue2003control} are also presented in Fig.~\ref{Fig:Re160M0_2Rotdp} for comparison. The results show that the pressure measured at $\theta=90^o$ is larger than that measured at $\theta=0^o$, indicating the lift dipole dominates the sound generation. The profile is not symmetric with respect to the line $\theta=0^o$ and $180^o$ as that in the stationary condition, due to the effects of rotation. The good agreement between the present results and the computational results from Ref.~\cite{inoue2003control} shows that the present method has an excellent ability to handle the acoustic simulations involving moving boundaries.

\begin{figure}
  \begin{center}
  \includegraphics[width=3.5in]{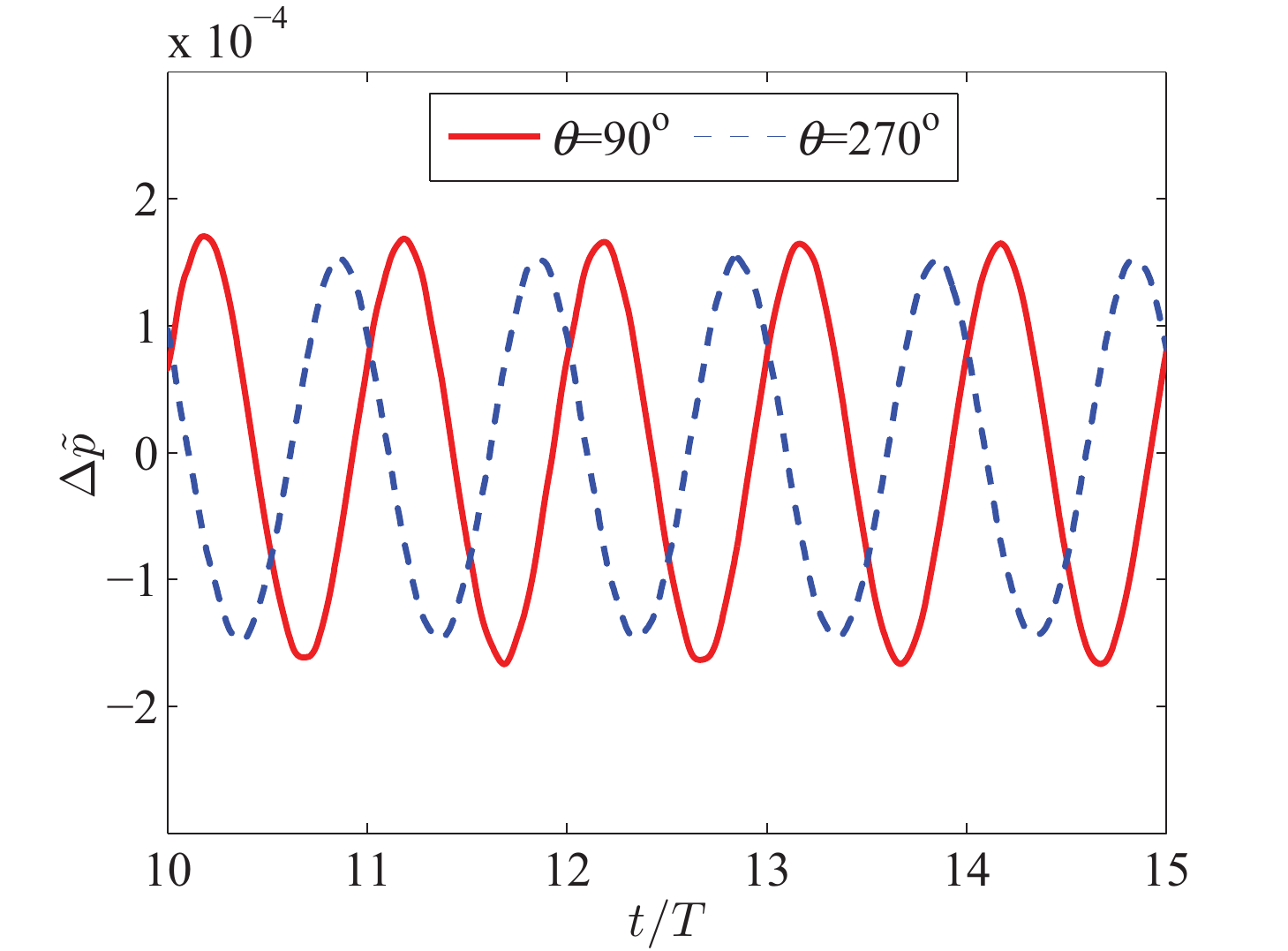}
  \end{center}
\caption{Sound generation by a rotating cylinder in a uniform flow: time histories of the fluctuating pressure $\Delta \tilde{p}$ measured at $r=75$, $\theta=90^o$ and $270^o$.}
\label{Fig:Re160M0_2Rotpt}
\end{figure}

Fig.~\ref{Fig:Re160M0_2Rotpt} shows the time histories of fluctuating pressure measured at $r=75$, $\theta=90^o$ and $270^o$. The results show that the pressure peaks measured at $\theta=90^o$ are higher than those at $\theta=270^o$, due to the fact that the anticlockwise rotation of the cylinder leads to asymmetrical lift and drag. This is also indicated by the root-mean-square of $\Delta \tilde{p}$ plotted in Fig.~\ref{Fig:Re160M0_2Rotdp}.

\subsection{Sound generation by an insect in hovering flight}
\begin{figure}
  \begin{center}
  \includegraphics[width=3.5in]{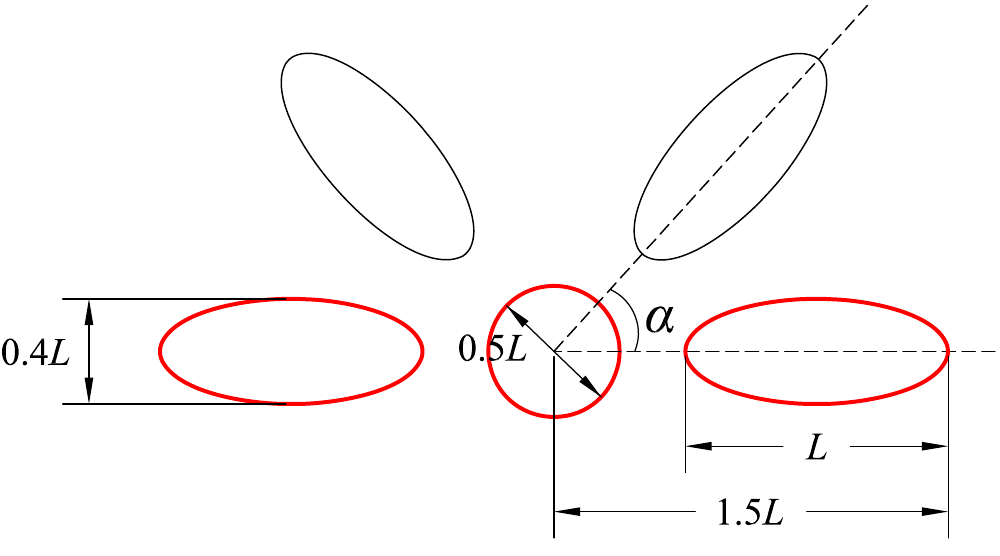}
  \end{center}
\caption{Schematic of the prescribed flapping motion of an insect.}
\label{Fig:mav_sch}
\end{figure}

In this section, an insect in hovering flight is considered to validate the current solver in handling moving body with relatively complex geometry configuration. The schematic of the problem is shown in Fig.~\ref{Fig:mav_sch}. A circular cylinder (body) with a diameter of $0.5L$, and two elliptic cylinders (wings) with the dimensions of $L$ and $0.4L$ are used to model the insect. The two wings flap symmetrically, and the flapping motion of the wings is prescribed by
\begin{equation}
\alpha(t)=\alpha_0[1+sin(2\pi f t)],
\label{eq:mav_premotion}
\end{equation}  
where the amplitude of the flapping angle $\alpha_0$ is $25.3^o$ and $f$ is the flapping frequency. The Reynolds number defined by $\rho_f U_{max} L/\mu$ is 200, where $U_{max}$ is the maximum wing tip velocity. The Strouhal number defined by $f L/U_{max}$ is 0.25, and the Mach number based on $U_{max}$ is 0.1. The computational domain extends from ($-100L$, $-100L$) to ($100L$, $100L$), with 50 points to resolve the wing length.

\begin{figure}
 \begin{center}
  \includegraphics[width=3.2in]{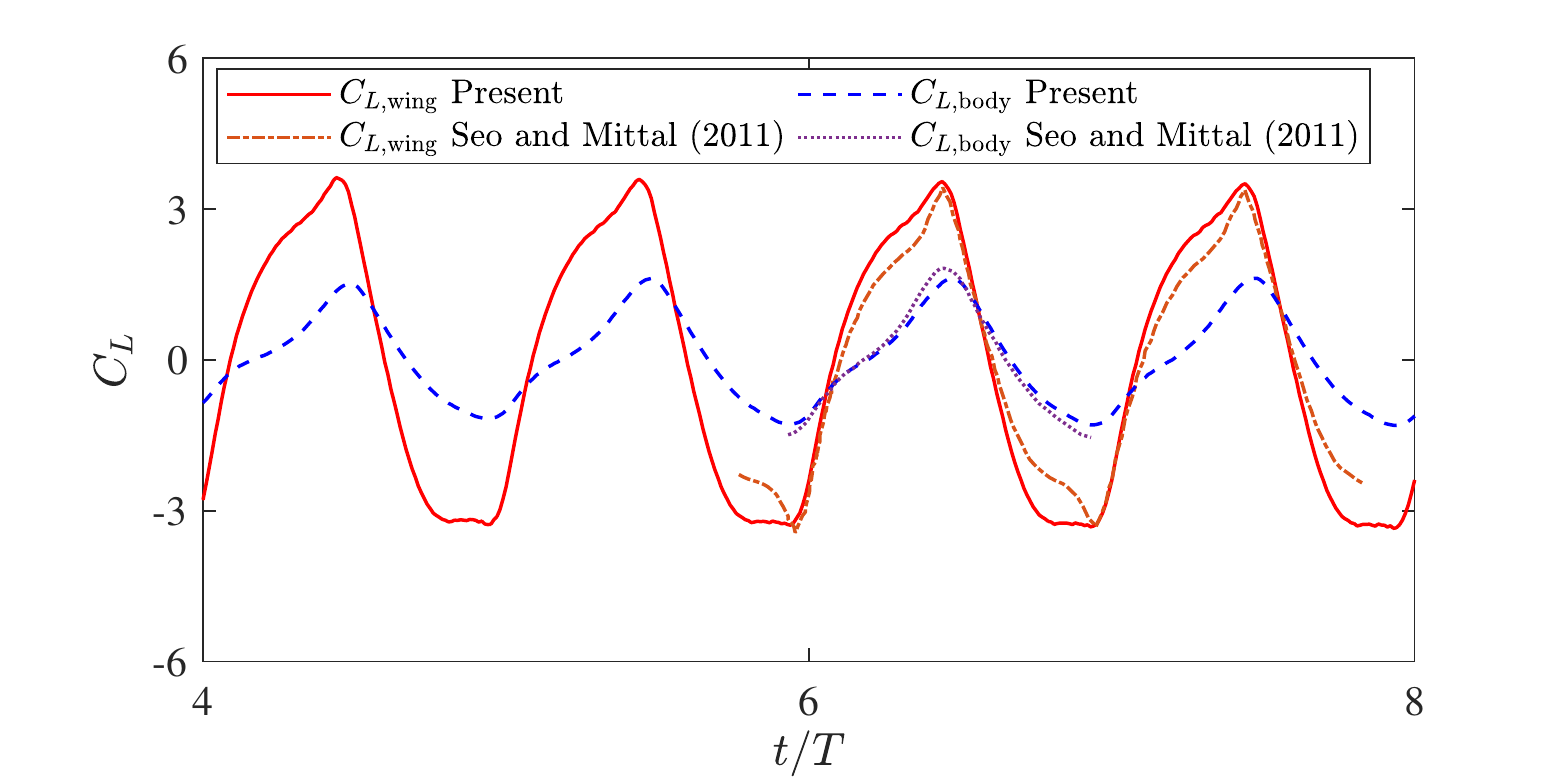}
  \end{center}
\caption{Sound generation by an insect in hovering flight: comparison of the lift generated by the wing and body.}
\label{Fig:mav_cl}
\end{figure}

\begin{figure}
  \begin{center}
  \hskip-3.0in (a) \hskip3.0in (b)\\
  
  \includegraphics[width=3.0in]{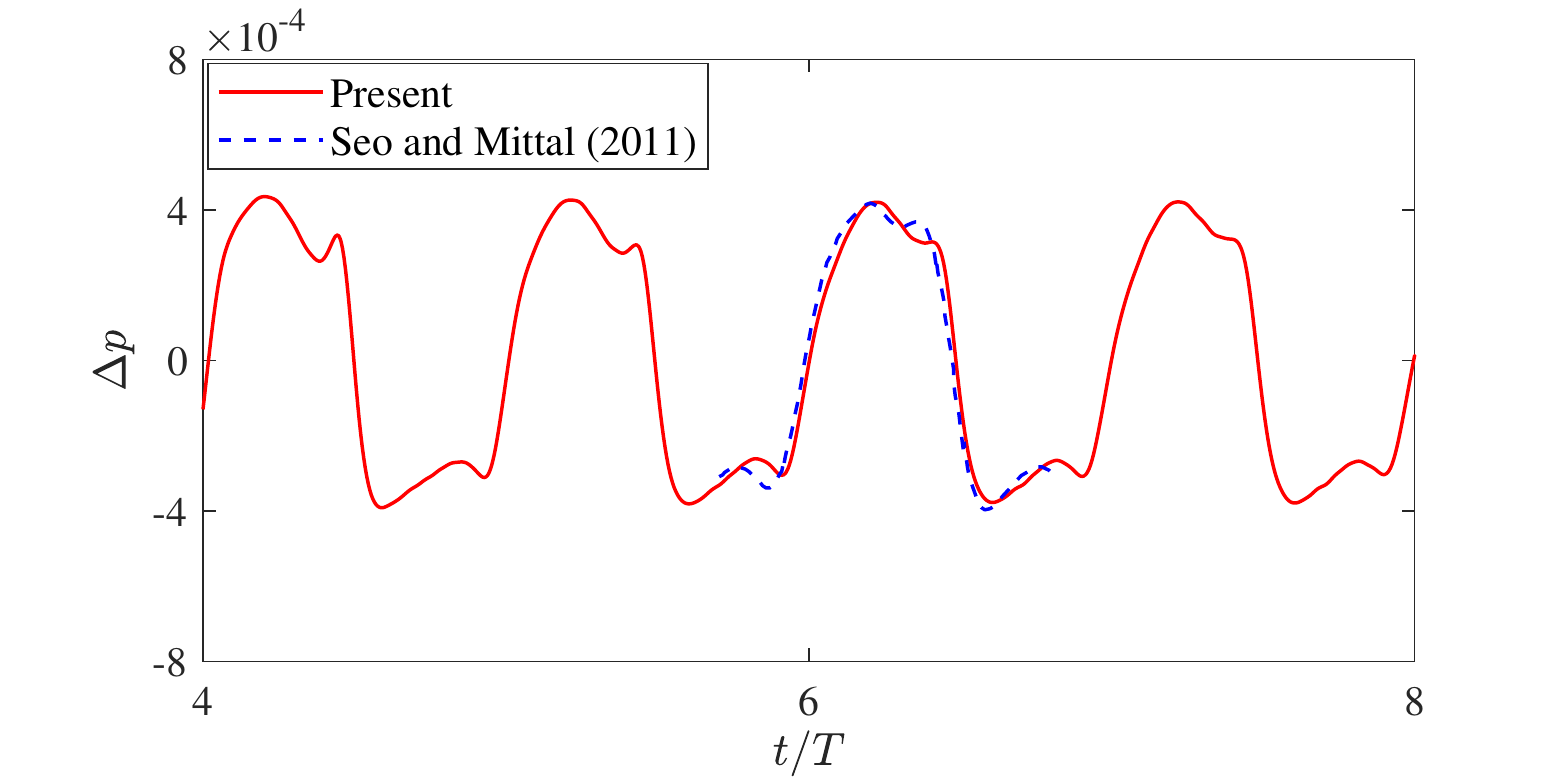}
  \hskip0.1in
  \includegraphics[width=3.0in]{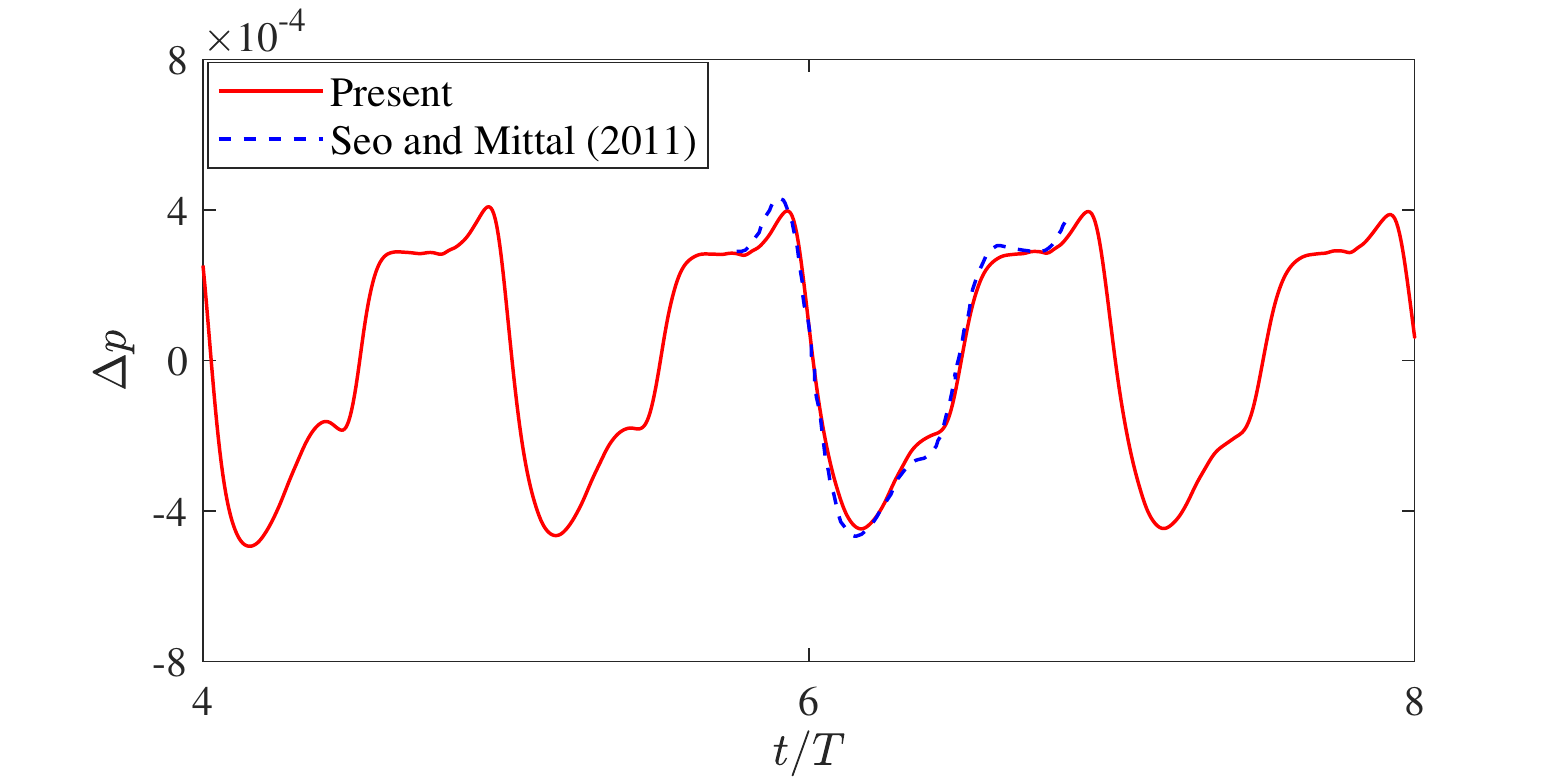}
  \end{center}
\caption{Sound generation by an insect in hovering flight: comparison of fluctuating pressure measured at (a) (0, $60L$) and (b) (0, $-60L$).}
\label{Fig:mav_d60_dp}
\end{figure}

\begin{figure}
  \begin{center}
  \includegraphics[width=3.5in]{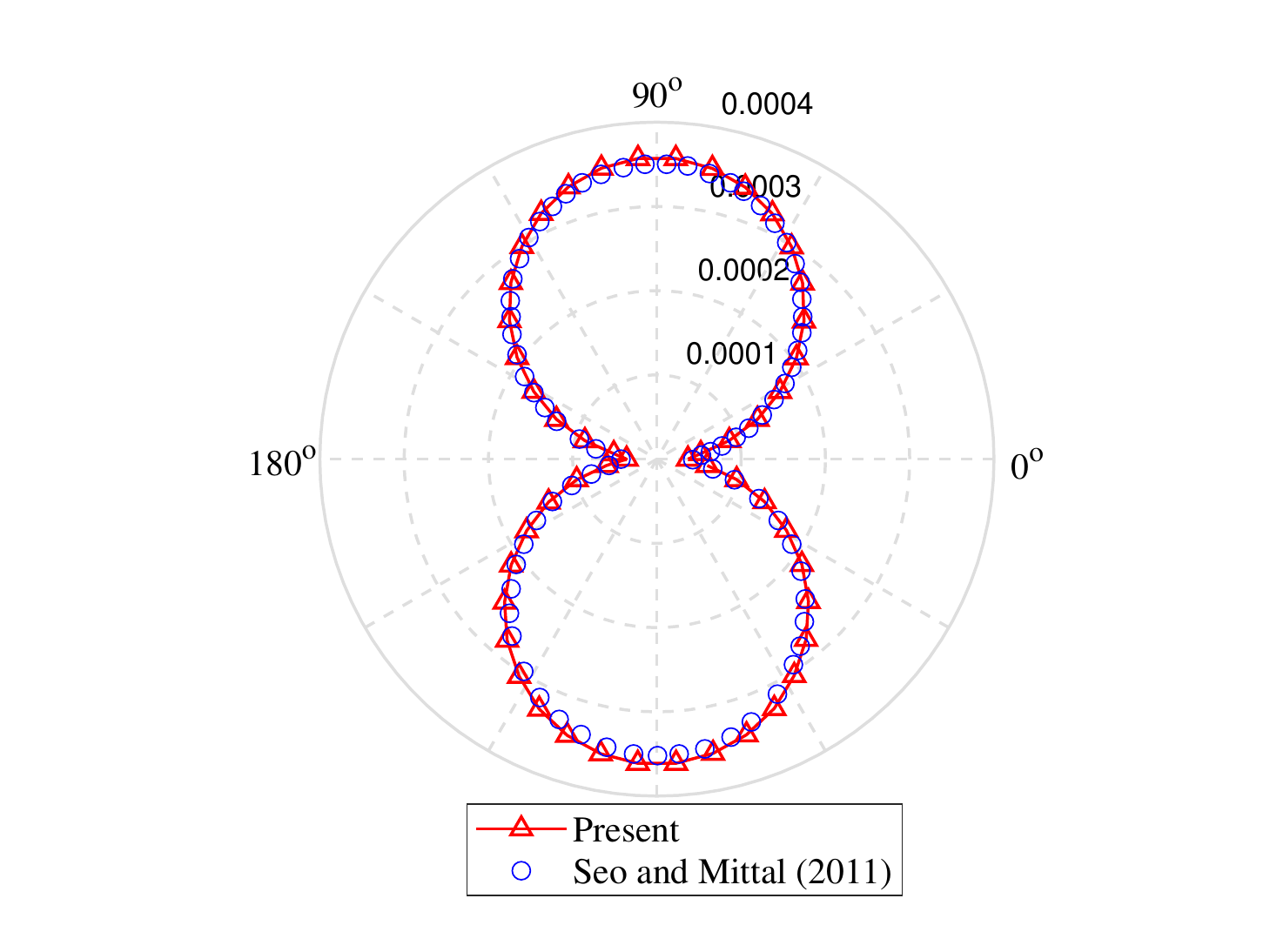}
  \end{center}
\caption{Sound generation by an insect in hovering flight: comparison of the root-mean-square of fluctuating pressure measured at a distance of $50L$.}
\label{Fig:mav_d50_dprms}
\end{figure}

\begin{figure}
 \begin{center}
  \includegraphics[width=3.2in]{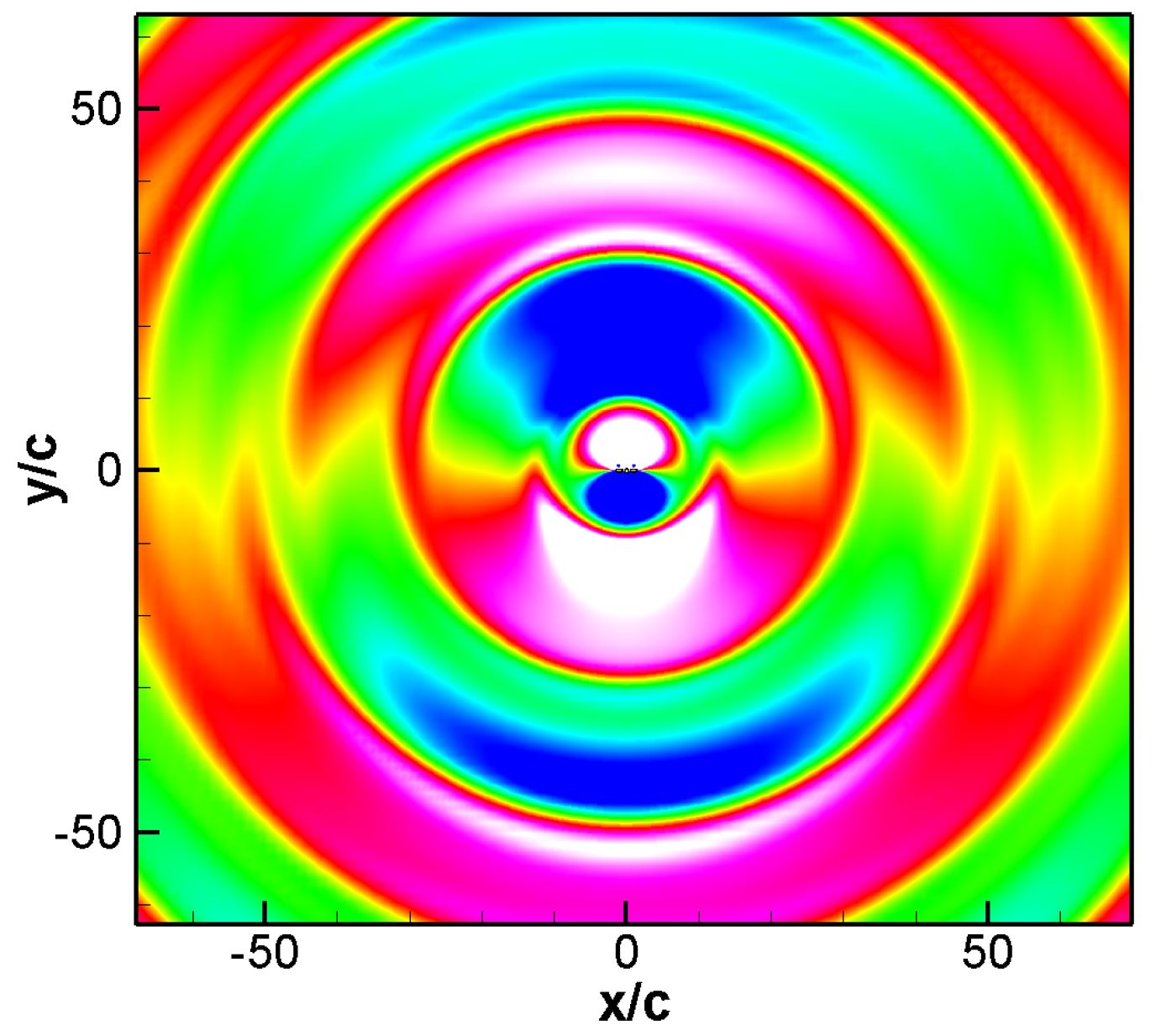}
  \hskip0.1in
  \includegraphics[width=3.0in]{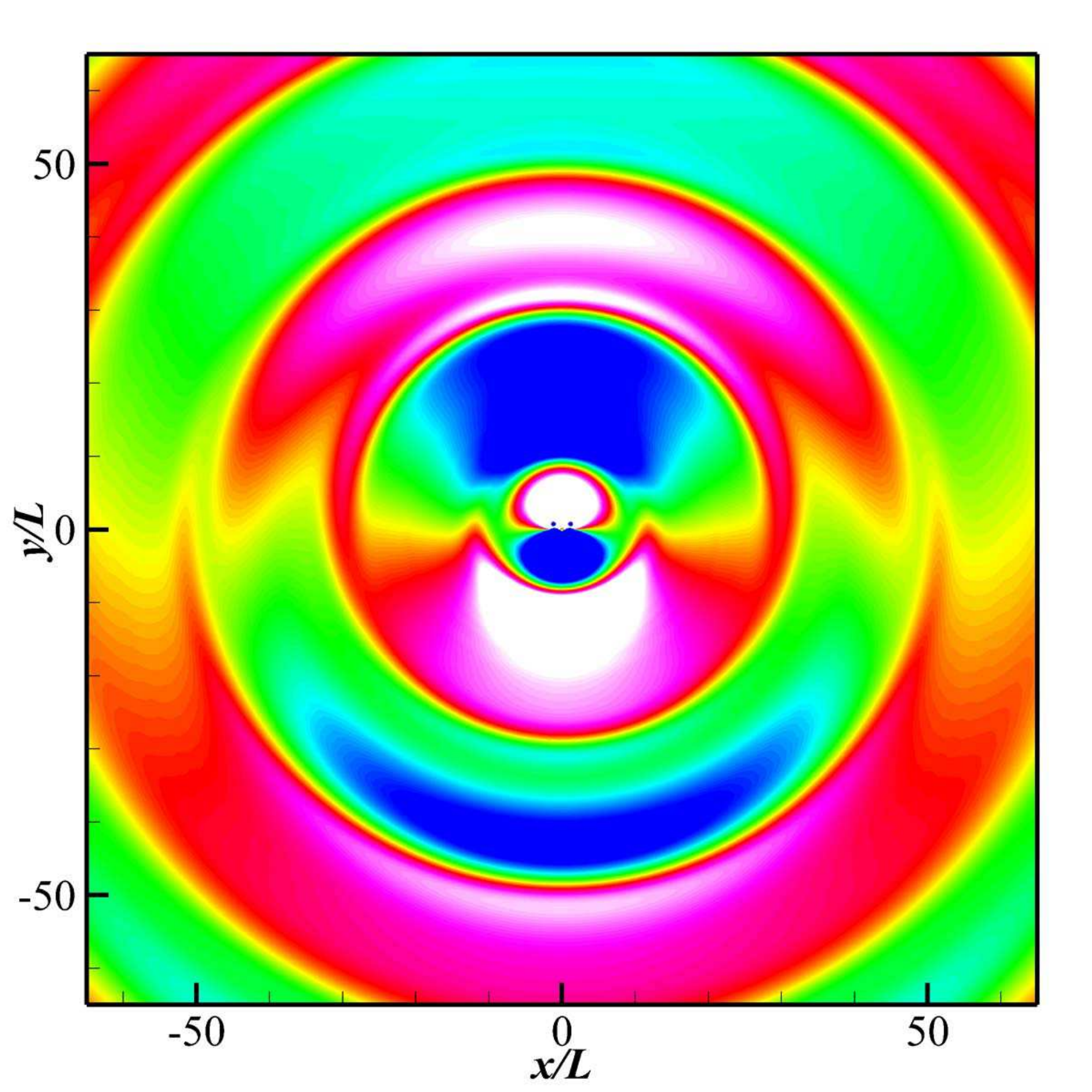}
  \end{center}
\caption{Sound generation by an insect in hovering flight: comparison of the instantaneous fluctuating pressure contours calculated by Seo and Mittal~\cite{seo2011computation} (left) and the present method (right). The contour level ranges from $-0.005\rho_f c^2$ (blue) to $0.005\rho_f c^2$ (red).}
\label{Fig:mav_pres_contour}
\end{figure}

A direct comparison of the lifts generated by the body and wing is shown in Fig.~\ref{Fig:mav_cl}. It is found that the present results agree well with the data from Ref.~\cite{seo2011computation}. Fig.~\ref{Fig:mav_d60_dp} shows that the fluctuating pressure measured at (0, $60L$) and (0, $-60L$) from present computation is also in a good agreement with the data from Ref.~\cite{seo2011computation}. Good agreements of the root-mean-square of fluctuating pressure measured at a distance of $50L$ shown in Fig.~\ref{Fig:mav_d50_dprms} confirm the good performance of the current method in handling acoustics involving complex moving geometries. Additionally, a qualitatively comparison of the fluctuating pressure is presented in Fig.~\ref{Fig:mav_pres_contour}.

\subsection{Deformation of a red blood cell induced by acoustic waves}
\begin{figure}
 \begin{center}
  \hskip-1.8in (a) \hskip1.8in (b) \hskip1.8in (c)

  \includegraphics[width=2.1in]{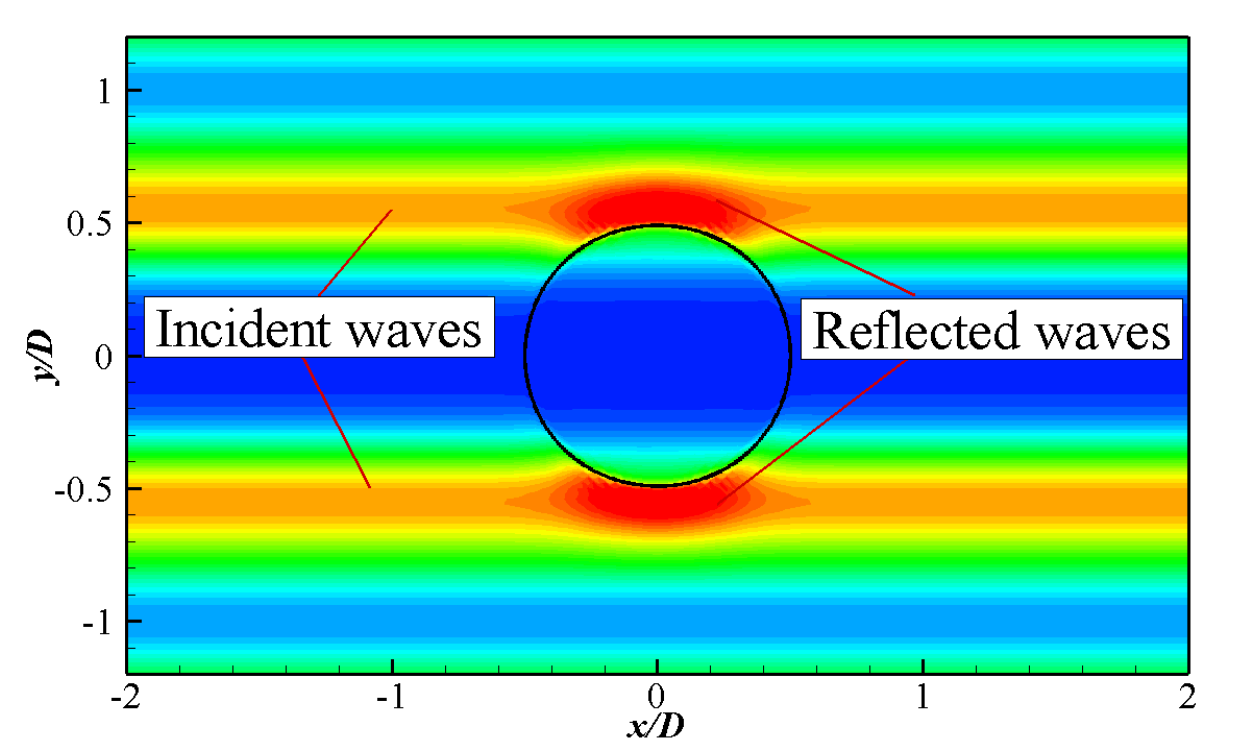}
  \includegraphics[width=2.1in]{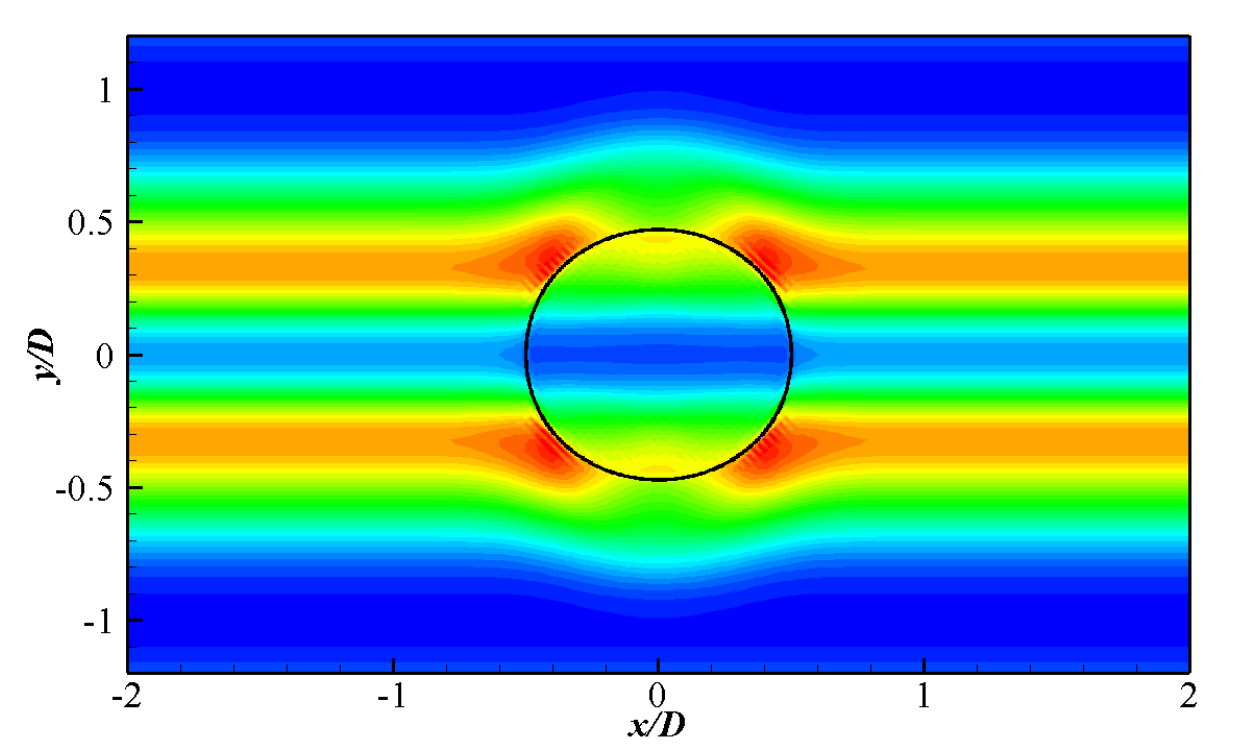}
  \includegraphics[width=2.1in]{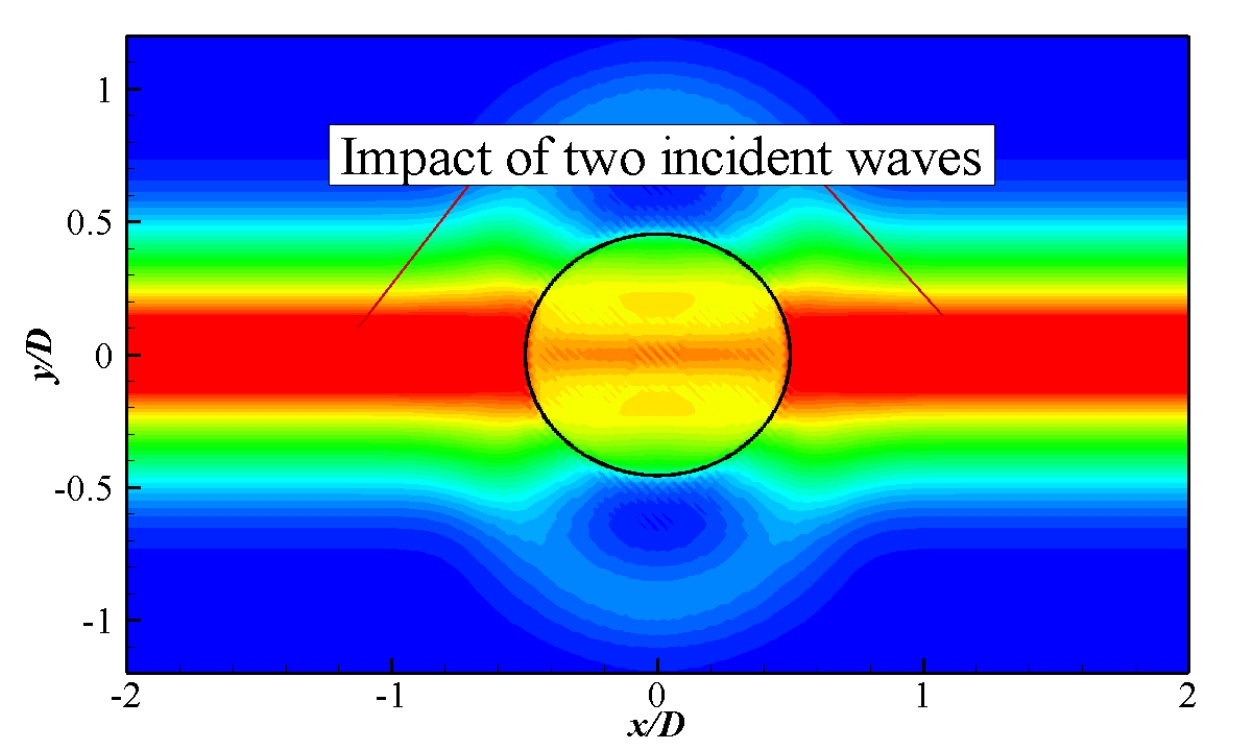}
  \end{center}
\caption{Deformation of a red blood cell induced by acoustic waves: instantaneous fluctuating pressure contour with $A^*=0.4$ at $tc/D=$ 0.4 (a), 0.6 (b) and 0.8 (c). The contour level ranges from 0 (blue) to $0.2\rho_f c^2$ (red).}
\label{Fig:rbc_contour}
\end{figure}

\begin{figure}
 \begin{center}
  \includegraphics[width=1.2in]{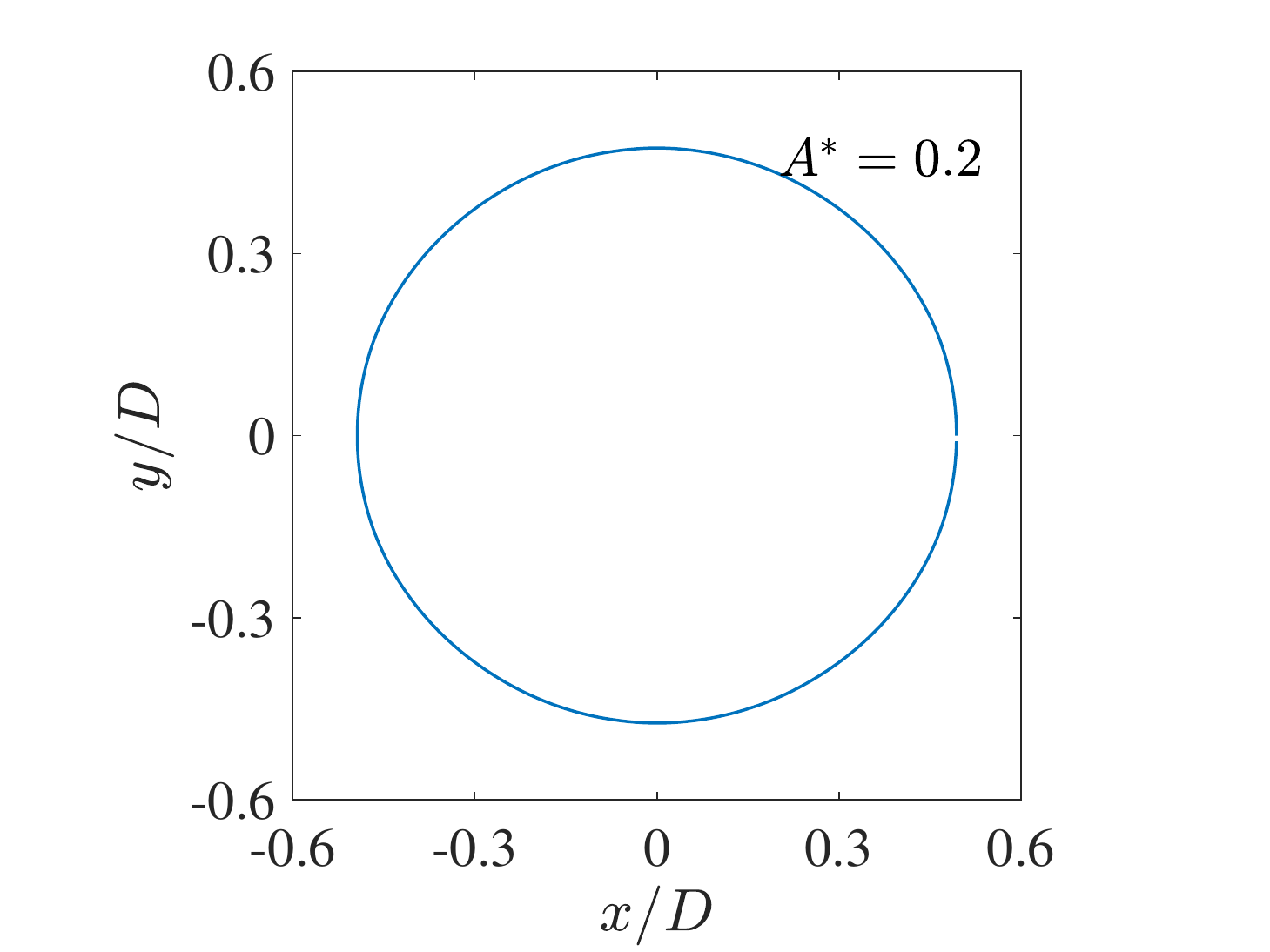}
  \includegraphics[width=1.2in]{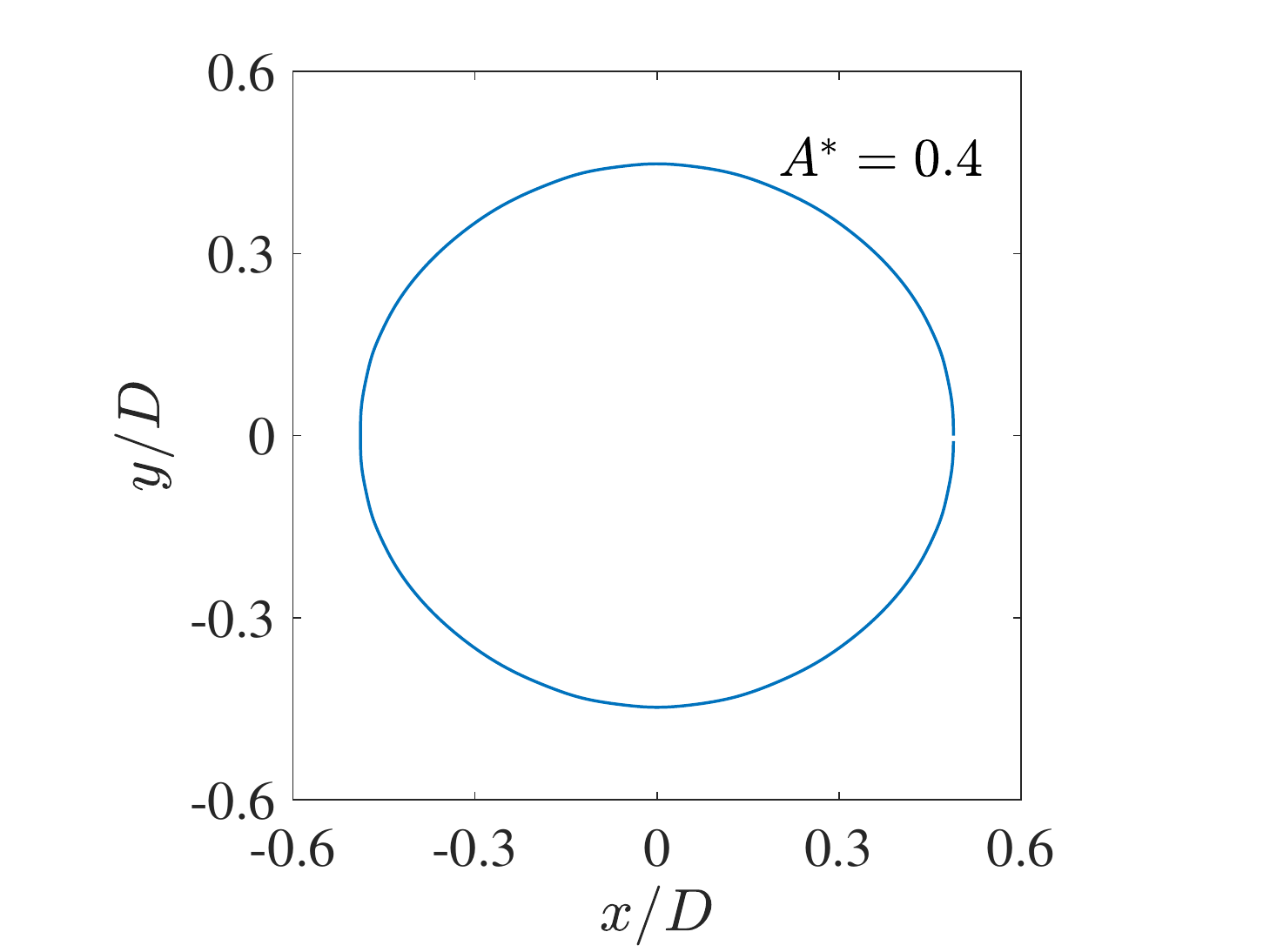}
  \includegraphics[width=1.2in]{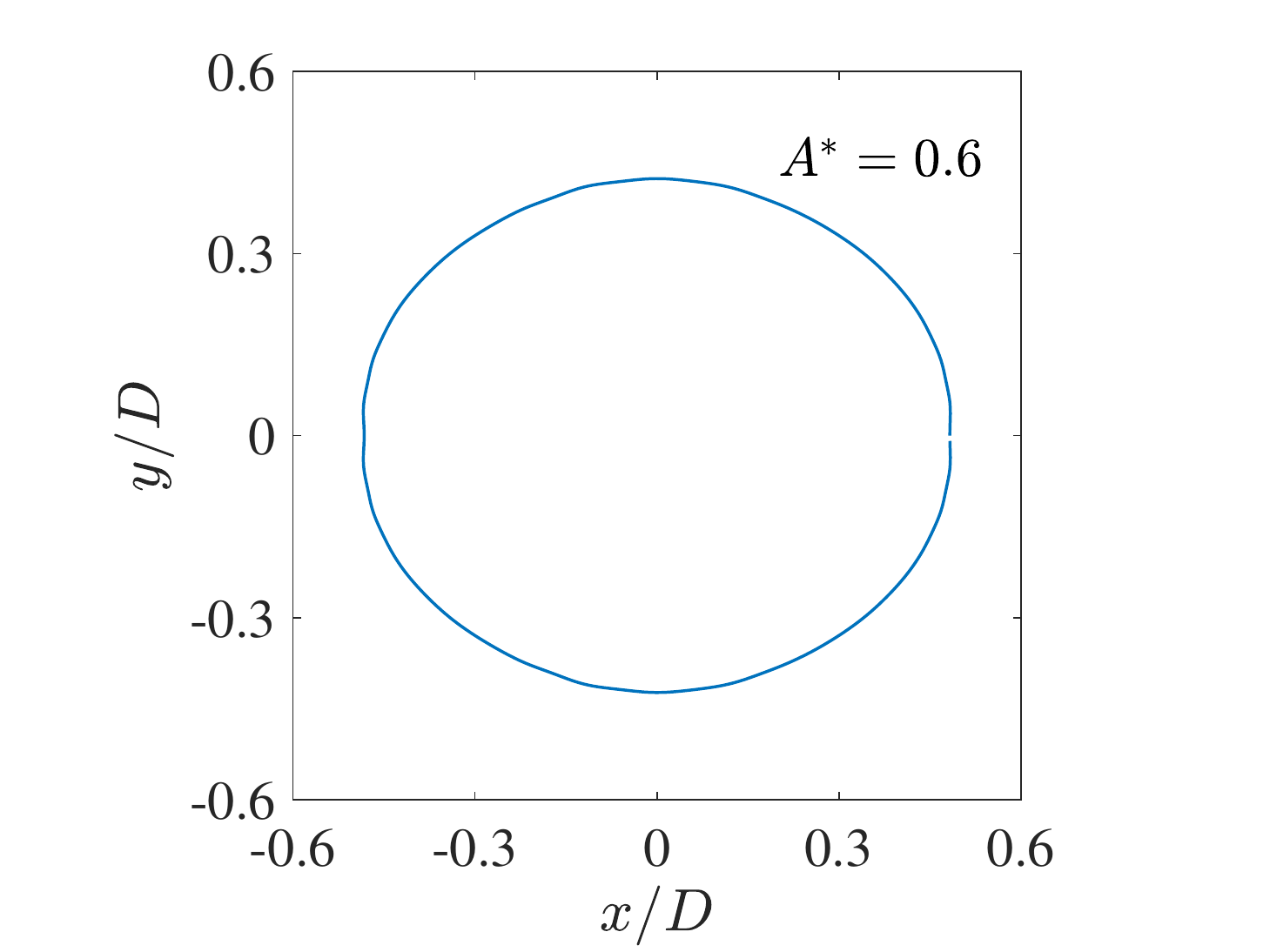}
  \includegraphics[width=1.2in]{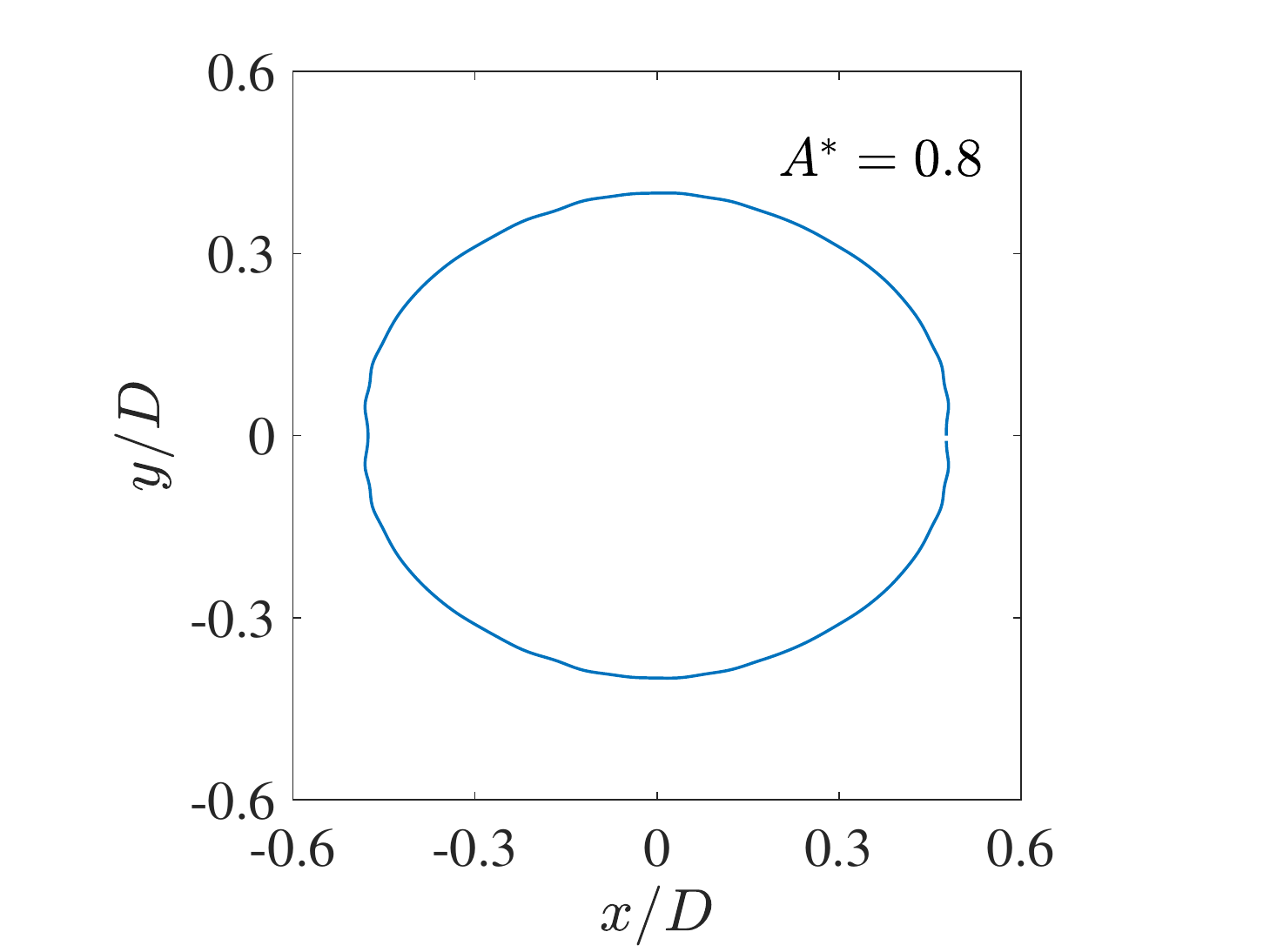}
  \includegraphics[width=1.2in]{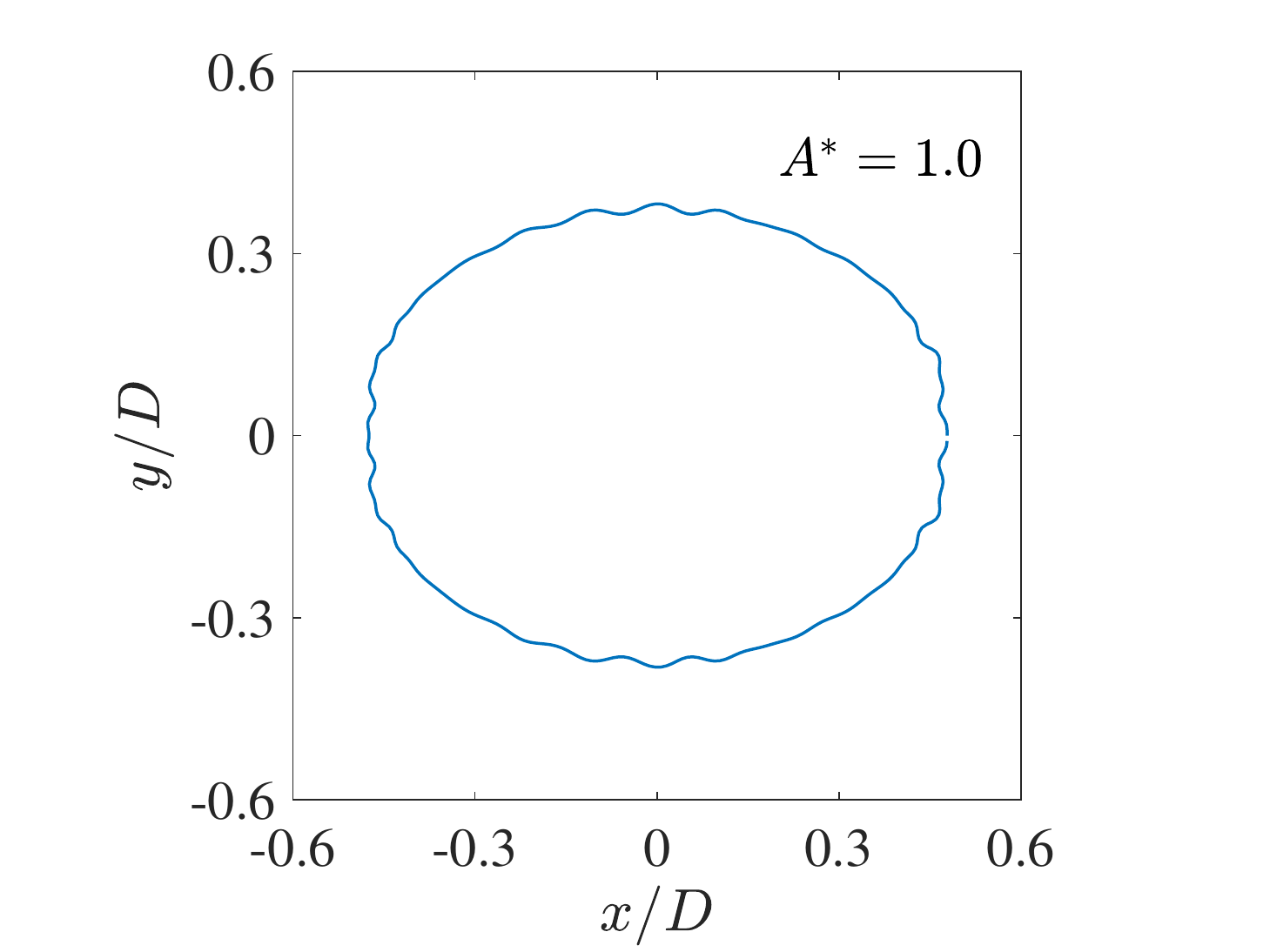}\\
  
  \includegraphics[width=6.0in]{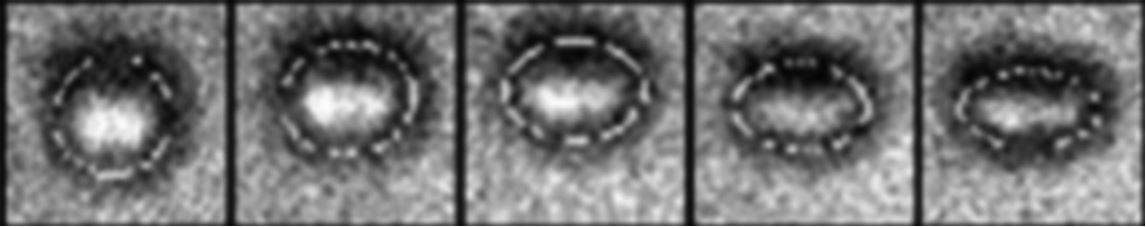}
  \end{center}
\caption{Deformation of a red blood cell induced by acoustic waves: qualitatively comparison of the deformations of the RBCs calculated by present solver (top) and those obtained by Mishra et al.~\cite{Puja2014Deformation} from experiments (bottom). }
\label{Fig:rbc_def}
\end{figure}

In this section, we consider the deformation of a red blood cell (RBC) induced by acoustic waves to validate the current method in handling fluid--structure--acoustics interactions. This problem was experimentally studied by Mishra et al.~\cite{Puja2014Deformation}. In the simulation, a localized pressure perturbation of Gaussian distribution is applied on the fluid domain, which is expressed as

\begin{equation}
P_a=A^*{\rm exp}[-{\rm ln}(2)\frac{(y-1)^2}{0.04}]+A^*{\rm exp}[-{\rm ln}(2)\frac{(y+1)^2}{0.04}],
\end{equation}
where $A^*=A/(\rho_f c^2)$ is the non-dimensional perturbation amplitude, and $c$ is the sound speed of the unperturbed fluid. The structure-to-fluid mass ratio is $m^*=\rho_s/(\rho_f D)=0.2$, where $D$ is the diameter of the cylindrical cell. The RBC is assumed to be elastic, and its non-dimensional stretching and bending rigidity are $K_S^*=K_S/(\rho_f c^2 D)=0.1$ and $E_B^*=E_B/(\rho_f c^2 D^3)=1.0\times10^{-6}$, respectively. The prestress is also applied on the RBC to hold it in a cylindrical shape initially. The pressure perturbation amplitudes ranging from 0.2 to 1.0 are examined. The instantaneous sound pressure contours at three instants for $A^*=0.4$ are shown in Fig.~\ref{Fig:rbc_contour}. Fig.~\ref{Fig:rbc_def} shows a qualitative comparison of the instantaneous deformation of the RBC. The results show that the shape of the RBC changes from a cylinder to an ellipse under the load of the symmetrically distributed pressure perturbations. As shown in Fig.~\ref{Fig:rbc_def}, the deformation increases with the pressure amplitude, which qualitatively agrees with the experimental results. In addition, the wavy deformations of the RBCs are observed for large $A^*$ (e.g. $A^*=1.0$). This phenomenon is the dynamic response of the RBC to the external load, and is consistent with the experimental observation.

\subsection{Acoustic waves scattered by a stationary sphere}
In this section, acoustic waves scattered by a stationary sphere is conducted to validate the present solver in capturing acoustic waves in three-dimensional space. In this case, a rigid sphere with a radius of $R$ is fixed in the fluid with its center at (0, 0). A periodic Gaussian pressure source in the fluid domain is applied
\begin{equation}
A_p=-A{\rm exp}\{-B {\rm log}(2)[(x-2)^2+y^2+z^2]\}{\rm cos}(\omega t),
\end{equation}
where $A=0.01$, $B=16$ and $\omega=2\pi$. All quantities are non-dimensionalized by the radius of the sphere, the ambient fluid density and sound speed. 

\begin{figure}
 \begin{center}
  \hskip-3.0in (a) \hskip3.0in (b)

  \includegraphics[width=3.2in]{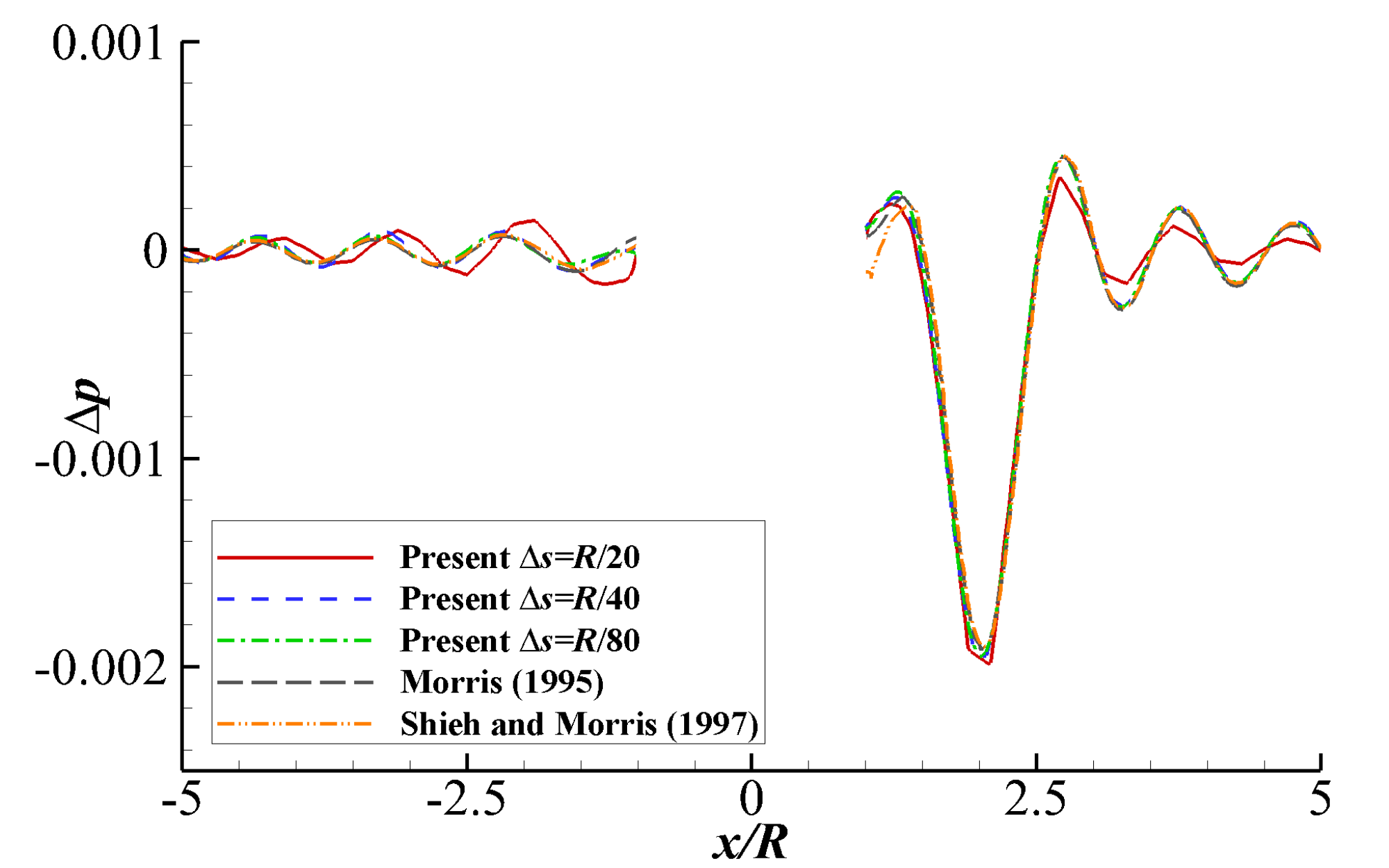}
  \hskip0.1in
  \includegraphics[width=3.2in]{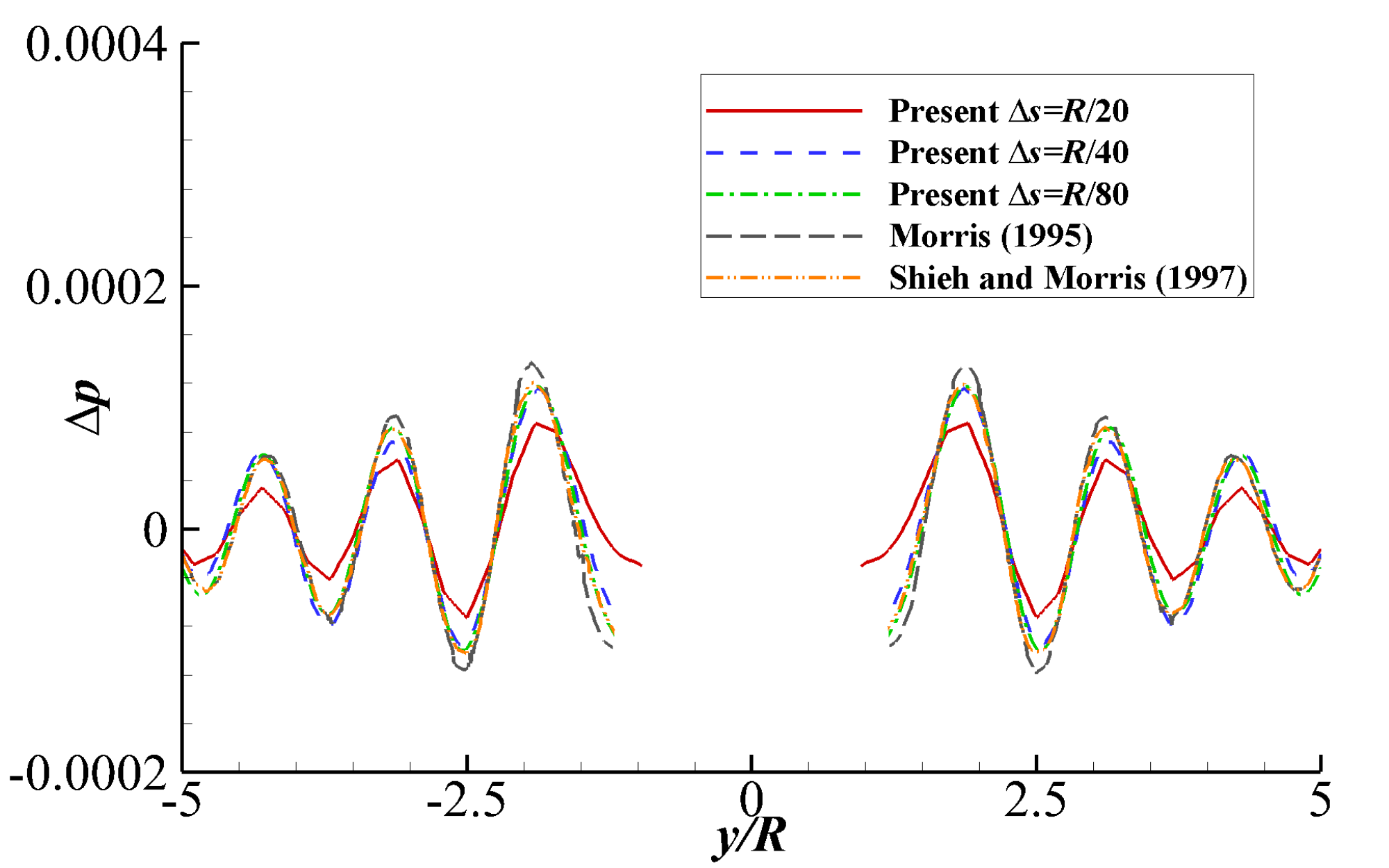}
  \end{center}
\caption{Acoustic waves scattered by a stationary sphere: comparison of the pressure fluctuation along $x$-direction on the line of $y=0, z=0$ (a) and $y$-direction on the line of $x=0, z=0$ with available data from Refs.~\cite{morris1995scattering,tam1997second}.}
\label{Fig:acoustic-sphere-dp}
\end{figure}

\begin{figure}
 \begin{center}

  \includegraphics[width=2.1in]{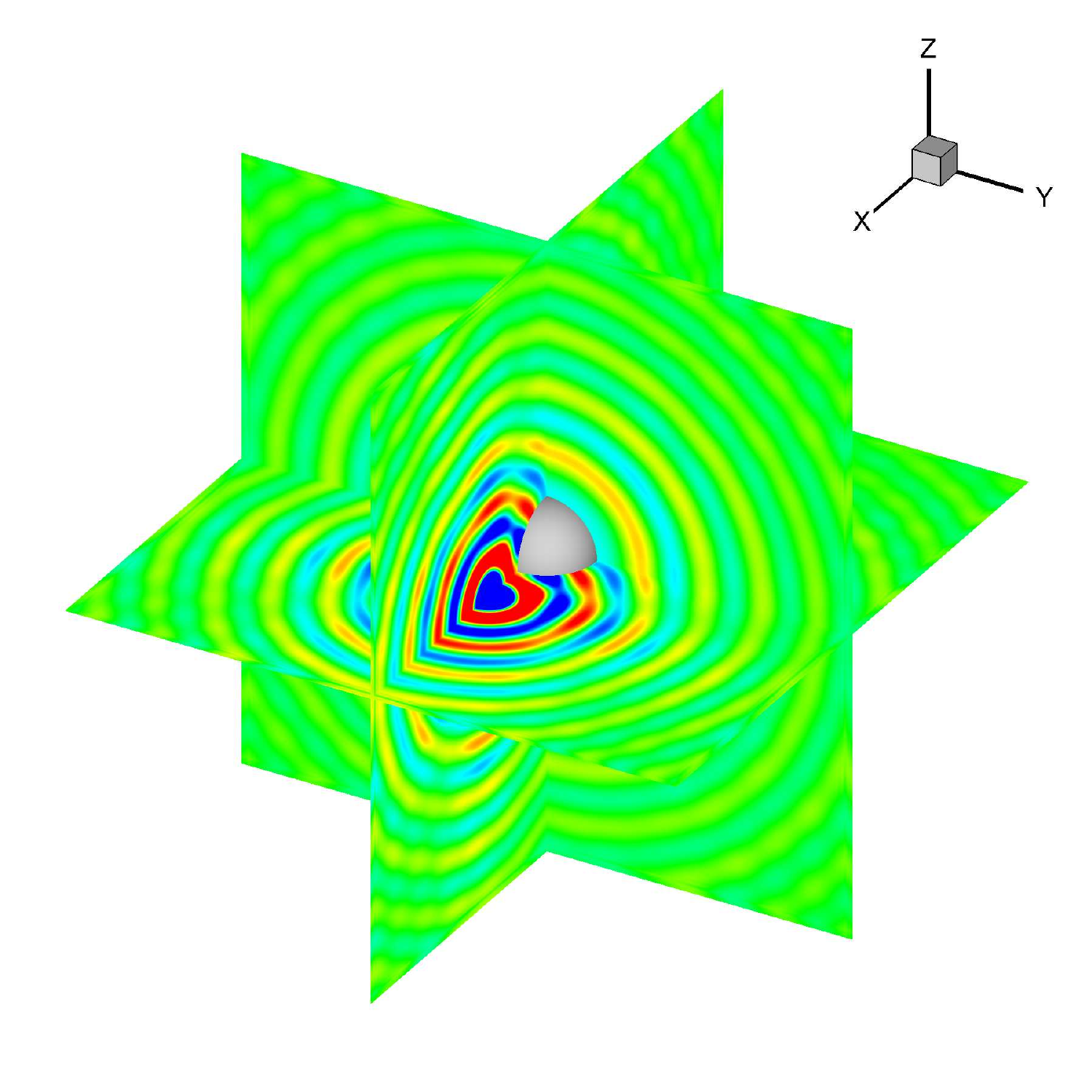}
  \end{center}
\caption{Acoustic waves scattered by a stationary sphere: a snapshot of the fluctuating pressure field at the mesh size of $R/80$. The contour level ranges from $-2\times10^{-4}$ (blue) to $2\times10^{-4}$ (red) with an interval of $4\times10^{-6}$.}
\label{Fig:acoustic-sphere-contour}
\end{figure}

Three mesh regions are used to improve the computational efficiency and conserve the accuracy as used in Section 3.2. Three mesh spacings of $R/20$, $R/40$ and $R/80$ within a domain of $2R\times2R\times2R$ covering the sphere are used. The far field mesh spacing is $R/10$. The sphere is discretized by triangular elements with a mesh spacing approximately equal to that of the fluid domain. The fluctuating pressure along $x$-direction on the line of $y=0, z=0$ and along $y$-direction on the line of $x=0, z=0$ are presented in Fig.~\ref{Fig:acoustic-sphere-dp} with data from Refs.~\cite{morris1995scattering,tam1997second}. As shown in this figure, the fluctuating pressure from the present simulation approaches to the results from the references when the mesh is refined. Good agreement is achieved when the finest mesh spacing of $R/80$ is used, showing that the present numerical method is able to predict acoustic waves in three-dimensional space. Fig.~\ref{Fig:acoustic-sphere-contour} shows three views of the fluctuating pressure field at the mesh spacing of $R/80$ when the periodic solution is obtained.

\begin{figure}
 \begin{center}
  \includegraphics[width=3.2in]{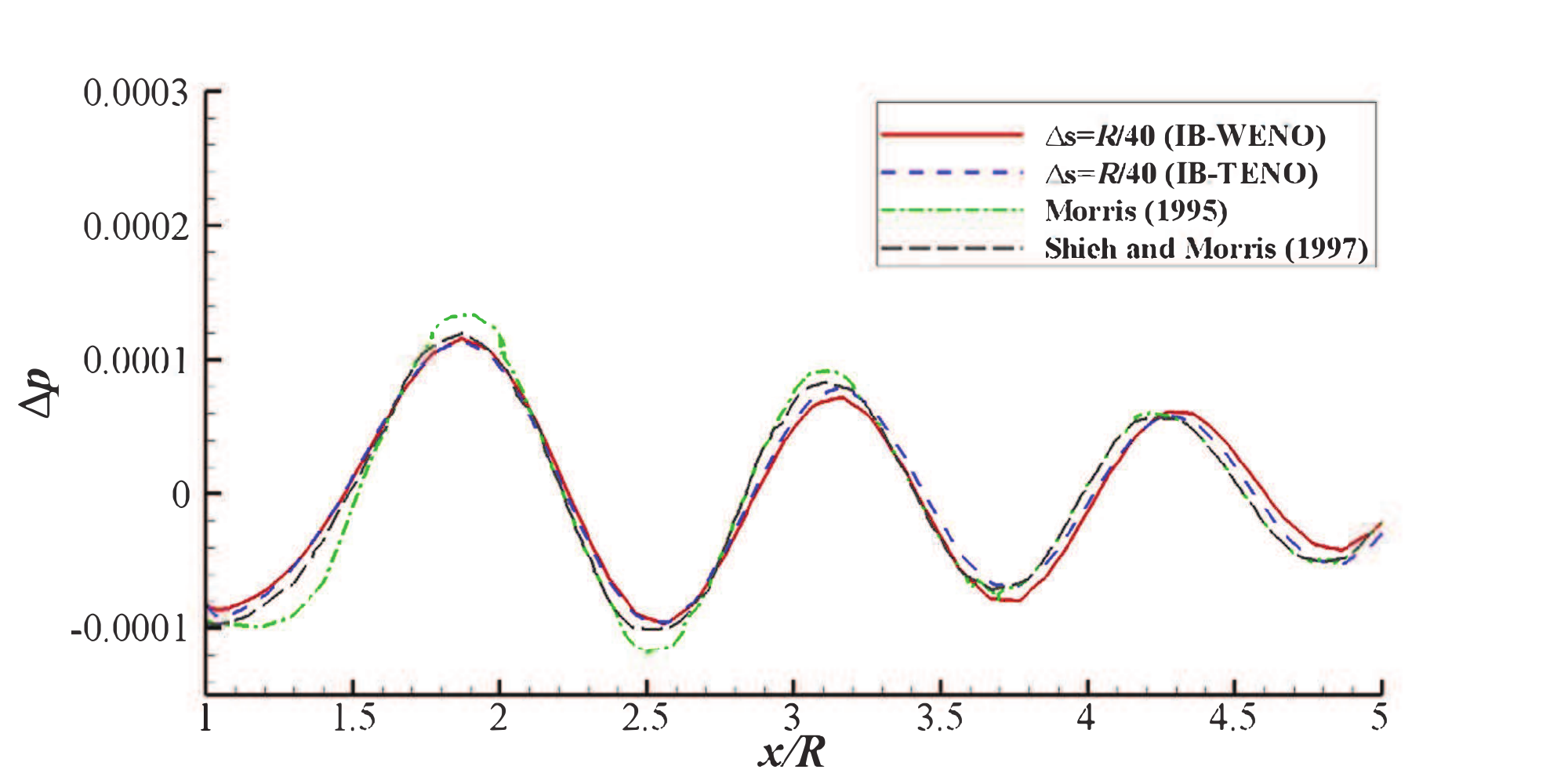}
  \end{center}
\caption{Acoustic waves scattered by a stationary sphere: comparison of the pressure fluctuation along $x$-direction on the line of $y=0, z=0$ calculated by WENO and TENO.}
\label{Fig:acoustic-sphere-dp_TENO}
\end{figure}

To compare the performance of the IB--WENO and the IB--TENO, Fig.~\ref{Fig:acoustic-sphere-dp_TENO} shows the pressure fluctuation along $x$-direction. It is found that the numerical dissipation of the IB-TENO is lower compared to the IB-WENO, as demonstrated by the peak values near $x/R=3.1$ and 4.7.

\section{Sound generation by flapping foils in a uniform flow}
Having conducted rigorous validations, the sound generated by flapping foils in forward flight and flapping foil energy harvester is numerically studied in the two-dimensional domain by considering the geometrical shape and the flexibility of the foil. Both the force and sound generations are compared and analyzed.
\subsection{Flapping foils in forward flight}
\subsubsection{Physical problem}
\begin{figure}
  \begin{center}
  \includegraphics[width=3.5in]{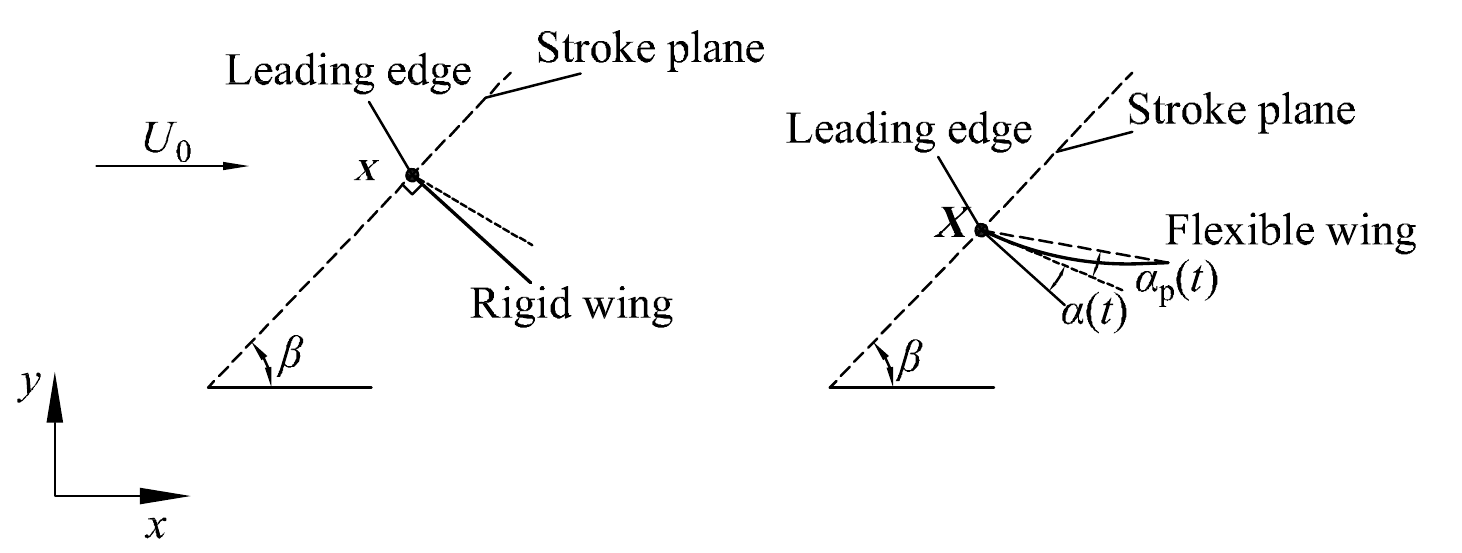}
  \end{center}
\caption{Schematic of a rigid (left) and flexible (right) foil flapping in a uniform flow.}
\label{Fig:rflapsch}
\end{figure}

In this section, the acoustic perturbations induced by flapping foils in forward flight are considered, as shown in Fig.~\ref{Fig:rflapsch}~\cite{tian2013force}. Here, the foil is clamped at the leading edge, and the clamping device undergoes a combined translational and rotational motions, given by~\cite{tian2013force}
\begin{equation}
\boldsymbol{X}_0(t)=\frac{A_0}{2} {\rm cos}(2 \pi f t) [{\rm cos}\beta, {\rm sin}\beta],\quad \alpha(t)=\frac{\alpha_m}{2}{\rm sin}(2 \pi f t+ \phi).
\label{eq:flapping_motion}
\end{equation}
where $A_0$ is the translational amplitude, $f$ is the flapping frequency, $\alpha_m$ is the rotation amplitude, and $\beta$ is the angle between the stroke plane and the horizontal plane. The phase difference ($\phi$) is 0 unless it is mentioned.

The non-dimensional parameters including the flapping amplitude, Reynolds number, inlet velocity, Mach number, mass ratio and frequency ratio that control this problem are given by
\begin{equation}
\frac{A_0}{L},\quad {\rm Re}=\frac{\rho_f UL}{\mu},\quad U_r=\frac{U_0}{U},\quad M=\frac{\pi U}{2c},\quad m^*=\frac{\rho_s}{\rho_f L},\quad \omega^*=\frac{2\pi f}{\omega_n},
\label{eq:flapping_para}
\end{equation}
where $L$ is the chord length, $U=2 f A_0$ is the average translational velocity of the leading edge, $f$ is the flapping frequency of the foil, $c$ is the sound speed of the fluid, $\rho_f$ is the fluid density, $\rho_s$ is the linear density of the foil, and $\omega_n=k_n^2/L^2\sqrt{E_B/\rho_s}$ with $k_n=1.8751$ (the frequency of the first natural vibration mode of the wing with fixed leading edge~\cite{landau1986theory,tian2013force}) and $E_B$ being the bending rigidity. Here, Re=100, $m*=5.0$ and $U_r=0.4$. The thrust, lift and power coefficients are defined as
\begin{equation}
C_T=\frac{2F_T}{\rho_f U^2 L},\quad C_L=\frac{2F_L}{\rho_f U^2 L},\quad C_p=\frac{-2 \int_0^L \boldsymbol{f}\cdot \boldsymbol{v} dl}{\rho_f U^3 L},
\label{eq:ctcl}
\end{equation}
where $F_T$ and $F_L$ are respectively the thrust and lift force acting on the foil by the ambient fluid, $\boldsymbol{f}$ is the hydrodynamic traction on the foil, and $\boldsymbol{v}$ is the velocity of the foil. The fluctuating pressure $\Delta \tilde{p}$ is defined in the same way in Section 3.2. All the pressures presented afterwards are scaled by the fluid density $\rho_f$ and sound speed $c$.

\subsubsection{Mesh convergence study}
\begin{figure}
 \begin{center}
  \hskip-3.0in (a)

  \includegraphics[width=5.0in]{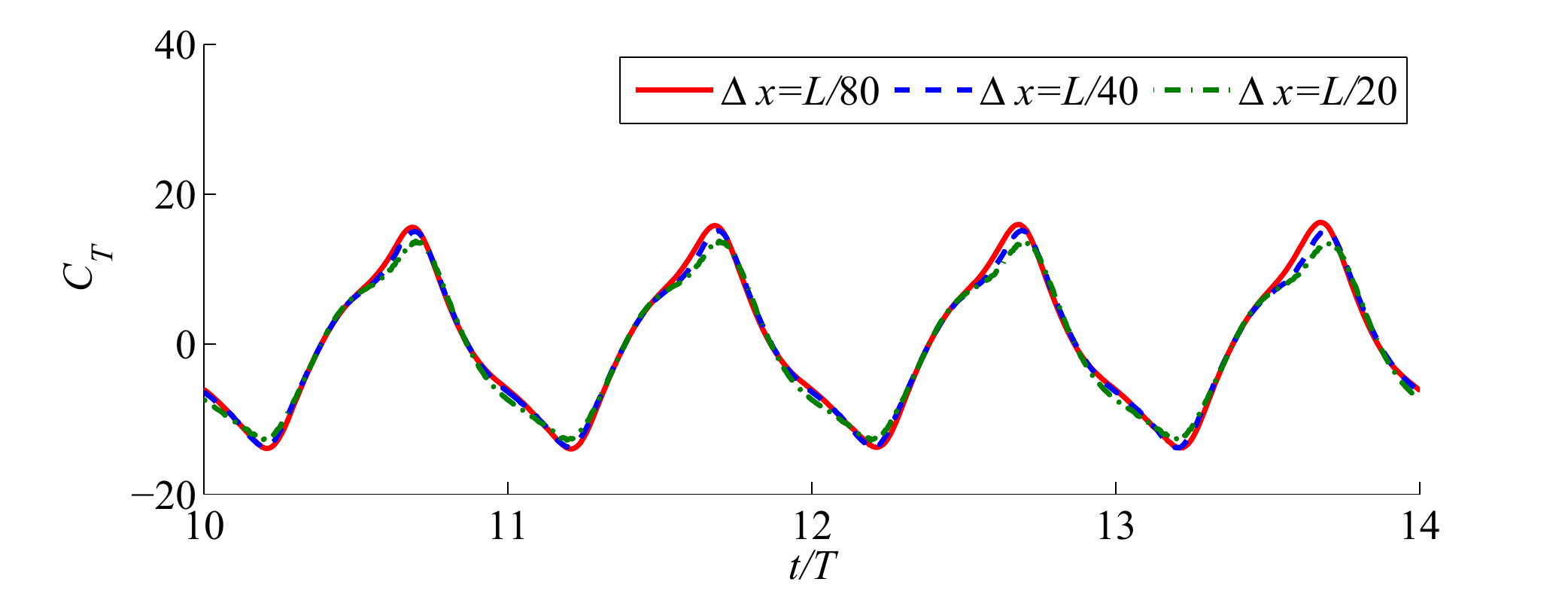}
  
  \hskip-3.0in (b)

  \includegraphics[width=5.0in]{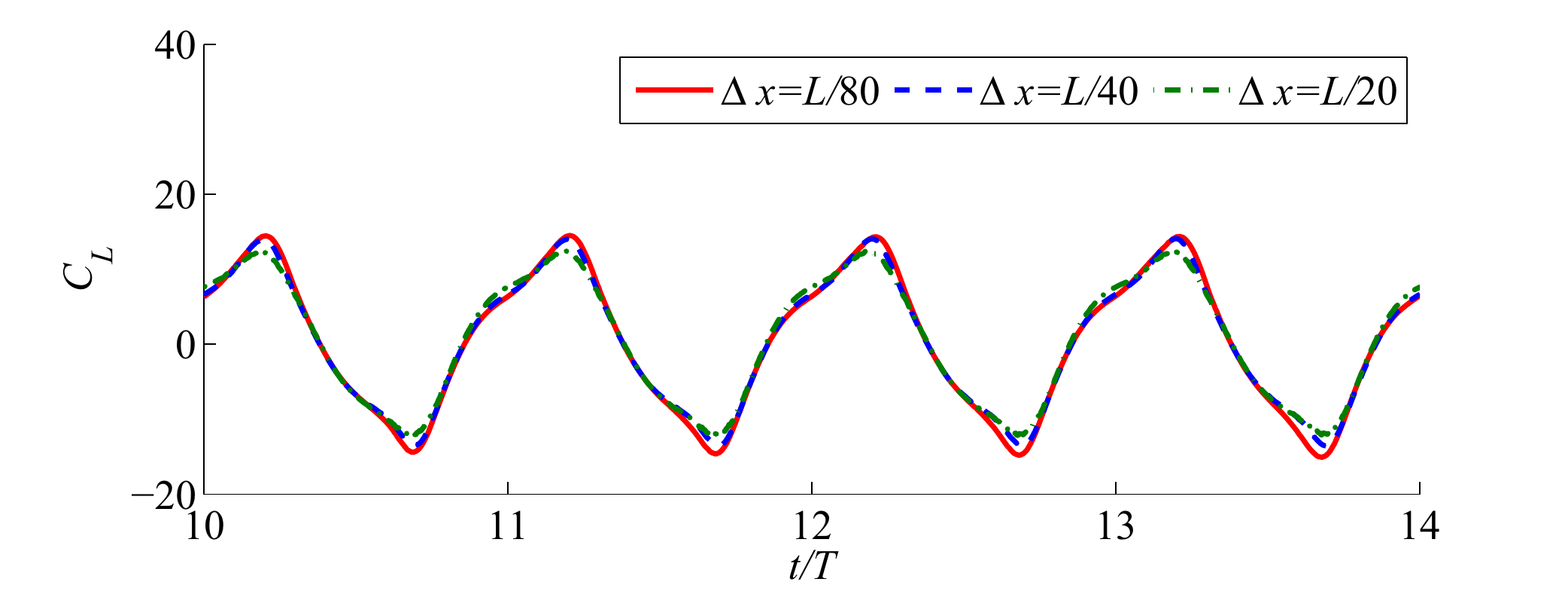}
  
  \hskip-3.0in (c)

  \includegraphics[width=5.0in]{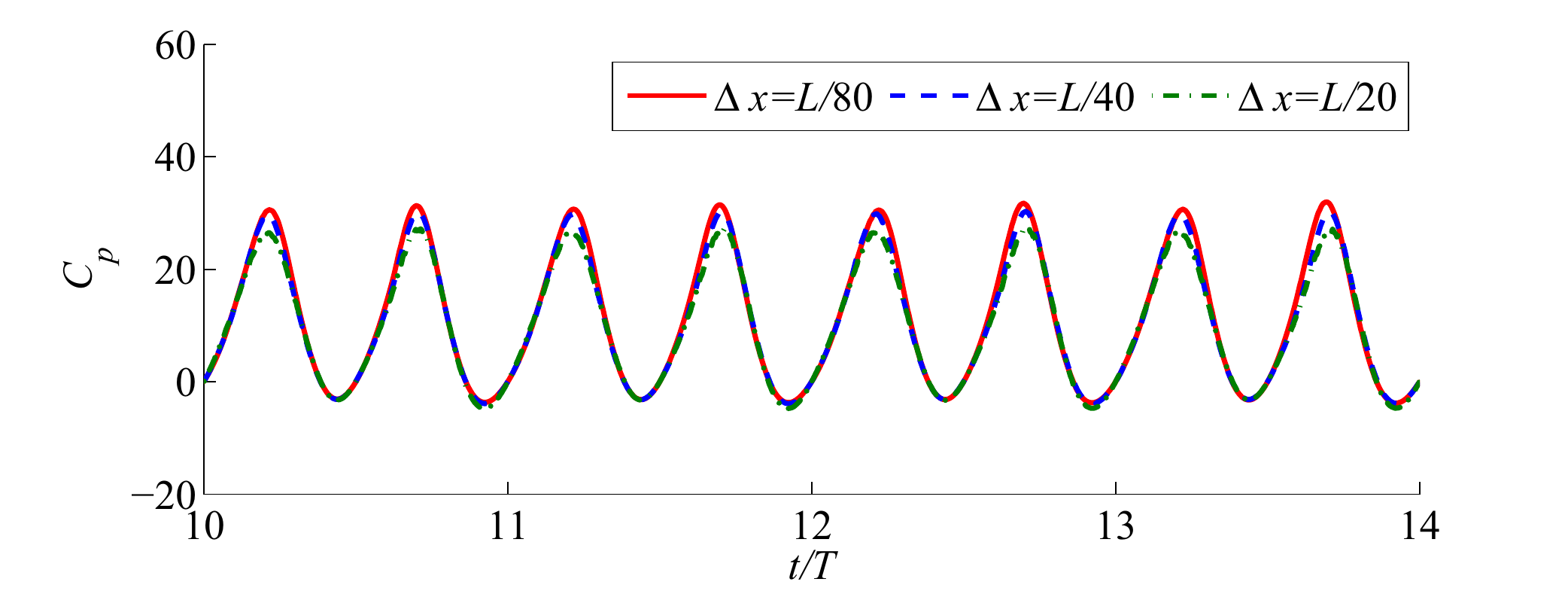}
  \end{center}
\caption{Flapping foils in forward flight: time histories of $C_T$ (a), $C_L$ (b) and $Cp$ (c) for Re=100, $M=0.1$, $U_r=0.4$, $A_0/L=1.25$ and $\alpha_m=0$.}
\label{Fig:ctlp-meshcong}
\end{figure}

First, mesh convergence is conducted to guarantee the reliability of the solver in modeling the acoustic perturbations induced by flapping foils. Non-uniform mesh scheme is adopted to improve the computational efficiency, where a uniform refined mesh region around the foil is used to enhance the accuracy. In the simulation, the computational domain extends from ($-36.25L$, $-36.25L$) to ($36.25L$, $36.25L$). Three cases are considered for the mesh convergence study: $\Delta x=L/80$, $L/40$ and $L/20$, where $\Delta x$ is the finest mesh spacing around the foil. Other non-dimensional parameters for these three cases are: $M=0.1$, $A_0/L=1.25$, $\beta=45^o$ and $\alpha_m=0^o$. In Fig.~\ref{Fig:ctlp-meshcong}, the time histories of  thrust, lift and power coefficients are presented. The results show that reliable results can be achieved with $\Delta x=L/40$.

\begin{figure}
  \begin{center}
  \includegraphics[width=5.0in]{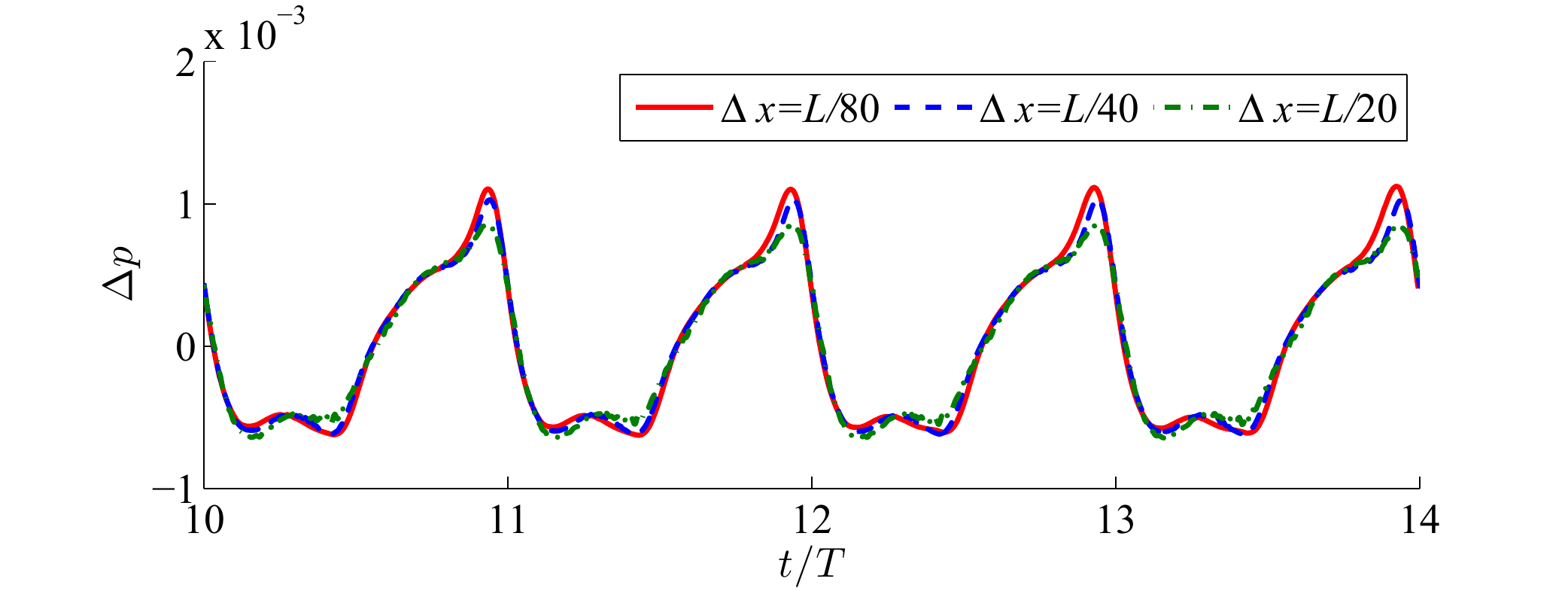}
  \end{center}
\caption{Flapping foils in forward flight: time histories of the fluctuating pressure measured at $r=10$, $\theta=90^o$ for Re=100, $M=0.1$, $U_r=0.4$, $A_0/L=1.25$ and $\alpha_m=0$.}
\label{Fig:dpmeshcong}
\end{figure}

Similar to Section 3.2, we define $r$ as the distance from the origin ( nondimensionalized by $L$) and $\theta$ as the circumferential angle. The histories of the fluctuating pressure measured at $r=10, \theta=90^o$ are presented in Fig.~\ref{Fig:dpmeshcong}, where it is found that the effects of the mesh difference are negligible for mesh size of $L/80$ and $L/40$. Therefore, $\Delta x=L/40$ is adopted in the later simulations. This figure also shows that the positive pressure wave peaks are significantly larger than the negative ones, which will be discussed in detail later.

\subsubsection{Propagation and decay of pressure waves}
\begin{figure}
 \begin{center}
  \hskip-3.0in (a) \hskip3.0in (b)

  \includegraphics[width=3.0in]{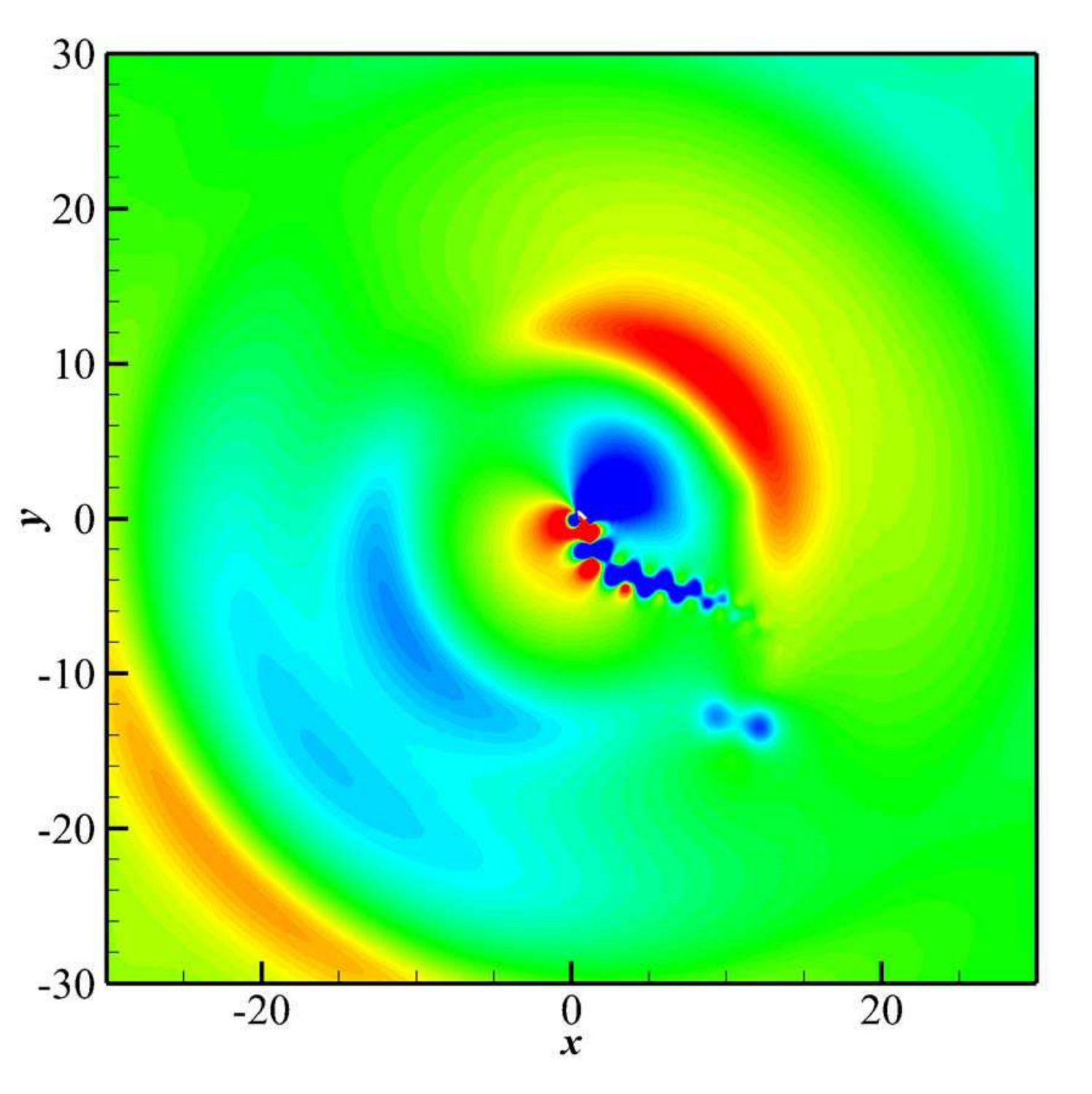}
  \hskip0.1in
  \includegraphics[width=3.0in]{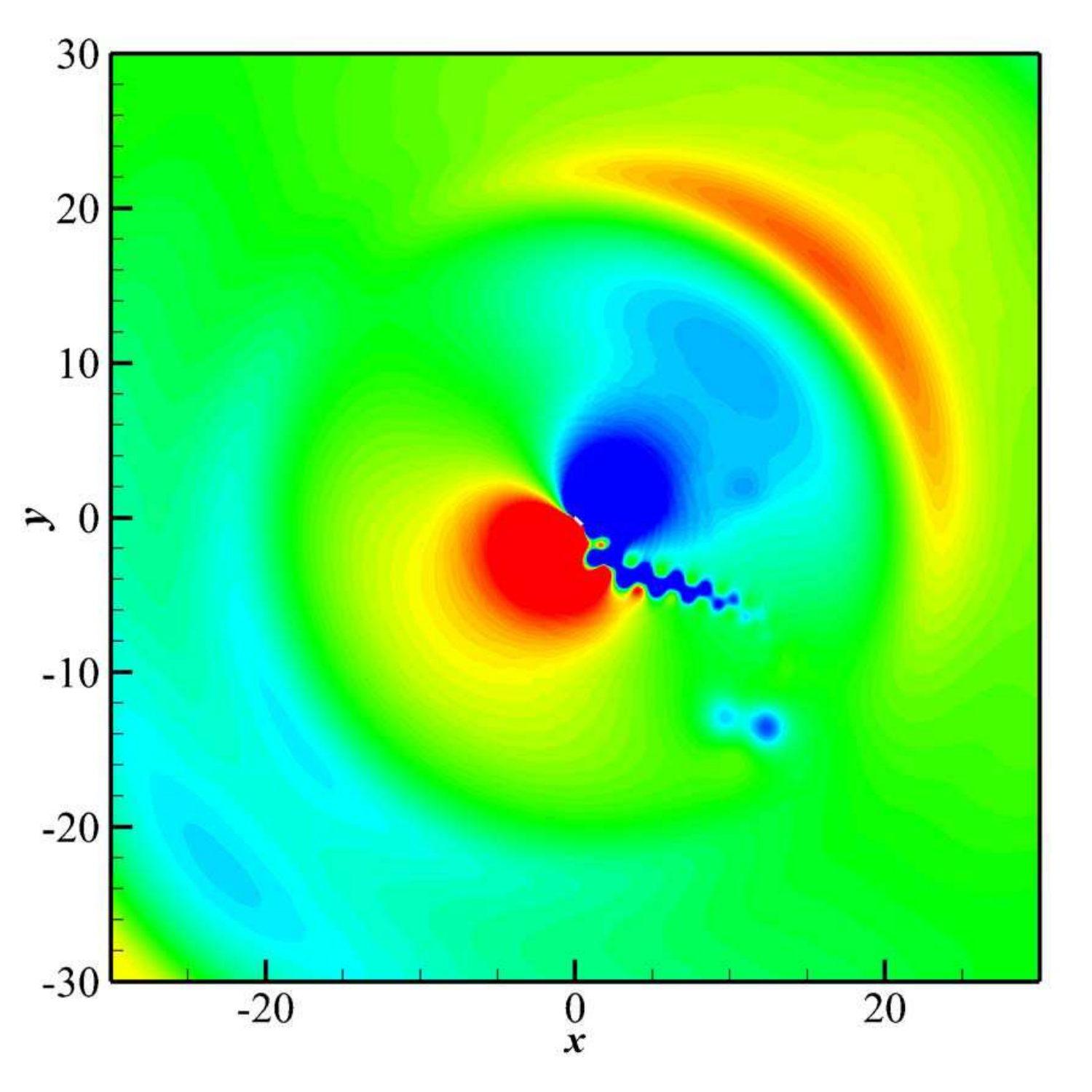}\\
  
  \hskip-3.0in (c) \hskip3.0in (d)

  \includegraphics[width=3.0in]{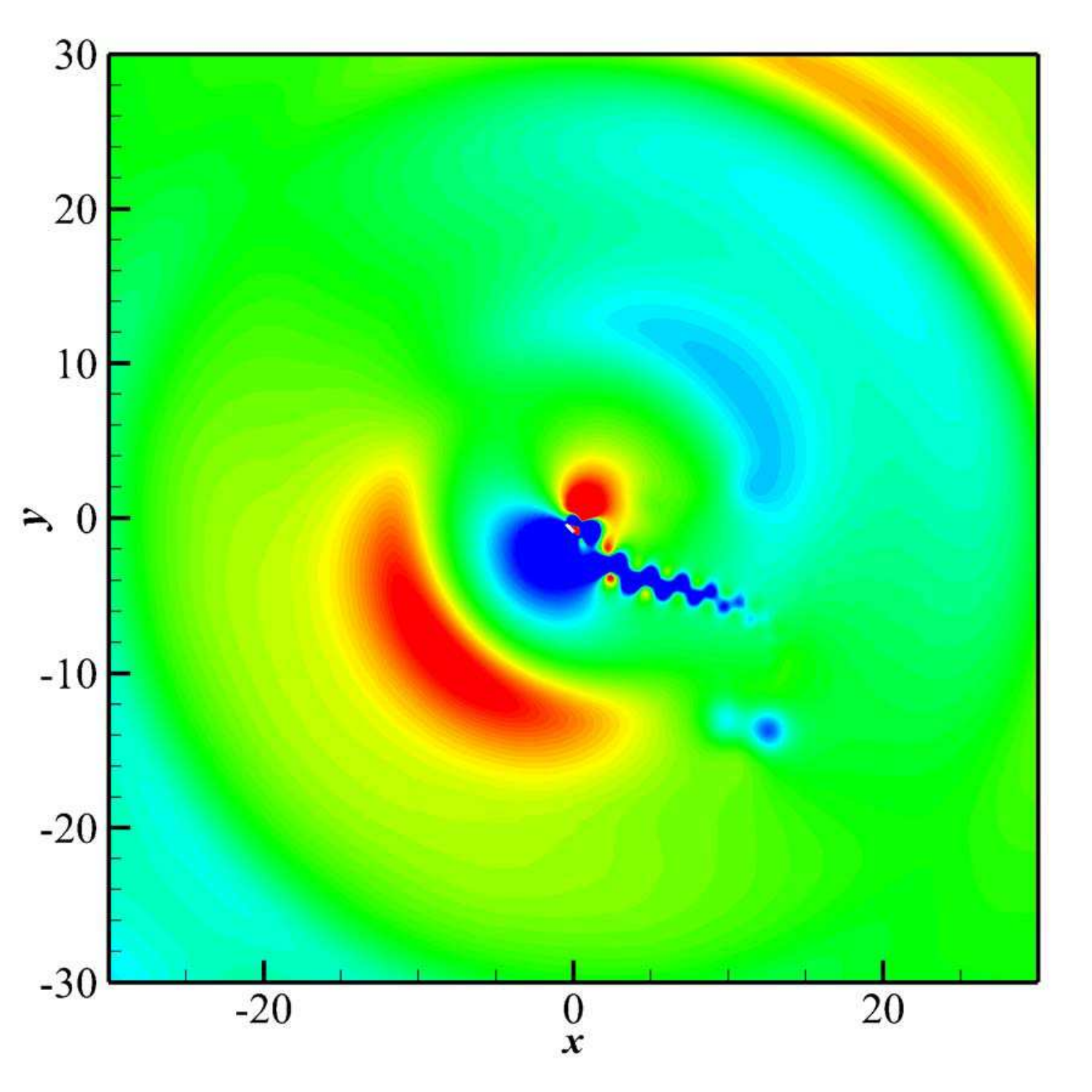}
  \hskip0.1in
  \includegraphics[width=3.0in]{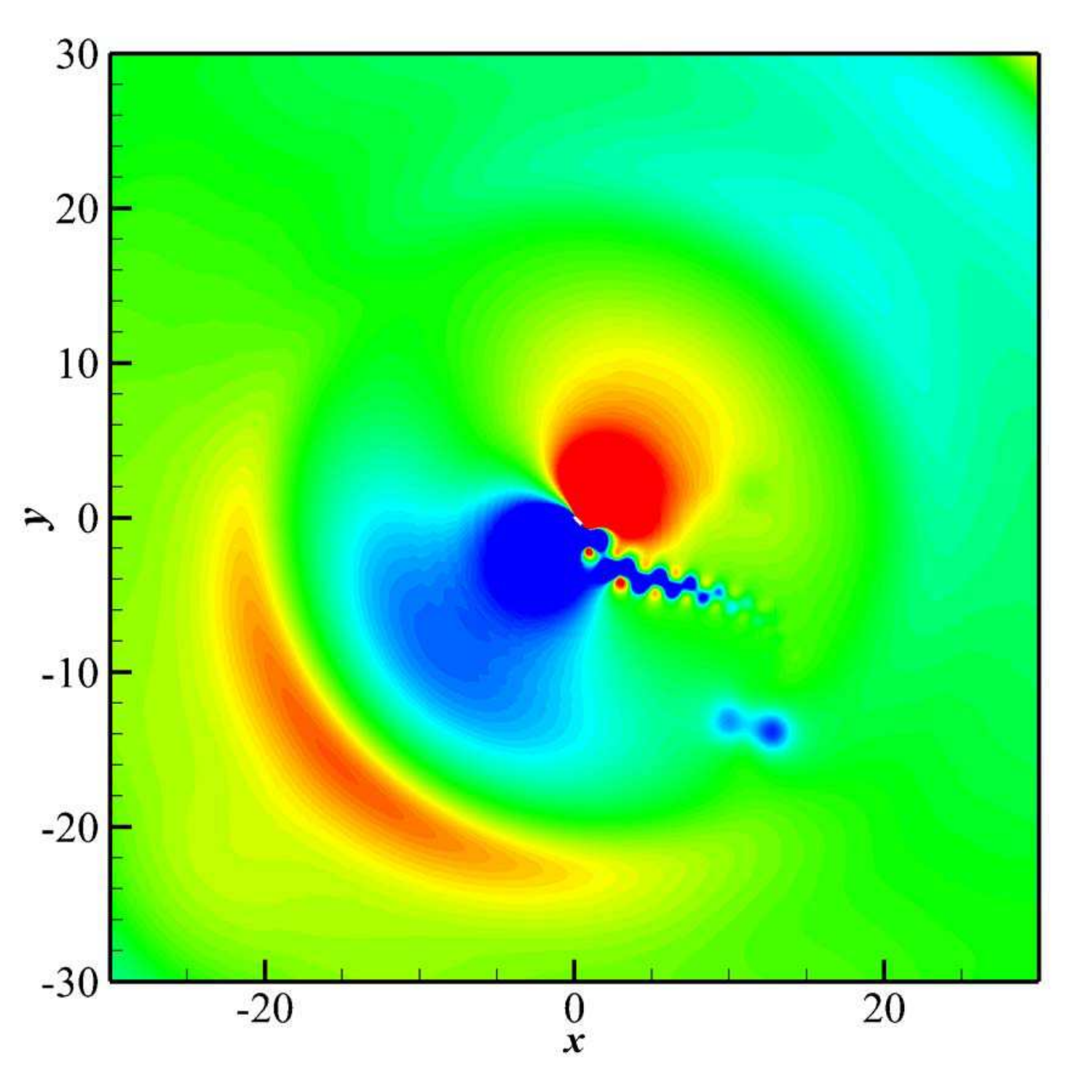}\\
  \end{center}
\caption{Flapping foils in forward flight: instantaneous contours of the fluctuating pressure $\Delta p$ at Re=100, $M=0.1$, $U_r=0.4$, $A_0/L=1.25$ and $\alpha_m=0$ at (a) $t = 0T/4$, (b) $t = T/4$, (c) $t = 2T/4$ and (d) $t = 3T/4$ with the finest mesh size of $L/40$. The contour level ranges from $-1.0\times10^{-3}\rho_f c^2$ (blue) to $1.0\times10^{-3}\rho_f c^2$ (red) with the interval of $4.0\times10^{-5}\rho_f c^2$.}
\label{Fig:rflappre}
\end{figure}

In Fig.~\ref{Fig:rflappre}, the instantaneous contours of the fluctuating pressure in a flapping period are presented. It is observed that during downstroke, negative pressure pulses are generated from the upper side of the foil, and positive pressure pulses are generated from the lower side of the foil, and vice versa. During a foil stroke, positive pressure is formed on the loading face due to fluid being displaced and negative pressure is formed on the opposite face mainly due to the leading-edge vortices. The positive and negative pressure exchange sides during the switch between the downstroke and the upstroke~\cite{geng2017effect}. A clear illustration of the foil-loading mechanism can be found in the work by Inada et al.~\cite{inada2009numerical}.

\begin{figure}
  \begin{center}
  \hskip-3.0in (a) \hskip3.0in (b)

  \includegraphics[width=3.25in]{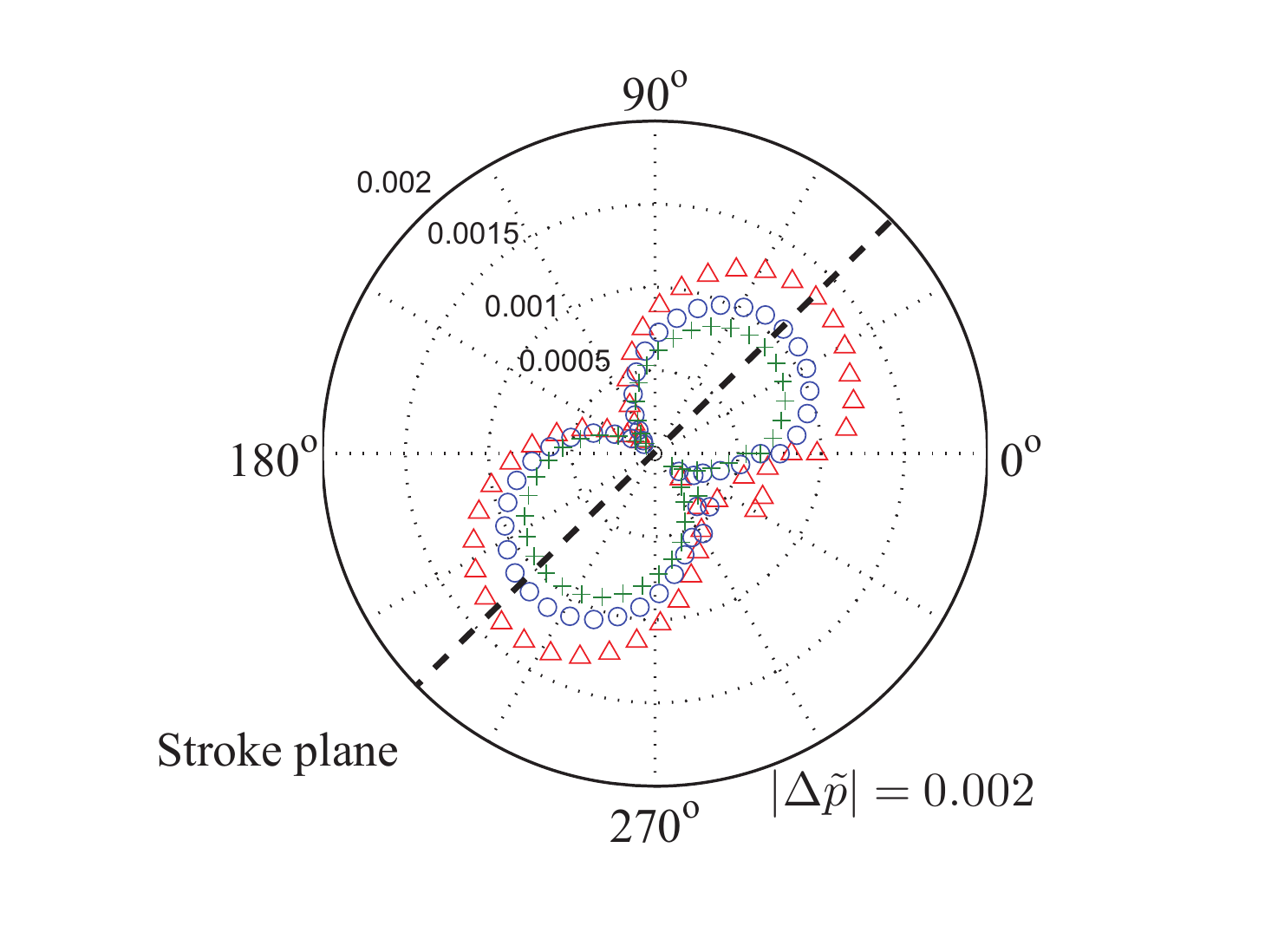}
  \hskip0.1in
  \includegraphics[width=3.25in]{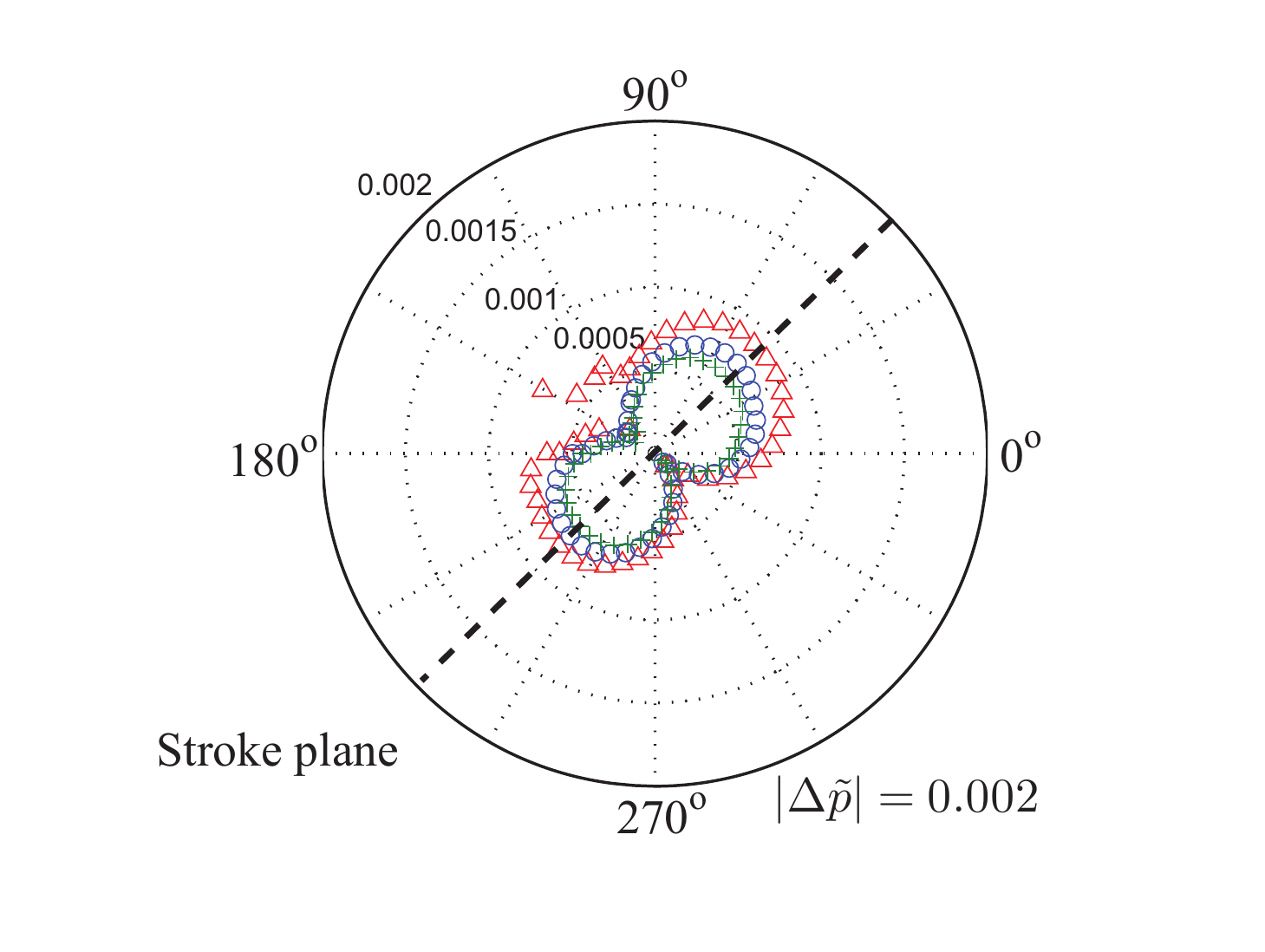}
  \end{center}
\caption{Flapping foils in forward flight: polar diagram of the fluctuating pressure peaks for Re=100, $M=0.1$, $U_r=0.4$, $A_0/L=1.25$ and $\alpha_m=0$. Positive (a) and negative (b) fluctuating pressure peaks at a distance of $10L$ ($\Delta$), $14L$ (o) and $18L$ ($+$). The dashed line indicates the direction of the stroke plane.}
\label{Fig:dpcirmeshcong}
\end{figure}

\begin{figure}
\begin{center}
  \hskip-1.8in (a) \hskip1.8in (b) \hskip1.8in (c)

  \includegraphics[width=2.0in]{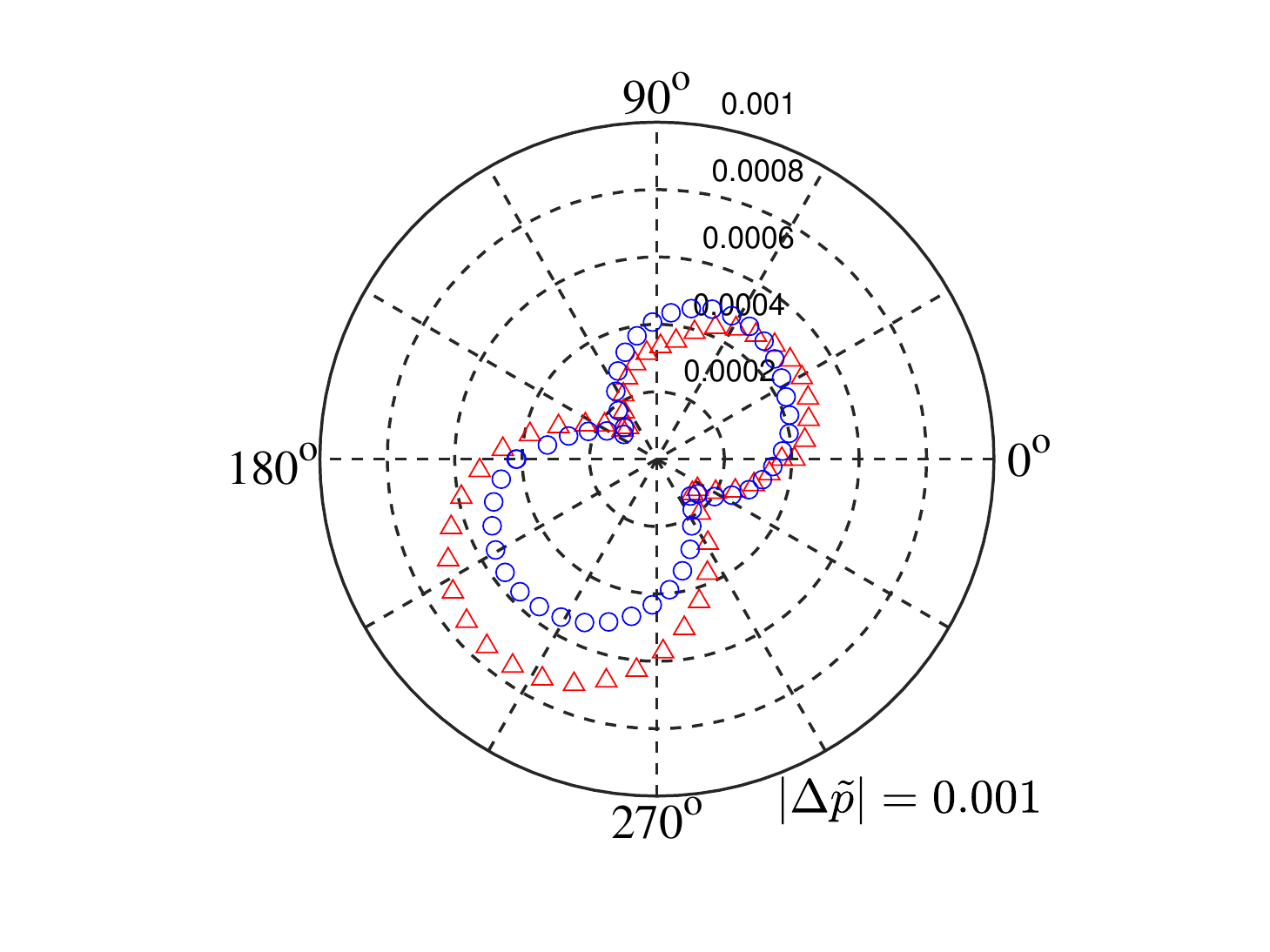}
  \hskip0.1in
  \includegraphics[width=2.0in]{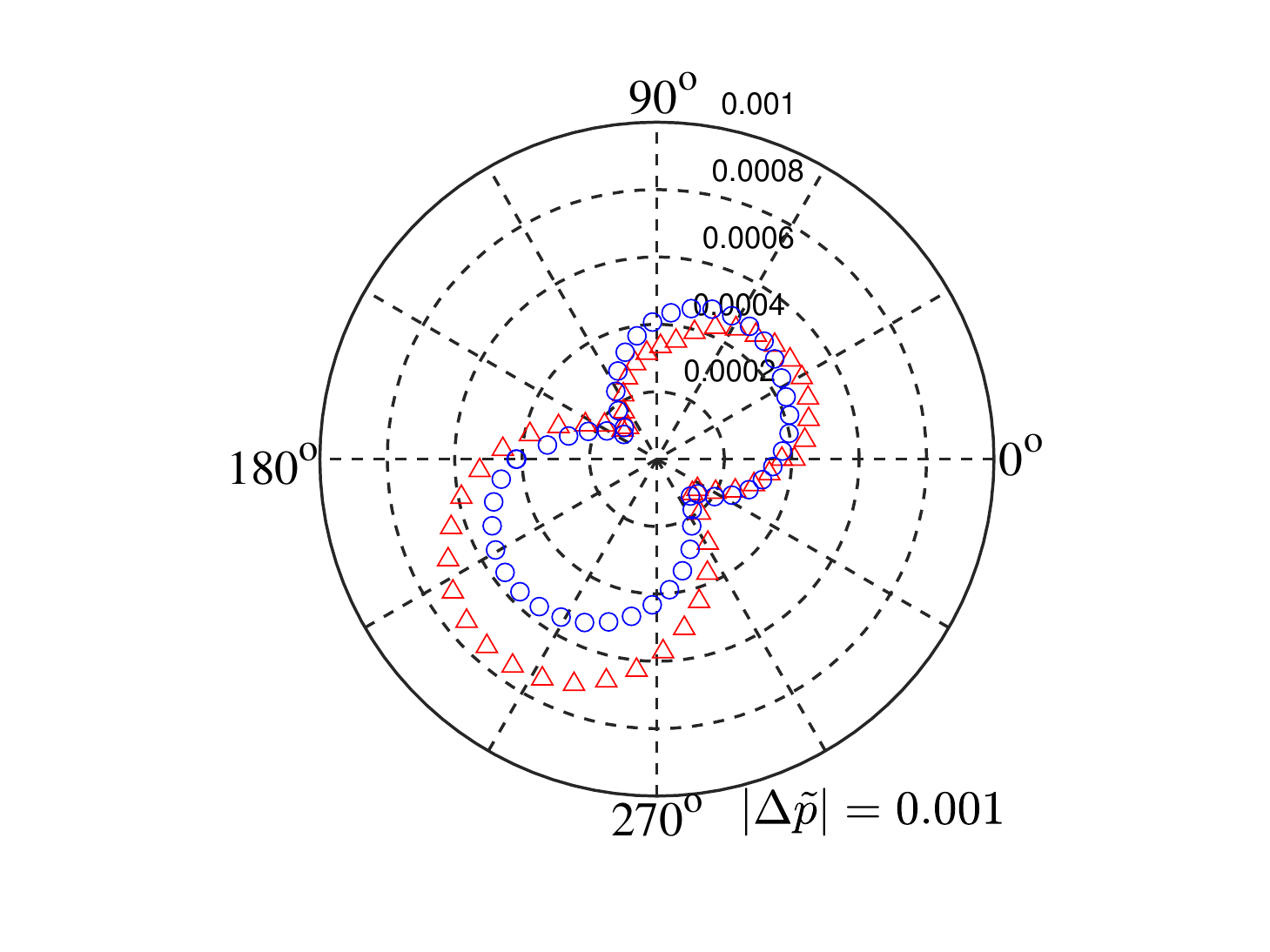}
  \hskip0.1in
  \includegraphics[width=2.0in]{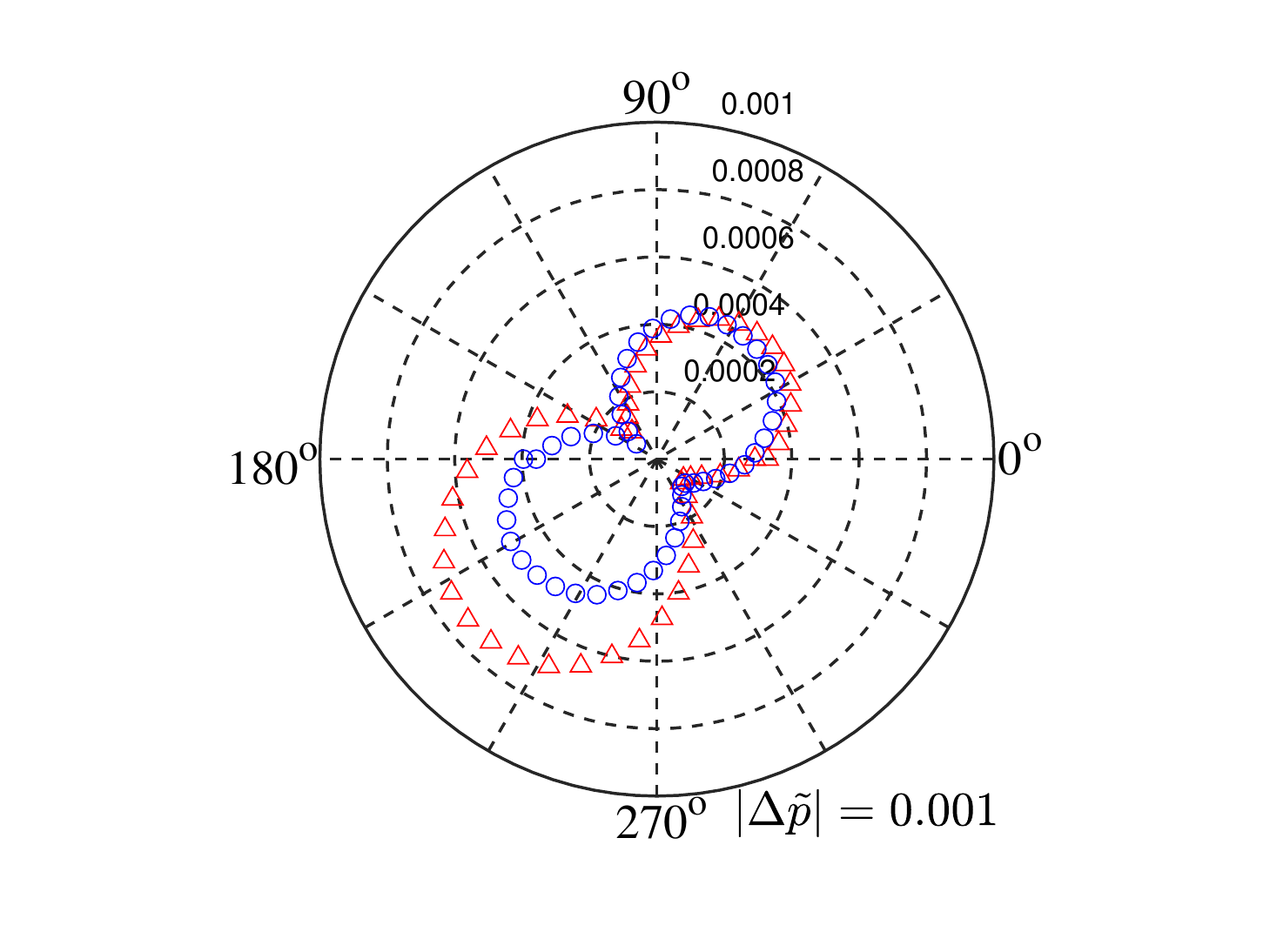}\\
  \end{center}
\caption{Flapping foils in forward flight: polar diagram of the positive ($\Delta$) and negative ($o$) fluctuating pressure peaks generated by a flapping wing under translational and rotational motions at a distance of $18L$: (a) $\phi=0^o$, (b) $\phi=45^o$ and (c) $\phi=90^o$.}
\label{Fig:posneg_peak_phase}
\end{figure}

The circumferential distribution of fluctuating pressure peaks are presented in Fig.~\ref{Fig:dpcirmeshcong}. It shows that the pressure peaks decay with the distance. In addition, the positive peaks are much larger than the negative ones, because as mentioned above, the positive and negative pressure fluctuations are generated by different mechanisms. Fig.~\ref{Fig:dpcirmeshcong} shows that the pressure peaks decay with the increasing distance, and the positive peaks are much larger than the negative peaks. In addition, Fig.~\ref{Fig:dpcirmeshcong} shows that the pressure peaks distribution is symmetrical about the stroke plane, indicated by the dashed line in the figure. It is also noted that some markers plotted in Fig.~\ref{Fig:dpcirmeshcong} are not smoothly distributed along the circumference. A reasonable explanation is that the oscillation is introduced by the vortices. When the vortex approaches to the probes, the recorded pressure will be perturbed. As the distance increases, the fluctuating pressure along the circumference tend to be smoother, as shown in Fig.~\ref{Fig:dpcirmeshcong}. This is due to the fact that the vortex effects reduce with the distance. The fluctuating pressure peaks generated by the flapping wing on the windward side ($180^o-270^o$) are slightly larger than those on the leeward side ($0^o-90^o$), as shown in Fig.~\ref{Fig:dpcirmeshcong}, due to the presence of the free stream. In order to figure out the influence of the phase difference ($\phi$) between the translational and rotational motions on the fluctuating pressure, combined translational and rotational motions of the flapping wing at $\phi = 0^o, 45^o$ and $90^o$ with $\alpha_m=\pi/4$ are simulated. Polar diagrams of the fluctuating pressure peaks are presented in Fig.~\ref{Fig:posneg_peak_phase}. It is found that the fluctuating pressure peaks on the windward side decrease slightly with the phase difference. Similar to the case without rotational motion, the positive fluctuating pressure peaks on the windward side are larger than the negative ones, due to the presence of the free stream. On the leeward side, the difference between the positive and negative fluctuating pressure peaks are not significant. However, the comparison between the flapping with and without translational motion, as shown in Figs.~\ref{Fig:dpcirmeshcong} and \ref{Fig:posneg_peak_phase}, shows that the combined translational and rotational motions can reduce the fluctuating pressure peaks, especially those on the leeward side.

\begin{figure}
 \begin{center}
  \hskip-3.0in (a) \hskip3.0in (b)

  \includegraphics[width=3.2in]{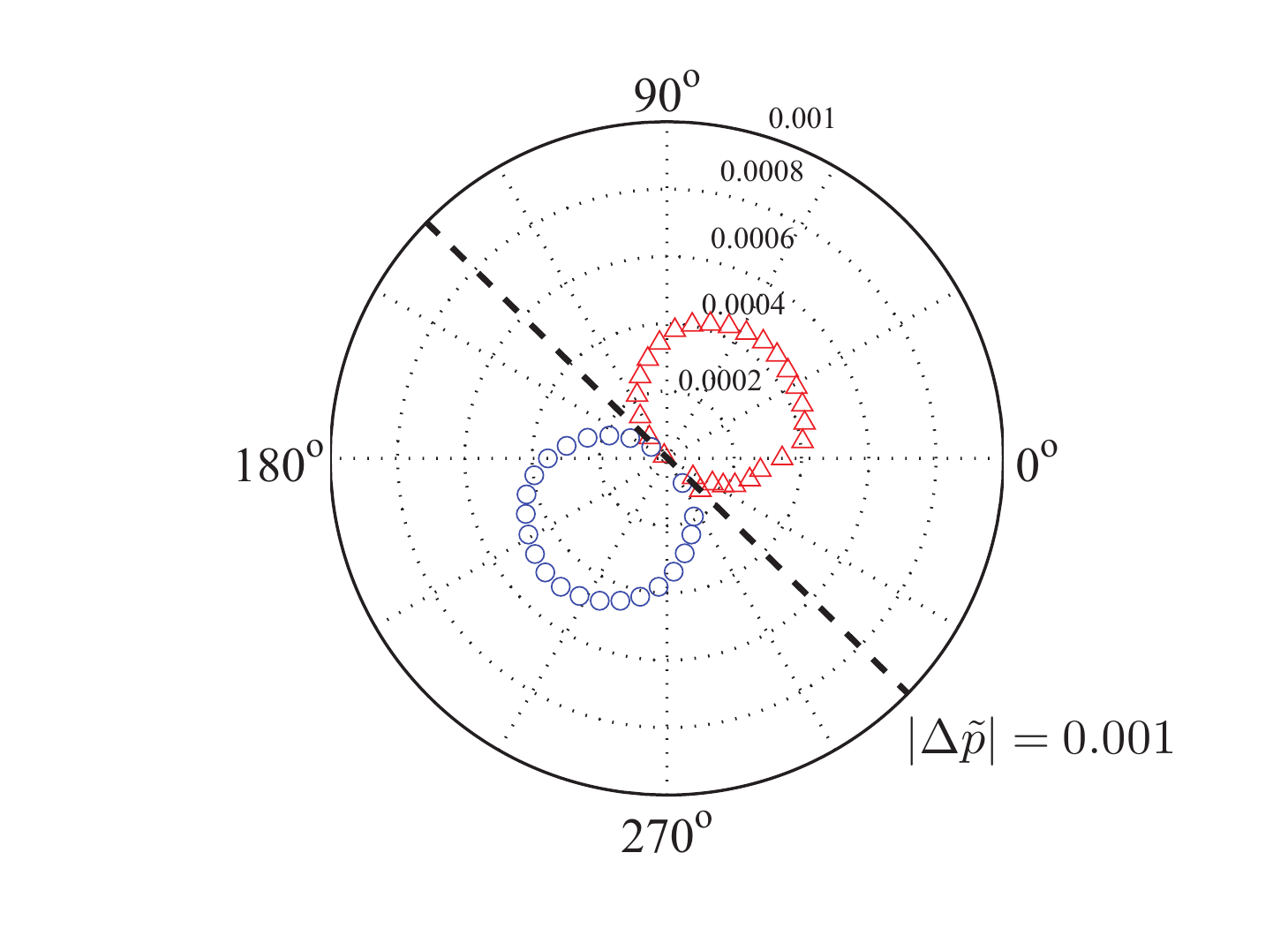}
  \hskip0.1in
  \includegraphics[width=3.2in]{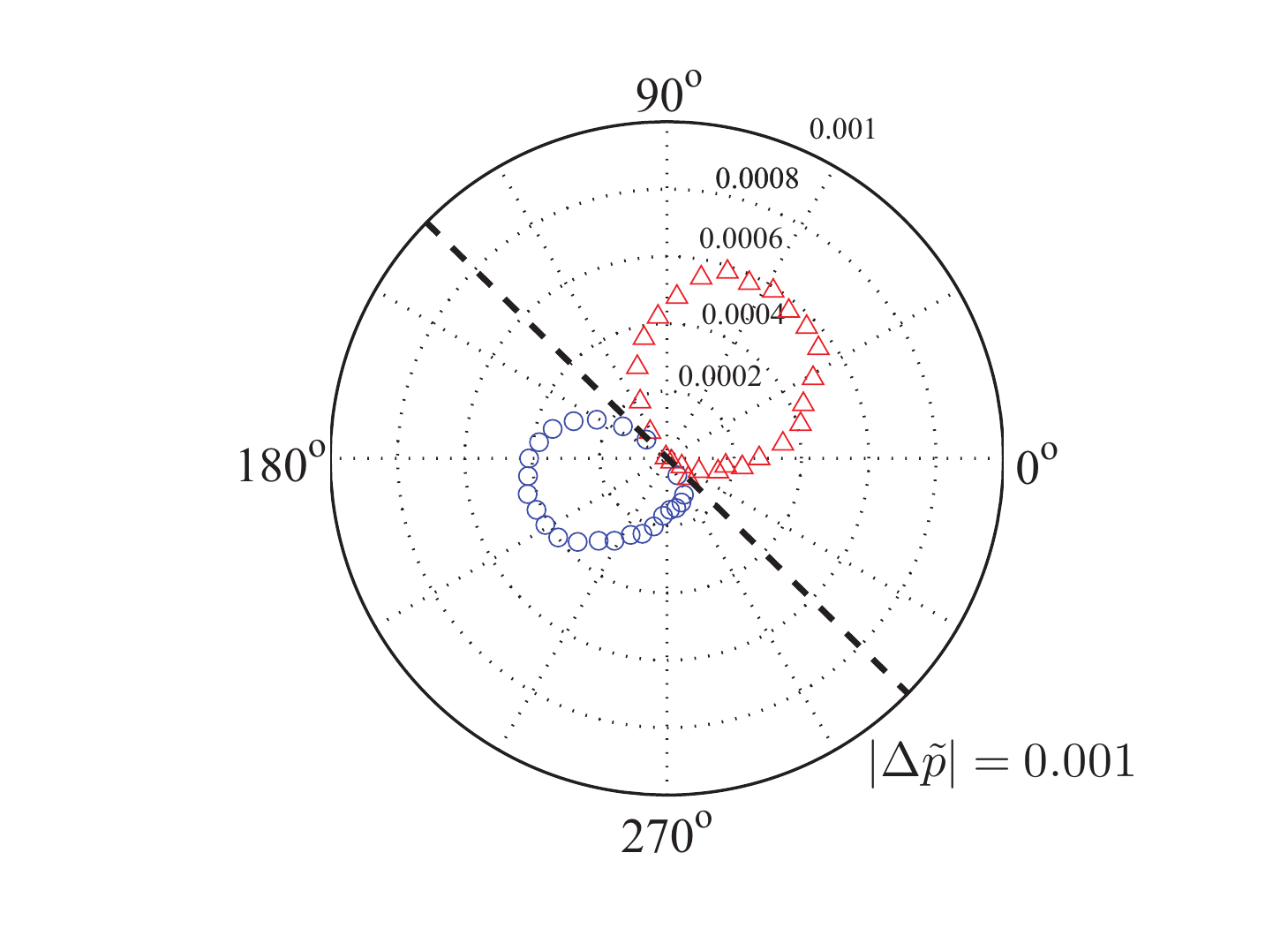}\\
  
  \hskip-3.0in (c) \hskip3.0in (d)

  \includegraphics[width=3.2in]{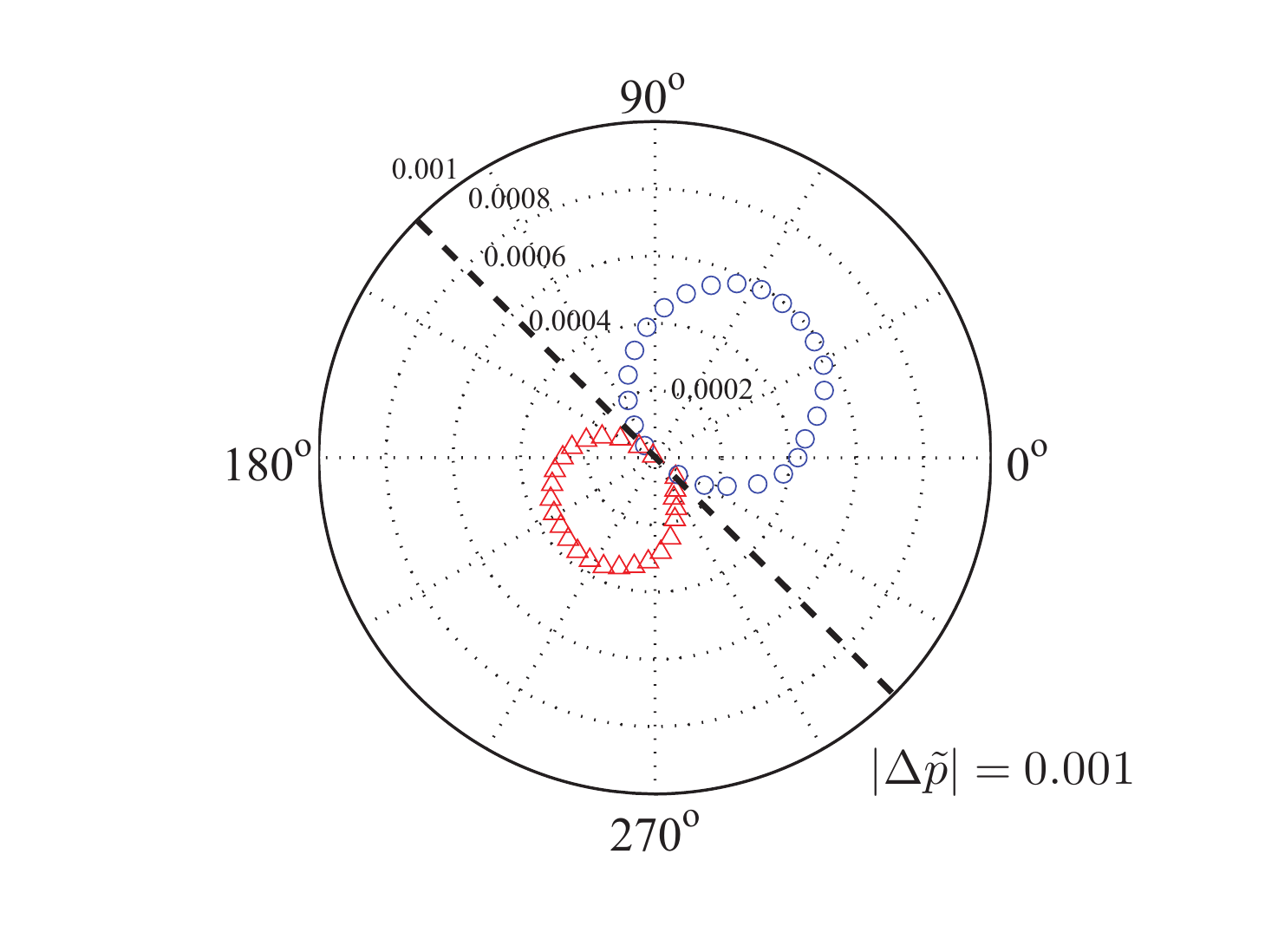}
  \hskip0.1in
  \includegraphics[width=3.2in]{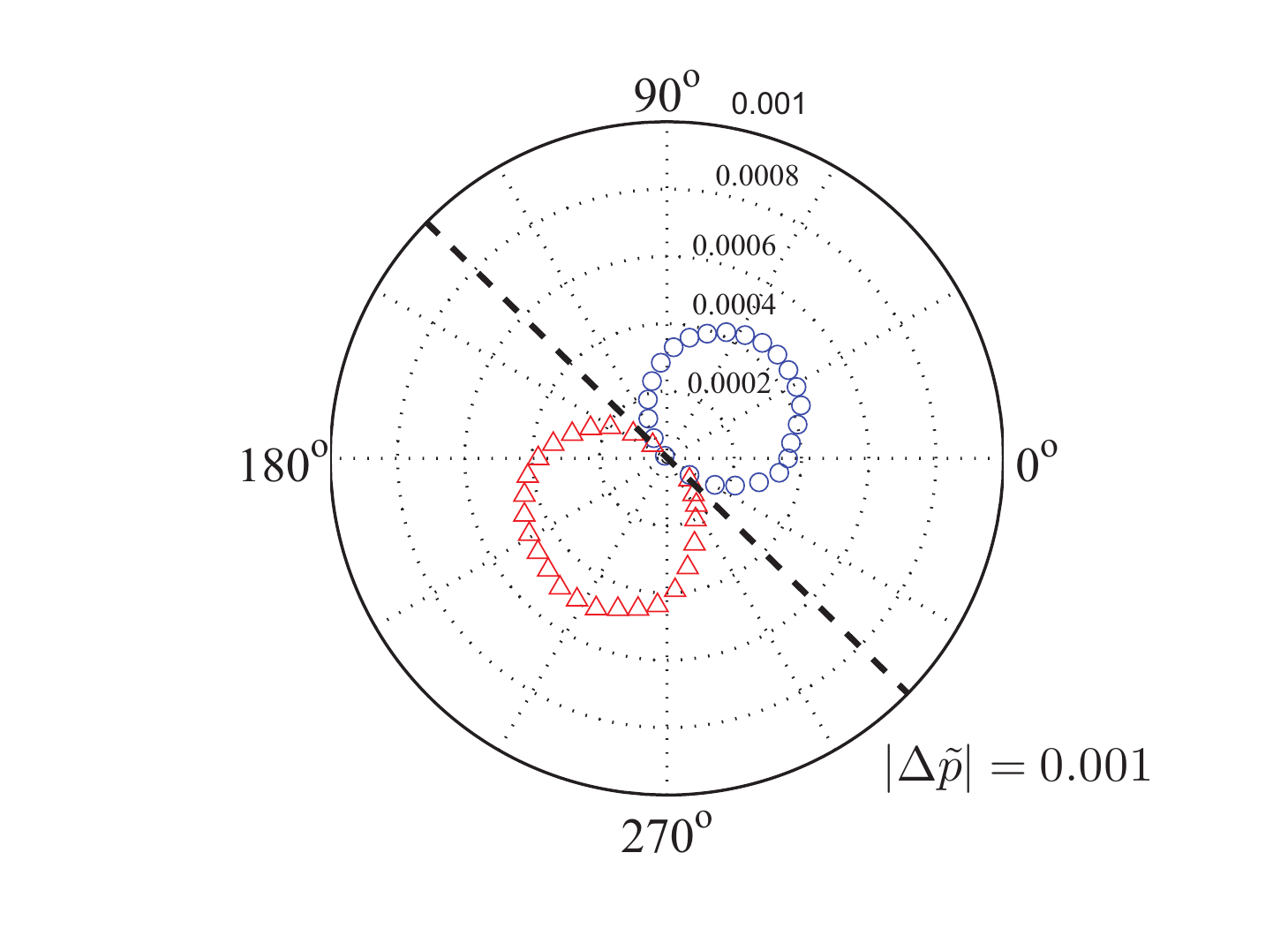}\\
  \end{center}
\caption{Flapping foils in forward flight: time variation of a polar diagram of $\Delta \tilde{p}$ for Re=100, $M=0.1$, $U_r=0.4$, $A_0/L=1.25$ and $\alpha_m=0$ measured at a distance of $18L$. (a) $t = 0T/6$, (b) $t = T/6$, (c) $t = 2T/6$ and (d) $t = 3T/6$. Where, $\Delta$ and o denote $\Delta \tilde{p}>0$ and $\Delta \tilde{p}<0$, respectively. The dashed line indicates the direction perpendicular to the stroke plane.}
\label{Fig:dpcirdirec}
\end{figure}

Shown in Fig.~\ref{Fig:dpcirdirec} are the polar plots of the fluctuating pressures $\Delta \tilde{p}$ at different instants from $0/6T$ to $3T/6$ in the 31st cycle. As the upstroke process is a inverse of the downstroke process, only a half period is presented. It is found that the directivity of the pressure waves fluctuates around the stroke angle $45^o$ at the flapping frequency. Fig.~\ref{Fig:dpcirdirec} also describes the whole sound generation process due to the flapping wing loading and the vortex shedding.

\begin{figure}
  \begin{center}
  \includegraphics[width=4.0in]{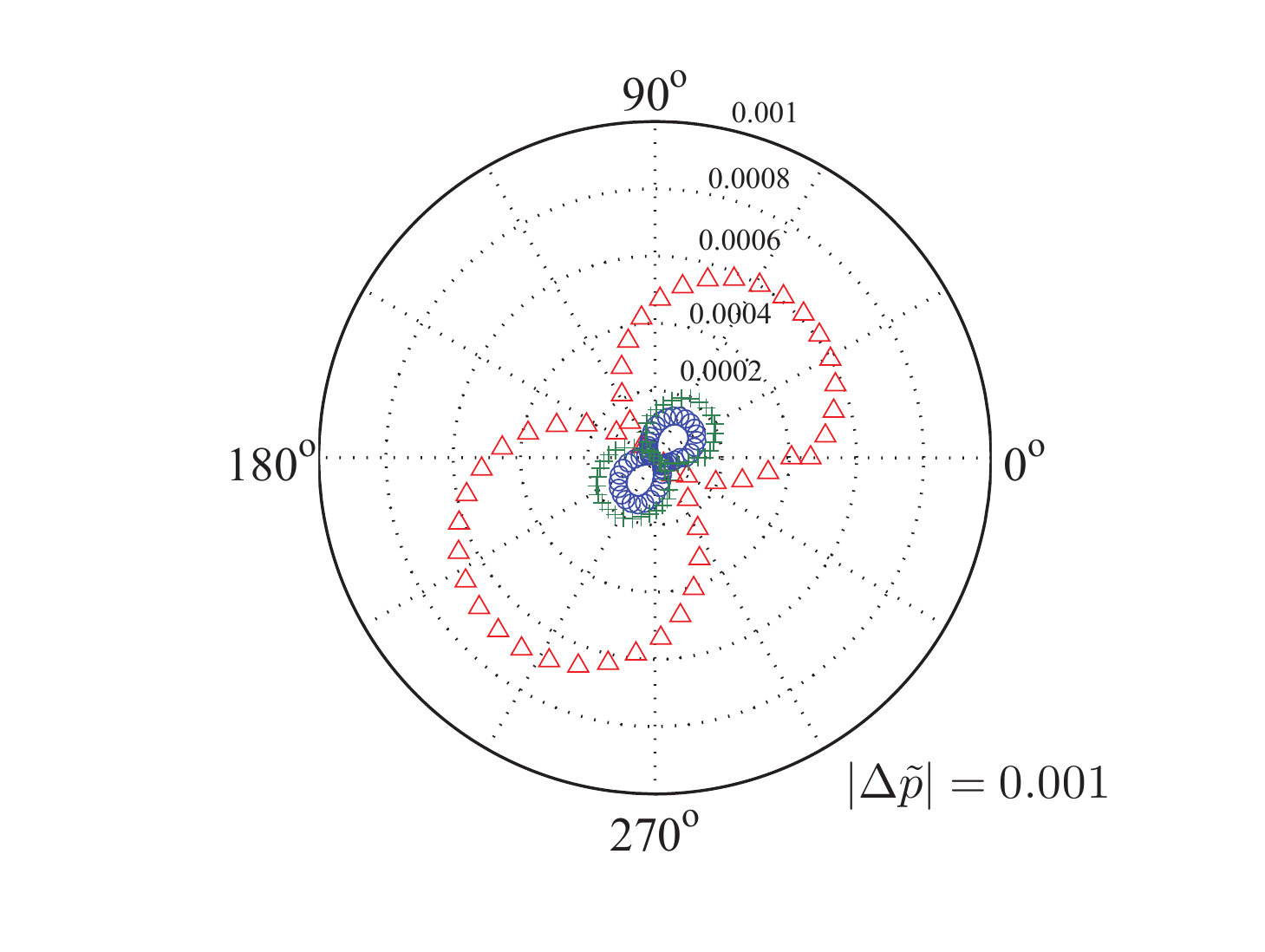}
  \end{center}
\caption{Flapping foils in forward flight: polar diagram of the fluctuating pressure peaks at $r=18$ with different frequencies. Where, $\Delta$, o and $+$ denote the frequency of $f$, $2f$ and $3f$, respectively.}
\label{Fig:dpcirfreq}
\end{figure}

After the analysis of the sound generation mechanism and amplitude, we would like to further analyze the frequency of the flapping foil-induced sound by using FFT to analyze the fluctuating pressure. The fluctuating pressure of ten cycles from $10T$ to $20T$ measured at the probes along $r=18$ are used in the frequency analysis. Fig.~\ref{Fig:dpcirfreq} shows the polar distribution of the fluctuating pressures at the frequency of $f$ (the flapping frequency), $2f$ and $3f$. It indicates that the sound pressure is dominated by the frequency of $f$. The peaks of the fluctuating pressures at the frequencies of $2f$ and $3f$ are much lower than those at the frequency of $f$. The fluctuating pressure at the frequencies over $3f$ are negligible.

\subsubsection{Comparison of the sound generated by flat plate and NACA0015 foil}
Here, two flat plates (a rigid and a flexible) and a rigid NACA0015 foil undergoing combined translational and rotational motion in a uniform flow, which widely exists in the insects wings~\cite{tian2013force}, are numerically simulated to study the influences of the foil flexibility and geometry.

\begin{figure}
 \begin{center}
  \hskip-1.8in (a) \hskip1.8in (b) \hskip1.8in (c)

  \includegraphics[width=2.0in]{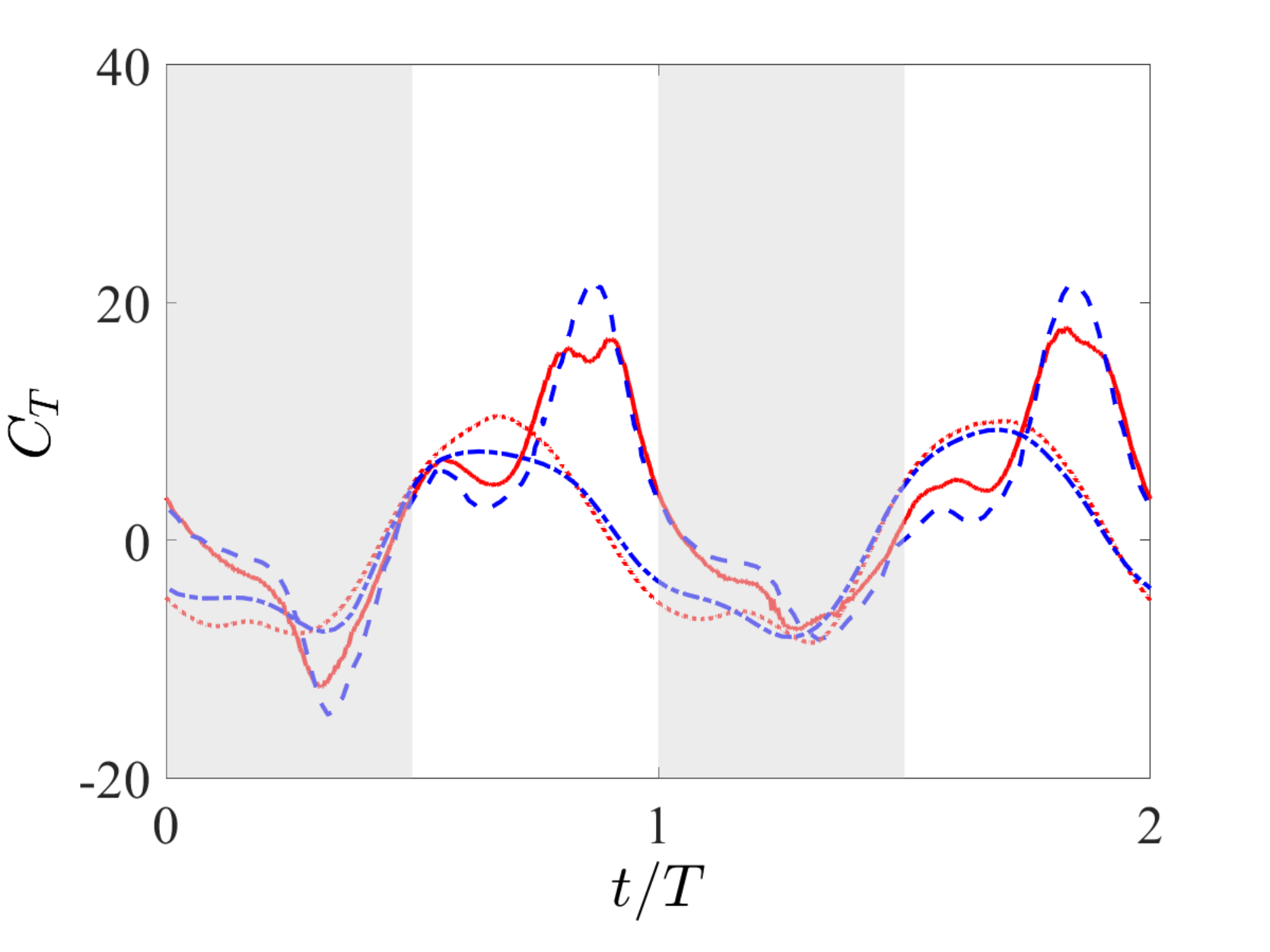}
  \includegraphics[width=2.0in]{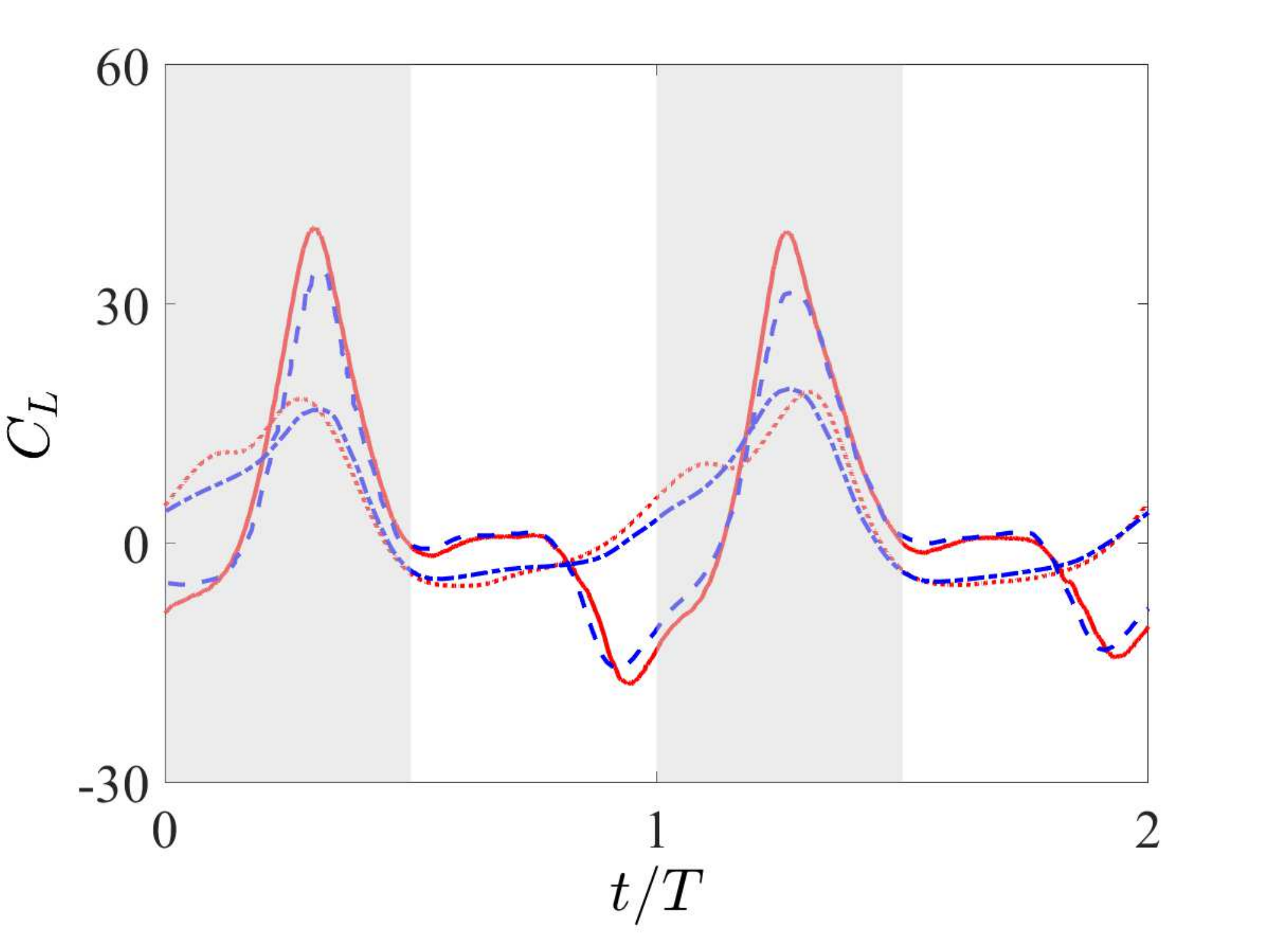}
  \includegraphics[width=2.0in]{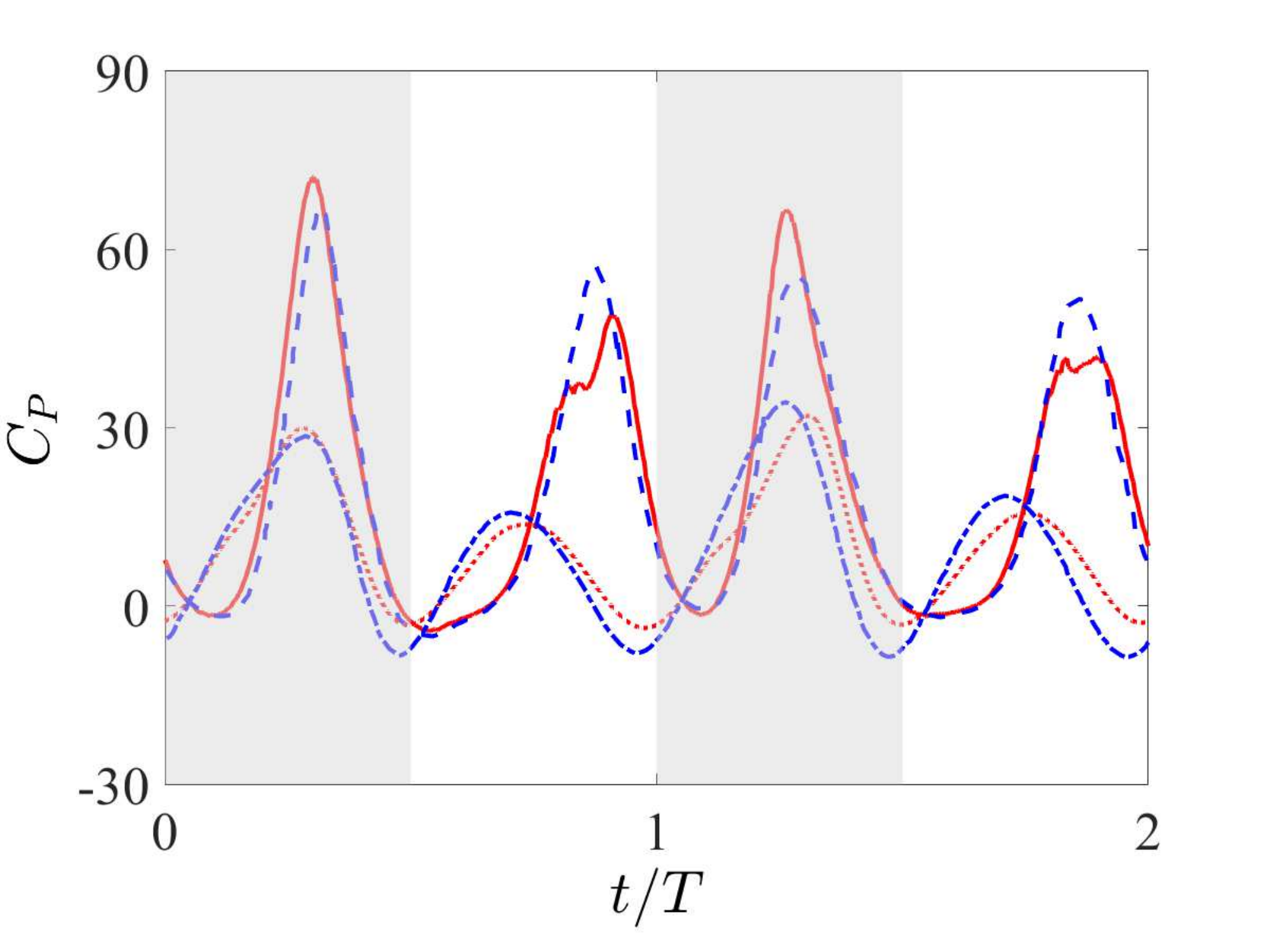}
  \end{center}
\caption{Flapping foils in forward flight: time histories of $C_T$ (a), $C_L$ (b) and $Cp$ (c) for Re=100, $M=0.1$, $U_r=0.4$, $A_0/L=1.25$ and $\alpha_m=\pi/4$. Where, solid line (present) and dashed line (Ref.~\cite{tian2013force}) denote the flexible foil, dotted line (present) and dash-dotted line (Ref.~\cite{tian2013force}) denote the rigid foil. The grey and white regions indicate downstroke and upstroke, respectively.}
\label{Fig:ctlp-rf}
\end{figure}

\begin{figure}
  \begin{center}
  \hskip-1.8in (a) \hskip1.8in (b) \hskip1.8in (c)

  \includegraphics[width=2.0in]{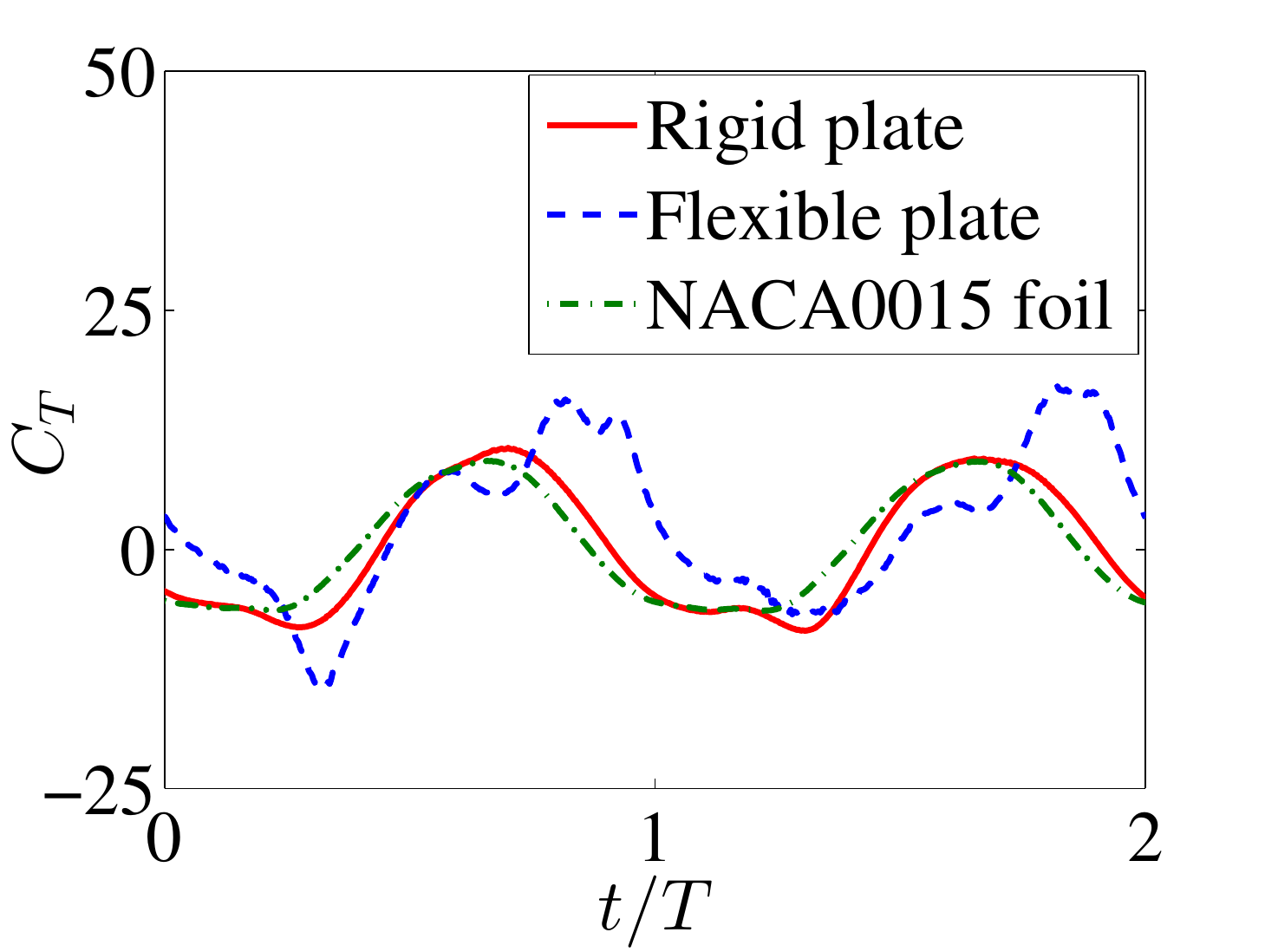}
  \hskip0.1in
  \includegraphics[width=2.0in]{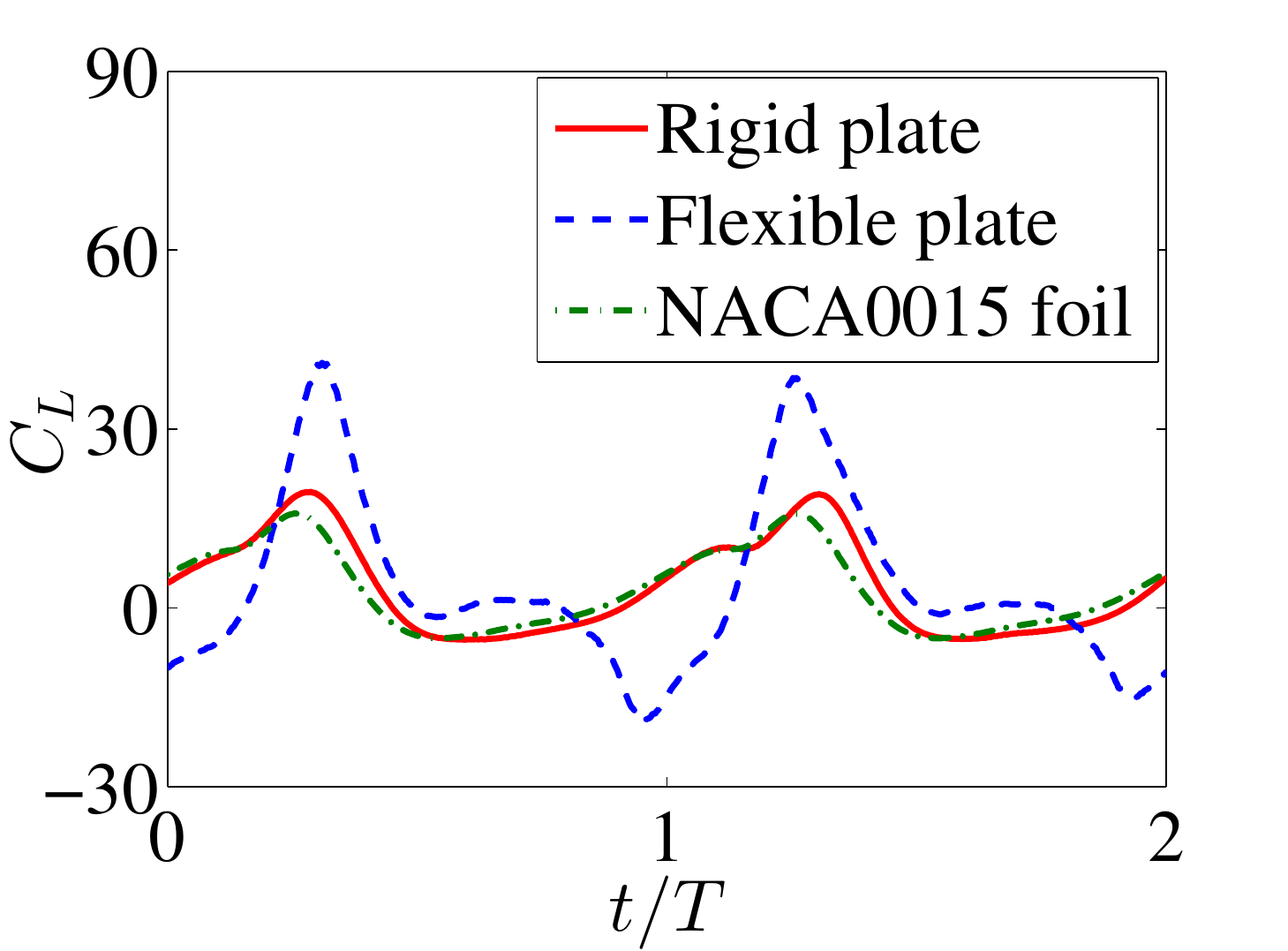}
  \hskip0.1in
  \includegraphics[width=2.0in]{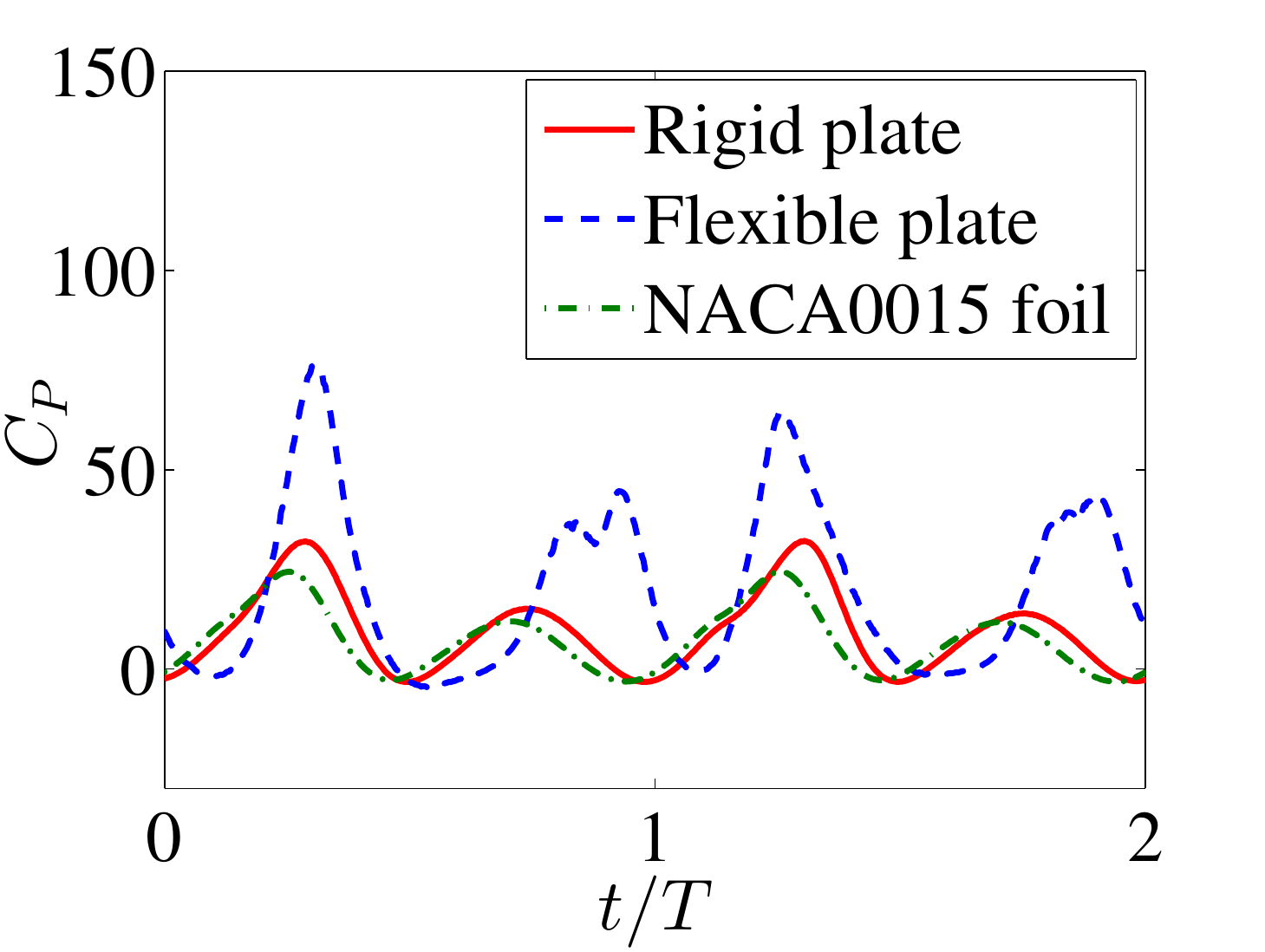}\\
  \end{center}
\caption{Flapping foils in forward flight: comparison of the drag coefficient (a), lift coefficient (b) and power coefficient (c) histories in a period. Here, Re=100, $M = 0.1$, $Ur = 0.4$, $A_0/L= 1.25$ and $\alpha_m=\pi/4$.}
\label{Fig:flapping_aerody}
\end{figure}

The motion of the foil is controlled by Eq.~\ref{eq:flapping_motion}. All the non-dimensional parameters are defined in Eq.~\ref{eq:flapping_para}. The amplitude of the rotating angle $\alpha_m=\pi/4$, other parameters for the flexible foils are: frequency ratio $\omega^*=0.6$ and $m^*=5.0$. The thrust, lift and power coefficients defined in Eq.~\ref{eq:ctcl} are presented in Fig.~\ref{Fig:ctlp-rf}, along with the numerical data from Ref.~\cite{tian2013force}. It is found the results from the current compressible solver agree well with the data obtained by Tian et al.~\cite{tian2013force} using an incompressible solver. Because the Mach number used in the current solver is low ($M=0.1$), the compressibility of the fluid is not significant. As seen from Fig~\ref{Fig:ctlp-rf}, the discrepancies between the current results and those from Ref.~\cite{tian2013force} for the flexible foil seem to be much more significant than those for the rigid foil. A plausible explanation is that the flexible foil induces higher fluid velocity around it, which makes the compressibility more significant. The histories of the thrust, lift and power coefficients in a period for the three cases are plotted in Fig.~\ref{Fig:flapping_aerody} for comparison. The results show that the flexibility of the plate has a remarkable influence on the force generation. The thrust and lift are enhanced significantly by the flexibility. However, the comparison between the rigid plate and NACA0015 foil indicates that the geometrical shape does not have significant effects on the aerodynamic characteristics at the conditions considered.

\begin{figure}
  \begin{center}
  \hskip-1.8in (a) \hskip1.8in (b) \hskip1.8in (c)

  \includegraphics[width=2.0in]{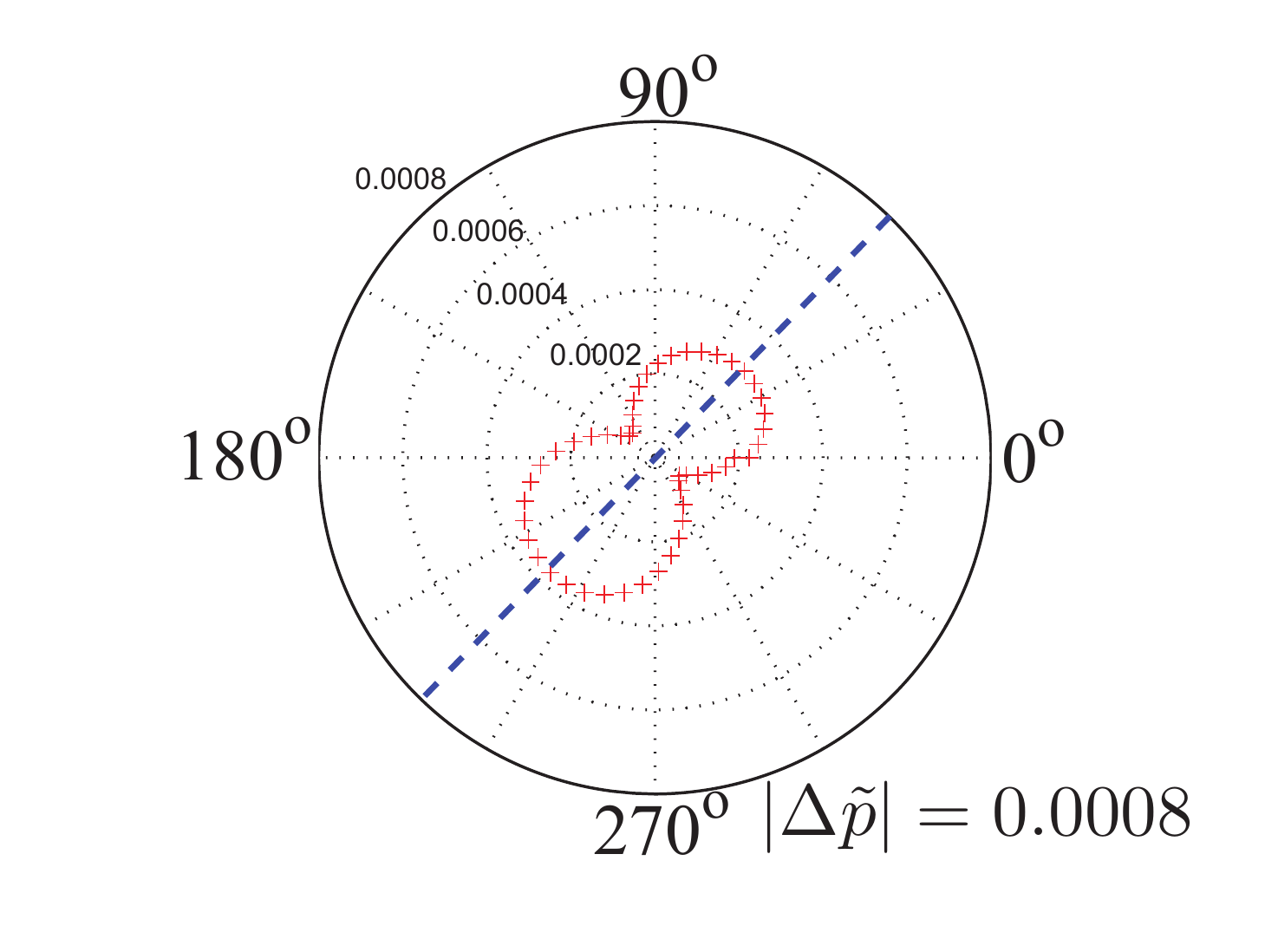}
  \hskip0.1in
  \includegraphics[width=2.0in]{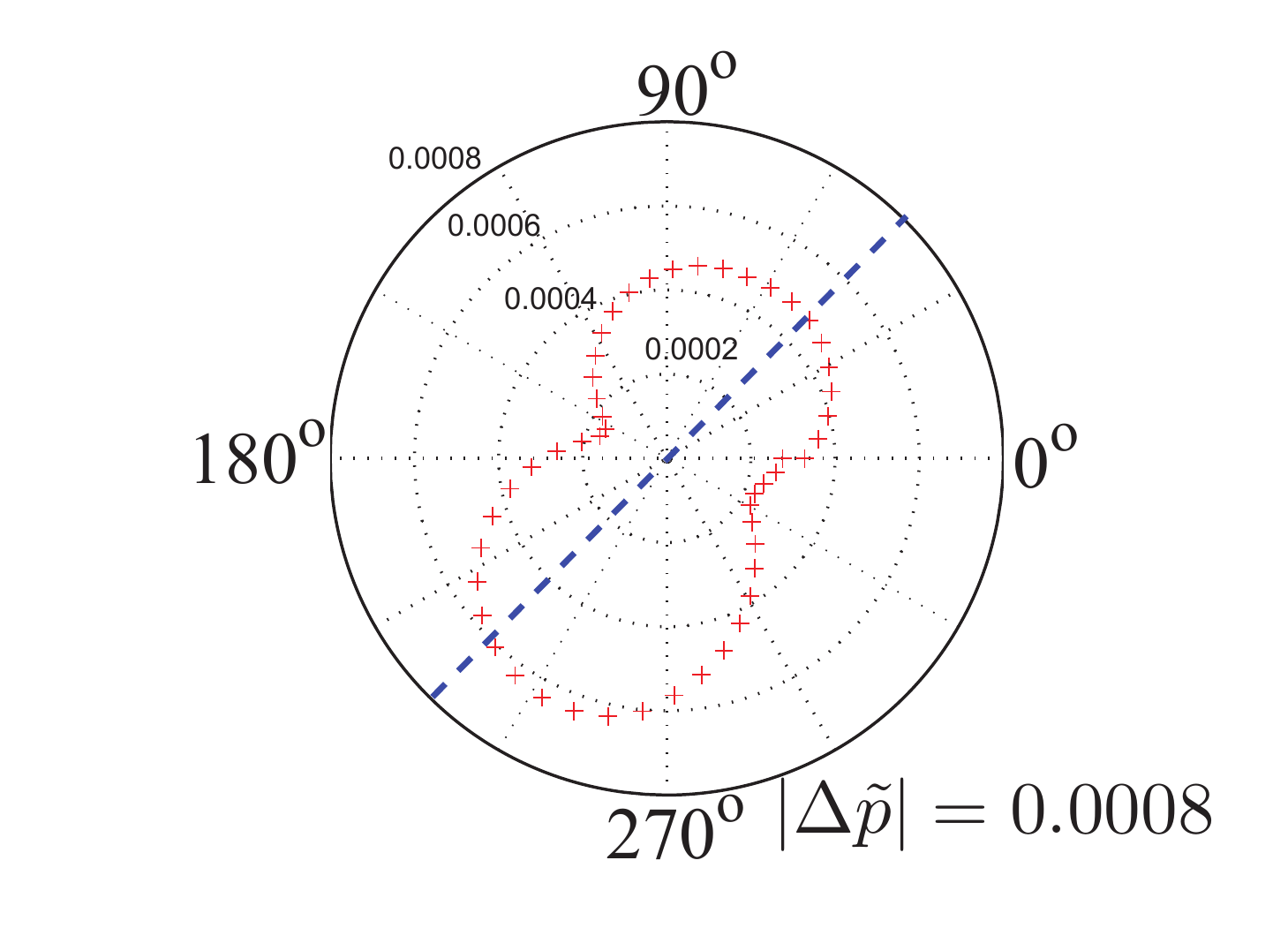}
  \hskip0.1in
  \includegraphics[width=2.0in]{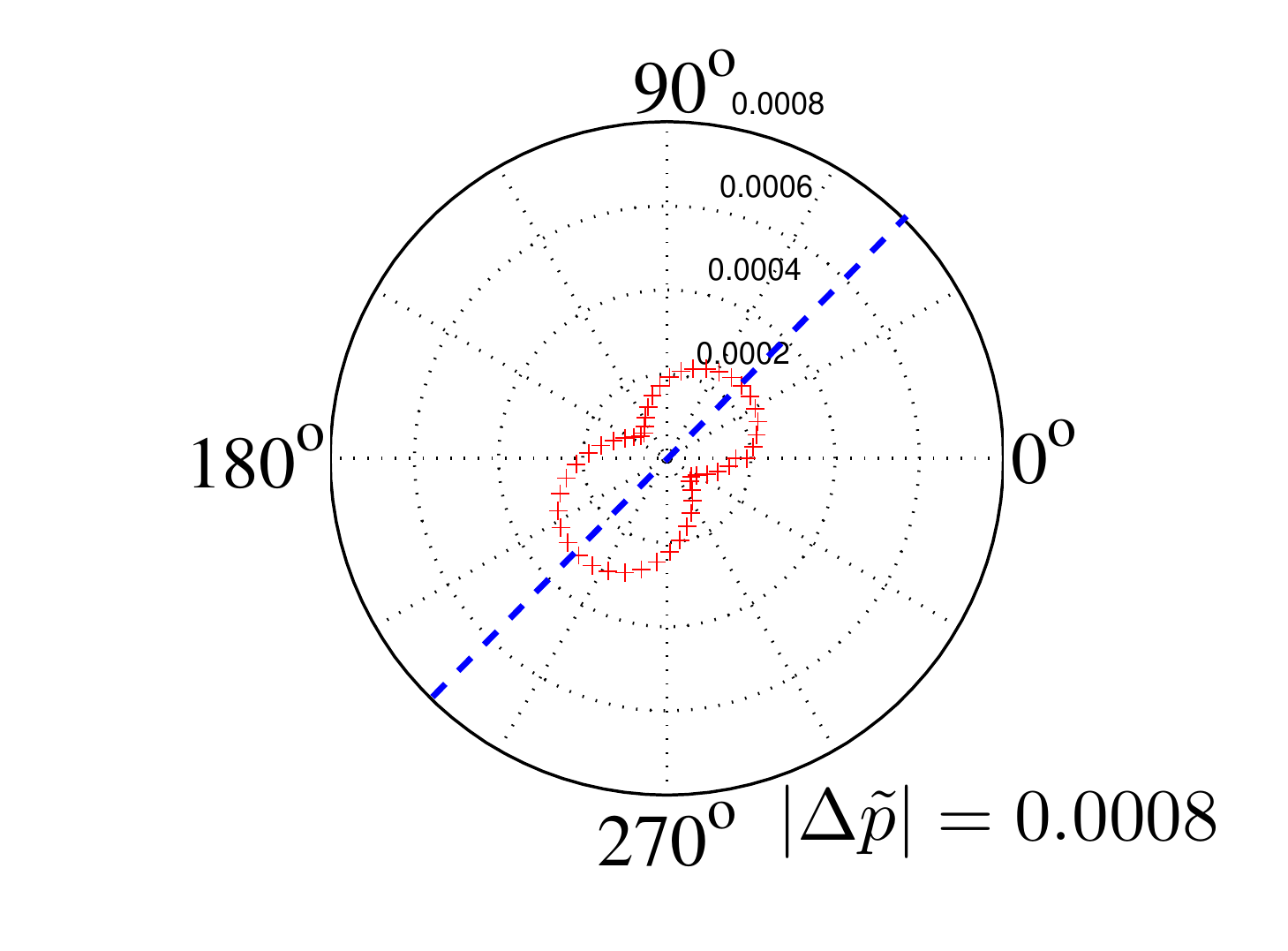}\\
  \end{center}
\caption{Flapping foils in forward flight: root-mean-square of $\Delta \tilde{p}$ measured at $r=30$: (a) rigid plate, (b) flexible plate and (c) NACA0015 foil. Re=100, $M = 0.1$, $Ur = 0.4$, $A_0/L= 1.25$ and $\alpha_m=\pi/4$. The dashed lines indicate the directions of the stroke planes.}
\label{Fig:rf_flapping_cir}
\end{figure}

\begin{figure}
  \begin{center}
  \hskip-3.0in (a) \hskip3.0in (b)

  \includegraphics[width=3.2in]{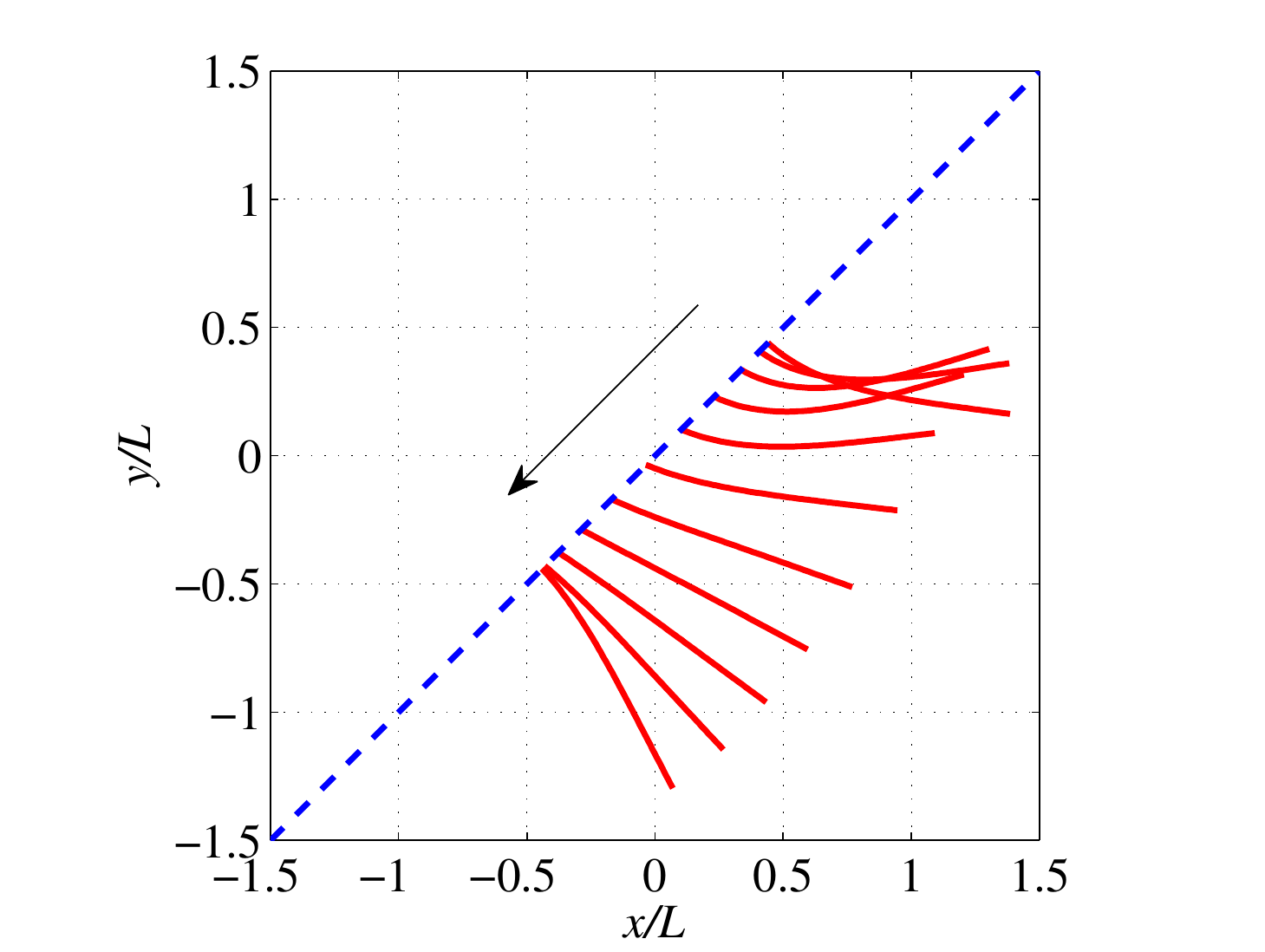}
  \hskip0.1in
  \includegraphics[width=3.2in]{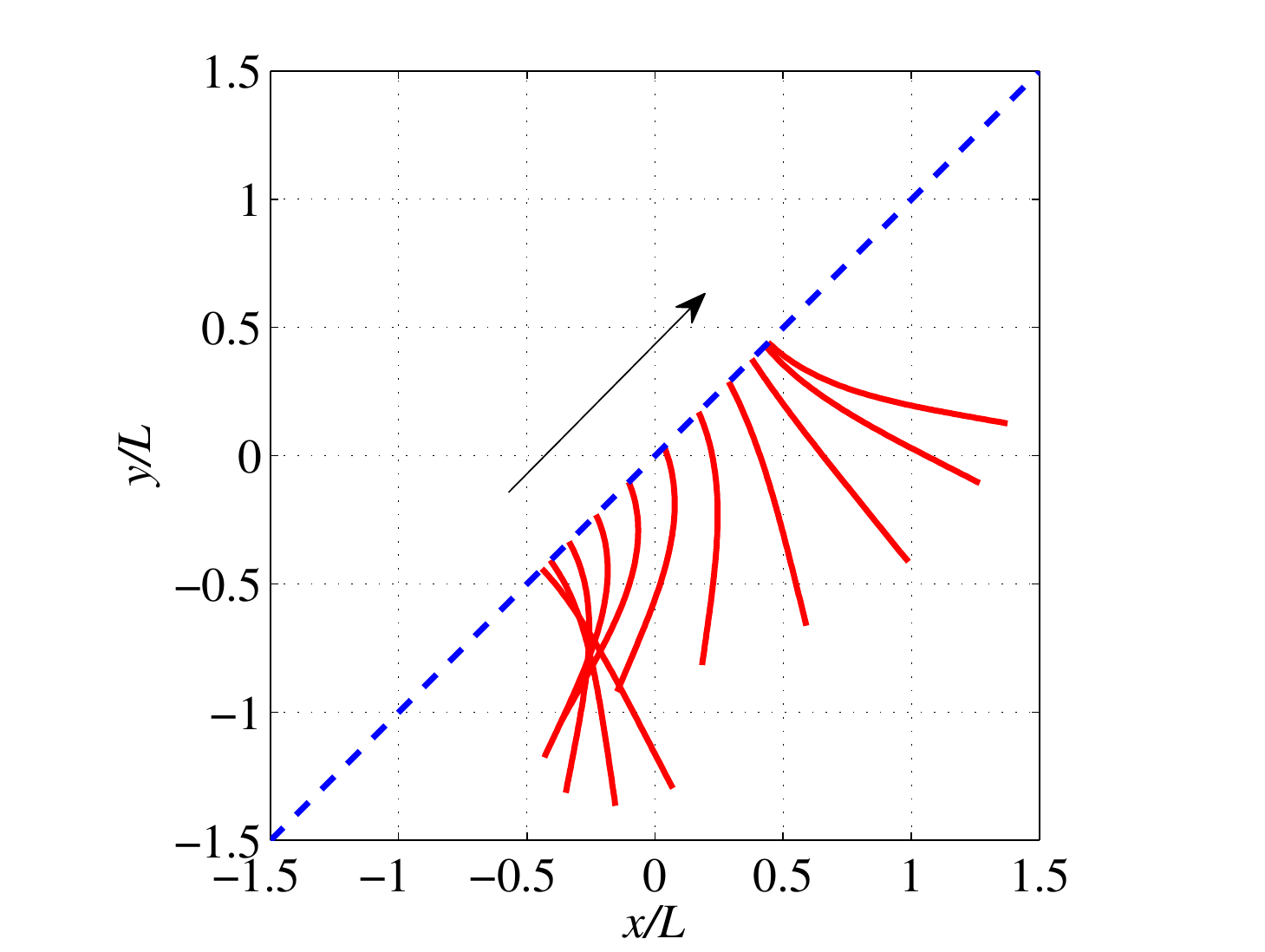}\\
  \end{center}
\caption{Flapping foils in forward flight: deformation of the flexible foil during (a) downstroke and (b) upstroke. Here, Re=100, $M = 0.1$, $Ur = 0.4$, $A_0/L= 1.25$, $\alpha_m=\pi/4$, $\omega^*=0.6$ and $m^*=5.0$. The dashed lines indicate the directions of the stroke planes.}
\label{Fig:deform_flapping}
\end{figure}

In order to compare the sound generation of rigid and flexible foils, the polar diagrams of the root-mean-square of $\Delta \tilde{p}$ measured at $r=30$ are presented in Fig.~\ref{Fig:rf_flapping_cir}. As shown in Fig.~\ref{Fig:rf_flapping_cir} (a) and (c), the fluctuating pressure generated by the rigid plate and NACA0015 foil distributes symmetrically about the stroke plane (indicated by the dashed line in Fig.~\ref{Fig:rf_flapping_cir} (a) and (c)). However, Fig.~\ref{Fig:rf_flapping_cir} (b) shows that the distribution of the fluctuating pressure generated by the flexible foil is asymmetrical. An evident shift (about $15^o$ anticlockwise) is observed, as shown in Fig.~\ref{Fig:rf_flapping_cir} (b). As analyzed in Ref.~\cite{tian2013force}, the deformation of the flexible plate during upstroke is higher than that during downstroke, due to the presence of free stream. Obviously, this sound shift is introduced by the asymmetrical deformation (as shown in Fig.~\ref{Fig:deform_flapping}) of the foil. Fig.~\ref{Fig:rf_flapping_cir} also shows that the flexible plate generates the largest fluctuating pressure, which is significantly larger than that generated by the rigid plate and NACA0015 foil. The results indicate that the flexibility influences the sound generation significantly. The effects of geometrical shape on the sound amplitudes are much smaller. In addition, it is found that the fluctuating pressure on the lower section is larger than that on the upper section. A reasonable explanation is that the lower section is located on the windward side and the upper section is on the leeward side.

\begin{figure}
  \begin{center}
  \hskip-1.8in (a) \hskip1.8in (b) \hskip1.8in (c)

  \includegraphics[width=2.0in]{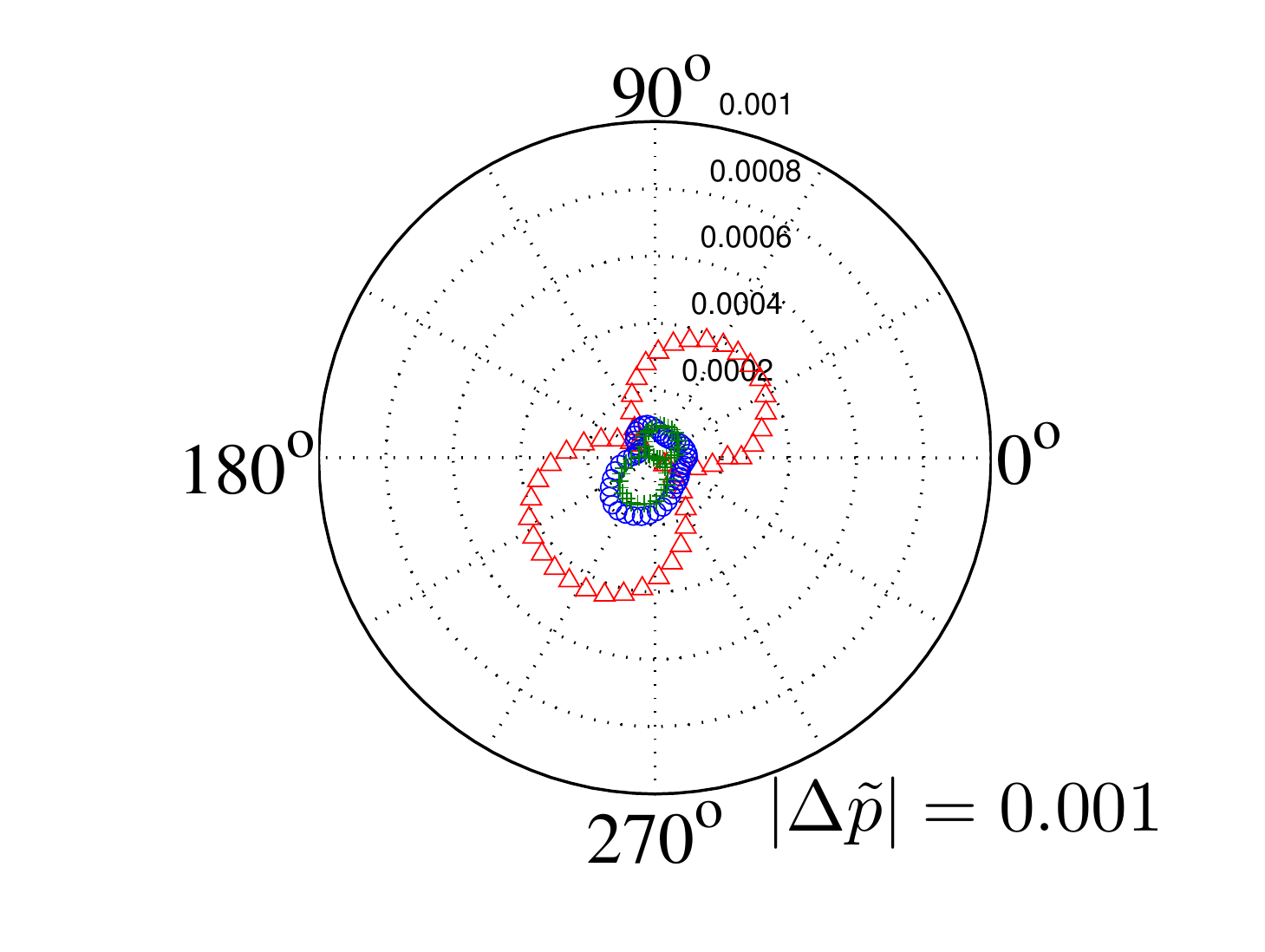}
  \hskip0.1in
  \includegraphics[width=2.0in]{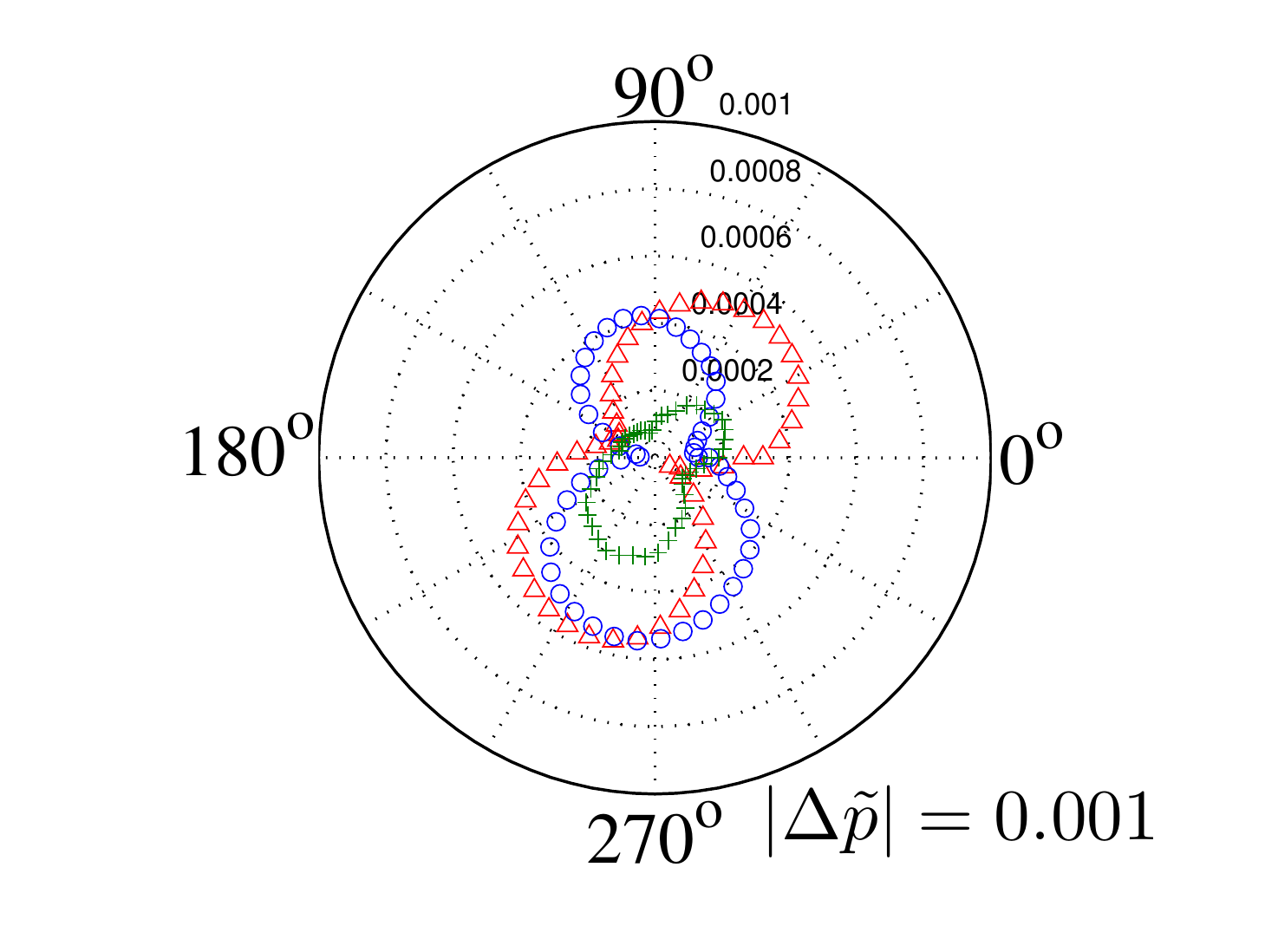}
  \hskip0.1in
  \includegraphics[width=2.0in]{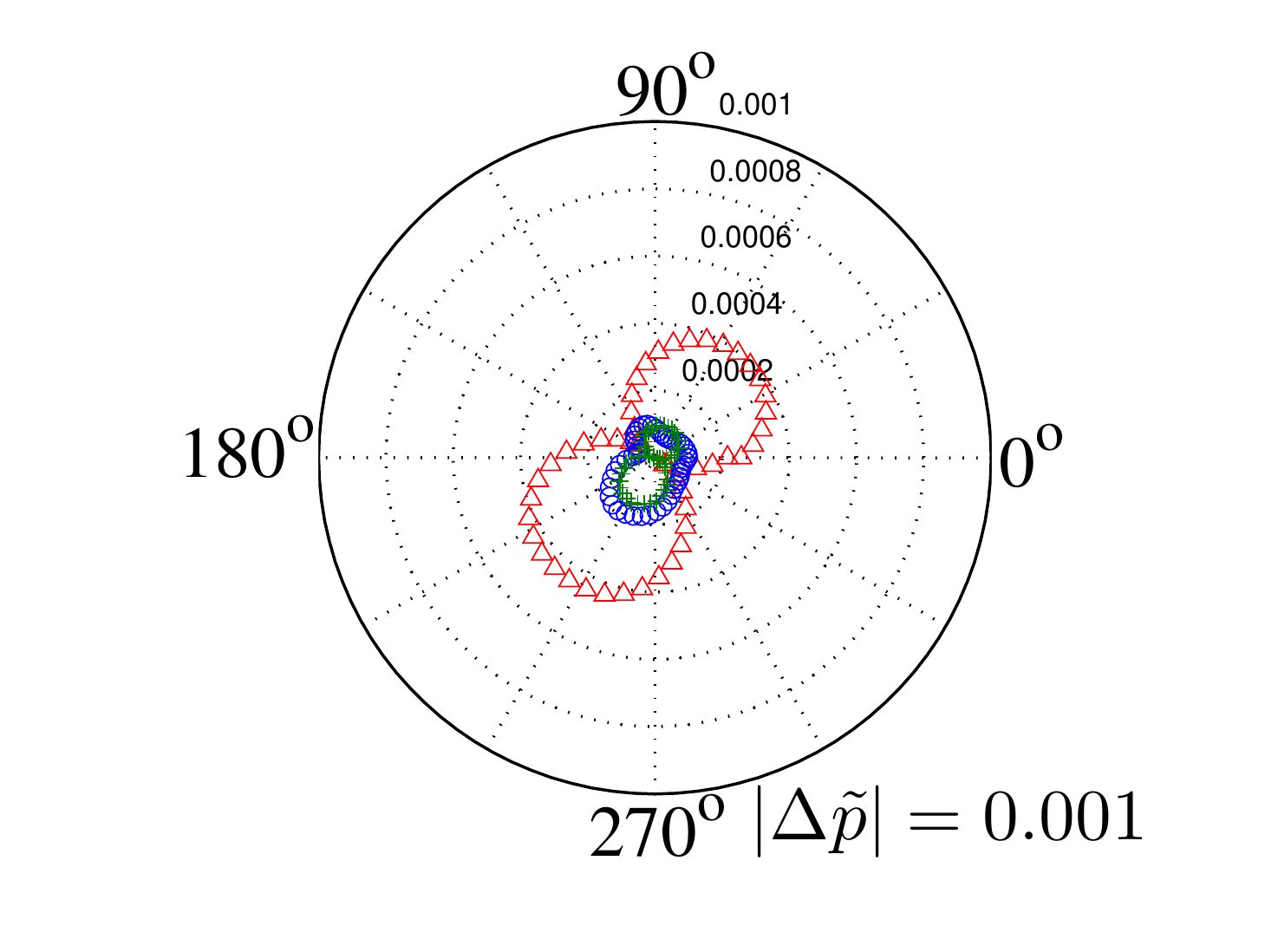}\\
  \end{center}
\caption{Flapping foils in forward flight: polar diagram of the fluctuating pressure peaks at $r=30$ with different frequencies: (a) rigid plate, (b) flexible plate and (c) NACA0015 foil. Here, Re=100, $M = 0.1$, $Ur = 0.4$, $A_0/L= 1.25$ and $\alpha_m=\pi/4$. Where, $\Delta$, o and $+$ denote the frequency of $f$, $2f$ and $3f$, respectively.}
\label{Fig:freq_flapping}
\end{figure}

The FFT is further used to study the frequency of the fluctuating pressure. According to the frequency analysis of the fluctuating pressure peaks, three main frequencies ($f$, $2f$ and $3f$) are captured in these three flapping foils. The fluctuating pressures at higher frequencies are negligible. The polar diagrams of the fluctuating pressure peaks at the three main frequencies are presented in Fig.~\ref{Fig:freq_flapping}. It is found that the frequency of $f$ dominates in the rigid flapping foil. However, Fig.~\ref{Fig:freq_flapping} (b) shows that both the frequency of $f$ and $2f$ dominate in the flexible flapping foil. As shown in Fig.~\ref{Fig:freq_flapping} (a) and (c), the fluctuating pressure for rigid foils  distributes symmetrically to the stroke plane as mentioned before. Fig.~\ref{Fig:freq_flapping} (b) shows that the fluctuating pressure at the frequency of $f$ generated by the flexible foil shift about $15^o$ anticlockwise compared with that generated by the rigid foil. However, the shift of the fluctuating pressure at the frequency of $2f$ approaches about $30^o$. This indicates that the propagation direction of the fluctuating pressure from a flexible foil at the frequency of $f$ and $2f$ exhibit remarkable differences, which can be used to identify the characteristics of the sound source. Additionally, the comparison between the rigid plate and NACA0015 foil also shows that the geometrical effects on the sound generation is not significant as mentioned in Section 4.1.4.

\subsection{Flapping foil energy harvester}
In this section, we further adopt the current numerical method to study the sound generated by a thin plate (without considering the thickness of the plate) and a NACA0015 foil (stream-lined shape) flapping as energy harvesters.

\begin{figure}
 \begin{center}
  \hskip-3.0in (a)

  \includegraphics[width=3.0in]{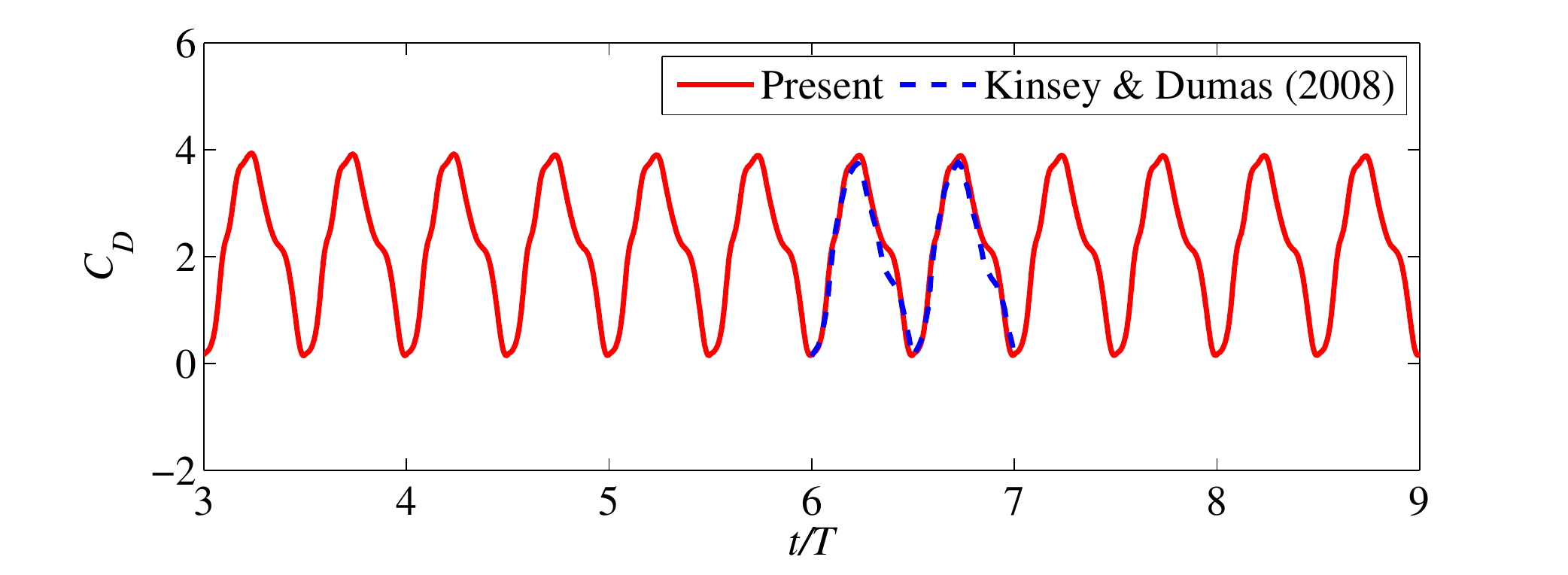}
  
  \hskip-3.0in (b)

  \includegraphics[width=3.0in]{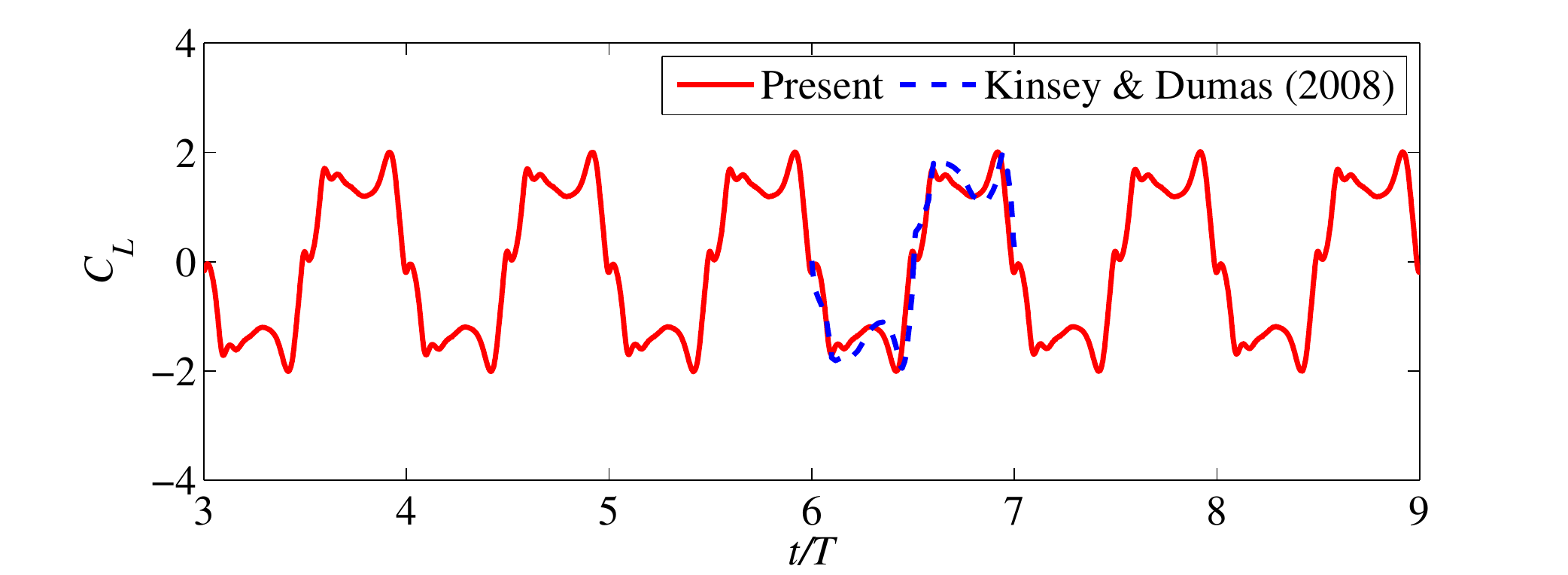}
  
  \hskip-3.0in (c)

  \includegraphics[width=3.0in]{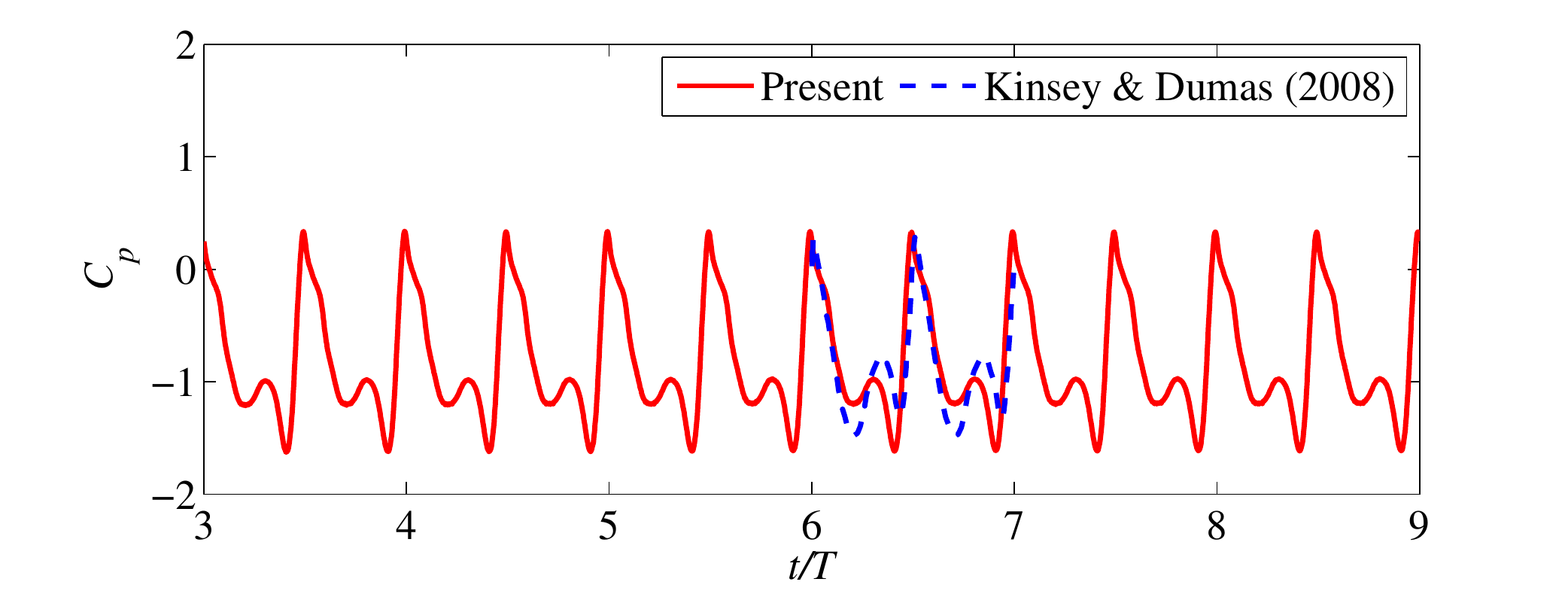}
  \end{center}
\caption{Flapping foil energy harvester: time histories of $C_D$ (a), $C_L$ (b) and $C_p$ (c) from a NACA0015 foil flapping in a uniform flow at Re=1100, $M=0.1$, $\alpha_m=152.6^o$, $A_0/L=2.0$ and $\beta=90^o$.}
\label{Fig:naca_clcd}
\end{figure}

A foil undergoing prescribed plunge and pitch governed by Eq.~\ref{eq:flapping_motion} in a uniform flow is considered. The characteristic parameters defined in Eq.~\ref{eq:flapping_para} are given as: $A_0/L=2.0$, $M=0.1$, $U_r=1.78$, $\alpha_m=152.6^o$ and $\beta=90^o$. The Reynolds number, defined by Re=$\rho_f U_0 L/\mu$ is 1100, and the distance from leading edge to the pitching axis is $L/3$. The details of the setup can be found in Refs.~\cite{kinsey2008parametric,tian2015fsi}. The computational domain based on a non-uniform Cartesian mesh has a size of $96L\times96L$, with a uniform area extending from ($-2.5L$, $-2.5L$) to ($2.5L$, $2.5L$), where the mesh spacing is $L/100$. Extensive preliminary tests have been conducted to make sure the results are independent of the mesh size and computational domain. The time histories of non-dimensional drag ($C_D$, defined as $-C_T$), lift ($C_L$) and power ($C_p$) extracted from the fluid obtained from the present simulations agree well with those available in Ref.~\cite{kinsey2008parametric} as shown in Fig.~\ref{Fig:naca_clcd}. 

\begin{figure}
  \begin{center}
  \hskip-1.8in (a) \hskip1.8in (b) \hskip1.8in (c)

  \includegraphics[width=2.0in]{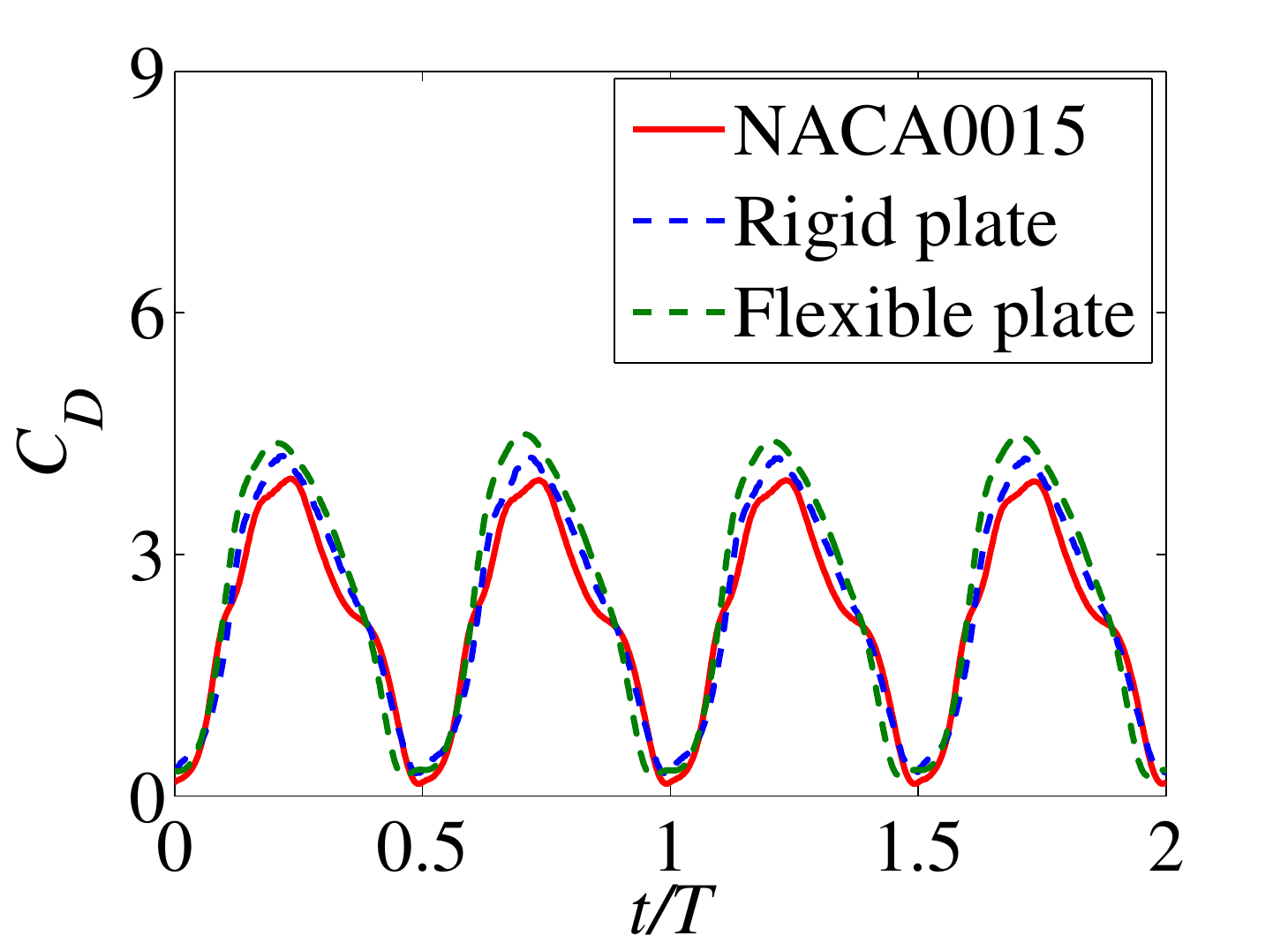}
  \hskip0.1in
  \includegraphics[width=2.0in]{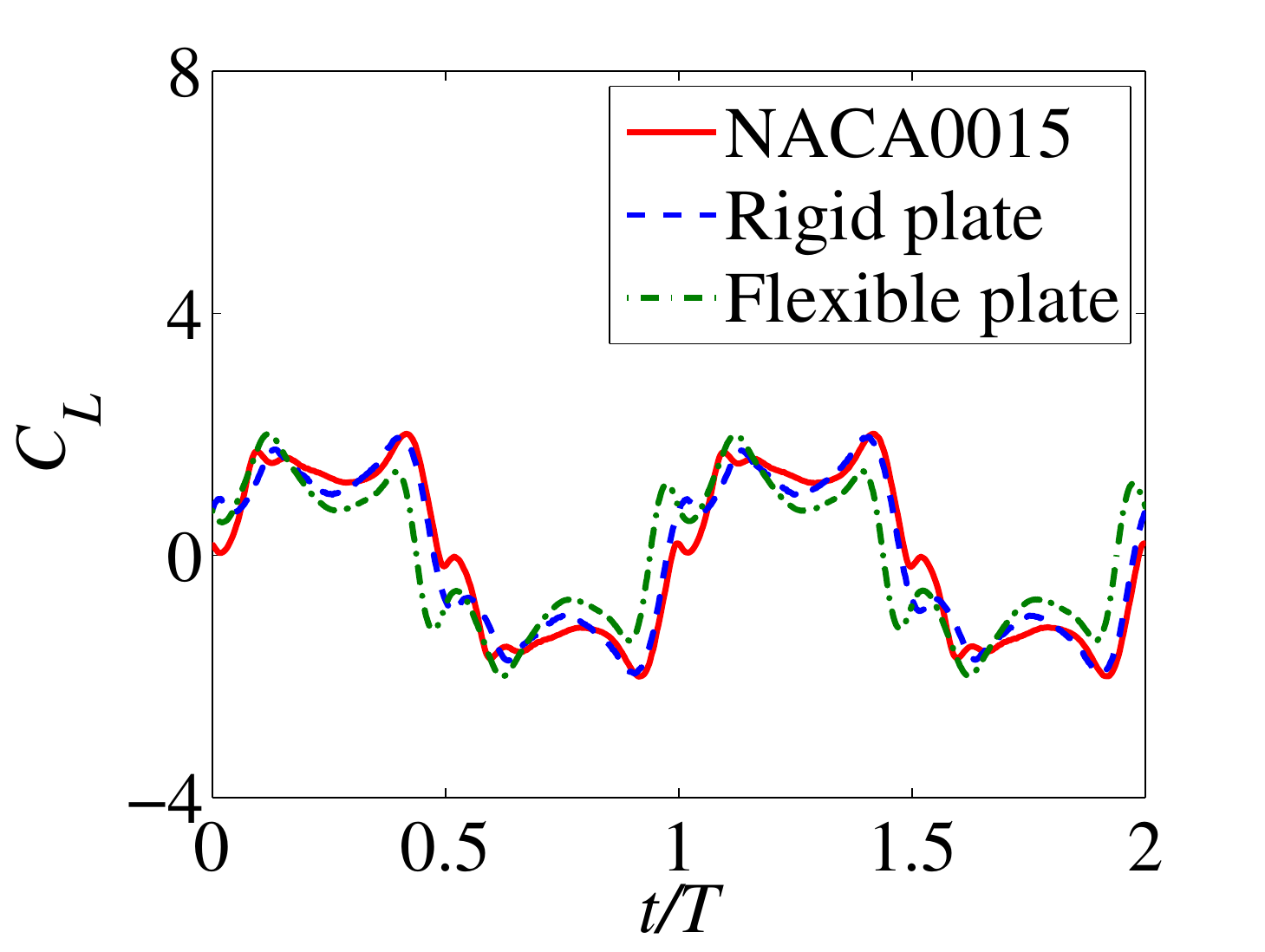}
  \hskip0.1in
  \includegraphics[width=2.0in]{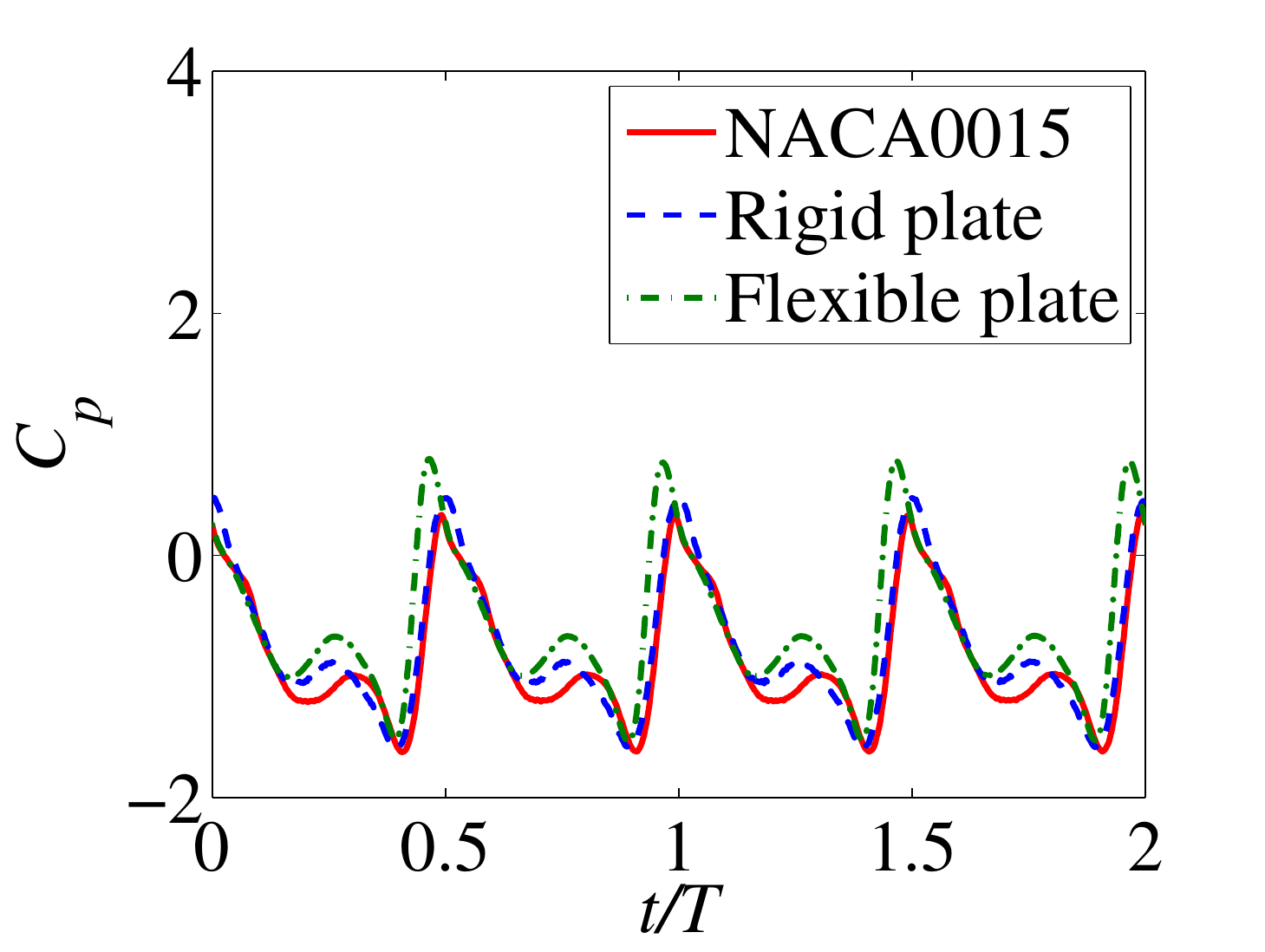}\\
  \end{center}
\caption{Flapping foil energy harvester: comparison of the drag coefficient (a), lift coefficient (b) and power coefficient (c) histories in a period. Here, Re=1100, $M=0.1$, $\alpha_m=152.6^o$, $A_0/L=2.0$ and $\beta=90^o$.}
\label{Fig:naca_flex_clcd}
\end{figure}

\begin{figure}
 \begin{center}
	\hskip-5.0in (a)	\\
  \includegraphics[width=1.5in]{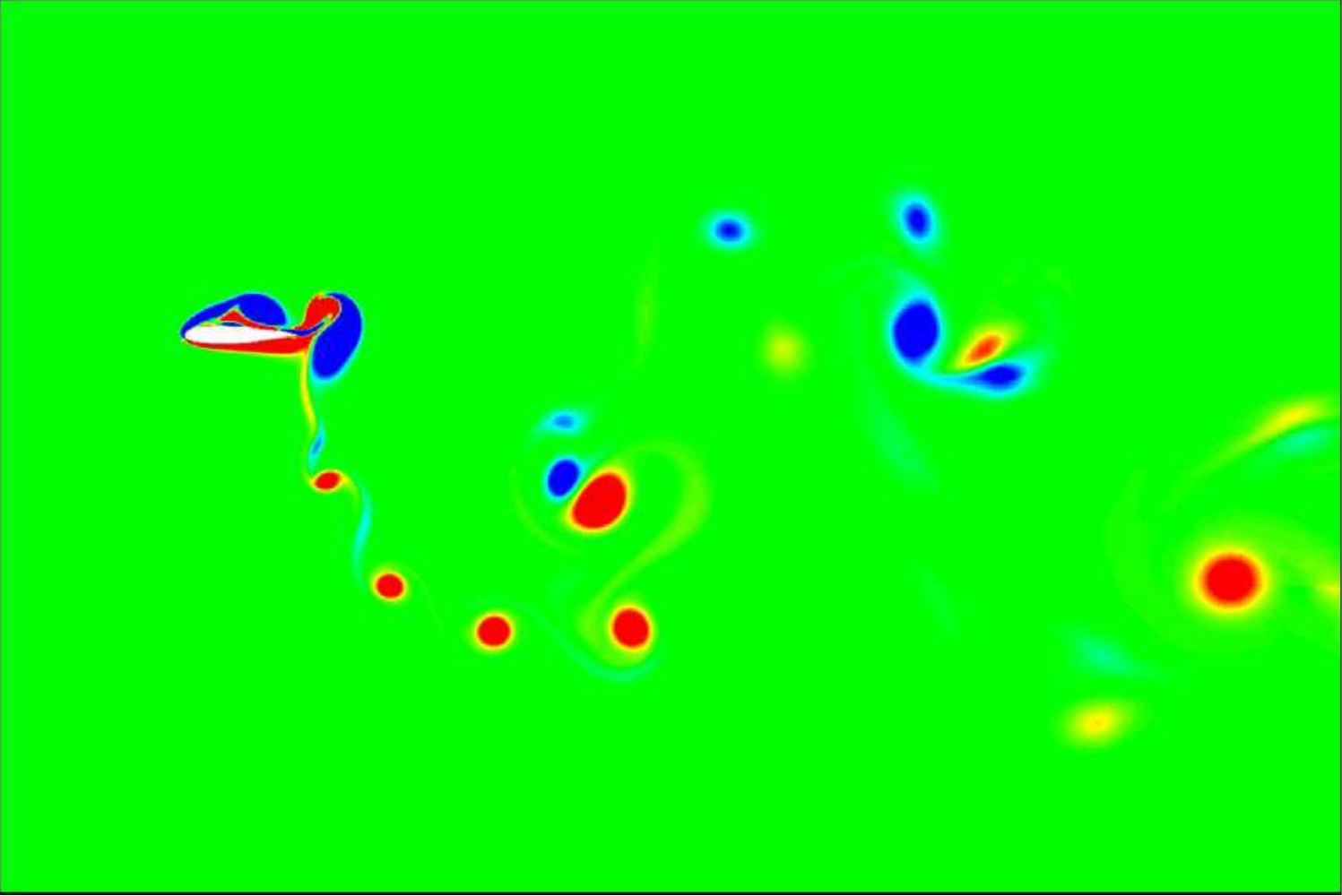}
  \includegraphics[width=1.5in]{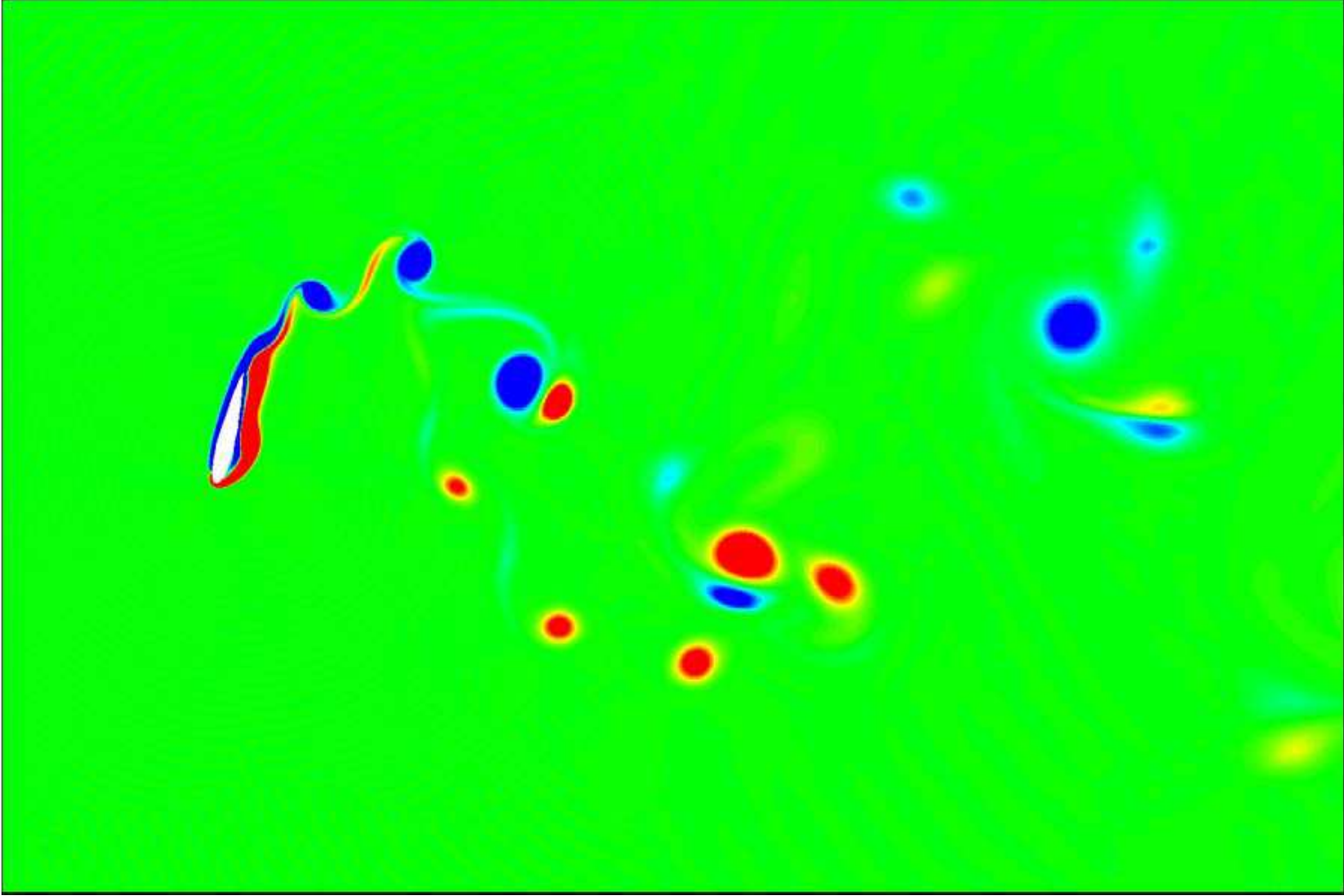}
  \includegraphics[width=1.5in]{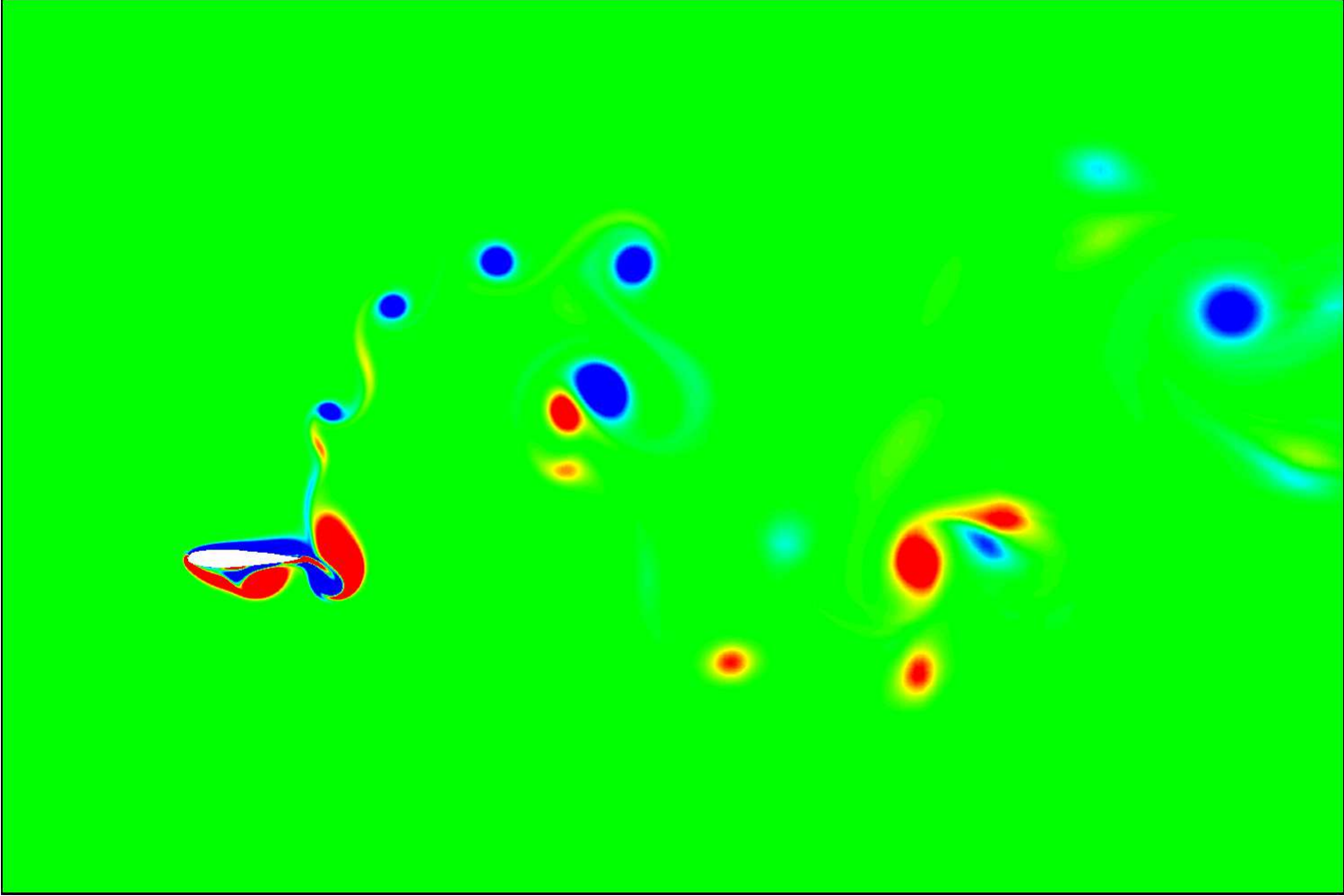}
  \includegraphics[width=1.5in]{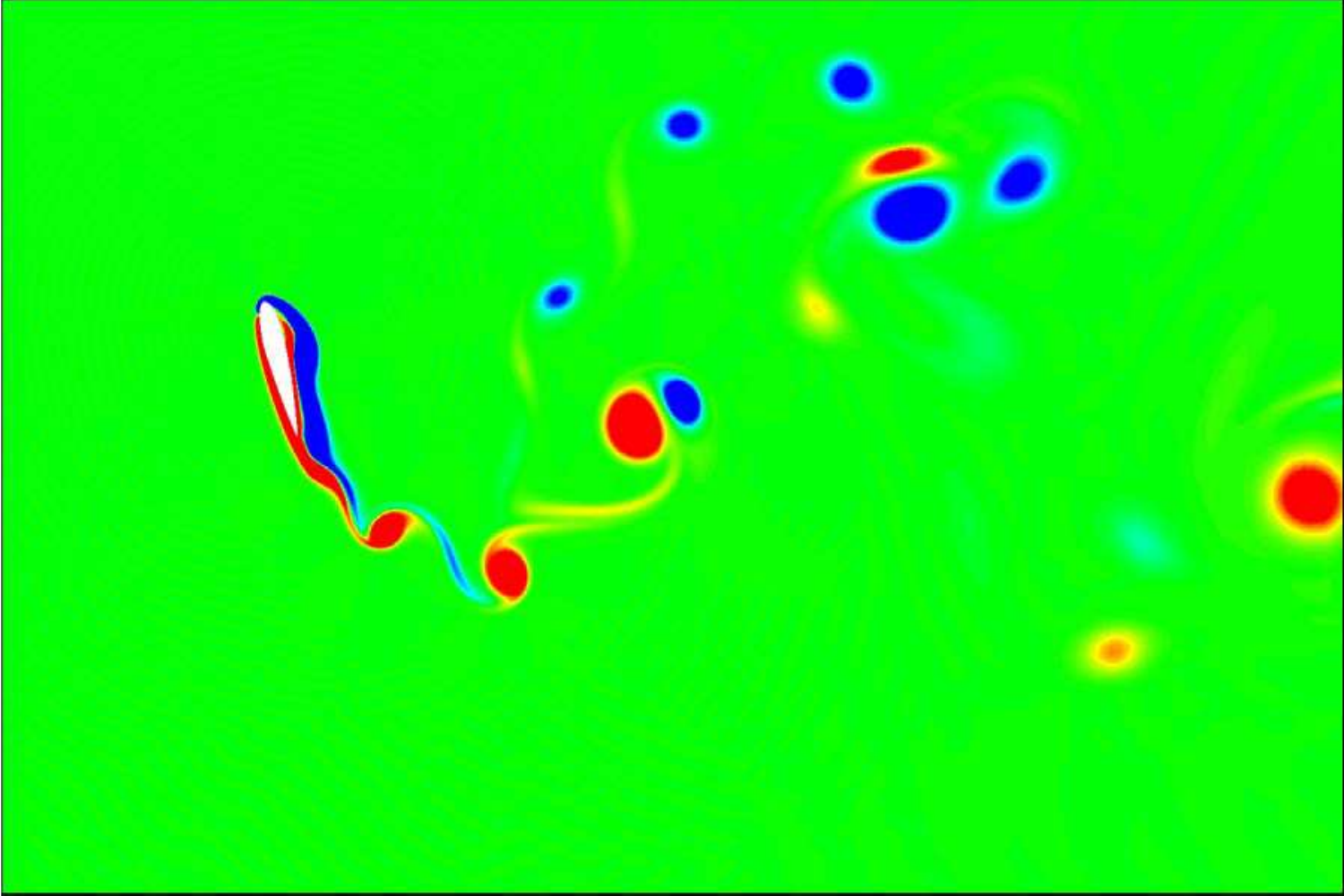}\\
   \hskip-5.0in (b)	\\
  \includegraphics[width=1.5in]{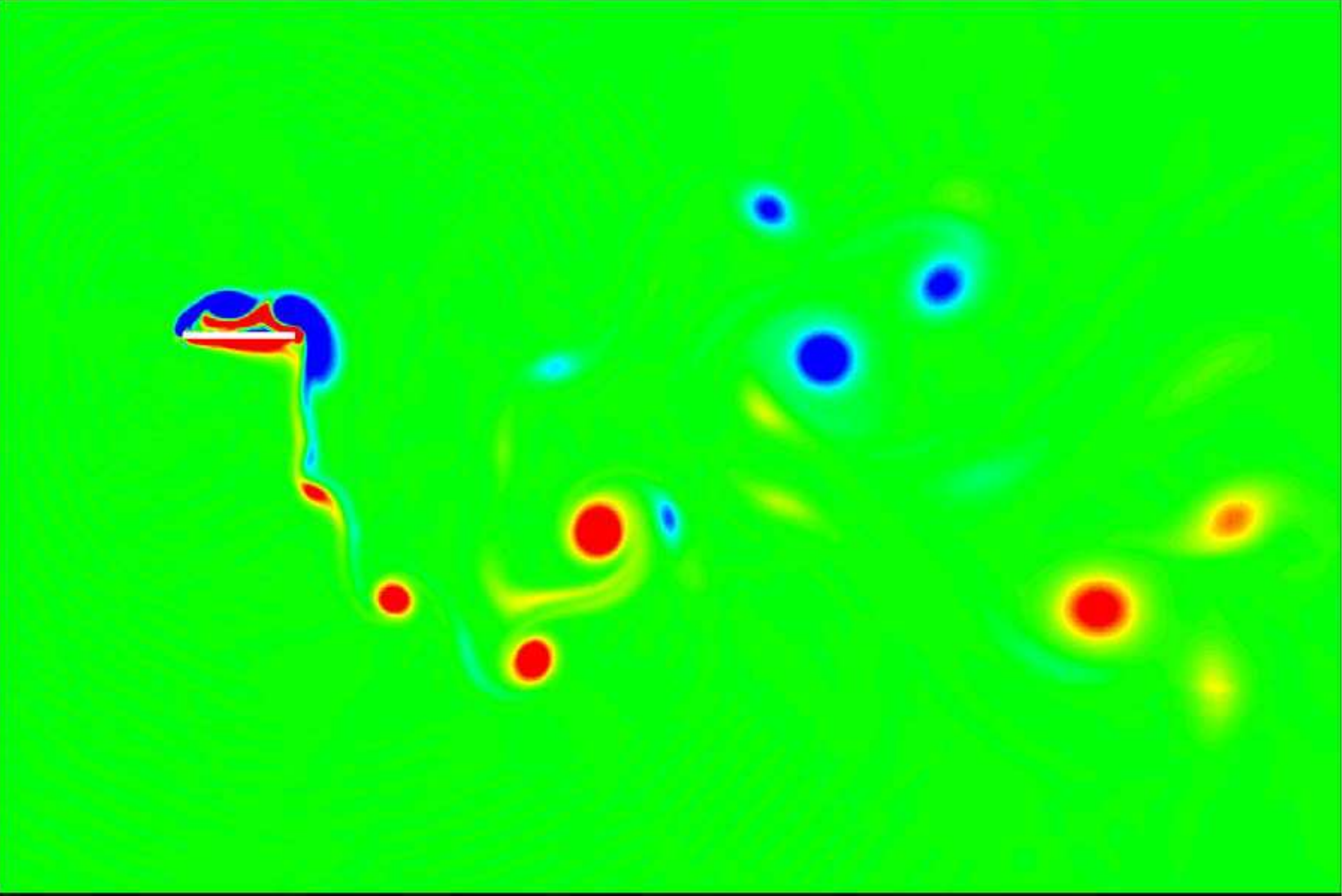}
  \includegraphics[width=1.5in]{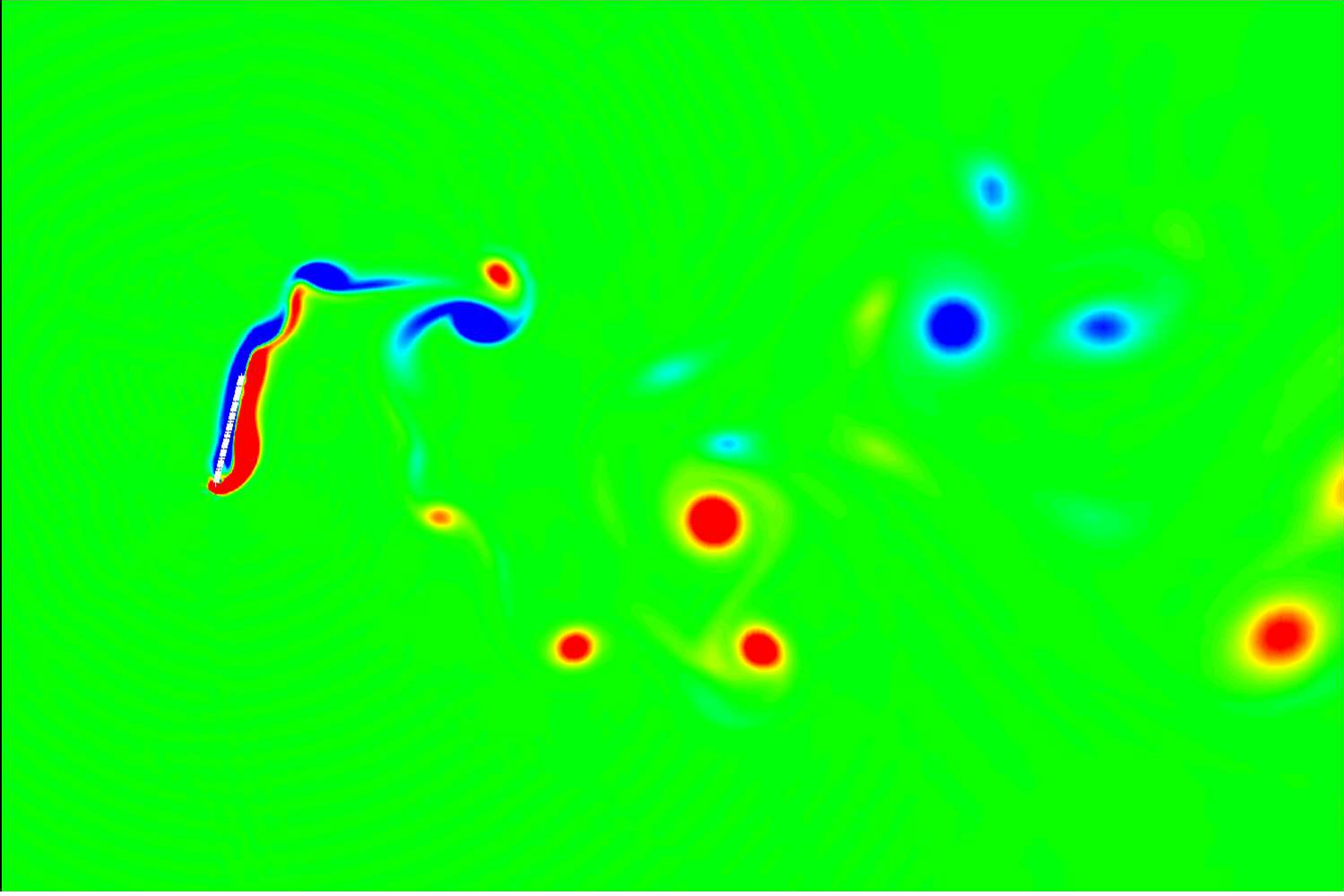}
  \includegraphics[width=1.5in]{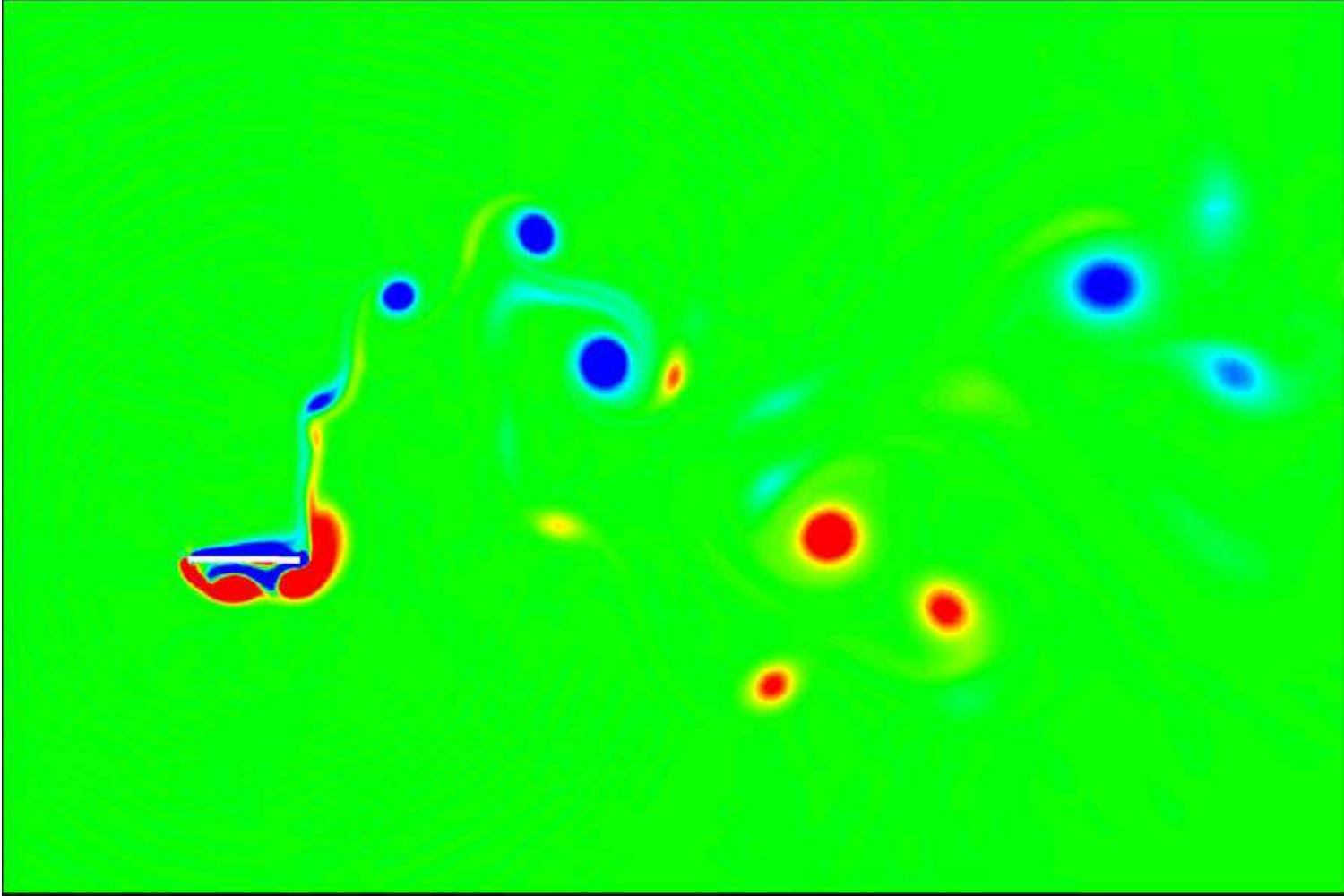}
  \includegraphics[width=1.5in]{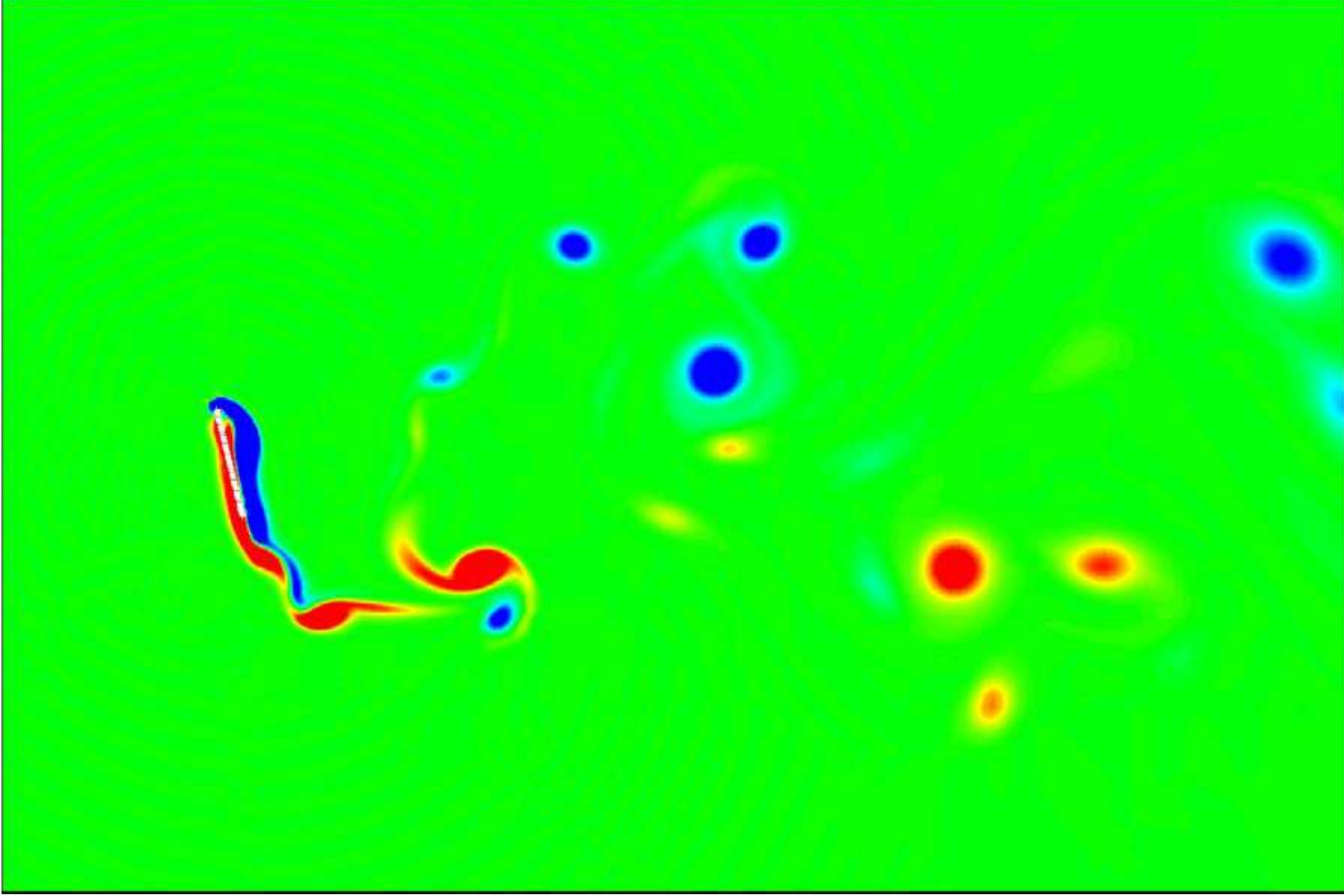}\\
  \hskip-5.0in (c)	\\
  \includegraphics[width=1.5in]{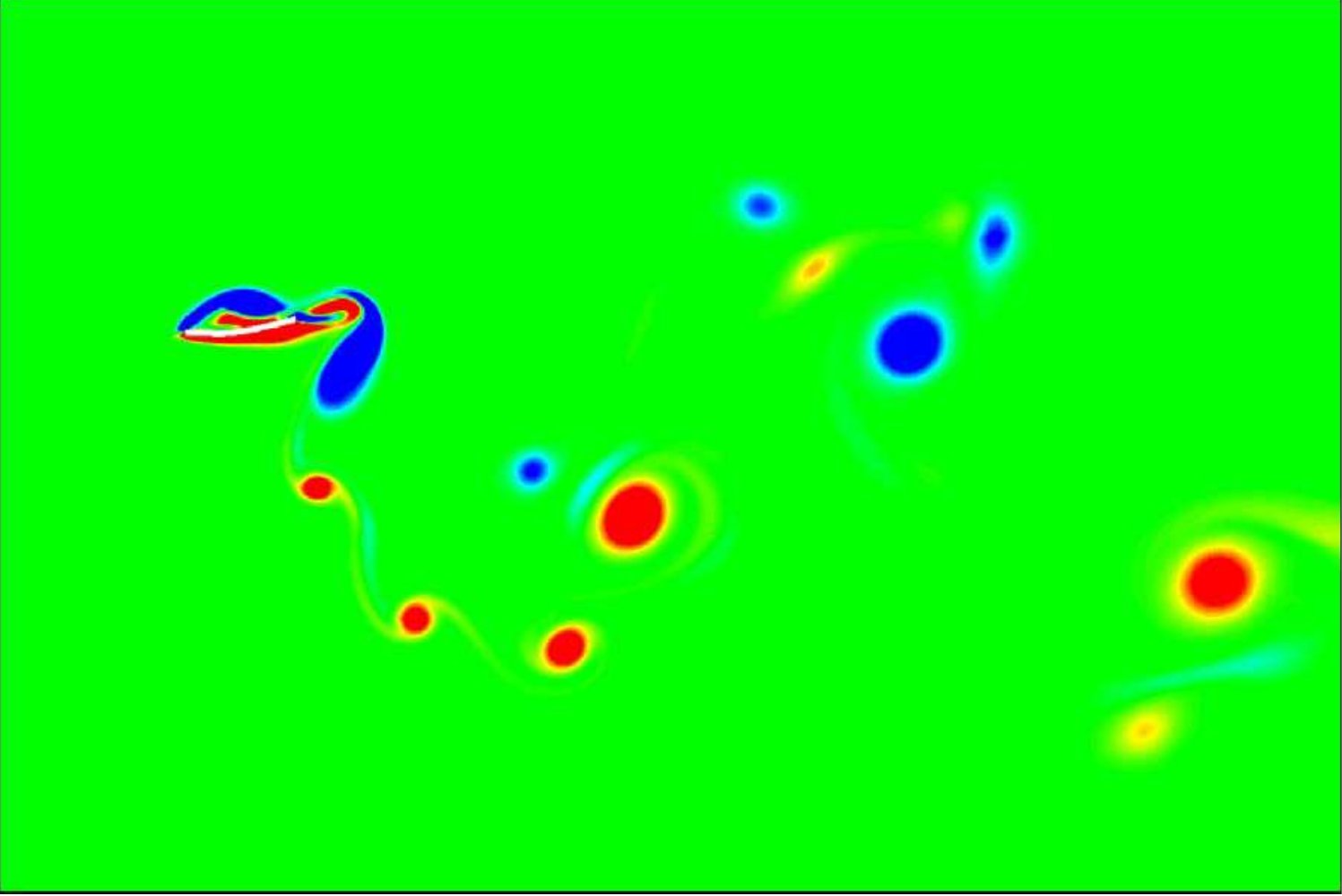}
  \includegraphics[width=1.5in]{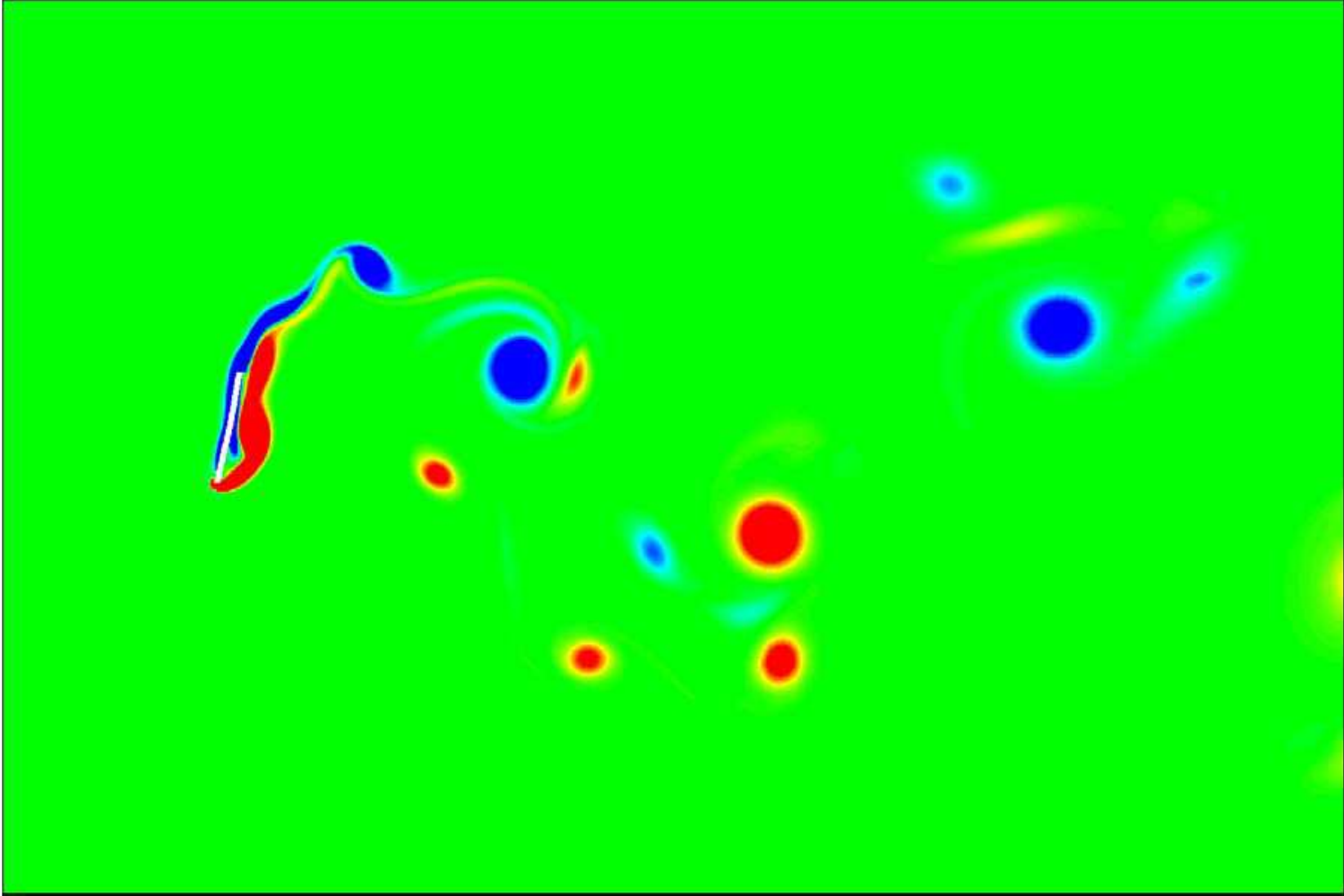}
  \includegraphics[width=1.5in]{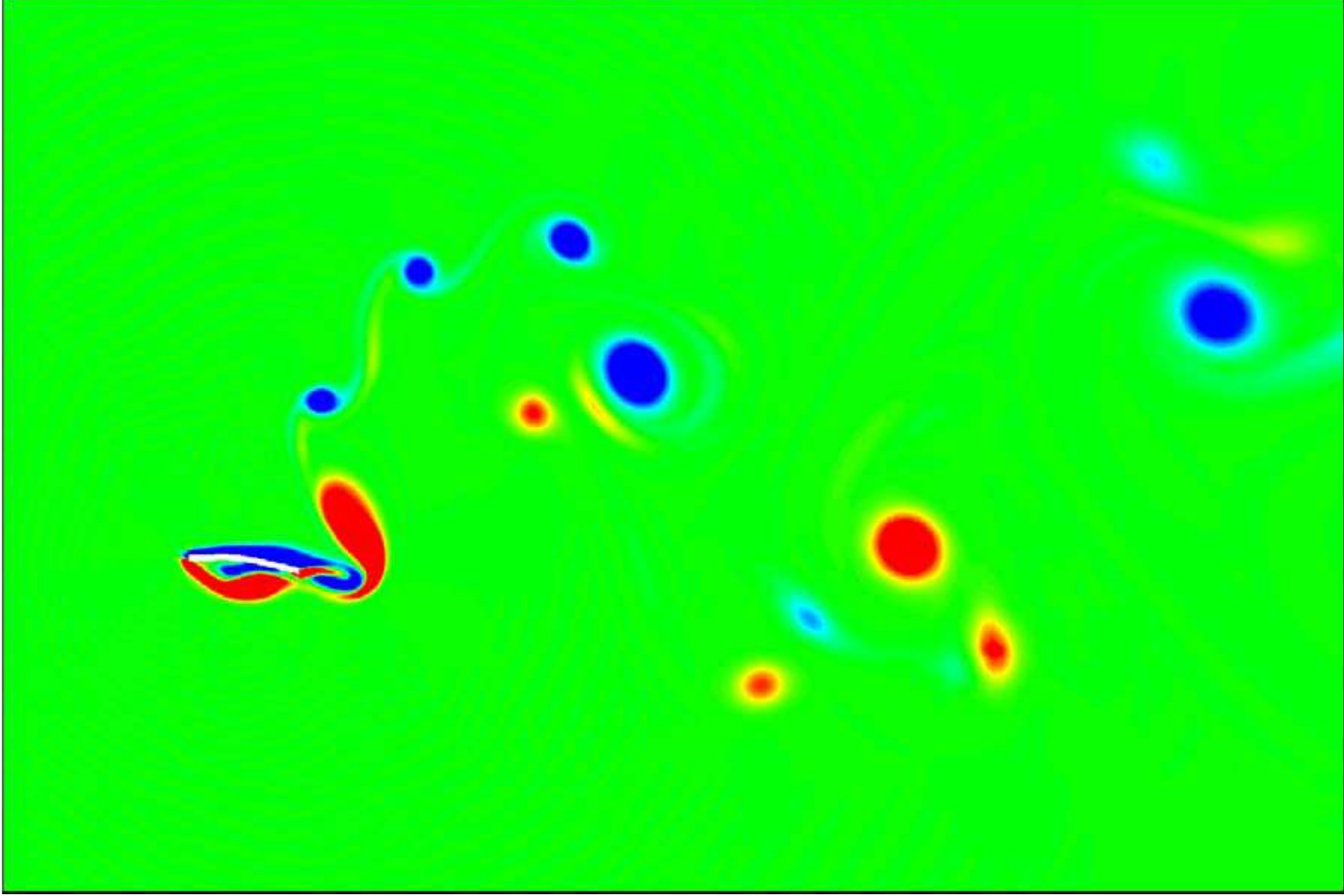}
  \includegraphics[width=1.5in]{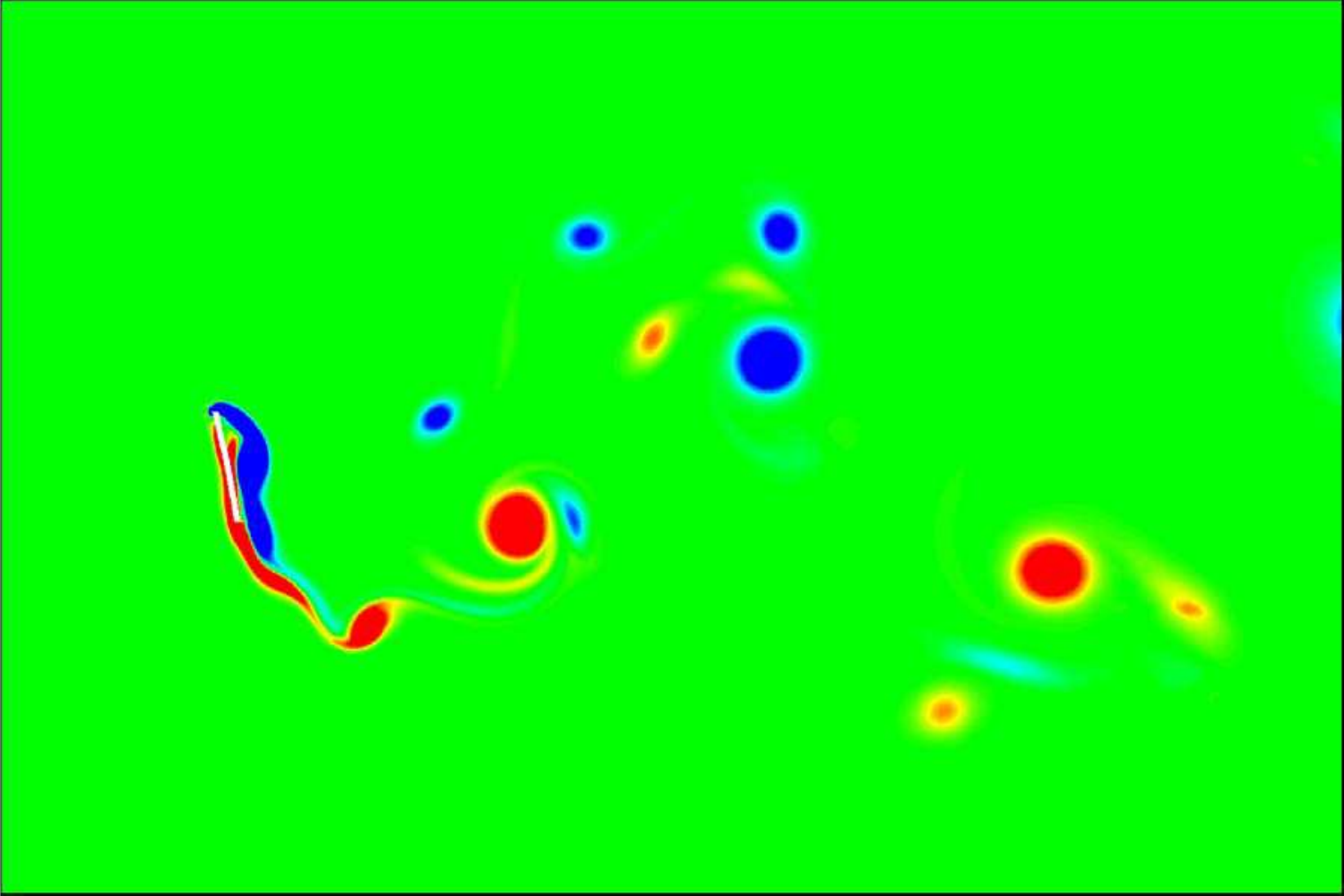}
  \end{center}
\caption{Flapping foil energy harvester: instantaneous vorticity contours in a period with an interval of $T/4$: (a) NACA0015 foil, (b) rigid plate and (c) flexible plate. Here, Re=1100, $M=0.1$, $\alpha_m=152.6^o$, $A_0/L=2.0$ and $\beta=90^o$. The contour level of the vorticity ranges from $-5U_0 /D$ (blue) to $5U_0 /D$ (red).}
\label{Fig:naca_vortex}
\end{figure}

Rigid and flexible plates without considering the thickness effect are also numerically simulated for comparison. All the non-dimensional parameters for the rigid plate are same as that for the NACA0015 foil. For the flexible plate, the mass ratio defined in Eq.~\ref{eq:flapping_para} is 5.0, the non-dimensional bending rigidity of the flexible plate defined as $E^*=E_B/(\rho_f U_0^2 L^3)$ is $1.0$. Fig.~\ref{Fig:naca_flex_clcd} presents the comparison of the drag, lift and power coefficients in a period obtained from the NACA0015 foil and the two plates. It is found that the tendencies of $C_D$, $C_L$ and $C_p$ are consistent, though small discrepancies are observed. Specially, Fig.~\ref{Fig:naca_flex_clcd} (a) shows that the plates induce larger drag than NACA0015 foil, and the flexibility of the plate also increases the drag coefficient. Fig.~\ref{Fig:naca_flex_clcd} (b) shows that the lift generation of the rigid plate and NACA0015 is very similar, the flexibility tends to generate lower lift amplitude. The mean power coefficient (defined by $-\bar{C}_p$) is generally used to evaluate the efficiency of the foil in energy harvester. Here, the power coefficients in five periods are used to calculate $-\bar{C}_p$. The mean power coefficient of the NACA0015 foil and rigid plate are 0.85 and 0.82, respectively, compared with that of only 0.58 for the flexible plate. The results show that NACA0015 foil is the most efficient in energy harvesting among the three foils. The flexibility of the foil in this case deteriorates the energy harvesting efficiency. The instantaneous vorticity contours in a period are presented in Fig.~\ref{Fig:naca_vortex}. Similar vortex fields can be observed  for the three cases, without remarkable differences.

\begin{figure}
  \begin{center}
  \hskip-1.8in (a) \hskip1.8in (b) \hskip1.8in (c)

  \includegraphics[width=2.0in]{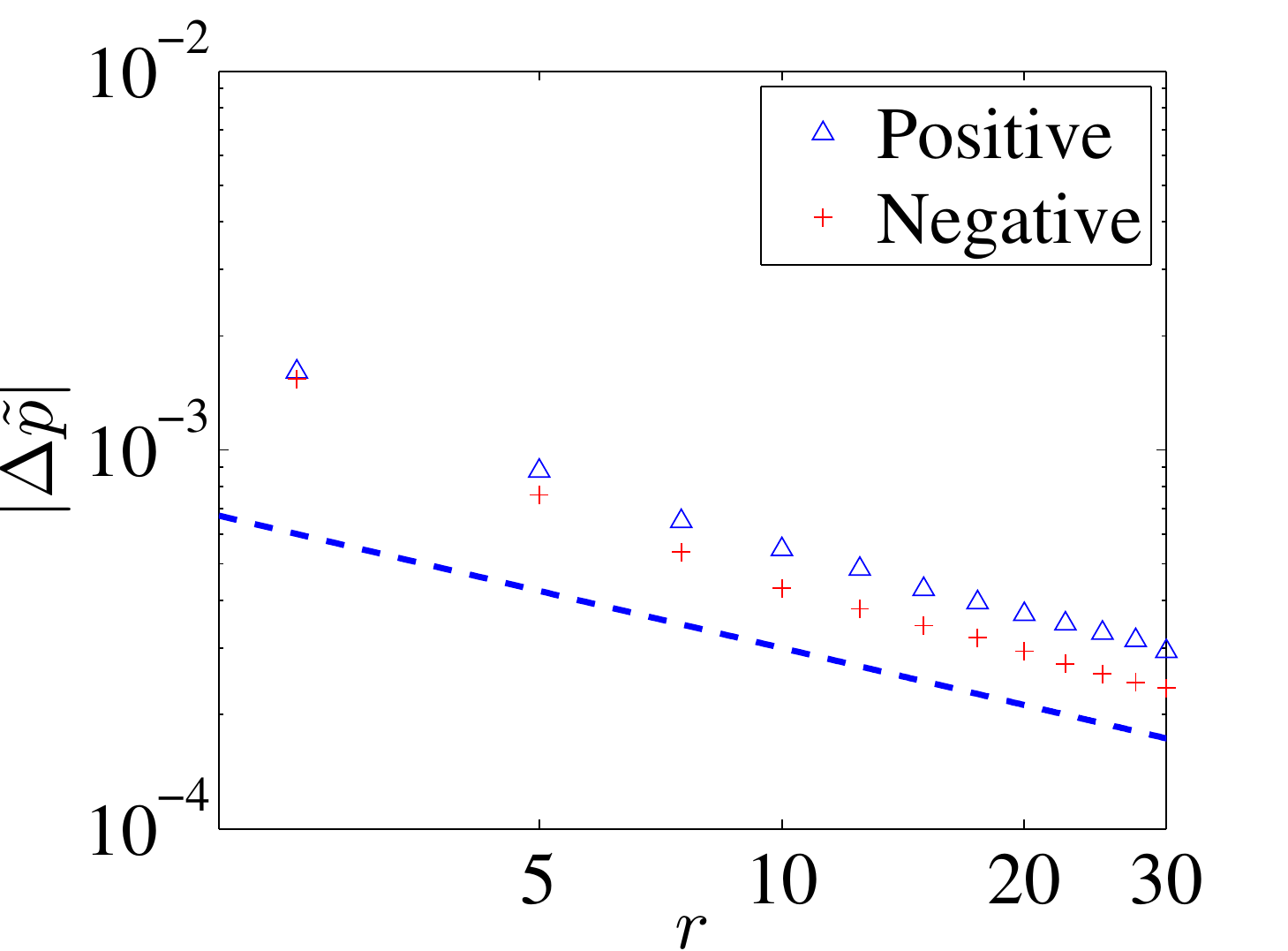}
  \hskip0.1in
  \includegraphics[width=2.0in]{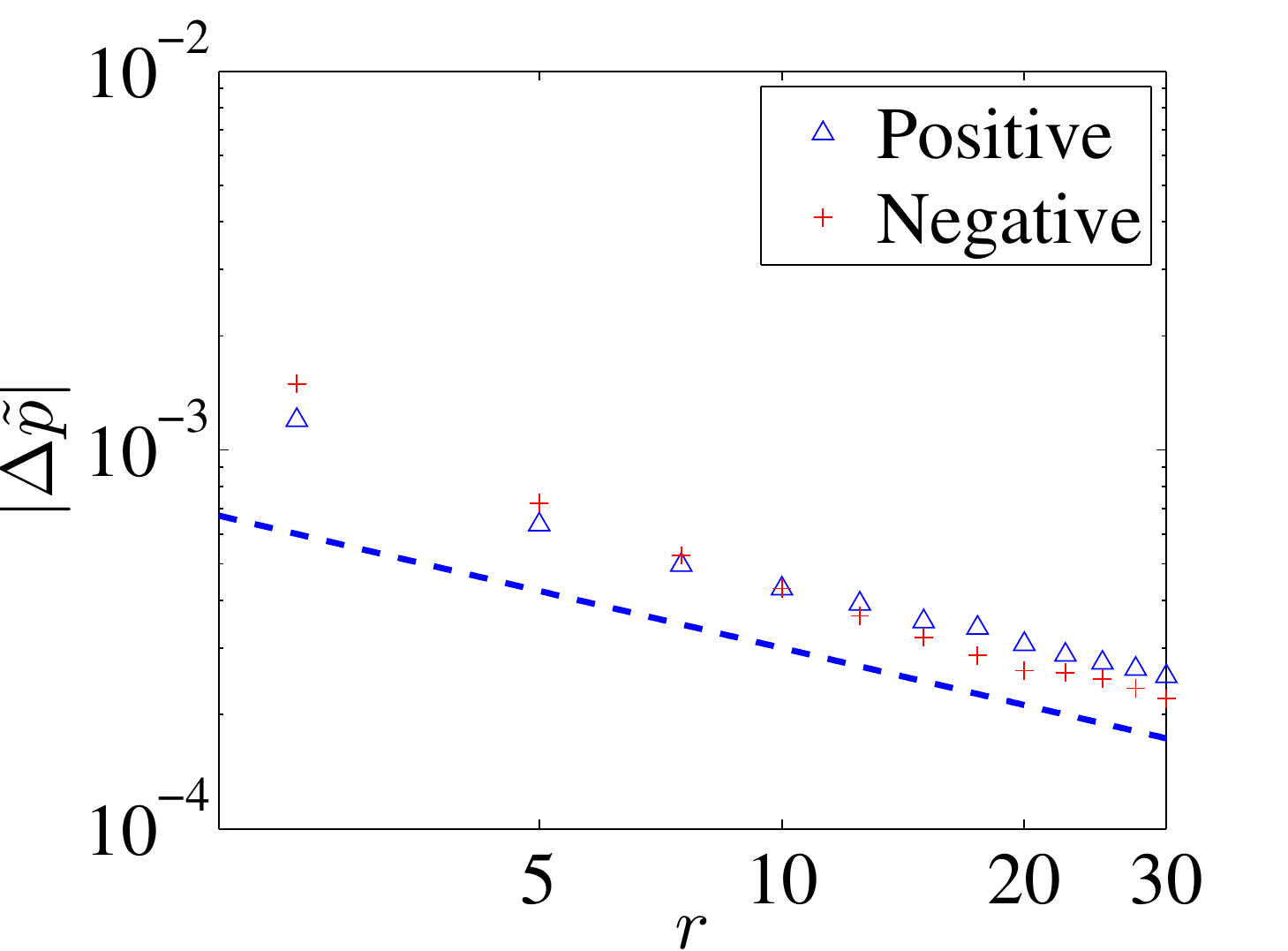}
  \hskip0.1in
  \includegraphics[width=2.0in]{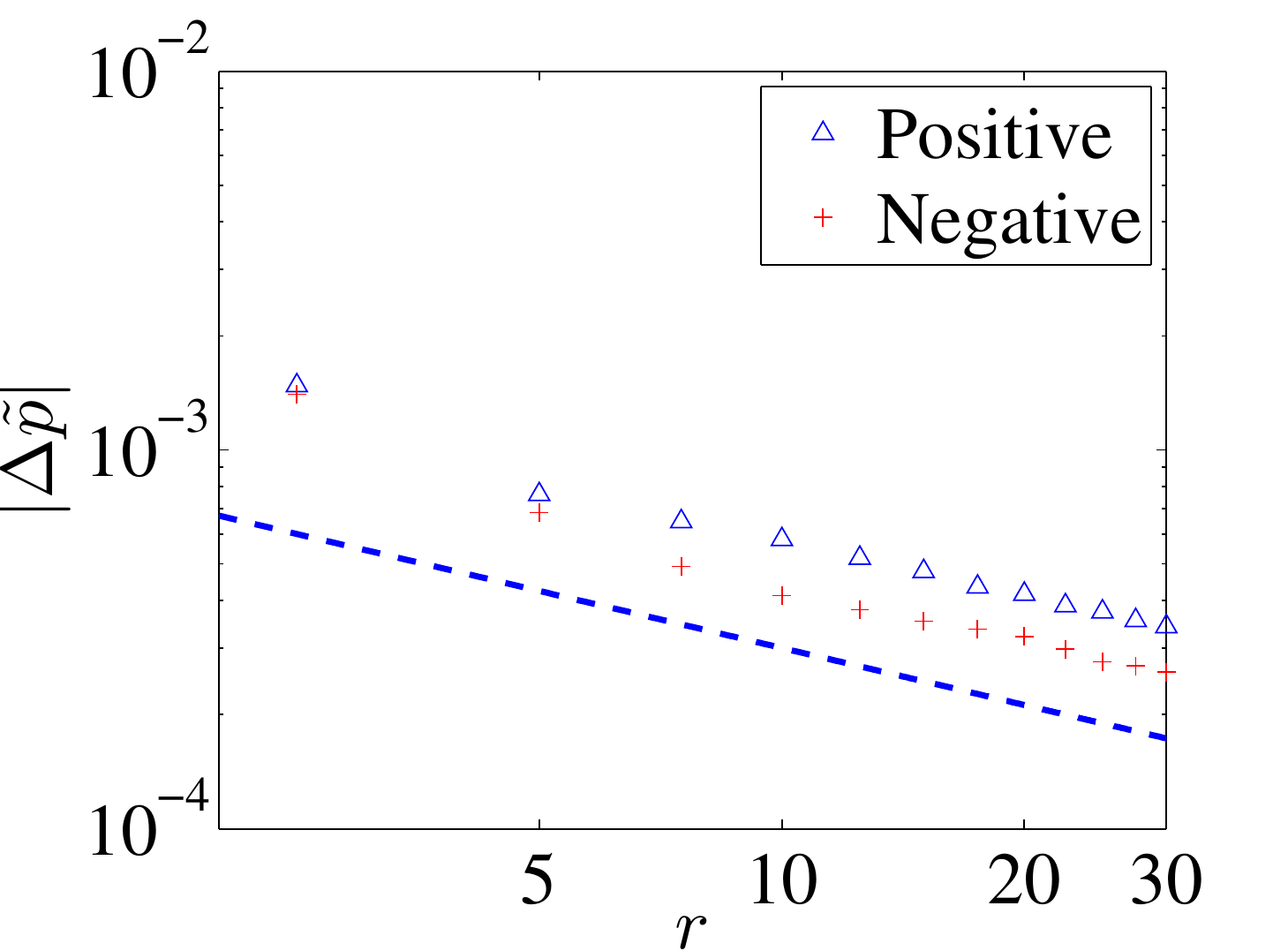}\\
  \end{center}
\caption{Flapping foil energy harvester: decay of pressure peaks measured at $\theta=90^o$: (a) NACA0015 foil, (b) rigid plate and (c) flexible plate. Here, Re=1100, $M=0.1$, $\alpha_m=152.6^o$, $A_0/L=2.0$ and $\beta=90^o$. The dashed line indicates that the pressure peaks tend to decay in proportion to $r^{-\frac{1}{2}}$.}
\label{Fig:naca_decay}
\end{figure}

Fig.~\ref{Fig:naca_decay} presents the decay of the fluctuating pressure (defined in Section 4.1) measured at $\theta=90^o$. It shows that the fluctuating pressure decays in proportion to the $r^{-\frac{1}{2}}$ in the intermediate and far fields, which agrees well with the theoretical result~\cite{Landau1987fluidmech}. In all cases, the positive fluctuating pressure is slightly larger than the negative one, which may indicate the loading process is stronger than the vortices shedding effects on the fluctuating pressure generation at the measured point. The comparison of Fig.~\ref{Fig:naca_decay} (b) and (c) shows that the flexibility of the plate increases the positive pressure at the measured point, which can be reasonably explained by the larger tip-displacement of the flexible plate.

\begin{figure}
  \begin{center}
  \hskip-1.8in (a) \hskip1.8in (b) \hskip1.8in (c)

  \includegraphics[width=2.0in]{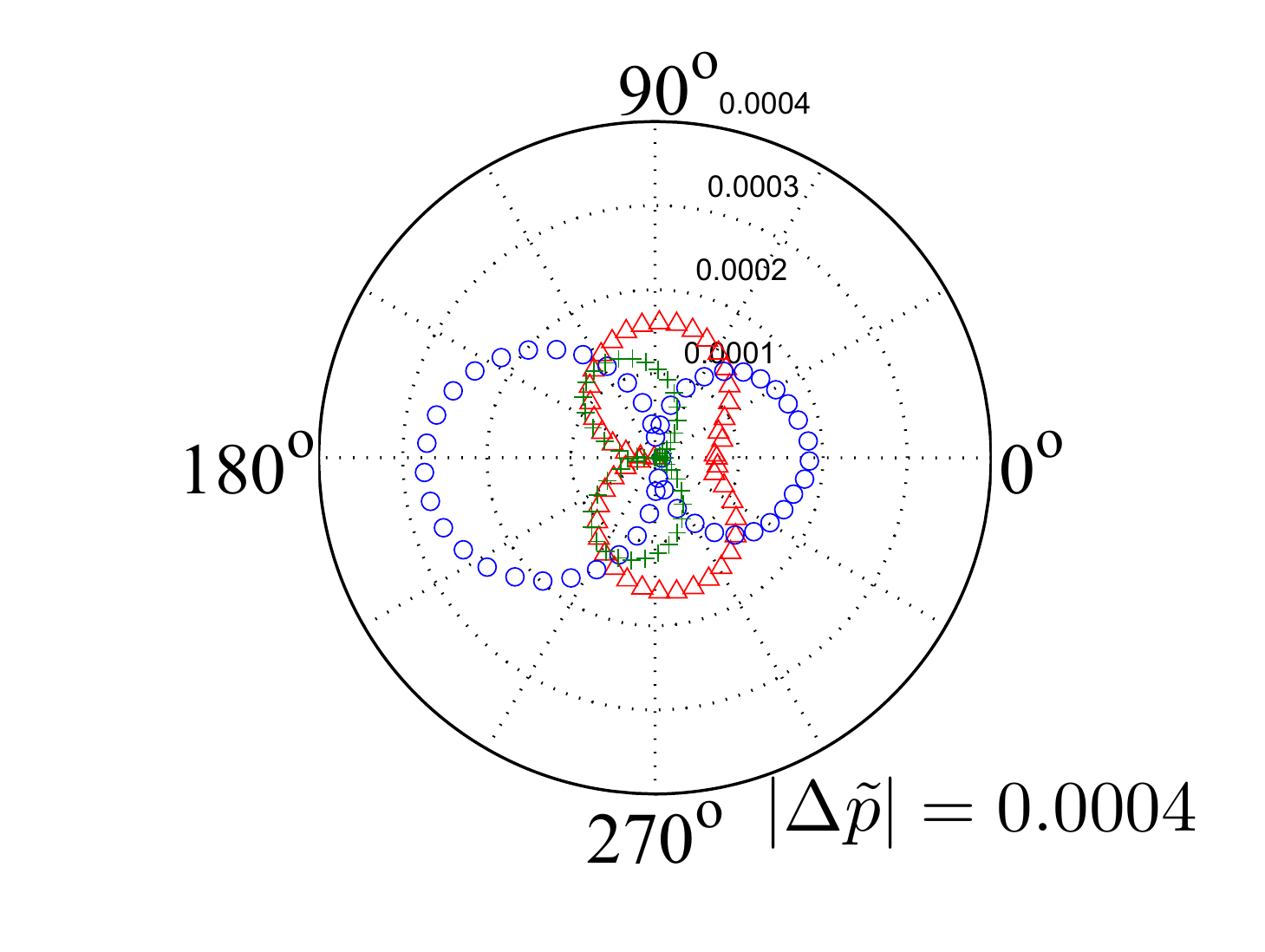}
  \hskip0.1in
  \includegraphics[width=2.0in]{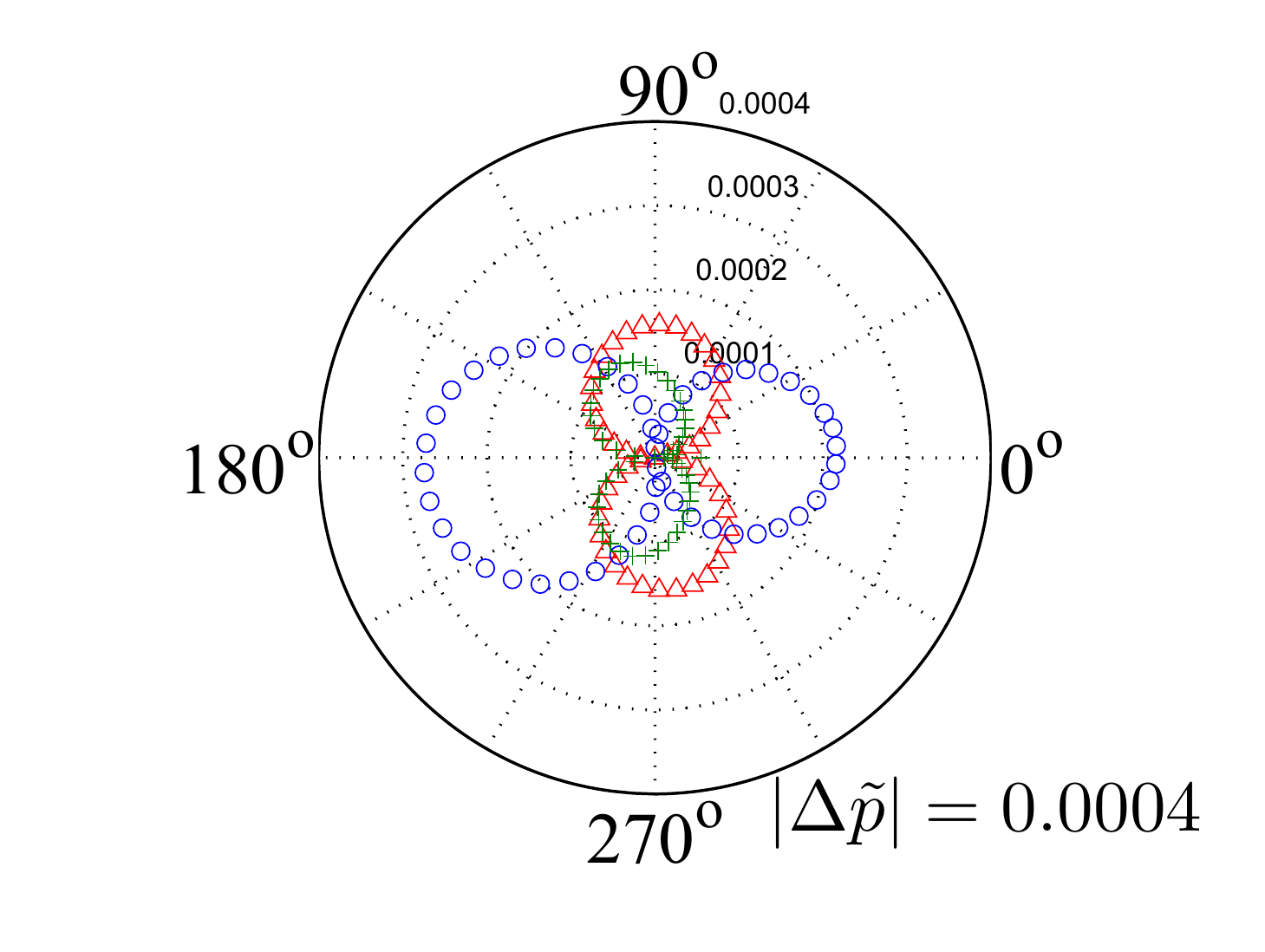}
  \hskip0.1in
  \includegraphics[width=2.0in]{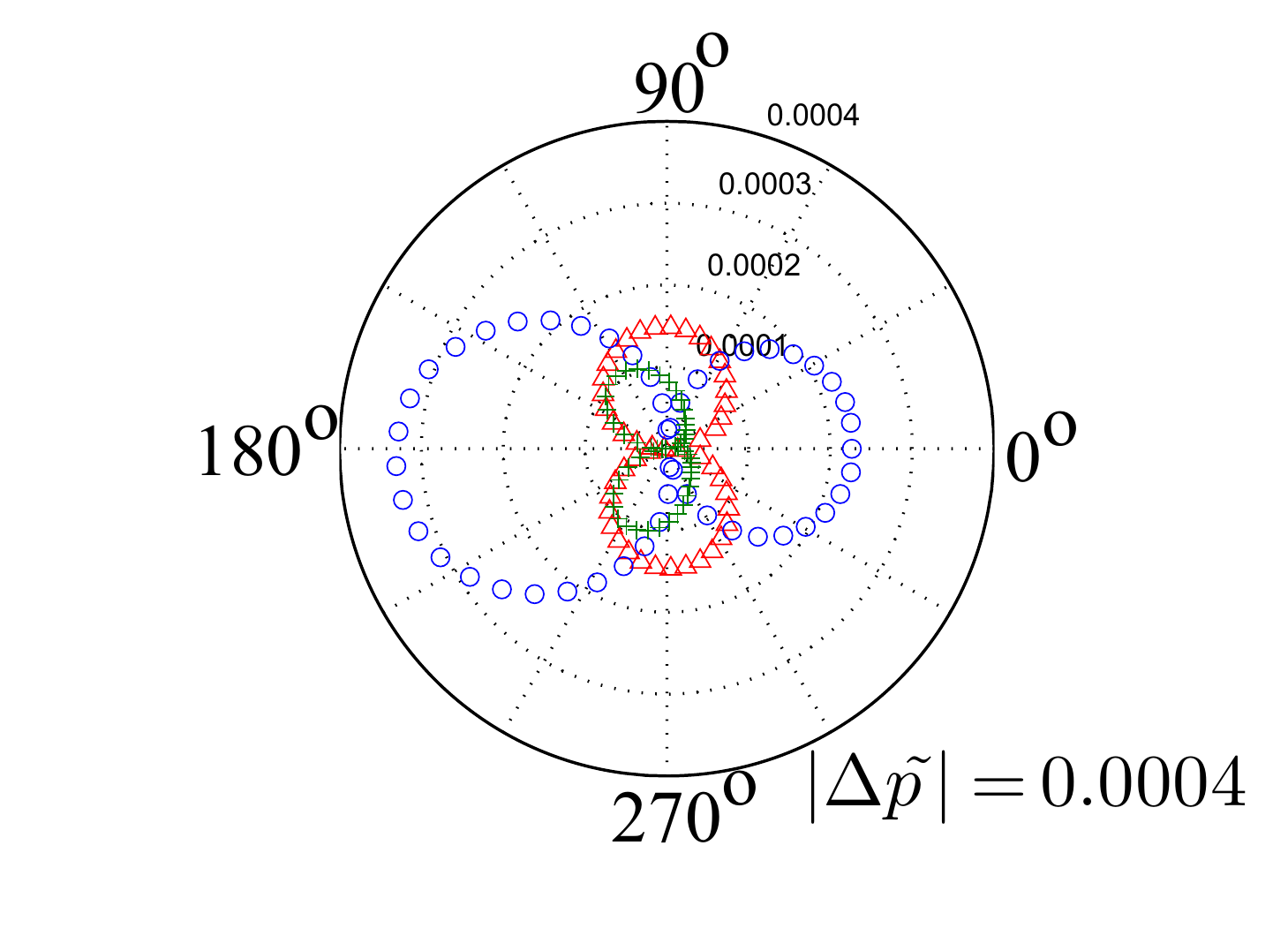}\\
  \end{center}
\caption{Flapping foil energy harvester: polar diagram of the fluctuating pressure peaks at $r=20$ with different frequencies: (a) NACA0015 foil, (b) rigid plate and (c) flexible plate. Here, Re=1100, $M=0.1$, $\alpha_m=152.6^o$, $A_0/L=2.0$ and $\beta=90^o$. Where, $\Delta$, o and $+$ denote the frequency of $f$, $2f$ and $3f$, respectively.}
\label{Fig:naca_freq}
\end{figure}

\begin{figure}
 \begin{center}
 \hskip-1.8in (a) \hskip1.8in (b) \hskip1.8in (c)	
 
  \includegraphics[width=2.0in]{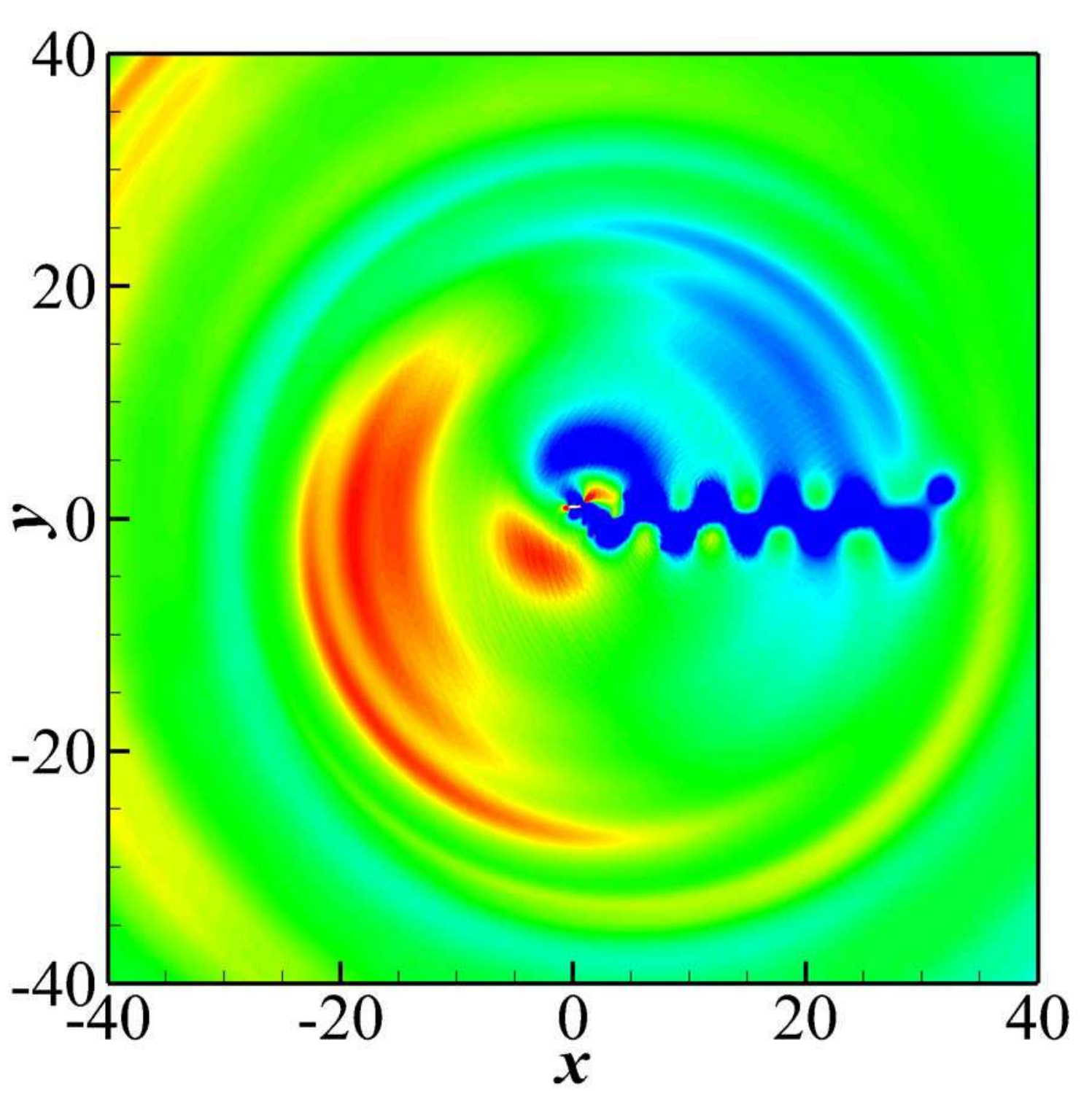}
  \includegraphics[width=2.0in]{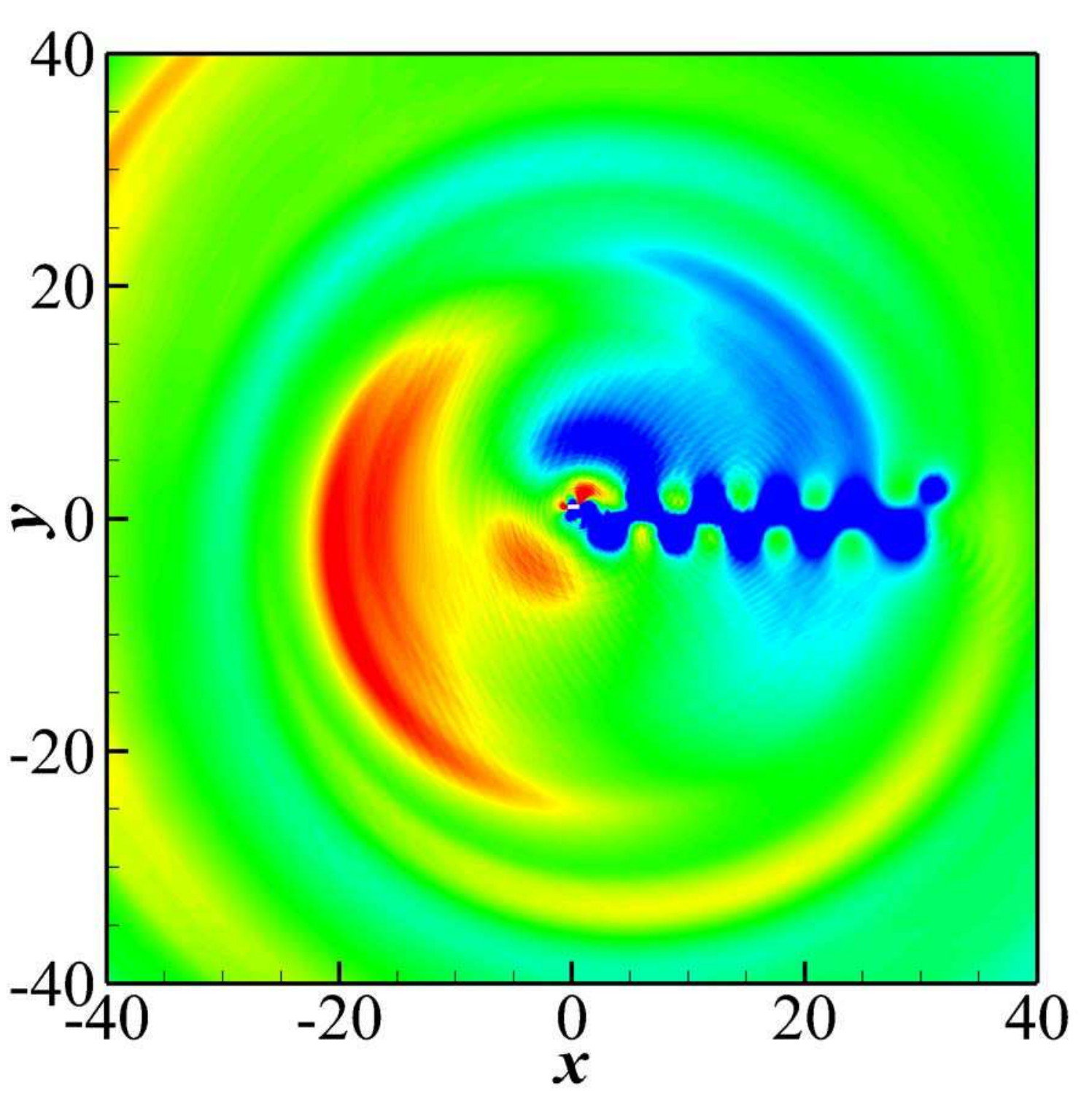}
  \includegraphics[width=2.0in]{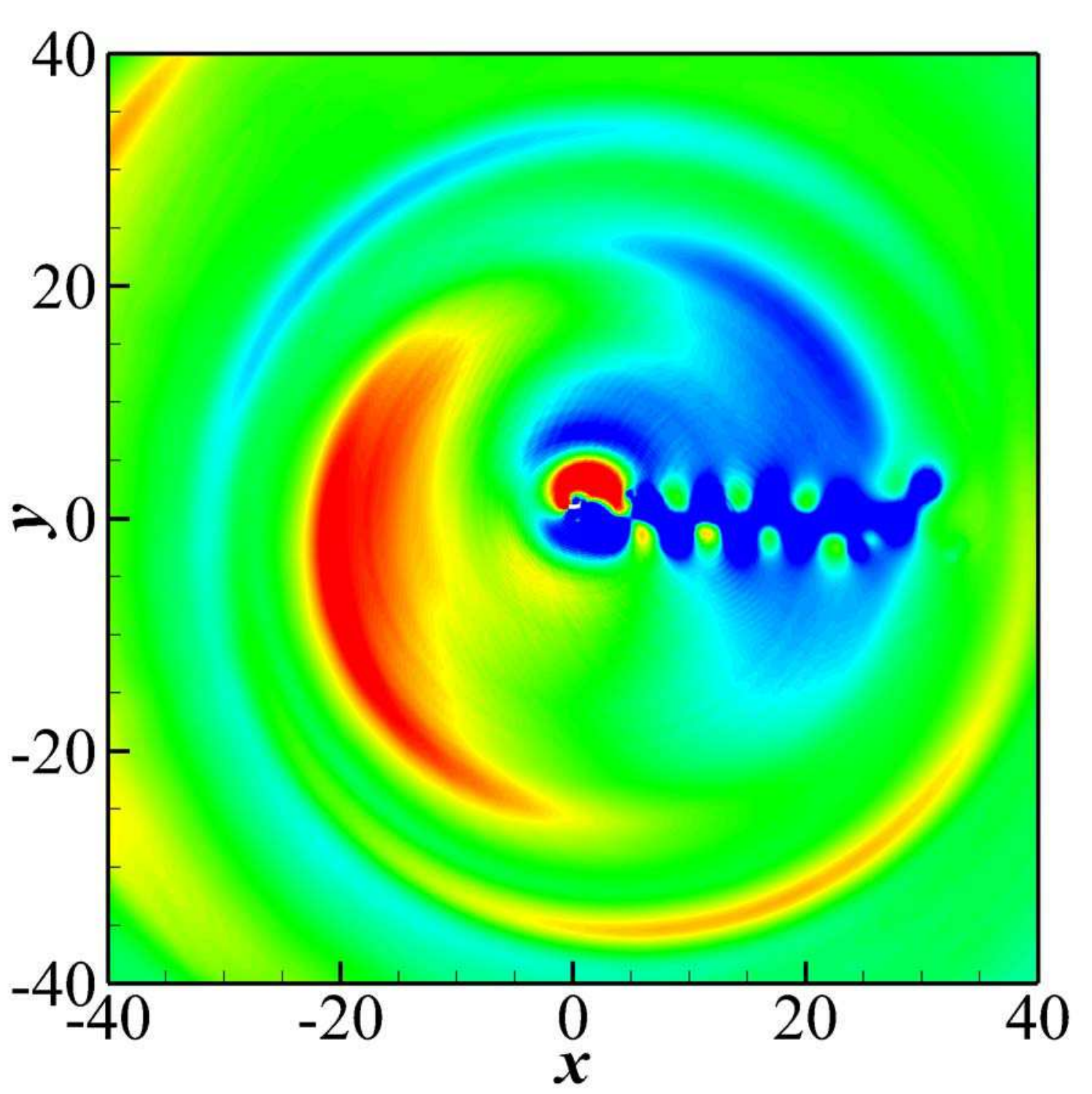}\\
  
  \includegraphics[width=2.0in]{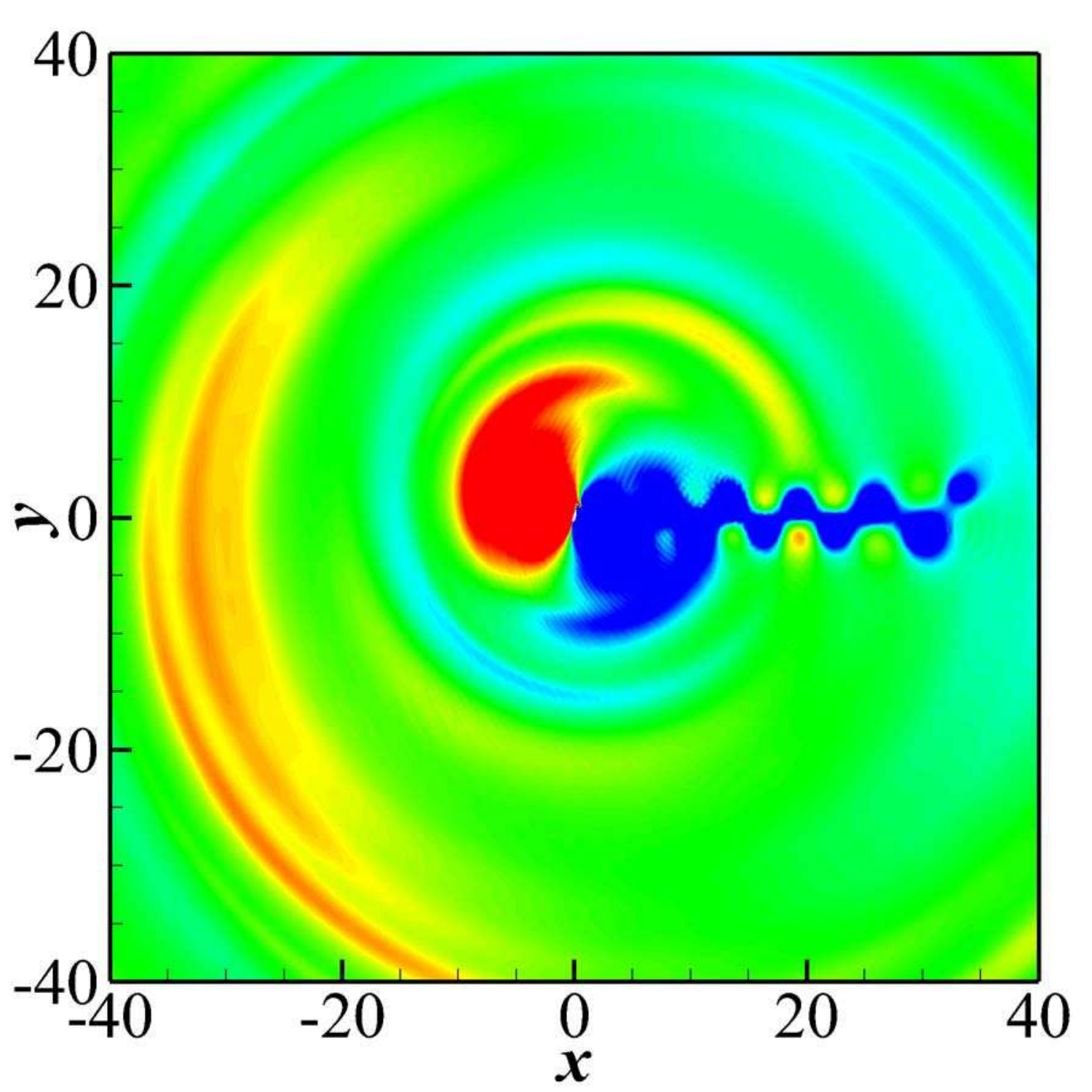}
  \includegraphics[width=2.0in]{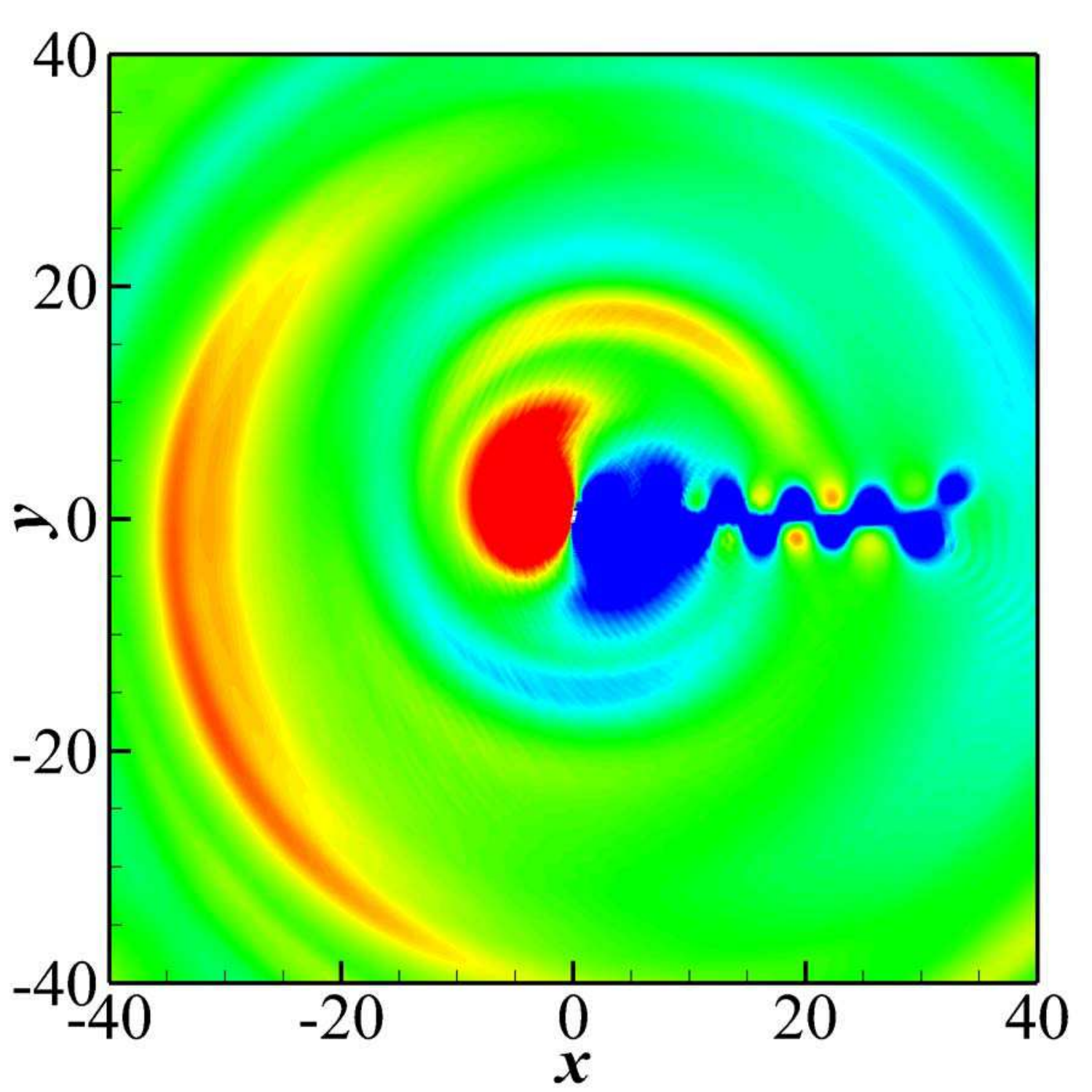}
  \includegraphics[width=2.0in]{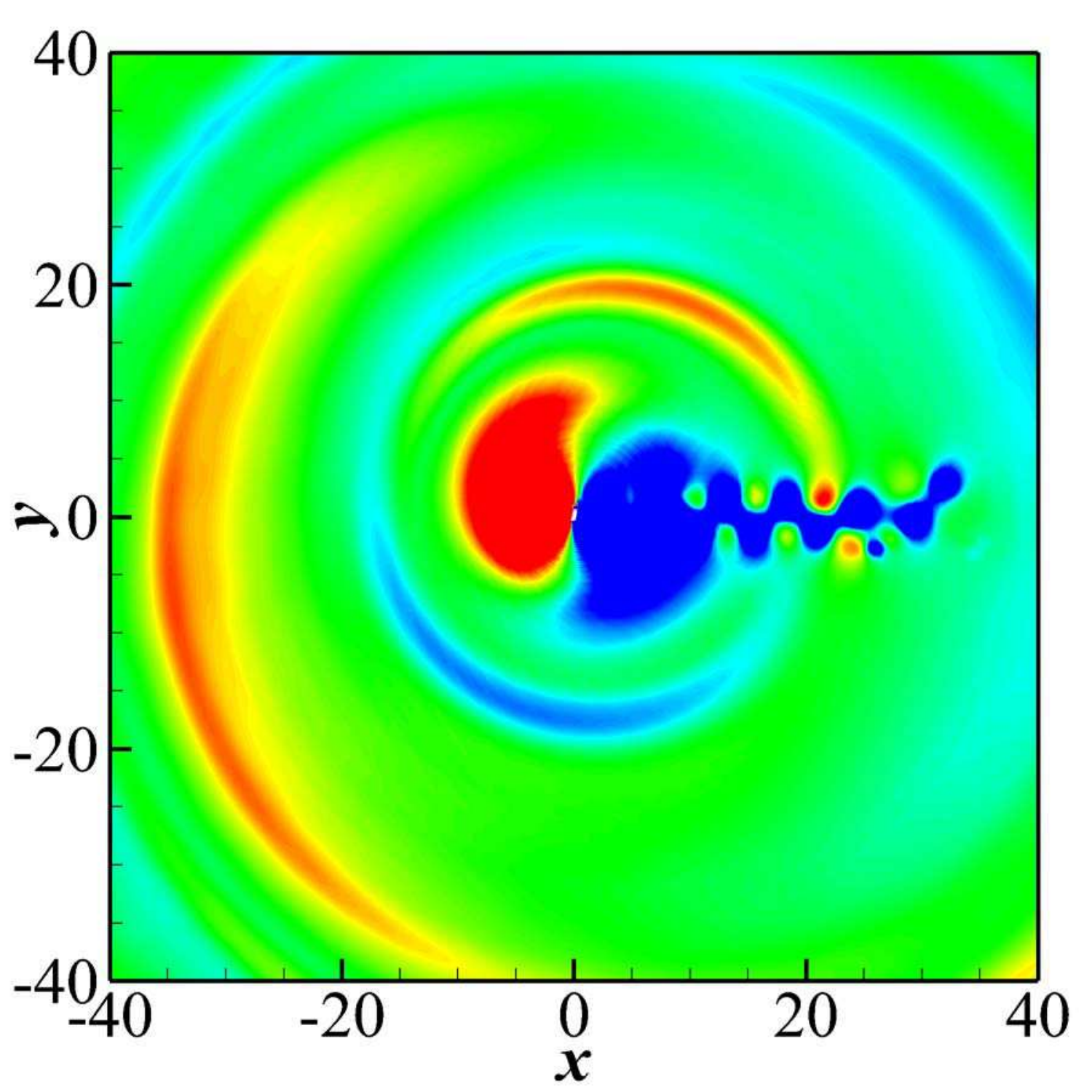}\\
  
  \includegraphics[width=2.0in]{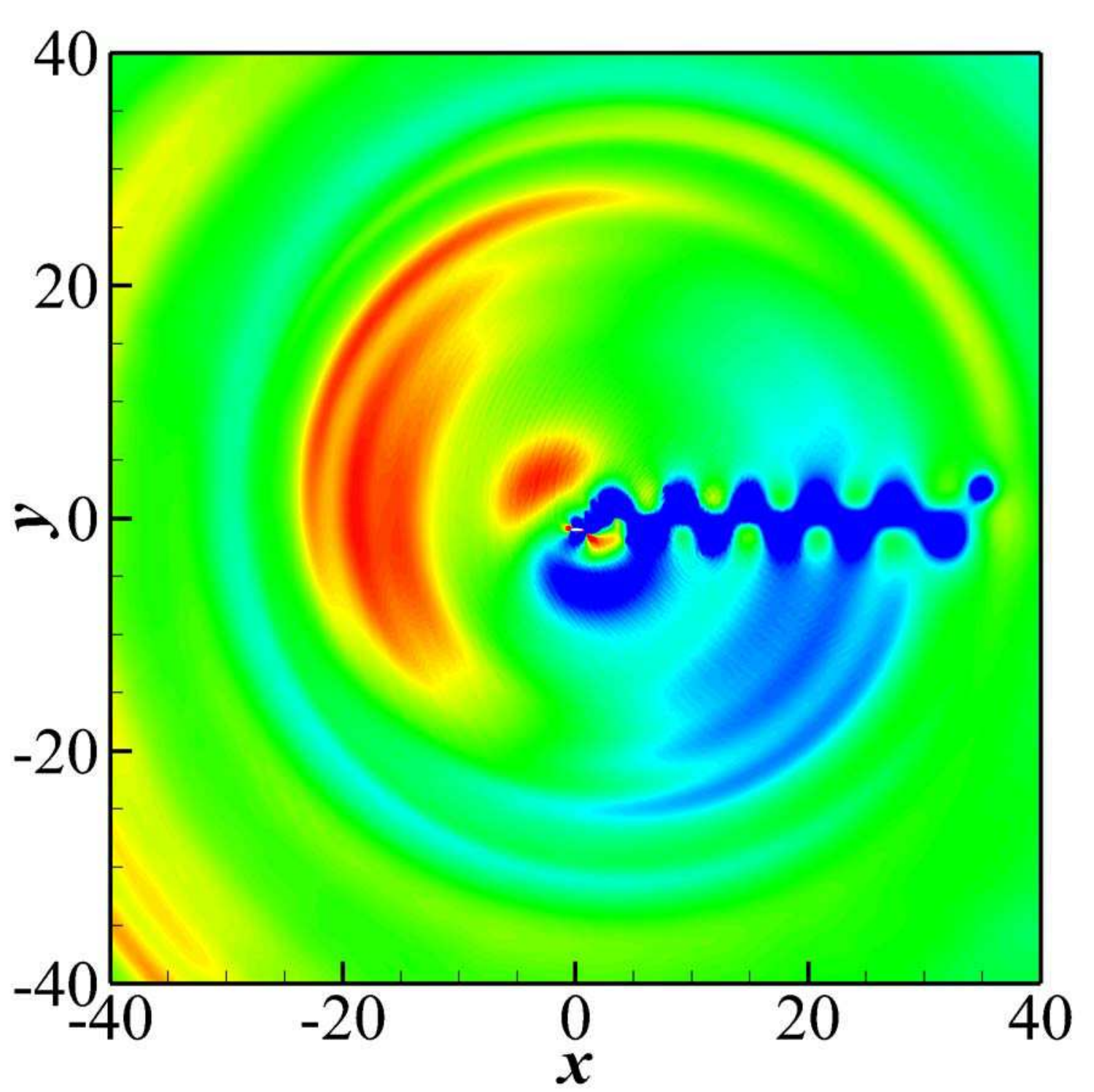}
  \includegraphics[width=2.0in]{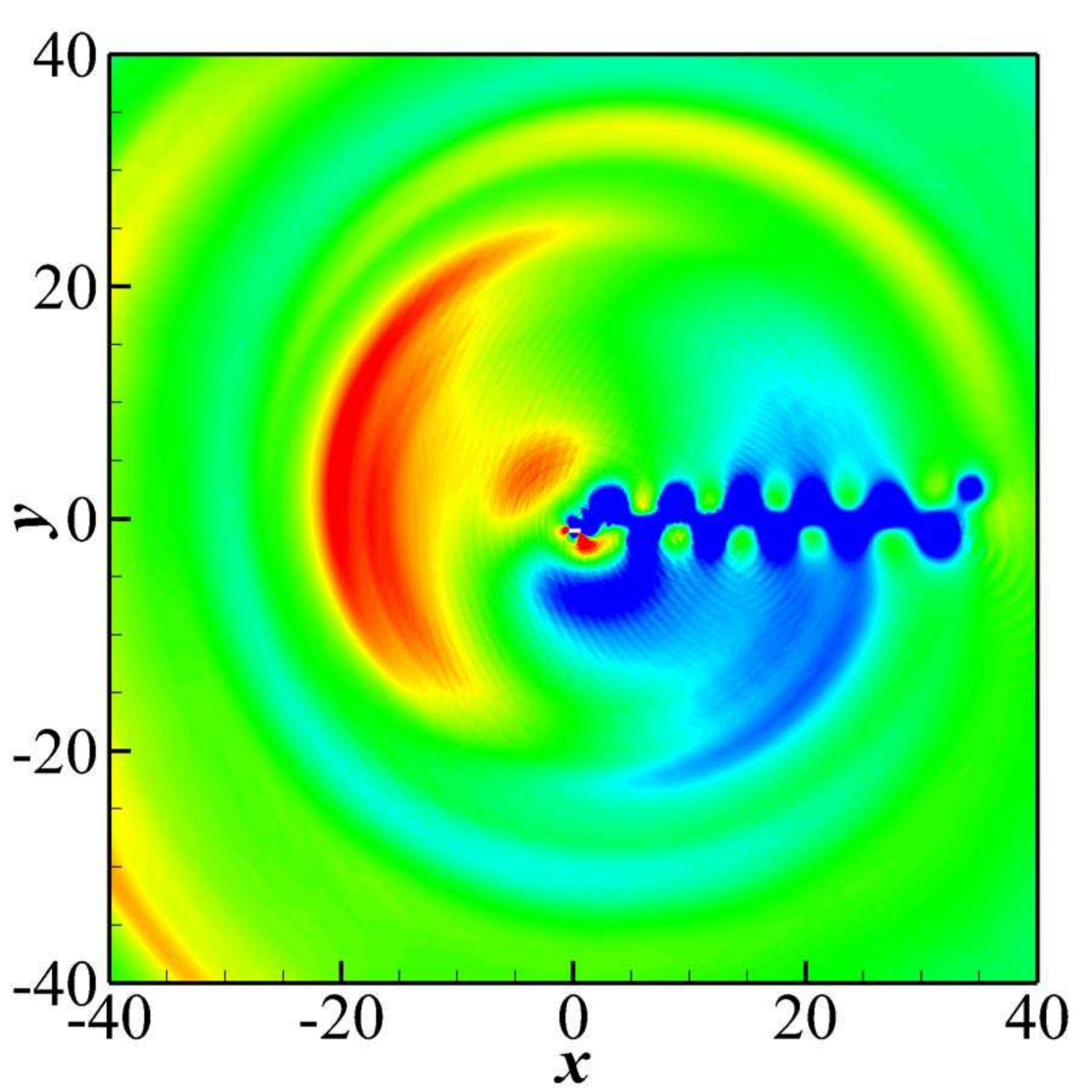}
  \includegraphics[width=2.0in]{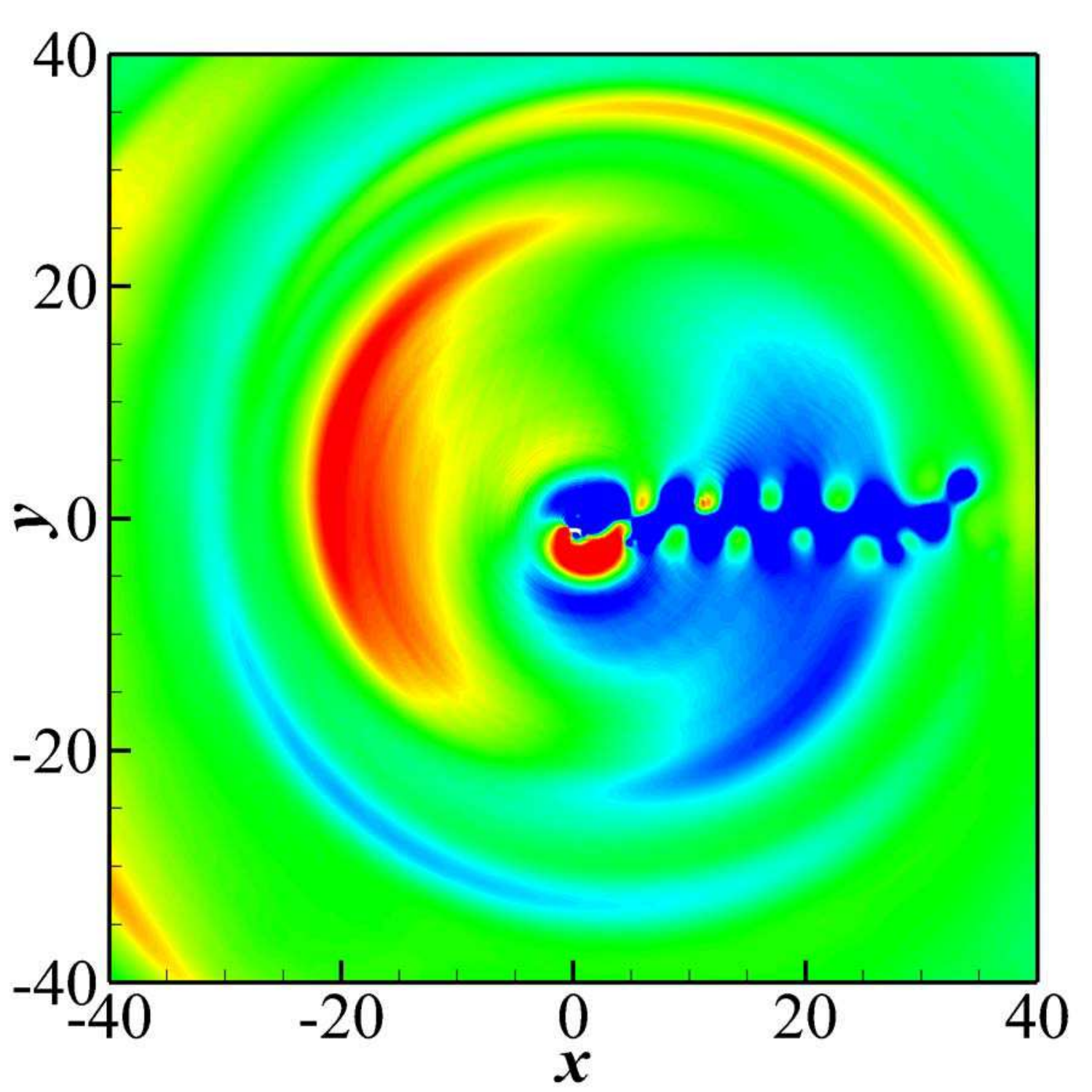}\\
  
  \includegraphics[width=2.0in]{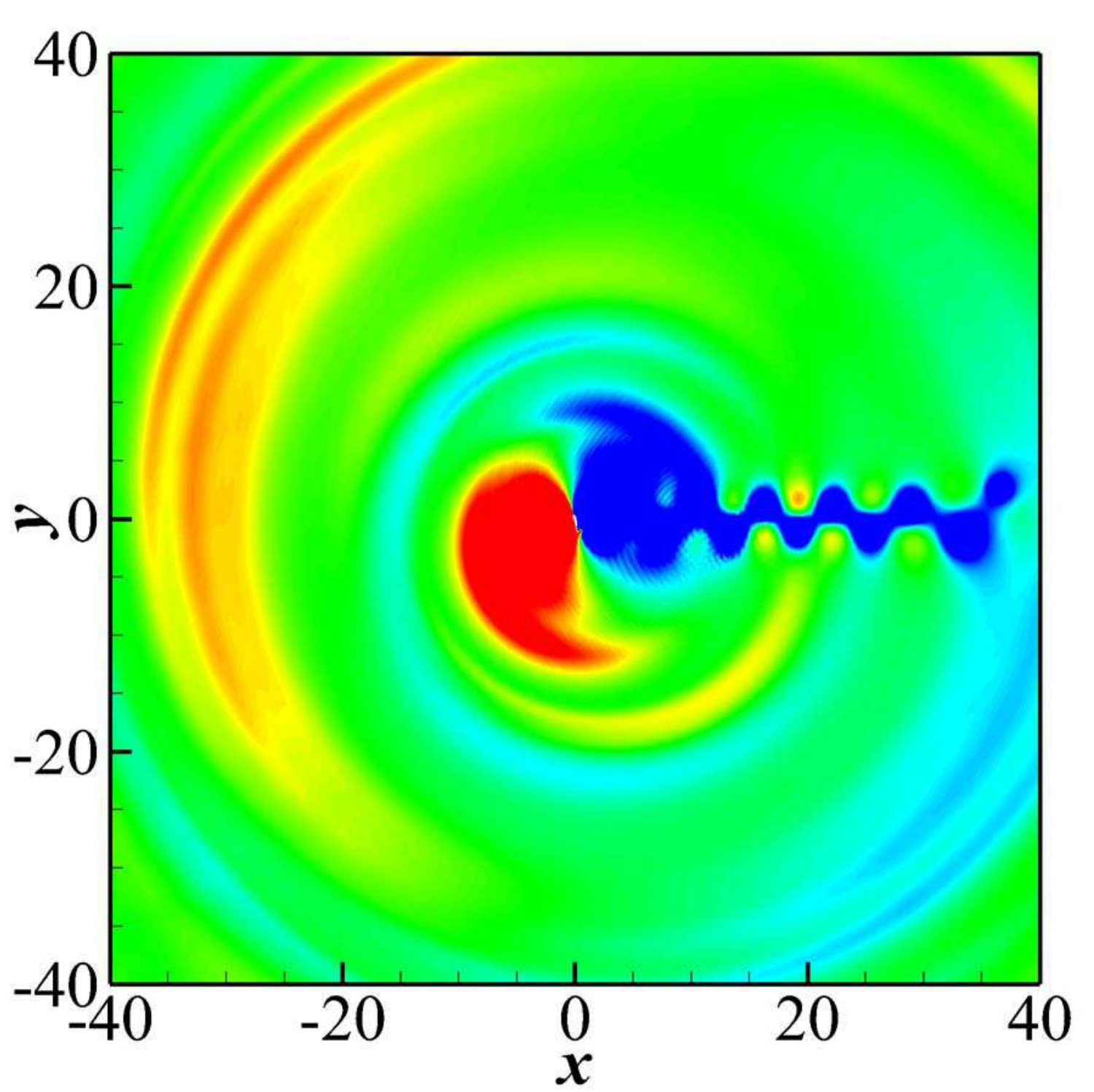}
  \includegraphics[width=2.0in]{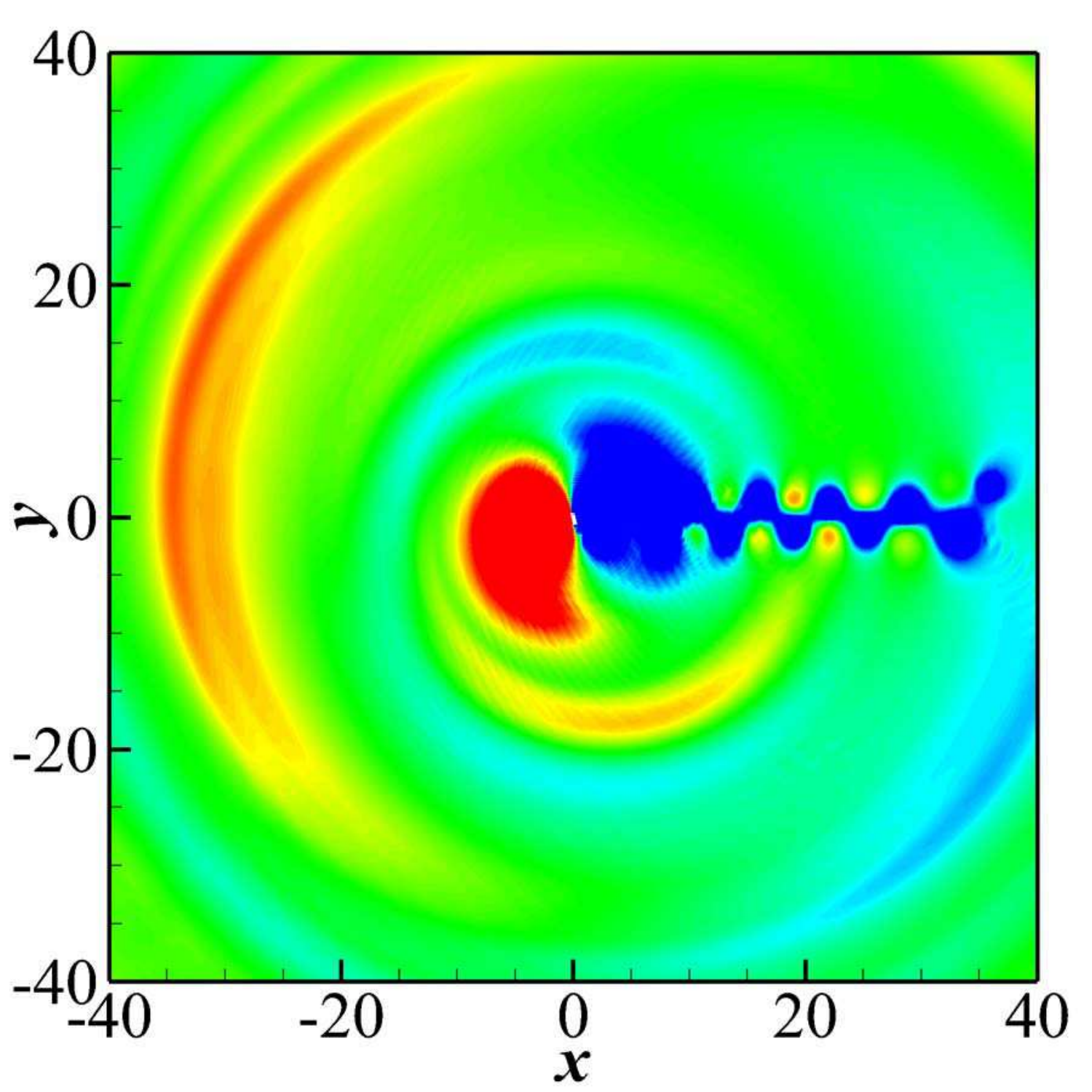}
  \includegraphics[width=2.0in]{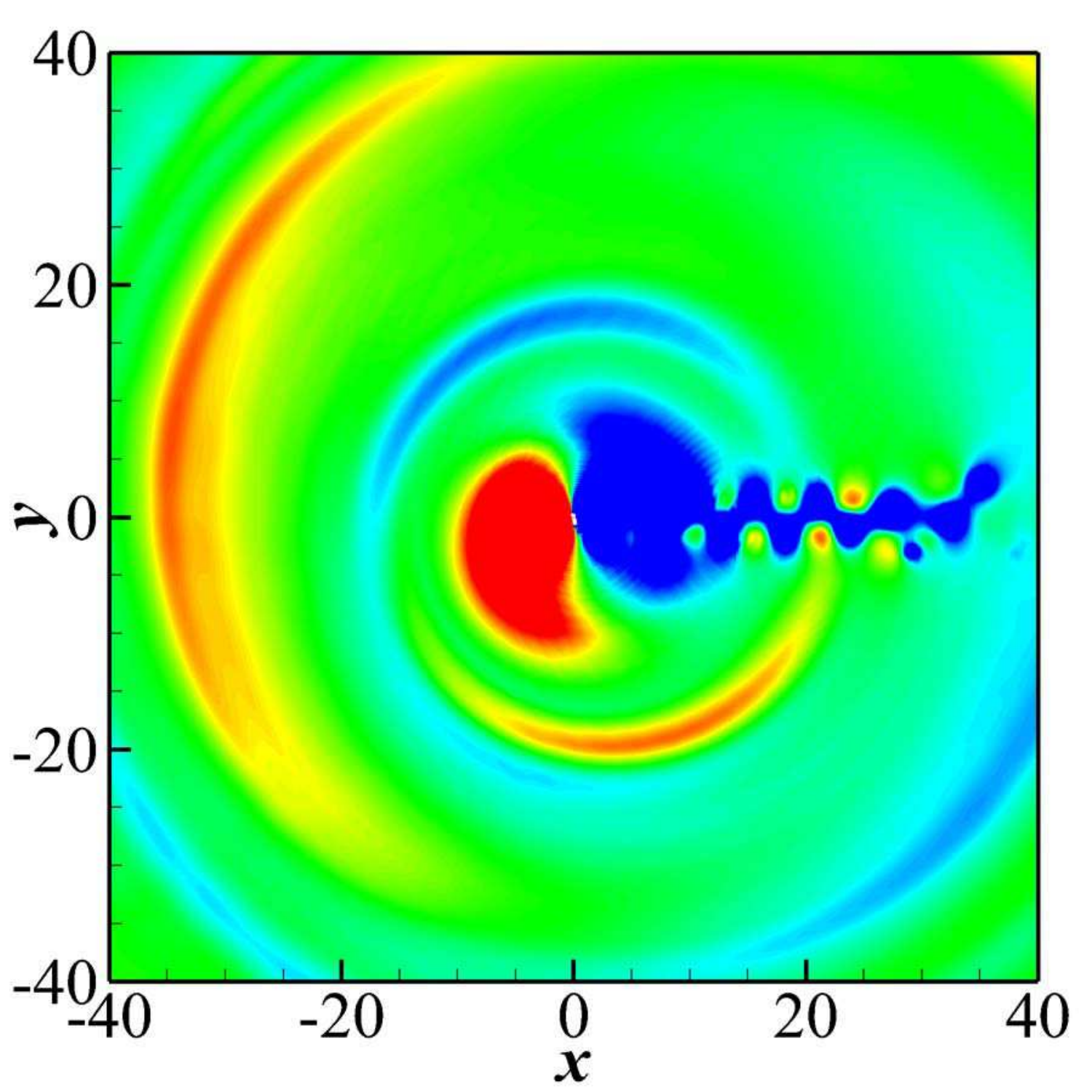}\\

  \end{center}
\caption{Flapping foil energy harvester: instantaneous contours of the fluctuating pressure $\Delta p$ induced by NACA0015 foil (a), rigid plate (b) and flexible plate (c) in a period with an interval of $T/4$. The contour level ranges from $-5.0\times10^{-4}\rho_f c^2$ to $5.0\times10^{-4}\rho_f c^2$ with an interval of $1.25\times10^{-5}\rho_f c^2$. Here, Re=1100, $M=0.1$, $\alpha_m=152.6^o$, $A_0/L=2.0$ and $\beta=90^o$.}
\label{Fig:naca_dp_contour}
\end{figure}

To illustrate the frequency characteristics of the fluctuating pressure generated by the foil, the FFT is used to analyze the fluctuating pressure. Polar diagrams of the fluctuating pressure peaks measured at $r=20$ with three frequencies $f$, $2f$ and $3f$ are shown in Fig.~\ref{Fig:naca_freq}. It is found that the sound at frequency of $f$ dominates in the vertical direction for all cases, while the fluctuating pressure at frequency of $2f$ dominates in the horizontal direction, and the fluctuating pressures at higher frequency are negligible. The fluctuating pressure in the vertical direction is dominated by the frequency of the lift ($f$, see Fig.~\ref{Fig:naca_freq} (b)). Similarly, the fluctuating pressure in the horizontal direction is dominated by the frequency of the drag ($2f$, see Fig.~\ref{Fig:naca_freq} (a)). Comparison of the fluctuating pressure generated by NACA0015 foil and rigid plate shows negligible differences indicating that the geometrical shape of the foil does not significantly affect the sound generation. It is also found that the fluctuating pressure at the frequency of $2f$ induced by the flexible plate is larger than that induced by NACA0015 foil and rigid plate, which agrees with the larger amplitudes of drag induced by the flexible plates, see Fig.~\ref{Fig:naca_flex_clcd} (a). The smaller fluctuating pressure at the frequency of $f$ and $3f$ induced by the flexible plate also agrees with the smaller lift amplitude generated by the flexible plate, see Fig.~\ref{Fig:naca_flex_clcd} (b). The instantaneous contours of the fluctuating pressure $\Delta p$ in a period are further presented in Fig.~\ref{Fig:naca_dp_contour}. It is found that the difference of the fluctuating pressure field generated by the NACA0015 foil and rigid plate is not significant. The one of the flexible plate tends to be different at $0T/4$ and $2T/4$, when the deformation of the plate is significant.

\begin{figure}
\begin{center}
  \includegraphics[width=3.5in]{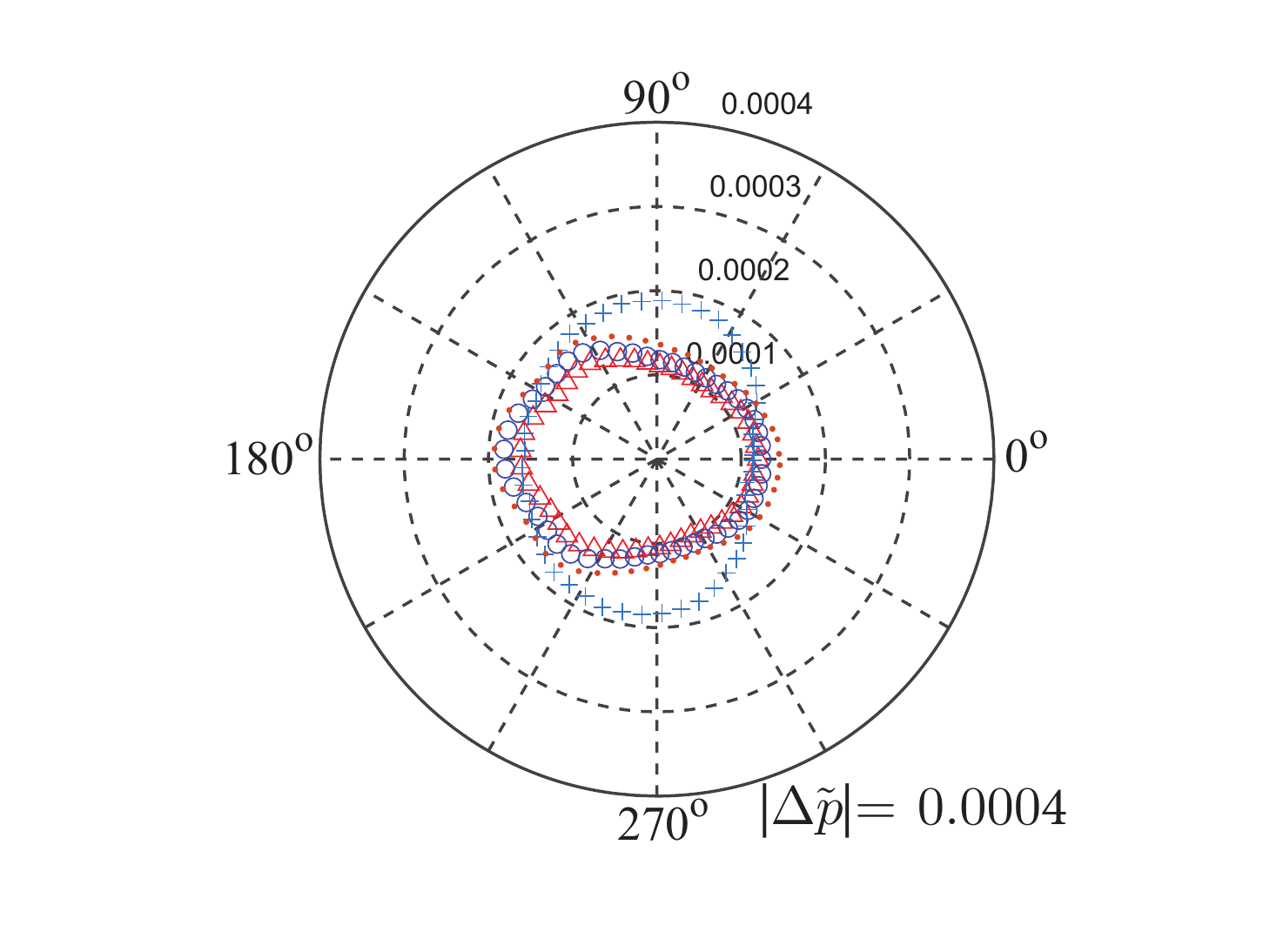}
  \end{center}
\caption{Flapping foil energy harvester: polar diagram of the root-mean-square of the fluctuating pressure at a distance of $37.5L$, where $\Delta$, o, $+$ and $\cdot$ denote $E_b^*$ at 5.0, 1.0, 0.5 and 0.25, respectively. Here, Re=1100, $M=0.1$, $\alpha_m=152.6^o$, $A_0/L=2.0$ and $\beta=90^o$.}
\label{fig:power_flex_dp_rms}
\end{figure}

Here, three additional cases at $E_b^*=0.25,$ 0.5 and 5.0 are considered to illustrate the effects of the flexibility on the sound generation. Fig.~\ref{fig:power_flex_dp_rms} shows the polar diagram of the root-mean-square of the fluctuating pressure. It shows that the fluctuating pressure generated by the flexible plates increase slightly with the flexibility. However, the flexible plate at $E_b^*=0.5$ generates a significantly larger fluctuating pressure, as the plate flaps near its natural frequency ($\omega^*$ defined in Eq.~\ref{eq:flapping_para} is 0.8).

\section{Conclusions}
In this paper, an immersed boundary method for fluid--structure--acoustics interactions involving large deformations and complex geometries is presented. The validations including acoustic waves scattered from a stationary cylinder, sound generation by a stationary and a rotating cylinder in a uniform flow, sound generation by an insect in hovering flight, deformation of a red blood cell induced by acoustic waves and acoustic waves scattered by a stationary sphere have been conducted. Results show that the current solver has a good performance in modelling fluid--structure--acoustics interactions involving large deformations and complex geometries. It indicates that the immersed boundary method handled by delta function is accurate enough for predicting the dilatation and acoustics. We further adopt the present solver to model fluid--structure--acoustics interactions of flapping foils in forward flight and flapping foil energy harvester. The numerical examples presented here can also enrich the limited database of fluid--structure--acoustics interactions.

In the computation of flapping foils in forward flight, the present simulation captures the propagation and the decay of the sound pressure accurately. Two main mechanisms to generate positive and negative pressure are observed. The present simulation shows that the positive pressure formed on the loading face is much larger than the negative pressure generated by the vortex shedding. The directivity of the pressure wave also presents fluctuations around the stroke plane. The flexibility of the foil generates larger thrust with higher fluctuating pressure, but the geometrical shape does not have significant influences on the force and sound generation. Based on the FFT analysis of the fluctuating pressures, the fluctuating pressures are dominated by the flapping frequency $f$ of the foil for the rigid and NACA0015 flapping foil. However, both $f$ and $2f$ components are significant for the flexible foil.

In the computation of flapping foil energy harvester, it is found that the fluctuating pressure generated by NACA0015 foil, rigid plate and flexible plate are similar in terms of the frequency. The lift dominates the pressure fluctuation in the vertical direction and the drag dominates the pressure fluctuation in the horizontal direction. Some differences of the fluctuating pressure are also observed and analyzed. The results show that the geometrical shape does not have significant effects on the force and sound generation, while the flexibility of the plate tends to deteriorate the power extraction. The current flexible plate also induces larger sound at the frequency of $2f$ and weaker sound at the frequencies of $f$ and $3f$.

While many merits have been demonstrated by the cases presented, it should be pointed out for fluid--structure--acoustics interactions at large Reynolds numbers where turbulence flows exist, large eddy simulation and wall models should be considered. We will address this challenge in the near future.


\section*{Acknowledgements}
 Dr. F.-B. Tian is the recipient of an Australian Research Council Discovery Early Career Researcher Award (project number DE160101098). This work was conducted with the assistance of resources from the National Computational Infrastructure (NCI), which is supported by the Australian Government.

\bibliographystyle{unsrt}  
\bibliography{paper} 

\end{document}